\documentclass[twocolumn,twocolappendix]{aastex631}
\usepackage{float}
\usepackage{xspace}
\usepackage{amsmath}
\usepackage{hyperref}    
\usepackage{xcolor}      

\newcommand{\OIII}{[\ion{O}{3}]\xspace}

\newcommand{\SII}{[\ion{S}{2}]\xspace}

\received{}
\revised{}
\accepted{}
\submitjournal{ApJ}

\shorttitle{PNLF-MUSE}
\shortauthors{Jacoby et al.}

\graphicspath{{./}{figures/}}

\begin{document}

\title{Towards Precision Cosmology With Improved PNLF Distances Using VLT-MUSE\\
II. A Test Sample from Archival Data
}

\correspondingauthor{George H. Jacoby}
\email{george.jacoby@noirlab.edu}

\author[0000-0001-7970-0277]{George H. Jacoby}   
\affiliation{NSF’s NOIRLab, 950 N. Cherry Ave., Tucson, AZ 85719, USA}

\author[0000-0002-1328-0211]{Robin Ciardullo}   
\affiliation{Department of Astronomy \& Astrophysics, The Pennsylvania State University, University Park, PA 16802, USA}
\affiliation{Institute for Gravitation and the Cosmos, The Pennsylvania State University, University Park, PA 16802}

\author[0000-0003-2451-739X] {Martin M. Roth}
\affiliation{Leibniz Institute for Astrophysics Potsdam (AIP), An der Sternwarte 16, 14482 Potsdam, Germany}

\author[0000-0001-7214-3009]{Magda Arnaboldi}
\affiliation{European Southern Observatory, Karl-Schwarzschild-Straße 2, 85748, Garching, Germany}

\author[0000-0003-4766-902X] {Peter M. Weilbacher}
\affiliation{Leibniz Institute for Astrophysics Potsdam (AIP), An der Sternwarte 16, 14482 Potsdam, Germany}




\begin{abstract}
Thanks to the MUSE integral field spectrograph on the VLT, extragalactic distance measurements with the \OIII $\lambda 5007$ planetary nebula luminosity function (PNLF) are now possible out to $\sim 40$~Mpc.  Here we analyze the VLT/MUSE data for 20 galaxies from the ESO public archive to identify the systems' planetary nebulae (PNe) and determine their PNLF distances.  Three of the galaxies do not contain enough PNe for a robust measure of the PNLF, and the results for one other system are compromised by the galaxy's internal extinction.  However, we obtain robust PNLF distances for the remaining 16 galaxies, two of which are isolated and beyond 30~Mpc in a relatively unperturbed Hubble flow.  From these data, we derive a Hubble Constant of $74.2\pm7.2$ (stat) $\pm3.7$ (sys) km-s$^{-1}$-Mpc$^{-1}$, a value that is very similar to that found from other quality indicators (e.g., Cepheids, the tip of the red giant branch, and surface brightness fluctuations). At present, the uncertainty is dominated by the small number of suitable galaxies in the ESO archival and their less than ideal observing conditions and calibrations. Based on our experience with these systems, we identify the observational requirements necessary for the PNLF to yield a competitive value for $H_0$ that is independent of the SN~Ia distance scale, and help resolve the current tension in the Hubble constant.


\end{abstract}



\section{Introduction} \label{sec:intro}

The technique of obtaining distances using the Planetary Nebulae Luminosity Function (PNLF) is now over three decades old \citep{Ciardullo+89}.  During this time, most planetary nebula (PN) surveys were conducted with narrow-band interference filters tuned to the wavelength of \OIII $\lambda 5007$ (and sometimes H$\alpha$) at the redshift of the targeted galaxy.  In practice, this meant performing PN photometry on images taken through $\sim 40$\,\AA\ wide bandpasses and accepting the background noise associated with the filter width.  This effectively limited the technique to objects within $\sim 15$~Mpc --- a range that, until recently, was similar to that of Cepheids and the tip of the red giant branch (TRGB) techniques.

Today, observations through wide-field integral-field spectrographs (IFS) have supplanted interference-filter based photometry for deep PNLF observations.  In particular, the Multi-Unit Spectroscopic Explorer (MUSE) optical IFS \citep{Bacon+10} has revolutionized PNLF measurements to both spiral and elliptical galaxies by coupling the large aperture and excellent image quality of the 8.2\,m Very Large Telescope (VLT) with the effective resolution of an $R \sim 2000$ spectrograph.  By reducing the sky + galaxy noise underlying each PN by more than a factor of 10 and allowing the simultaneous measurement of several spectral lines (such as \OIII $\lambda 5007$, H$\alpha$, and \SII $\lambda\lambda 6716,6731$), MUSE has facilitated the identification of large numbers of extragalactic PNe with almost no contamination from interloping objects, such as \ion{H}{2} regions, supernova remnants (SNRs), and background emission-line galaxies.  The result has been PNLF measurements to several dozen spiral and elliptical galaxies within $\sim 20$~Mpc \citep[e.g.,][]{Spriggs+21, Scheuermann+22}. Moreover, \citet{Roth+21} have shown that through the careful use of a Differential Emission Line Filter (DELF) technique, precision PN photometry is now possible out to distances as far away as $\sim 40$~Mpc. 

This changes the landscape for the PNLF method.  When the PNLF was restricted to distances of $\lesssim20$~Mpc, it could only be used as an inter-method cross-check that might detect deficiencies or systematic offsets between Population~I standard candles, such as Cepheids, and techniques that work best in Population~II systems, i.e., the TRGB and surface brightness fluctuations (SBF\null).  However, by enabling PN photometry out to $\sim 40$~Mpc, MUSE allows the PNLF to be competitive with the very deepest Cepheid \citep{Riess+22} and TRGB \citep{Jang+17} measurements obtained from space.  Moreover, at a distance of $\sim 40$~Mpc, the typical peculiar velocity of a galaxy is only $\sim 10\%$ that of the Hubble flow.  Thus, PN photometry in a sample of $\sim 50$ galaxies beyond $\sim 30$~Mpc can place a meaningful constraint on the Hubble constant that is independent of the SN Ia calibration.


In this paper, we apply the photometric methods described in \citet[][hereafter Paper~I]{Roth+21} and the analysis techniques detailed in \citet{Chase+23} to further illustrate the ability of the PNLF to yield accurate distances to galaxies well beyond the Local Supercluster.  We do this by using the archival MUSE data cubes of 20 representative galaxies spanning a wide range of distances, absolute magnitudes, and Hubble types.  We also examine possible systematic errors associated with the PNLF and examine strategies to improve the precision of the method. In short, this paper explores the range over which the PNLF can be used effectively to derive reliable extragalactic distances.

In Section~\ref{sec:galaxies}, we describe our criteria for selecting promising PNLF targets from the MUSE archive. In Section~\ref{sec:observations}, we review our method for identifying faint PN candidates in the target galaxies, removing interloping objects such as \ion{H}{2} regions and supernova remnants, and measuring the PNe's apparent \OIII $\lambda 5007$ magnitudes.  In Section~\ref{sec:pnlf_fitting}, we outline our method of deriving distances from the PNLF, including the case where unrecognized PN superpositions may be present in the dataset.   In Section~\ref{sec:results}, we present our PNLF distances and provide brief commentaries on the unique challenges presented by each galaxy.  In Section~\ref{sec:discuss}, we compare our results to those of other techniques and assess the reliability/repeatability of the distances. 

\section{Galaxy Selection} 
\label{sec:galaxies}

The ESO science archive contains a large number of galaxies with MUSE observations.  For this paper, we examined the database as of 05 December 2021 to identify a set of galaxies amenable to PNLF measurements that also provide an adequate and diverse sample to demonstrate the applicability of the technique.  We then refined our list based on the following criteria:


\begin{enumerate}
    \item Good image quality (i.e., a point-source full-width at half maximum (FWHM) $< 1\farcs 0$): The dominant source of noise for any extragalactic PN measurement is the background. Consequently, the identification and measurement of PNe against the continuum light of a host galaxy is a strong function of image quality, especially in the bright central regions where the highest density of PNe are located.  Moreover, as the seeing degrades, the likelihood that two, closely separated PNe will the recorded as a single, bright object increases; this effect can distort the observed shape of the PNLF and lead to a systematic error in the galaxy's derived distance and its uncertainty \citep[e.g.,][]{Chase+23}.  Thus, in order to obtain a large statistically complete sample of PNe with a minimal number of image blends, one generally requires excellent seeing.  Image quality was therefore of paramount importance for our choice of galaxies.
    
    \item Effective exposure time ($t > 3600$~s):  PNLF distances are defined by the apparent magnitude of the rapid cutoff seen at the bright end of the PN luminosity function. A high-precision measurement of this cutoff magnitude ($m^*$) requires that the PN sample be reasonably complete at least $\sim 0.5$~mag fainter than this cutoff, and preferably to $m^* + 1$.  In $\sim 1\arcsec$ seeing, the exposure time needed for MUSE to detect such objects in a galaxy $\sim 20$\,Mpc away is roughly one hour.  
    
    \item Likely distance ($D < 30$~Mpc): In almost all cases, a galaxy's recessional velocity, coupled with a bulk-motion model \citep[e.g.,][]{Lynden-Bell+88, Tonry+00, Shaya+17} provides a rough estimate of its distance; for a subset of galaxies, very good distance estimates exist from the measurement of resolved (or semi-resolved) stars. If this distance is too great and if the exposure time is not long enough to compensate, the data will not reach deep enough to observe the PNLF cutoff. Since one goal of this study is to compare MUSE PNLF distances to distances obtained from other indicators, we focused on targets that promised to yield good results, i.e., galaxies within $\sim 30$~Mpc.  We analyze more distant systems only if the exposure times are long and the image quality is excellent; in Paper~I, for example, we included NGC 474 at $\sim 37$\,Mpc, as its halo was observed for 10 hr in $0\farcs 78$ seeing.
    
    \item Surveyed luminosity 
    ($M_V < -19$):  Bright planetary nebulae are rare objects:  from the fuel consumption theorem, an $M_V \sim -21.2$ galaxy will create only one PN per year \citep{Renzini+86, Buzzoni+06}, and less than $\sim 2\%$ of all PNe are in the top $\sim 1$~mag of the PN luminosity function.  As a result, in order for the top magnitude of the PNLF to be well-populated, one needs to survey at least $M_V \sim -19.5$ of galactic light \citep[e.g.,][]{Ciardullo+05,Buzzoni+06}. PN surveys of dwarf galaxies and systems with only limited MUSE coverage will, at best, produce distances with large error bars.
    
    \item Preference for having Cepheid, TRGB, or SBF distances in the literature: There are exactly zero galaxies whose distances are known \textit{a priori}; the two systems that come closest to satisfying this criterion are the LMC \citep[via eclipsing binaries and the SN 1987A light echo;][] {Pietrzynski2019, Panagia+91} and NGC\,4258 \citep[via the geometry of its megamaser;][]{Reid2019}. The best one can do to validate the results of any distance indicator is to compare its results to those obtained from other methods.  Still, this type of analysis only demonstrates consistency, not correctness. At this time, Cepheid and TRGB distances lead the way in terms of perceived accuracy, with SBF results close behind. Thus, to achieve the goal of validating the accuracy of the PNLF, we prefer to include targets with distances measured using well-established techniques. 
    
    \item Preference for hosting Type Ia supernovae: There are two ways to view this selection criterion. If a galaxy hosts a Type Ia supernova, then a PNLF distance can be used to calculate the supernova's peak luminosity, and thereby improve the calibration of SN~Ia as a distance indicator. Alternatively, we can use the existing SN Ia measurements to provide another check on the PNLF distances. Unfortunately, at the time of our galaxy selection, there were only three known Type Ia SNe (in two galaxies) that had MUSE archival observations which satisfied our other criteria (though more may exist in the future).
\end{enumerate}

Table~\ref{tab:galaxylist} gives the resultant list of the galaxies.  The group is quite diverse, and includes a central cD galaxy (NGC\,1399), a low-luminosity AGN with two, apparently dark matter-less satellite dwarfs (NGC\,1052), a Seyfert 2 galaxy with a megamaser (NGC\,4418), two spiral galaxies undergoing interactions (NGC\,4038/39 and NGC\,1385), and several normal elliptical, lenticular, and spiral systems.  Such diversity allows us to test the methods detailed in Paper~I in different environments and provides a guide to the range of galaxies amenable to future PNLF measurements.

\begin{deluxetable*}{llccccccl}
\tabletypesize{\scriptsize}
\tablecaption{List of Galaxies Analyzed
\label{tab:galaxylist} } 
\tablehead{
    &   &\colhead{GSR Velocity}  &                    &                 &\multicolumn{3}{c}{Range of Distances (Mpc)}         &\\
\colhead{Galaxy}  &\colhead{Type\tablenotemark{\footnotesize a}}  &\colhead{(km~s$^{-1}$)\tablenotemark{\footnotesize b}} &\colhead{$E(B-V)$\tablenotemark{\footnotesize c}} &\colhead{$M_B${\tablenotemark{\footnotesize d}}}  &\colhead{Cepheid\tablenotemark{\footnotesize e}}  &\colhead{TRGB\tablenotemark{\footnotesize e}} &\colhead{SBF\tablenotemark{\footnotesize e}}    &\colhead{SN Ia Name} }
\startdata
NGC  253          &Sc             & 243                    & 0.0161             &$-20.0$       &           & 3.3 -- 3.9                       &                 & \\
NGC 1052          &E3/S0          & 1473                   & 0.0228             &$-19.8$         &                    &               & 18.0 -- 20.6    & \\
NGC 1326          &SBa            & 1242                   & 0.0163             &$-19.8$         &                    &               &                 & \\
NGC 1351          &S0$_1$/E6      & 1396                   & 0.0115             &$-18.3$         &                    &               & 19.2 -- 22.6    & \\
NGC 1366          &S0$_1$         & 1120                   & 0.0139             &$-17.5$         &                    &               & 18.6 -- 21.2    & \\
NGC 1385          &Sc             & 1400                   & 0.0173             &$-19.8$         &                    &               &                 & \\
NGC 1399          &E1 (cD)        & 1301                   & 0.0109             &$-20.3$         &                    &               & 17.6 -- 21.4    & \\
NGC 1404          &E2             & 1823                   & 0.0097             &$-19.9$         &                    &       18.1$^f$, 18.7$^g$        & 17.9 -- 22.2    & SN2007on, SN2011iv \\
NGC 1419          &E0$_p$         & 1442                   & 0.0112             &$-17.5$         &                    &               & 19.2 -- 22.9    & \\
NGC 1433          &SBb            & 928                    & 0.0078             &$-19.7$         &                    & 9.0           &                 & \\
NGC 1512          &SBb$_{\rm p}$  &  746                   & 0.0091             &$-19.0$     &        & 11.4 -- 11.9                   &                 & \\
NGC 2207          &Sc             & 2577                   & 0.0748             &$-21.9$         &                    &               &                 & SN1975A \\
NGC 3501          &Scd            & 1057                   & 0.0200             &$-18.9$         &                    &               &                 & \\
NGC 4038/9        &Sc$_{\rm p}$   & 1548                   & 0.0398             &$-21.1$         & 18.1 -- 21.5       & 20.0 -- 21.6, 21.7$^f$   &                 & SN2007sr \\
NGC 4365          &E3             & 1170                   & 0.0184             &$-20.5$         &                    &               & 20.4 -- 23.6    & \\
NGC 4418          &S0/a           & 2027                   & 0.0202             &$-19.0$           &                    &               &                 & \\
NGC 4472          &E1/S0$_1$      & 913                    & 0.0191             &$-21.8$         &                    &               & 14.5 -- 17.8    & \\
NGC 5248          &Sbc            & 1126                   & 0.0210             &$-21.1$         &                    &               &                 & \\
NGC 6958          &S0$_1$         & 2730                   & 0.0385             &$-20.5$         &                    &               &                 & \\
MCG-06-08-024     &E4$_p$         & 1692                   & 0.0101             &$-16.1$         &                    &               &19.2 -- 19.5                & \\
\enddata
\tablenotetext{b}{From \citet{RC3-1991}}
\tablenotetext{c}{From \citet{Schlafly+11}}
\tablenotetext{d}{All total $B$ magnitudes estimated from the Nearby Galaxies Catalog \citep{Tully-cat} and the NASA/IPAC Extragalactic Database (NED), except for NGC\,4418, which is derived from \citet{RC3-1991} and our own distance modulus of $(m-M)_0 = 32.59$.}
\tablenotetext{e}{Compiled by the NASA/IPAC Extragalactic Database, except where noted}
\tablenotetext{f}{From \citet{Anand2022}}
\tablenotetext{g}{From \citet{Hoyt+21}}

\end{deluxetable*}


\section{Observations and Reductions}
\label{sec:observations}
For our analysis, we extracted publicly available, fully-reduced MUSE data cubes of the galaxies listed in Table~\ref{tab:galaxylist} from the ESO archive \citep{Romaniello+18}.  These data, which were obtained between 25 June 2014 and 05 December 2020, were taken under a variety of observing conditions, with exposure times ranging between 621 and 9600~sec, and seeing (as listed in the archive) between $0\farcs 51$ and $1\farcs 27$.  Because MUSE consists of 24 identical integral-field spectrographs, which, in its wide-field mode, covers only a $1\arcmin \times 1\arcmin$ field of view, most galaxies in our list were only partially surveyed via a small number of pointings.  In many cases, these pointings overlapped, allowing (at least in theory), a consistency check of each data cube's flux calibration.

\begin{deluxetable}{lccc}
\tablecaption{Archival Data Cubes
\label{tab:MUSE-pointings}  }
\tablehead{
\colhead{Pointing}  &\colhead{Date of Obs} &\colhead{Exp Time (s)} &\colhead{Archive Seeing}  }
\startdata
\hline
NGC 253  - P1  & 07 Nov 2018  & 1820 & $0\farcs 89$  \\
NGC 253  - P2  & 29 Jul 2019  & 1833 & $0\farcs 83$  \\
\hline
NGC 1052 - P1  & 05 Sep 2019  & 1685 & $0\farcs 61$ \\
\hline
NGC 1326 - P1  & 19 Nov 2017  & 3354 & $1\farcs 11$ \\
\hline
NGC 1351 - P1  & 15 Nov 2017  & 3379 & $0\farcs 69$ \\
NGC 1351 - P2  & 05 Oct 2016  & 4680 & $0\farcs 93$ \\
\hline
NGC 1366 - P1  & 09 Aug 2019  & 2445 &$ 1\farcs 04$ \\
\hline
NGC 1385 - P1  & 31 Dec 2019  & 2360 &$0\farcs 74$ \\
NGC 1385 - P2  & 05 Dec 2020  & 2417 &$1\farcs 03$ \\
NGC 1385 - P3  & 21 Jan 2020  & 2424 &$0\farcs 63$ \\
NGC 1385 - P4  & 17 Jan 2019  & 2418 &$0\farcs 72$ \\
NGC 1385 - P5  & 20 Jan 2020  & 2414 &$0\farcs 67$ \\
\hline
NGC 1399 - M   & 02 Oct 2014  & 954 & $0\farcs 81$ \\
\hline
NGC 1404 - P1  & 22 Nov 2017  & 3287 &$0\farcs 70$ \\
NGC 1404 - P2  & 14 Nov 2017  & 1687 &$0\farcs 80$ \\
\hline
NGC 1433 - P1  & 22 Nov 2019  & 2580 & $0\farcs 75$ \\
NGC 1433 - P2  & 21 Nov 2019  & 2580 & $0\farcs 73$ \\
NGC 1433 - P3  & 20 Nov 2019  & 2580 & $1\farcs 05$ \\
NGC 1433 - P4  & 05 Oct 2019  & 2580 & $0\farcs 78$ \\
NGC 1433 - P5  & 26 Aug 2016  & 3840 & $0\farcs 76$ \\
NGC 1433 - P6  & 06 Oct 2019  & 2580 & $0\farcs 70$ \\
NGC 1433 - P7  & 21 Dec 2019  & 2580 & $0\farcs 82$ \\
NGC 1433 - P8  & 22 Dec 2019  & 2580 & $0\farcs 87$ \\
NGC 1433 - P9  & 23 Dec 2019  & 2580 & $0\farcs 64$ \\
\hline
NGC 1512 - P1  & 30 Dec 2018  & 2406 & $0\farcs 79$ \\
NGC 1512 - P2  & 17 Feb 2018  & 3000 & $1\farcs 75$ \\
NGC 1512 - P3  & 18 Feb 2018  & 2402 & $0\farcs 92$ \\
NGC 1512 - P4  & 30 Dec 2018  & 2307 & $0\farcs 58$ \\
NGC 1512 - P5  & 21 Sep 2017  & 3280 & $0\farcs 67$ \\
NGC 1512 - P6  & 19 Feb 2018  & 2402 & $0\farcs 65$ \\
NGC 1512 - P7  & 10 Jan 2019  & 2384 & $1\farcs 27$ \\
NGC 1512 - P8  & 10 Jan 2019  & 2312 & $0\farcs 78$ \\
NGC 1512 - P9  & 10 Jan 2019  & 2300 & $0\farcs 61$ \\
\hline
NGC 2207 - P1  & 19 Feb 2019  & 2238 & $0\farcs 83$ \\
NGC 2207 - P2  & 17 Nov 2017  & 3338 & $0\farcs 65$ \\
NGC 2207 - P3  & 18 Nov 2017  & 3892 & $0\farcs 83$ \\
\hline
NGC 3501 - M   & 30 Dec 2016  & 9600 & $1\farcs 51$ \\
\hline
NGC 4038/9 - C1   & 23 Apr 2015  & 4951 & $0\farcs 67$ \\
NGC 4038/9 - C3   & 11 May 2015  & 4811 & $0\farcs 58$ \\
NGC 4038/9 - C4   & 13 May 2015  & 4985 & $0\farcs 51$ \\
NGC 4038/9 - C5   & 21 May 2015  & 5117 & $0\farcs 82$ \\
NGC 4038/9 - C9   & 24 Apr 2017  & 2584 & $0\farcs 72$ \\
NGC 4038/9 - C10  & 12 May 2016  & 2592  &$0\farcs 74$ \\
NGC 4038/9 - C11a & 23 Apr 2017  & 2498 & $0\farcs 79$ \\
NGC 4038/9 - C11b & 25 Apr 2017  & 2523 & $0\farcs 74$ \\
NGC 4038/9 - C12a & 01 Feb 2016  & 2700  & $1\farcs 22$ \\
NGC 4038/9 - C12c & 01 Feb 2016  & 2700  & $1\farcs 09$ \\
NGC 4038/9 - C12e & 04 Feb 2016  & 2700  & $0\farcs 86$ \\
\hline
NGC 4365 - P1  & 13 Feb 2015  & 2343 &$0\farcs 83$ \\
\hline
NGC 4418 - P1  & 30 Jan 2020  & 5999 &$0\farcs 76$ \\
\hline
NGC 4472 - M1  & 14 Apr 2015  & 626 &$0\farcs 72 $ \\
NGC 4472 - M2  & 10 Jul 2015  & 623 &$0\farcs 82 $ \\
NGC 4472 - M3  & 12 Apr 2015  & 621 &$0\farcs 93 $ \\
\hline
NGC 5248 - P1  & 04 Apr 2016  & 3411 & $0\farcs 76$ \\
\hline
NGC 6958 - P1  & 18 Sep 2017  & 1906 & $0\farcs 96$ \\
\hline
MCG-06-08-024 - P1  & 27 Dec 2016  & 4821 &$0\farcs 67$ \\
\enddata
\end{deluxetable}

To identify and measure the PNe in these galaxies, we used the Differential Emission-line Filtering technique, which is described in detail in Paper~I. This method, which is the IFS equivalent of the traditional on-band/off-band procedure commonly used with interference filters, maximizes the contrast of emission-line objects over the background while minimizing spaxel-to-spaxel flat-field residuals associated with the instrument.  We outline our steps below. With DELF, each PN effectively is observed through its own optimized and tuned 3.75--6.25\,\AA\ narrow-band filter.

After downloading a data cube from the archive, we create a 125\,\AA\ wide intermediate-bandwidth ``continuum'' image from the co-addition of 100 contiguous data cube layers centered $\sim 90$\,\AA\ redward of \OIII $\lambda 5007$ at the redshift of the galaxy.  We then subtract a scaled version of this image from each of the cube's 15 layers surrounding \OIII $\lambda 5007$ to create 15 ``difference images,'' one for each 1.25\,\AA\ data slice.

Next, to identify the PNe, we stepped through the 15 difference images, summing up three adjacent images at a time, and visually inspecting the frames for emission-line sources.  Since the wavelength resolution of MUSE at $\sim 5000$\,\AA\ is roughly 2.5\,\AA\ (i.e., twice the dispersion of the data cube), any point-like object appearing on three consecutive frames was considered a PN candidate.

Once these candidates were identified, their spectra were extracted in a small aperture, typically with a radius of 3 to 5 spaxels ($0\farcs 6$ to $1\farcs 0$), depending on the seeing. The \OIII\ emission lines were then examined for evidence of asymmetric, double-peaked, or broadened line profiles.  If a line profile was well-behaved, its total flux was obtained by modeling the emission with a Gaussian model; otherwise, an interactive tool was employed to deblend the Gaussian associated with the PN in question from other spectral components (usually arising from diffuse gas or other superposed sources). 

Aperture corrections were then found by identifying bright point sources in the field, co-adding 160 layers of the original data cube centered at redshifted $\lambda 5007$, and measuring the stars' curves-of-growth.  The total flux of each object was then translated into an \OIII $\lambda 5007$ magnitude, $m_{5007}$ via
\begin{equation}
m_{5007} = -2.5 \log F_{5007} - 13.74
\label{eq:mdef}
\end{equation}
where $F_{5007}$ is the \OIII line flux in ergs~cm$^{-2}$~s$^{-1}$.

Finally, the full spectrum of each object was inspected to exclude \ion{H}{2} regions \citep[using the ratio of H$\alpha$ to \OIII;][]{Herrmann+08}, supernova remnants \citep[based on the strength of \SII relative to H$\alpha$;][]{Kreckel+17}, and background Ly$\alpha$ emitting galaxies \citep[from the line profile and the lack of other spectral features;][]{Kudritzki+00, Arnaboldi+02}. Furthermore, any object with only a single line was rejected because it cannot be confirmed as a PNe; these were always quite faint.  A full description of this process is given in Paper~I.

\vspace{50pt}

\section{Obtaining a PNLF Distance}
\label{sec:pnlf_fitting}

\subsection{Historical approach}

The shape of the planetary nebula luminosity function is not universal: in old stellar populations, the slope at faint magnitudes changes from galaxy to galaxy \citep[e.g.,][]{Longobardi+13,Hartke+17,Hartke+20, Bhattacharya+21}, while in star-forming systems, the PNLF is known to contain multiple inflection points \citep{Jacoby+02, Reid+10, Ciardullo2010}.  However, at the extreme bright end of the function, the PNLF's behavior is remarkably invariant, and consistent with an abrupt truncation of an underlying power law.  Moreover, measurements of roughly a dozen galaxies within $\sim 10$~Mpc show that for systems more metal-rich than the Large Magellanic Cloud, the absolute magnitude where this truncation takes place is extraordinarily insensitive to stellar population \citep[e.g.,][]{Ciardullo2013}.  Thus, the traditional method of deriving a PNLF distance involves fitting the observed distribution of apparent \OIII $\lambda 5007$ magnitudes to a function that captures the PNLF's truncation, such as 
\begin{equation}
N(M) \propto e^{0.307 M} \{ 1 - e^{3 (M^* - M)} \},
\label{eq:pnlf}
\end{equation}
where $M^* = -4.54$ is the \OIII $\lambda 5007$ absolute magnitude of the brightest possible planetary. This procedure proved extremely successful for galaxies within $\sim 10$~Mpc, but beyond this distance, the values obtained from the PNLF were generally $\sim 0.2$~mag smaller than those from Cepheids and the SBF method.  This offset suggested the presence of one or more systematic errors \citep[e.g.,][]{Ferrarese+00, Ciardullo2022}, such as the inclusion of contaminating objects (i.e., SNRs, \ion{H}{2} regions, and background galaxies) in the PN samples, or an incorrect expression for the shape of the PNLF cutoff. 

\subsection{Recent Improvements with MUSE}

The MUSE analyses by \citet{Spriggs+21} and \citet{Scheuermann+22} have pointed to another possible error in the PNLF: the projection of two separate PNe onto a single spatial (and spectral) resolution element. Although a chance alignment of two rare objects would seem improbable, Paper~I demonstrated that photometric blends occur more often than previously realized.  When this happens, a source may appear to have the spectrum of a normal PN, but with an \OIII\ magnitude that is up to 0.75~mag brighter than $M^*$.  This can distort the observed PN luminosity function and lead to an incorrect (underestimated) value for both the PNLF distance and its uncertainty. 

\citet{Chase+23} detailed the procedures needed to include the possibility of image blends in a PNLF analysis.  In brief, if the \OIII fluxes of two superposed PNe are uncorrelated, the probability density function (PDF) for their summed \OIII flux is simply the convolution of the single object PDF with itself.  At any position in a galaxy, the probability of observing a PN with a given magnitude $m$ is therefore given by the weighted sum of two PDFs:  one for single objects (equation~\ref{eq:pnlf}) and one for superpositions.  Under the assumption that the distribution of PNe within a galaxy follows that of the underlying light, the weights of the two PDFs can be found by applying Poisson statistics to the number of PNe expected to be found on a single resolution element of data.  This expectation value can then be estimated from the surface brightness of the galaxy, the total number of PNe observed, the velocity dispersion of the galaxy's stars, the image quality of the data, and the spectral resolution of the instrument.  

Note that in our analysis, the expected PDF for every PN candidate depends not only on the form of the empirical function (equation~\ref{eq:pnlf}) but also on the galactic surface brightness and velocity dispersion underlying the PN's location.  Thus, one cannot simply fit the data with a single curve: a galaxy's best-fit distance must be derived by maximizing the combined likelihoods of the entire PN sample based on their individual PDFs.  When PN blends are a possibility, histograms and cumulative distributions of PN magnitudes are useful visualization tools, but should not be used as the basis for a quantitative analysis.




\subsection{Implementation of the Correction for PN Superpositions}
\label{subsec:blends}

As stated in the previous two paragraphs, a PNLF distance analysis which includes the possibility of blends requires having some idea of the surface brightness distribution of the galaxy, the velocity dispersion of the stars underlying the position of each PNe, and the spectral and spatial resolution of the data (i.e., how close can two PNe be before we cannot tell there are two objects at one location). These values do not necessarily need to be precise, as there is some degeneracy between the parameters.  However, PNLF analyses that do not include these factors can produce distances (and associated uncertainties) that are underestimated.

The most important parameter for modeling the effect of blends on the PNLF is the total amount of galaxy light sampled by a PN survey and the galactic surface brightness underlying the position of each PN\null.  These data can either be estimated from previously published surface photometry, or from continuum measurements off the MUSE data cubes themselves.  Of secondary importance is the stellar line-of-sight velocity dispersion at each PN's position in the galaxy.  Such data are not available for all galaxies, and even then, the precision of the  measurements is generally limited.  Fortunately, PNLF distances are generally not very sensitive to this parameter, and any error in this quantity is easily subsumed into the overall error budget of the calculation.  Finally, any analysis which includes the possibility of PN superpositions must include the spatial and spectral resolution of the data cube at the wavelength of the observed emission line.  Simulations performed by \citet{Chase+23} show that to a good approximation, two adjacent PNe can only be spatially resolved in a MUSE data cube if their angular separation is greater than half the seeing FWHM \textit{or} their velocity difference is greater than $\Delta v \approx 200$~km~s$^{-1}$.


\section{Results}
\label{sec:results}







In the following subsections, we present basic data for our program galaxies, including a summary of detected PN candidates, any peculiar challenges associated with the analysis, and the results of fitting the empirical PNLF of equation~(\ref{eq:pnlf}) to the data while including the possibility of PN superpositions \citep{Chase+23}.  The galaxies selected include both early and late-type systems, and their PN populations range from more than 200 detected objects, down to just a handful of PN candidates.   Throughout the analysis, we assume the Milky Way foreground extinction model of \citet{Schlafly+11} and $A_{5007} = 3.47 E(B-V)$ \citep{Cardelli+89}.

The PNLF analyses presented below all involve solving for two variables:  the distance modulus, which, in graphical form, shifts the fitted curve along the $x$-axis, and a (log) normalization, which offsets the curve in $y$.  Since the fuel-consumption theorem predicts that a stellar population's rate of PN production per unit bolometric luminosity should be almost independent of its age, metallicity, and initial-mass function \citep{Renzini+86, Buzzoni+06}, we express our normalizations in terms of $\alpha_{2.5}$, the number of PNe predicted to be within 2.5~mag of $M^*$ (assuming the shape given by equation~\ref{eq:pnlf}) scaled to the amount of galaxy bolometric luminosity sampled in the survey.  Effectively, this means estimating the total amount of $V$-band light falling onto the MUSE IFS, applying a bolometric correction of B.C. $= -0.85$ \citep{Buzzoni+06}, and extrapolating the shape of the luminosity function from the survey's completeness limit to a depth that is 2.5~mag fainter than the PNLF cutoff.  When expressed in this way, most galaxies have values of $\alpha_{2.5}$ between $\sim 5$ and $\sim 50$~PNe~$L_{\odot}$ \citep{Ciardullo2010}. 

Our PNLF analyses carry two important caveats.  The first is that the data cubes used in this work are archival, and most were obtained for purposes unrelated to our goals. Thus, many are not ideal for distance scale analysis.  For example, in order to measure the total fluxes of faint planetary nebulae, one needs bright stars in the field of view, as these allow a precise measurement of the data cube's point spread function (PSF) and stellar aperture corrections.  Many high galactic latitude MUSE observations do not have such stars.  Similarly, since not all MUSE programs require the photometric conditions necessary for distance scale work, it is likely that the flux calibrations of some archival data cubes are more accurate than others. While we do attempt to include these uncertainties in our error budget, in some cases, the exact values are hard to quantify.

The second caveat is that all our distances are based on fitting the observed PN magnitudes to the function defined by equation~(\ref{eq:pnlf}) with a constant value for the function's cutoff of $M^*-4.54$.  It is possible that this expression does not capture the true shape of the PN magnitude distribution; for example, if the \OIII luminosity of a PN can exceed $M^*$ --- either due to high central star luminosity or a fortuitous sight-line through the surrounding circum-nebular extinction -- then the PNLF distances may be systematically underestimated.  In addition, evidence suggests that $M^*$ fades in the metal-poor populations that may be found in low-luminosity galaxies and in the outer halos of larger systems \citep[e.g.,][]{Ciardullo+92, Dopita+92, Richer+08, Ciardullo2012, Bhattacharya+21}.  In the absence of a metallicity correction, the PNLF technque would then overestimate the distance to the system.  We comment on this latter possibility when necessary.

On the basis of our findings, we present our conclusions at the end of the paper regarding the prerequisites for targeted PNLF studies with MUSE\null.  Unless otherwise referenced, the basic data for the selected objects were retrieved from the NED and SIMBAD databases.

Each galaxy is depicted with a broadband RGB image, which indicates where the archival MUSE fields are located, a greyscale ``off-band'' image, which displays the light extracted in a 50~\AA\  bandpass longward of the redshifted \OIII and H$\beta$ lines, and an exemplary ``difference'' image formed by subtracting the off-band image from the sum of three MUSE data cube layers centered on \OIII\ $\lambda 5007$ at the redshift of the galaxy. These latter thumbnail images can be retrieved with full resolution through hyperlink access for immediate download.

\clearpage

\begin{figure*}[t!]
\hspace{20mm}
\href{https://cloud.aip.de/index.php/s/GXTtyGnwE3a7b38}{\bf \colorbox{yellow}{Off-band}}
\hspace{37mm}\href{https://cloud.aip.de/index.php/s/2MdSHb2nTKn6Z2C}{\bf \colorbox{yellow}{Diff}} \\
\includegraphics[width=52mm, bb=100 0 620 600,clip]{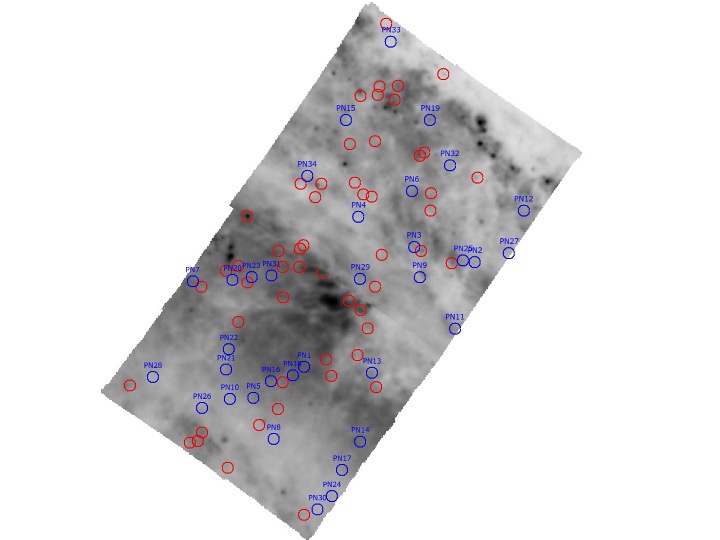}
\includegraphics[width=52mm, bb=100 0 620 600,clip]{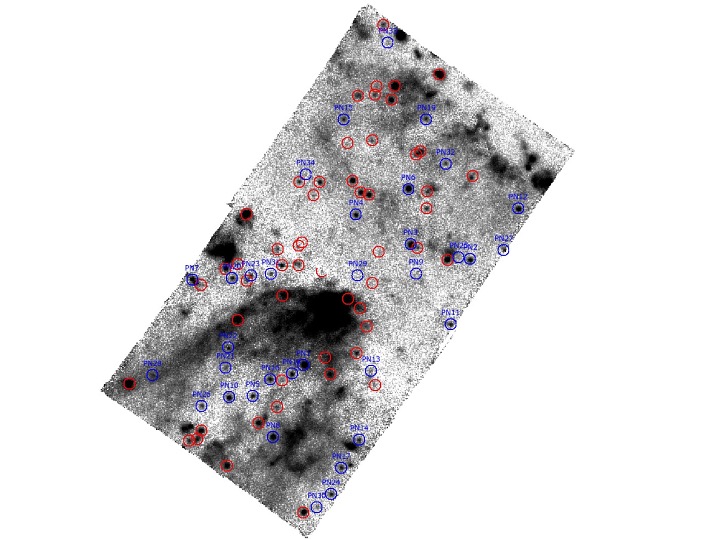}
\hspace{2mm}
\includegraphics[width=90mm,bb=0 0  2000 1400,clip]{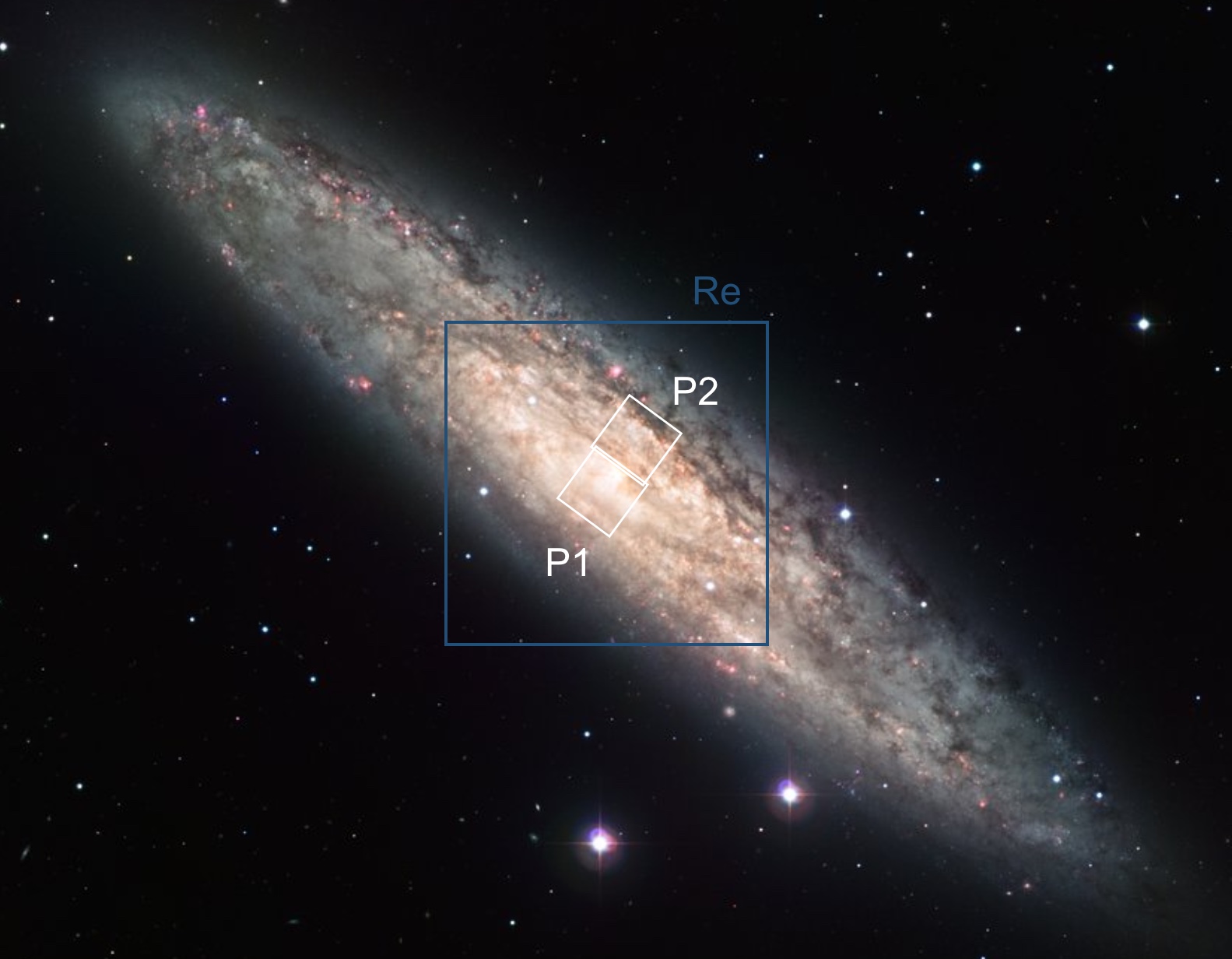}
\caption{NGC\,253. Left: thumbnail images of an off-band and difference image derived from the MUSE data cubes.  Our PNe candidates are highlighted in blue, while emission-line interlopers are shown in red. \href{https://cloud.aip.de/index.php/s/44ggaMYMXkfJ37D}{\colorbox{yellow}{VIDEO}}  Right: a broadband image showing the locations of pointings P1 and P2 (credit: ESO).
\label{fig:NGC253_FChart}}
\end{figure*}

\subsection{NGC 253 \label{subsec:NGC253}}
NGC\,253 is an SAB(s)c starburst galaxy and the most massive member of the Sculptor group.  At the time of our analysis, two MUSE data cubes of the galaxy's central regions were available in the ESO archive (IDs ADP.2018-11-22T21:29:46.157, ADP.2019-08-24T09:53:08.548, PI: L. Zschaechner, Program ID 0102.B-0078); a much more comprehensive set of observations has since been released and will be the subject of a separate paper.  The exposure times of the two data cubes considered here are similar (1820~s for P1 and 1833~s for P2), and our measurements of the image quality at 5007~\AA\ are similar ($1\farcs 06$ and $0\farcs 95$ for P1 and P2, respectively).  The locations of these pointings are shown in Figure~\ref{fig:NGC253_FChart}. 

The only previous PNLF study of NGC\,253's PNLF is by \citet{Rekola+05}, who used the classical on/off-band imaging technique to find 14 PNe in the field outlined in gray in the right panel of Fig.~\ref{fig:NGC253_FChart}.  These authors also modeled the effect of dust extinction in the galaxy; while one generally does not expect the PNLF to be greatly affected by galactic internal extinction --- the scale height of PNe should be several times that of the dust \citep[e.g.,][]{Feldmeier+97, Gonzalez-Santamaria+21, Guo+21} --- the high inclination of NGC\,253 ($i=74^{\circ}$) and the galaxy's prominent dust lanes are clearly an issue. \citet{Rekola+05} concluded that the effect of extinction on their sample of NGC\,253 PNe is quite small, and derived a PNLF distance to the system of $(m-M)_0 = 27.62^{+0.16}_{-0.26}$ for a foreground reddening of $E(B-V) = 0.019$.

Our initial list of point-like \OIII sources in NGC\,253's P1 and P2 fields contained 48 and 42 objects, respectively. However, after excluding objects based on their spectra (i.e., eliminating \ion{H}{2} regions and SNRs via the line strengths of H$\alpha$ and \SII), the list of PN candidates dropped to 19 and 15 sources.  Unfortunately, less than a dozen of these objects have magnitudes in the top $\sim 1$~mag of the PN luminosity function.

The derivation of NGC\,253's PNLF distance presents a number of challenges.  The first comes from the diffuse ionized gas (DIG) distributed throughout the field.  This gas, along with the galaxy's \ion{H}{2} regions and supernova remnants, is responsible for the large number of interlopers in our original list of PN candidates (red circles in the left panels of Fig.~\ref{fig:NGC253_FChart}) and for the possible existence of systematic errors in the background subtraction.  As a result, the relatively small formal photometric errors associated with the brightest PNe ($\sim 0.01$~mag) do not necessarily account for the true uncertainties in our measurements. To compensate for this, we included an additional 5\% error into all our photometry.

\begin{figure}[h!]
\includegraphics[width=0.473\textwidth]{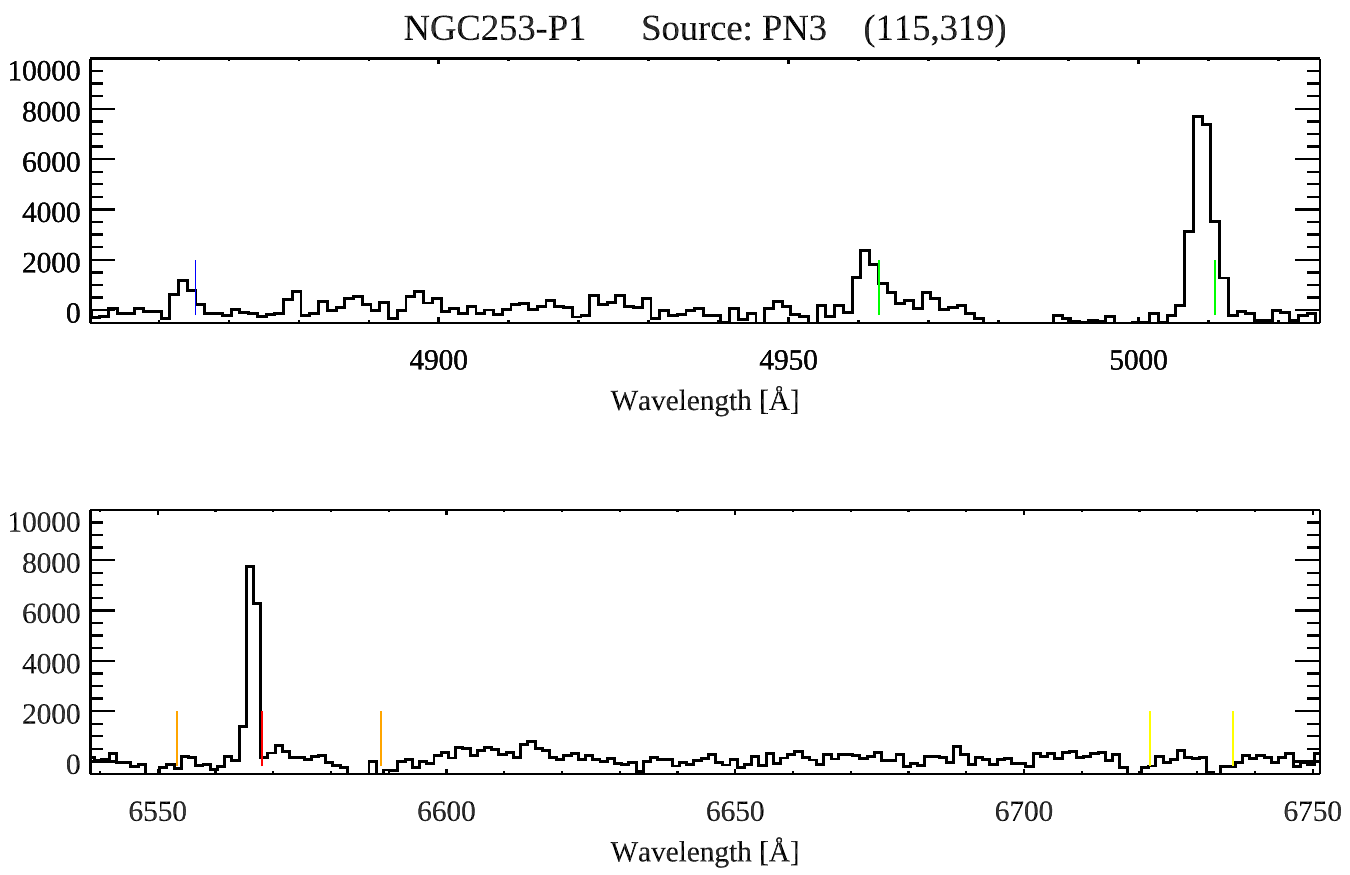}
\caption{Blue and red regions of the spectrum of the bright, highly extincted planetary PN3 found in pointing P1 of NGC\,253.  The $y$-axis is in units of  $10^{-20}~$erg\; cm$^{-2}$\;s$^{-1}$\;\AA$^{-1}$. The extreme H$\alpha$/H$\beta$ ratio indicates an extinction of $A_{5007} \sim 2.1$ mag.
\label{fig:NGC253_P1_PN3}}
\end{figure}

A second issue arises from the possibility of internal galactic extinction.  As evidenced by their Balmer decrements, many of the PNe in NGC\,253 are heavily obscured (e.g.,  Figure~\ref{fig:NGC253_P1_PN3}).  Simplistic scale-height based arguments \citep[e.g.,][]{Feldmeier+97, Gonzalez-Santamaria+21, Guo+21} suggest that internal extinction should not affect most PNLF measurements, but if NGC\,253's dust is extincting the PNe, the effect should be removed before deriving a distance.  Alternatively, since NGC\,253 is a starburst object, it is also likely that many of its PNe have evolved from relatively high-mass progenitors.  The circumnebular extinction produced by these objects  should \textit{not} be touched, since the component is an important contributor to the location of the PNLF cutoff \citep{Ciardullo2012, Davis+18b}.  Unfortunately, since there is no easy way to disentangle the two sources of extinction, one must simply live with the possibility that internal galactic extinction, might be affecting the results.  If so, then our formal PNLF distance will be an overestimate.  (This argument applies to all the dusty galaxies analyzed in this paper.)

Finally, we note that the P1 and P2 fields of NGC\,253 do not possess any bright reference stars with which to determine the observations' aperture corrections.  This is not an uncommon issue with MUSE archival data, since the field-of-view of the instrument is only $1\arcmin \times 1\arcmin$ (see Paper~I).  Our aperture corrections were therefore estimated using bright, compact star clusters as a reference.  For distant galaxies, the use of such systems for point-source photometry should be valid, but at the distance of the Sculptor group, some of the clusters may be marginally resolved.  If so, their use as ``point sources'' could introduce a zero-point error into our photometry and artificially brighten the PNLF\null.  Additionally, since there is little overlap between the two MUSE fields, the errors associated with the aperture corrections may result in the photometry of the two fields having different zero points.  The systematic error introduced by this effect is discussed in \ref{subsec:NGC1433}.


\begin{figure}
\includegraphics[width=0.473\textwidth]{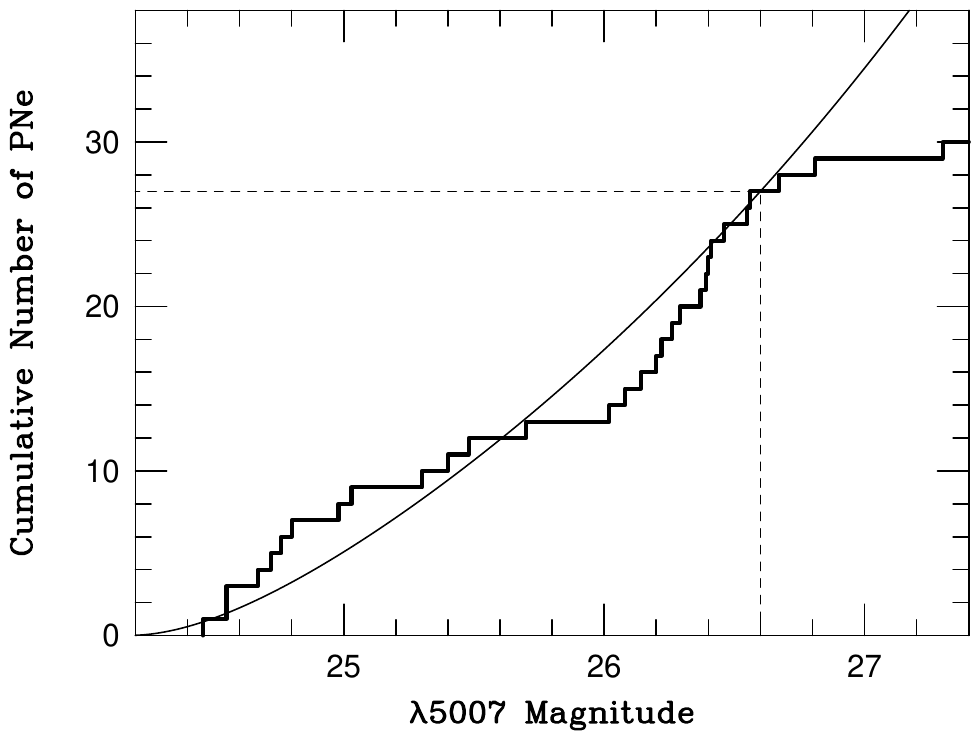}
\caption{The cumulative luminosity function for PNe in NGC\,253.  The dark line represents the observed data; the curve is equation~(\ref{eq:pnlf}) shifted to the most-likely apparent distance modulus.  The dashed line shows the location where incompleteness and the PNLF's star-formation ``dip'' cause the curve to overpredict the number of PNe.  Data brighter than this are consistent with being drawn from the empirical function.
\label{fig:NGC253_PNLF_v3}}
\end{figure}

Since our PN sample contains only $\sim 10$ objects in the top magnitude of the luminosity function, a plot showing the differential distribution of these points is not very revealing.  Consequently, Figure~\ref{fig:NGC253_PNLF_v3} plots the cumulative distribution of the PN magnitudes. (We will present cumulative distributions in later cases, as well, when the PN count is low.)  The rapid departure from the distribution predicted by equation~(\ref{eq:pnlf}) beyond $m_{5007} \sim 26.6$ is not principally due to incompleteness.  Rather, it comes from a ``dip'' in the luminosity function $\sim 1.5$ mag below the bright-end cutoff. This feature, which was first reported by \citet{Jacoby+02} for the PNLF of the SMC, is characteristic of most star forming systems, though the exact location and strength of the dip varies with stellar population.  \citep[For a qualitative explanation of the phenomeon, see][]{Ciardullo2010}.  At brighter magnitudes, the PN magnitude distribution follows the predictions of the empirical law quite well (i.e., a K-S test cannot rule out the hypothesis that the PNe are drawn from that distribution), but with so few objects in the top $\sim 1$~mag of the PNLF, the exact position of the function's cutoff cannot be fixed with any precision. 

We used the analysis program of \citet{Chase+23} to derive NGC\,253's PNLF distance.  To estimate the likelihood of any PN superpositions, we used the $H$-band surface photometry of \citet{Forbes+92} to measure the amount of galaxy light at each position in the data cube and converted these IR magnitudes to the $V$-band using the galaxy's integrated optical and IR colors \citep{Aaronson+77, Jarrett+03}.   We then assumed that the spectral lines of any possible PN blend would be unresolved, as the line-of-sight velocity dispersion in a disk galaxy such NGC\,253 should be much less than MUSE's spectral resolution.  We then computed the galaxy's most likely distance modulus, under the assumption of a foreground Milky Way reddening of $E(B-V) = 0.016$ \citep{Schlafly+11}.  

\begin{figure}[h]
\includegraphics[width=0.473\textwidth]{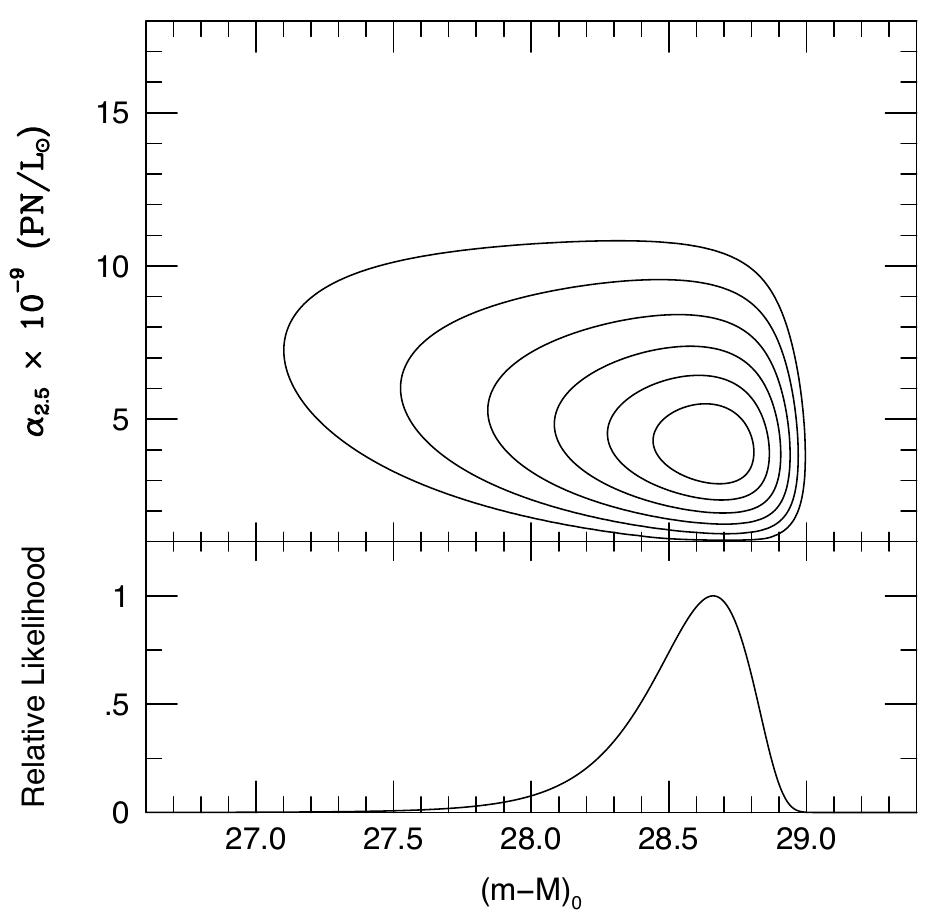}
\caption{The results of the maximum likelihood analysis for NGC\,253.  The top panel shows the likelihood contours drawn at $0.5\sigma$ intervals, with the abscissa giving the galaxy's true distance modulus, and the ordinate showing the number of PNe within 2.5~mag of $M^*$ normalized to the amount of galactic bolometric light contained in the two MUSE data cubes.  The lower panel marginalizes over the this variable.  The distribution is quite wide, due to the dearth of PNe in the top $\sim 1$~mag of the luminosity function. 
\label{fig:NGC253_contours}}
\end{figure}

Figure~\ref{fig:NGC253_contours} shows the results of our analysis.  With so few objects in the top magnitude of the luminosity function, the random error associated with our solution is substantial.  If we fit NGC\,253's PNLF down to a limiting magnitude $m_{5007} = 26.6$, we obtain a galaxy distance modulus of $28.66^{+0.12}_{-0.28}$, or $5.4^{+0.3}_{-0.6}$~Mpc.  This number is about a magnitude more distant than the PNLF distance found by \citet{Rekola+05} from a different set of PNe identified at larger radii in the galaxy.  (Unfortunately, there are no PNe common to both data sets.)  It is also a magnitude further than the galaxy's TRGB distances of $\sim27.5$ \citep[e.g.,][]{Mouhcine+05, Dalcanton+09, Radburn-Smith+11}.  



The errors quoted above do not include the possible systematic offsets introduced by our very limited choice of aperture correction reference stars and our general lack of knowledge of the internal reddening.  The former number is difficult to quantify with the present dataset, while evidence for the latter is indirect.  Although \citet{Rekola+05} concluded that internal extinction likely did not strongly affect their sample of NGC\,253 PNe, our MUSE pointings lie much closer to the center of the galaxy, where the extinction is likely to be larger.  Moreover, the best-fit value of $\alpha_{2.5}$ displayed in Figure~\ref{fig:NGC253_contours} is slightly lower than expected for PN measurements in a spiral galaxy \citep{Ciardullo2010}, suggesting that dust may be having an effect on PN detections.


The two MUSE data cubes analyzed in this work cover only a small fraction of the galaxy's $\sim 3 \times 10^{10} L_{\odot}$ of $B$-band light. Thus, our result is just a feasibility test. However, since 2022 a more complete survey of NGC\,253 has been published in the ESO archive. These MUSE pointings include proper aperture correction stars, and should contain enough PNe to produce a PNLF statistical error comparable to that of the galaxy's TRGB distance measurement. We will address this opportunity in a future paper.

\subsection{NGC 1052 \label{subsec:NGC1052}}

\begin{figure*}[ht]
\hspace{12mm}
\href{https://cloud.aip.de/index.php/s/GYipFQtwa3XSgHN}{\bf \colorbox{yellow}{Off-band}}
\hskip110pt \href{https://cloud.aip.de/index.php/s/Qm3B34SjXbepA9q}{\bf \colorbox{yellow}{Diff}} \\
\includegraphics[width=50mm,bb=0 0  900 900,clip]{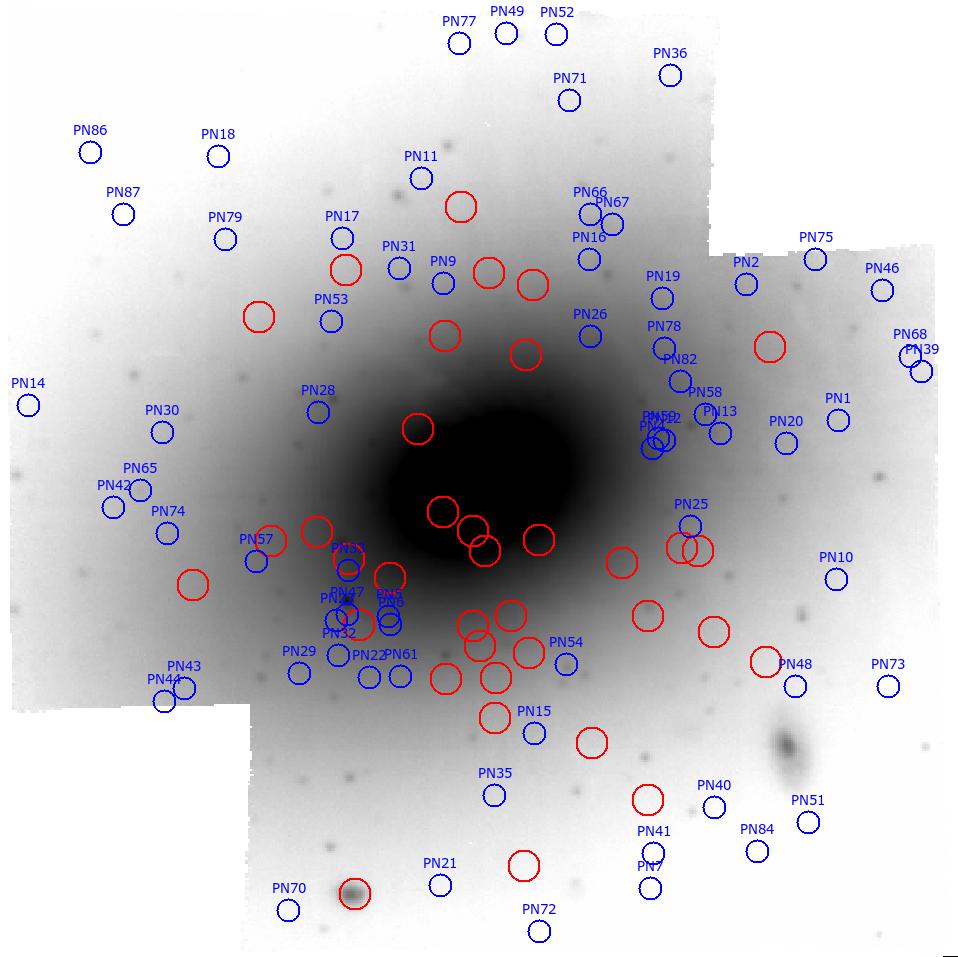}
\includegraphics[width=50mm,bb=0 0  900 900,clip]{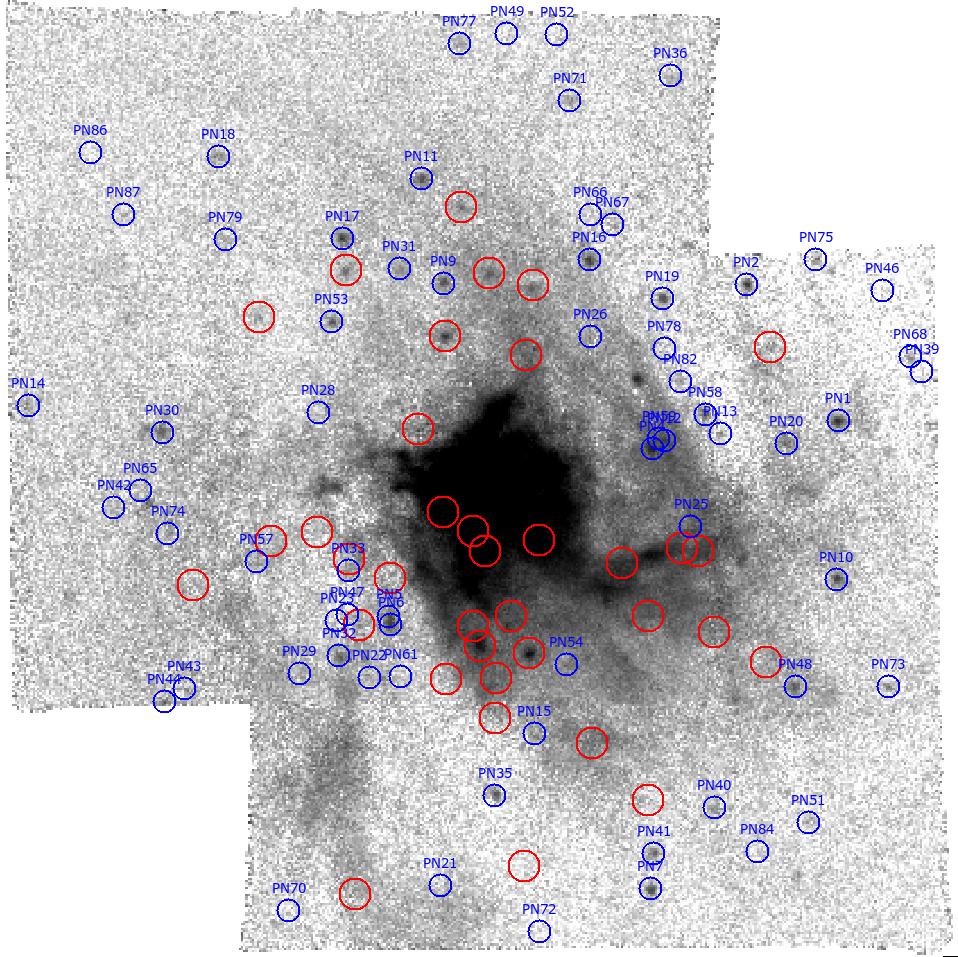}
\includegraphics[width=80mm,bb=0 0  2100 1600,clip]{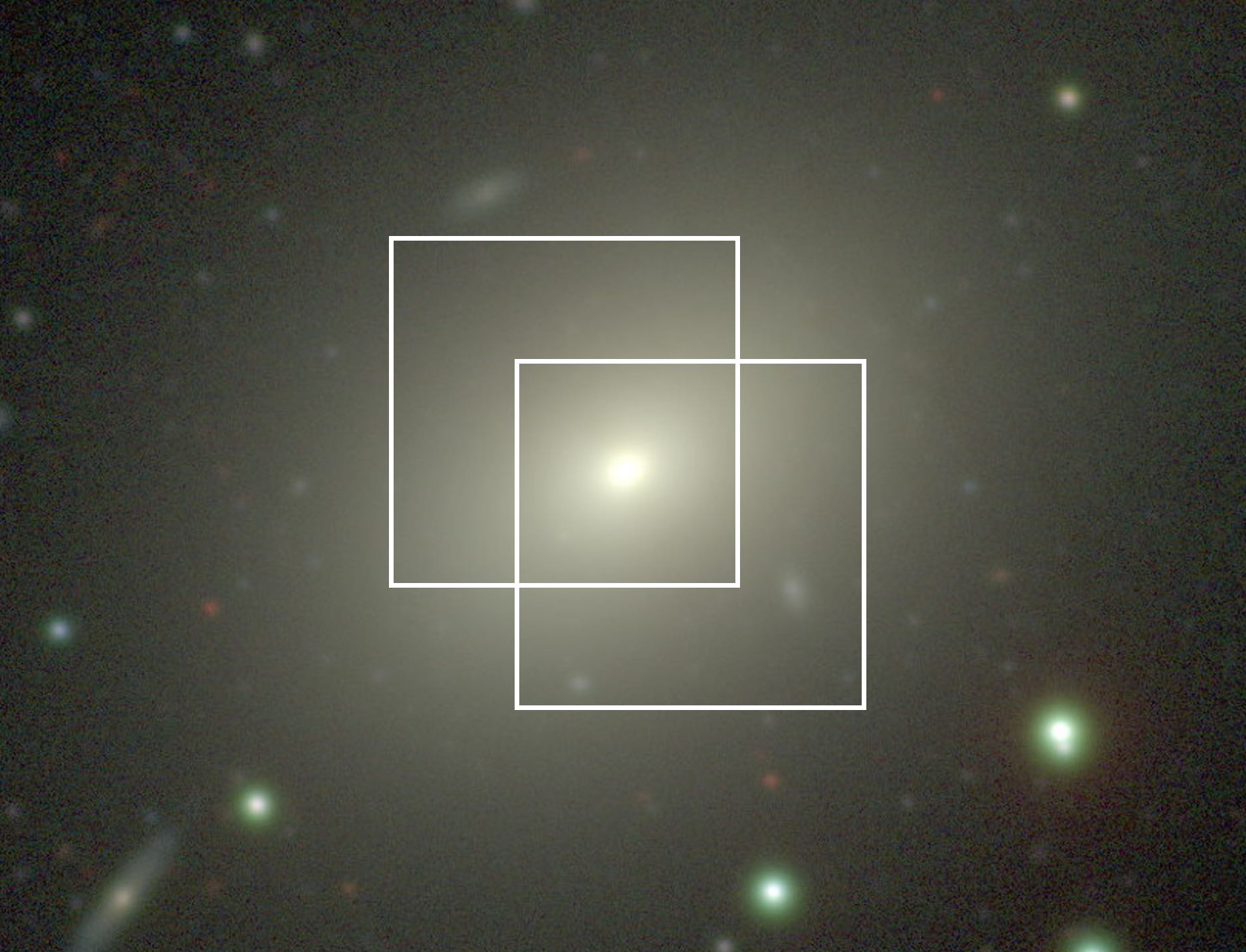}
\caption{NGC\,1052. Left: thumbnail off-band and difference images derived from the MUSE data cubes. Our PNe candidates are highlighted in blue, while emission-line interlopers are shown in red.  High-resolution images can be obtained by clicking on the hyperlink titles. \href{https://cloud.aip.de/index.php/s/6RQSjcK4tFxLYMj}{\colorbox{yellow}{VIDEO}}. Right: Broadband image of the galaxy with a mosaic of the two MUSE pointings outlined in white.  (Credit: Carnegie-Irvine Galaxy Survey (CGS)). \label{fig:NGC1052_FChart}}
\end{figure*}

NGC\,1052 is a bright elliptical (E4) galaxy with a radio jet, a LINER-type nucleus, and a large-scale outflow that has been studied with the Wide-Field IFU spectrograph (WiFeS) on the ANU 2.3-m telescope \citep{Dopita+15}.  Recently, the system has garnered considerable attention, as two of NGC\,1052's satellites, DF2 and DF4 appear to be ultra-diffuse dwarf galaxies with little or no dark matter \citep{VanDokkum+18}.  Since these dark matter estimates depend on the system's distance, an independent measure of this quantity is of considerable interest.

The ESO archive provides a data cube derived from the merger of two MUSE pointings with an effective exposure time of 1685~s and a measured seeing at 5007~\AA\ of $0\farcs 80$ (ESO archive ID ADP.2019-10-05T19:19:48.724, PI: L. Hernandez-Garcia, Program ID: 0103.B-0837).  As shown in  Figure~\ref{fig:NGC1052_FChart}, the data span the galaxy's minor axis, and overlap in the area of its bright nucleus  The ``diff'' image, shown in Fig.~\ref{fig:NGC1052_FChart} highlights the outflow, which is quite bright, even in the high excitation line of \OIII $\lambda 5007$.   

\begin{figure}[ht]
\includegraphics[width=0.473\textwidth]{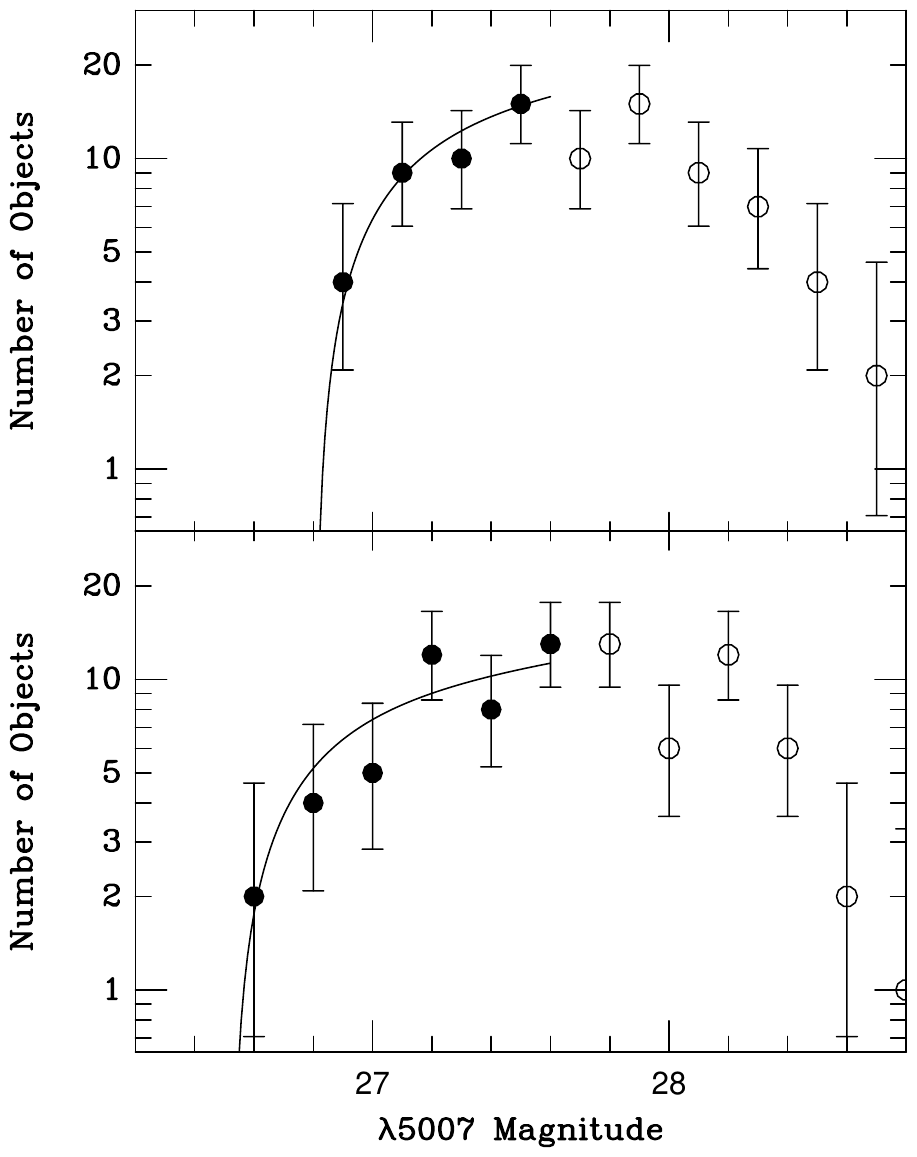}
\caption{Top:  The observed PNLF of NGC\,1052 binned into 0.2~mag intervals, with error bars displaying  $1\sigma$ confidence intervals \citep[see][]{Gehrels86}. Open circles represent data beyond the completeness limit of the survey.  The top panel shows the PNLF where the PN emission has been de-blended from the DIG using the methods described in Paper~I; the lower panel shows the same function when the sky estimates are based on simple aperture photometry.  The curves show the best-fit PNLF: without deblending, the most-likely apparent distance modulus is 0.2 mag smaller. 
\label{fig:NGC1052_PNLF}}
\end{figure}

\begin{figure}[ht]
\includegraphics[width=0.473\textwidth]{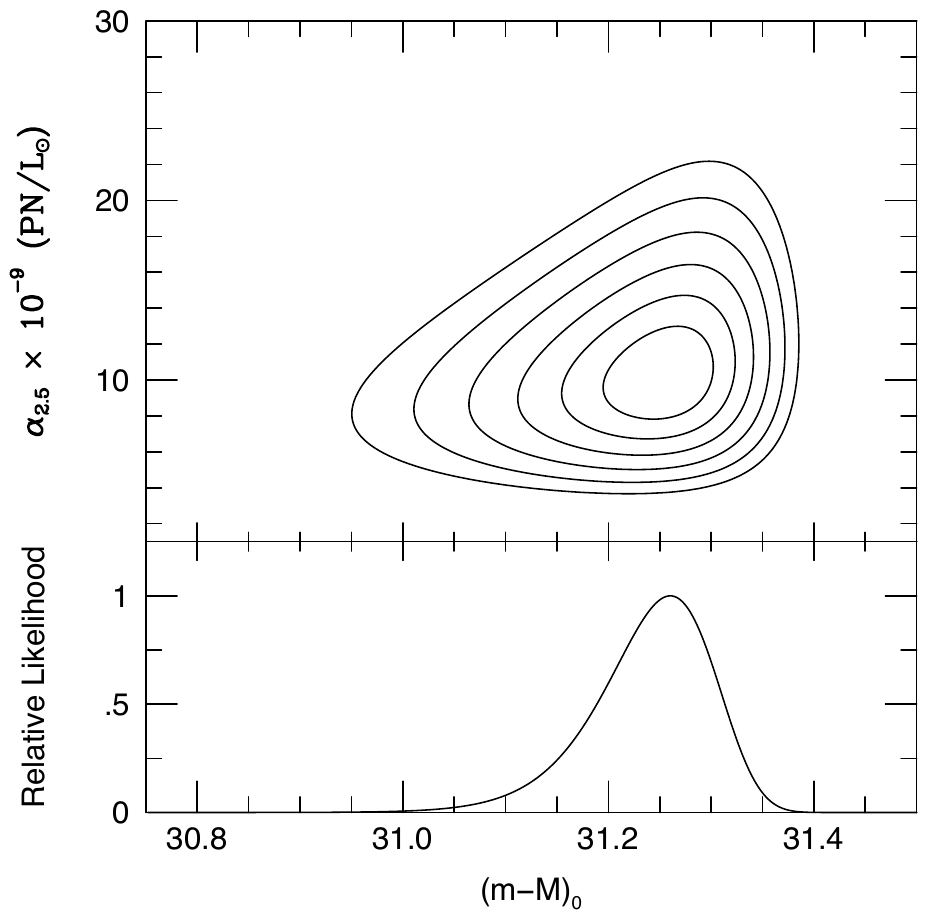}
\caption{The results of the maximum likelihood solution for NGC\,1052's distance modulus.  The top panel shows the likelihood contours (drawn at $0.5\sigma$ intervals), with the abscissa giving the galaxy's true distance modulus and the ordinate showing the number of PNe within 2.5~mag of $M^*$, normalized to the amount of bolometric light sampled.  The lower panel marginalizes over the latter variable.  
 \label{fig:NGC1052_contours}}
\end{figure}

Our initial examination of NGC\,1052's MUSE data cube found 100 PN candidates, with the brightest 20 having photometric errors less than 0.05 mag.  However, we rejected 13 of these candidates, as their spectra are inconsistent with that of an \OIII-bright PNe \citep[e.g.,][]{Herrmann+08, Kreckel+17}.  These are identified in Fig.~\ref {fig:NGC1052_FChart} with red circles.  In addition, as the distribution of objects in Fig.~\ref {fig:NGC1052_FChart} demonstrates, PN detections in the galaxy's inner $11\arcsec$ are problematic, due to the extremely bright emission from the region's diffuse gas.  The exclusion of this region removed one object from our sample, leaving 86 PNe suitable for analysis, and $\sim 50$ in the critical top $\sim 1$~mag of the PNLF.

One major challenge associated with measuring NGC\,1052's PNLF is contamination by the galaxy's diffuse emission-line gas.  As illustrated by the difference image of Fig.~\ref{fig:NGC1052_FChart}, the DIG in NGC\,1052 creates a bright, complex background that compromises the photometry of faint PNe, even in the high-excitation \OIII line.  Fortunately, in most cases, the PN's radial velocity is different enough from that of the DIG that the two components can be de-blended quite easily using the MUSE spectra (see Paper~I\null).  The exception is in the nuclear region of the galaxy, where the luminosity of the outflow overwhelms that of the point-sources.

The importance of NGC\,1052's DIG is illustrated in Figure~\ref{fig:NGC1052_PNLF}.  The bottom panel shows the galaxy's PNLF as measured using simple point-source photometry, where the background is estimated using an annulus surrounding each source.  The top panel shows the PNLF where the contribution of the diffuse emission-line gas has been removed by carefully examining each object's spectrum and de-blending the PN's \OIII emission from that of the DIG using the methodology of Paper~I.  Clearly, the latter set of photometry is a better fit to equation~(\ref{eq:pnlf}) and produces a distance that is $\sim 0.25$~mag more distant.  This highlights another advantage of MUSE PNLF measurements: narrow-band photometry would have been unable to disentangle the two sources of \OIII emission.

To fit NGC~1052's PNLF, we determined the amount of galactic $V$-band light at every position in the MUSE data cube using the $R$-band surface photometry of \citet{Jedrzejewski+87} and an assumed color of $(V-R) = 0.93$ \citep{Persson+79}.  We find that, after excluding the central $11\arcsec$ of the galaxy, roughly $V \sim 11.1$ of flux is contained within the two MUSE pointings.  We then estimated the stellar velocity dispersion underlying each PN's position using the long-slit spectroscopy of \citet{Binney+90}, and computed the most likely distance to the galaxy assuming a foreground extinction of $E(B-V) = 0.023$ \citep{Schlafly+11}.  Our solution using these numbers is illustrated in Figure~\ref{fig:NGC1052_contours}.  


As shown in the figure, equation~(\ref{eq:pnlf}) fits the observed PNLF extremely well, and there is no evidence for any overluminous objects.  Our PNLF distance to NGC\,1052 is $(m-M)_0 = 31.26^{+0.04}_{-0.07}$, ($17.9^{+0.3}_{-0.6}$~Mpc), where the uncertainties do not include the systematic error associated with  MUSE's flux calibration and our photometric aperture correction.  The former uncertainty is generally of the order of $\sim 3\%$ \citep{Weilbacher+20}; to estimate the latter, we examined the photometry of the four, point-like sources contained in the overlap region of the two NGC\,1052 MUSE fields.  These sources are relatively faint, with between 17175 to 37780 counts (in the units as explained in Appendix B), and the scatter of their inferred aperture corrections is $\sigma = 0.075$~mag.  If we use the standard deviation of the mean to defined the expected error on the data cube's aperture correction, then this systematic component is $\sim 0.04$~mag.

(We do note that the point-like sources used to define the data cube's PSF are likely not Milky Way stars, but globular clusters belonging to NGC\,1052.  However, this should not effect our measurement of the aperture correction.  If we scale M31's globular cluster system \citep{Barmby+01} to the distance of NGC\,1052, then a median globular cluster in the galaxy should have a half-light radius of only $0\farcs 02$, and the angular size of the M31's largest cluster would be $R_e \sim 0\farcs 31$.  This is much smaller than the $0\farcs 80$ seeing of the data cube.)

If we fold in the systematic errors associated with the MUSE flux zero-point, the data cube's aperture correction, and the likely error on the reddening, then our estimate for the galaxy's distance modulus becomes $(m-M)_0 = 31.26^{+0.07}_{-0.08}$ (for an $M^*$ value of $-4.54$).  This distance is essentially identical to that found from the \textit{HST}/NICMOS measurement of the galaxy's near-infrared surface brightness fluctuations \citep[$(m-M)_0 = 31.28\pm0.27$;][]{Jensen+03}. It is also consistent with the distances of the galaxy's two ultra-diffuse satellite dwarfs, as inferred by first taking the ratio of their $I$-band surface brightness fluctuations to those of the dwarf galaxies of the M96 group, and then anchoring the zero point to the megamaser distance of NGC\,4258  \citep{VanDokkum+18}.  Interestingly, direct TRGB measurements of these dwarfs give significantly larger values for the group's distance, with $(m-M)_0 = 31.72 \pm 0.12$ \citep{Shen+21} and $31.50 \pm 0.18$ \citep{Danieli+20}.  These distance moduli are excluded by our measurements.  Finally, we note that \citet{Fensch+19} report the identification of 3 PNe in the dwarf NGC\,1052-DF2\null. However, since  none of these objects are likely to have magnitudes near the bright-end cutoff of the PNLF, their detection does not provide a useful constraint on the galaxy's distance. 

\begin{figure*}[ht]
\hspace{19mm}
\href{https://cloud.aip.de/index.php/s/Aco4ErDEsHdFCoz}{\bf \colorbox{yellow}{Off-band}}
\hspace{37mm}
\href{https://cloud.aip.de/index.php/s/A2gDEx8Z9zD4q9p}{\bf \colorbox{yellow}{Diff}} \\
\includegraphics[width=49mm,bb=0 0  650 750,clip]{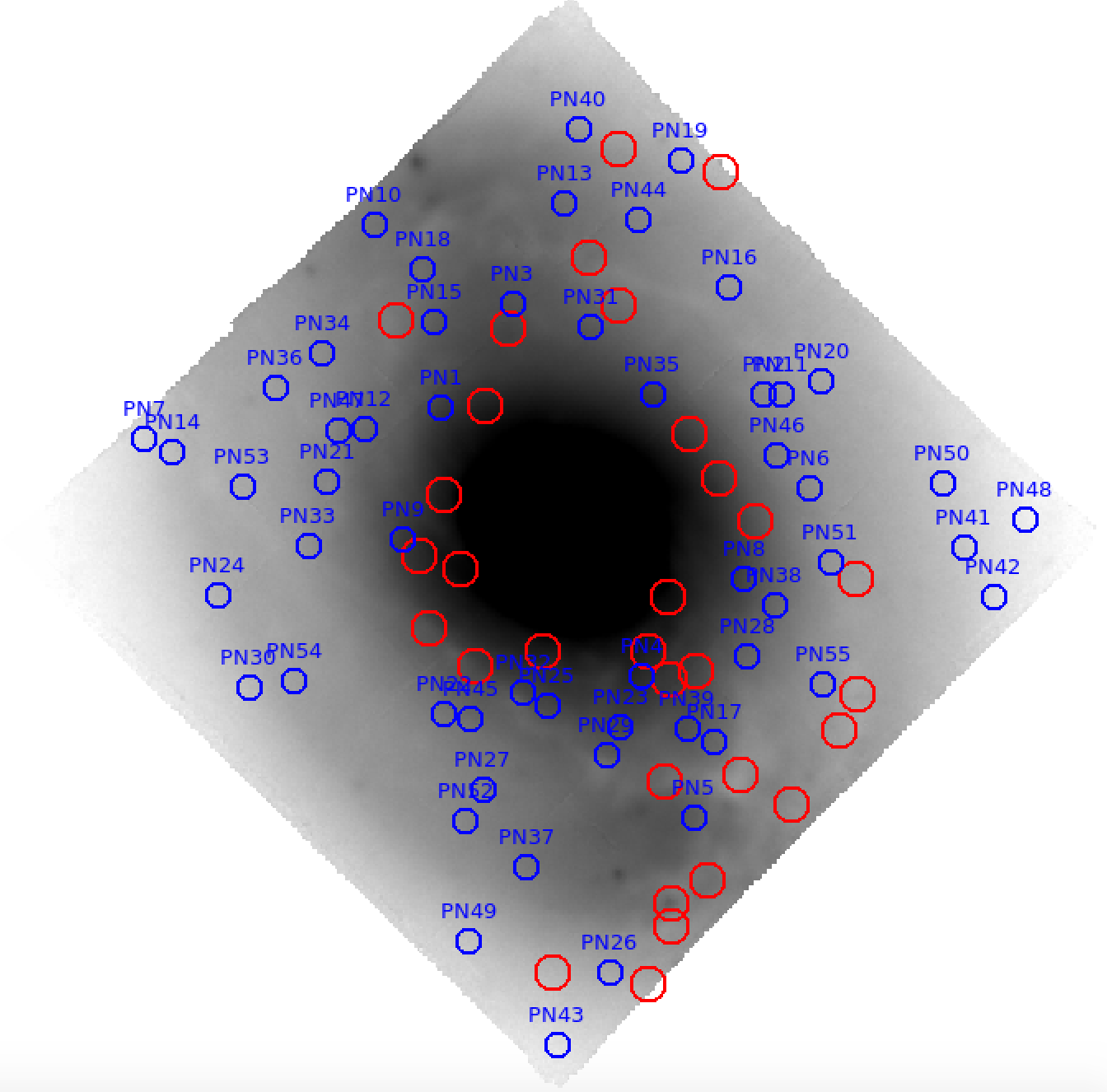}
\hspace{3pt}
\includegraphics[width=49mm,bb=0 0  650 750,clip]{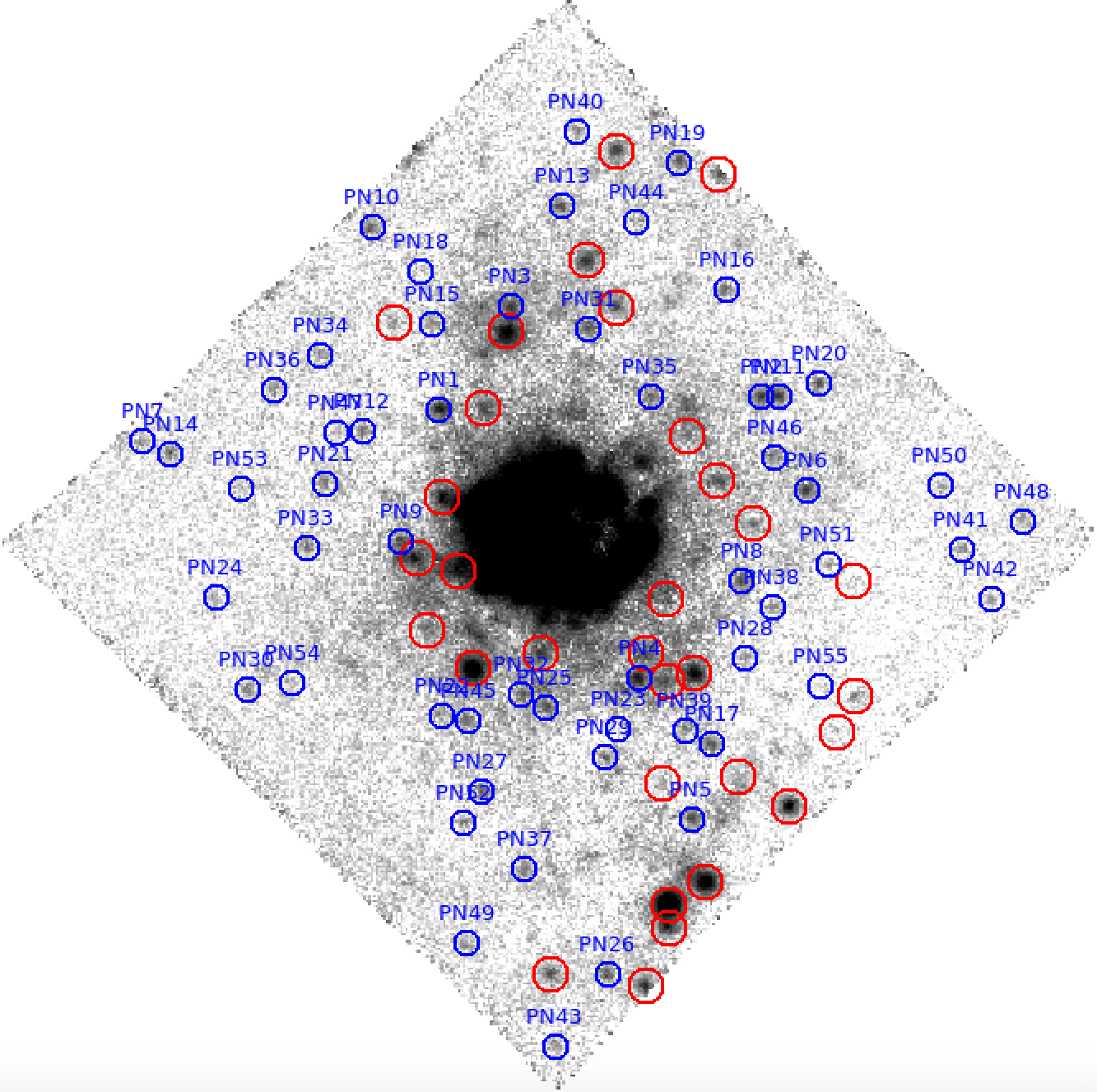}
\hspace{4pt}
\includegraphics[width=77mm,clip]{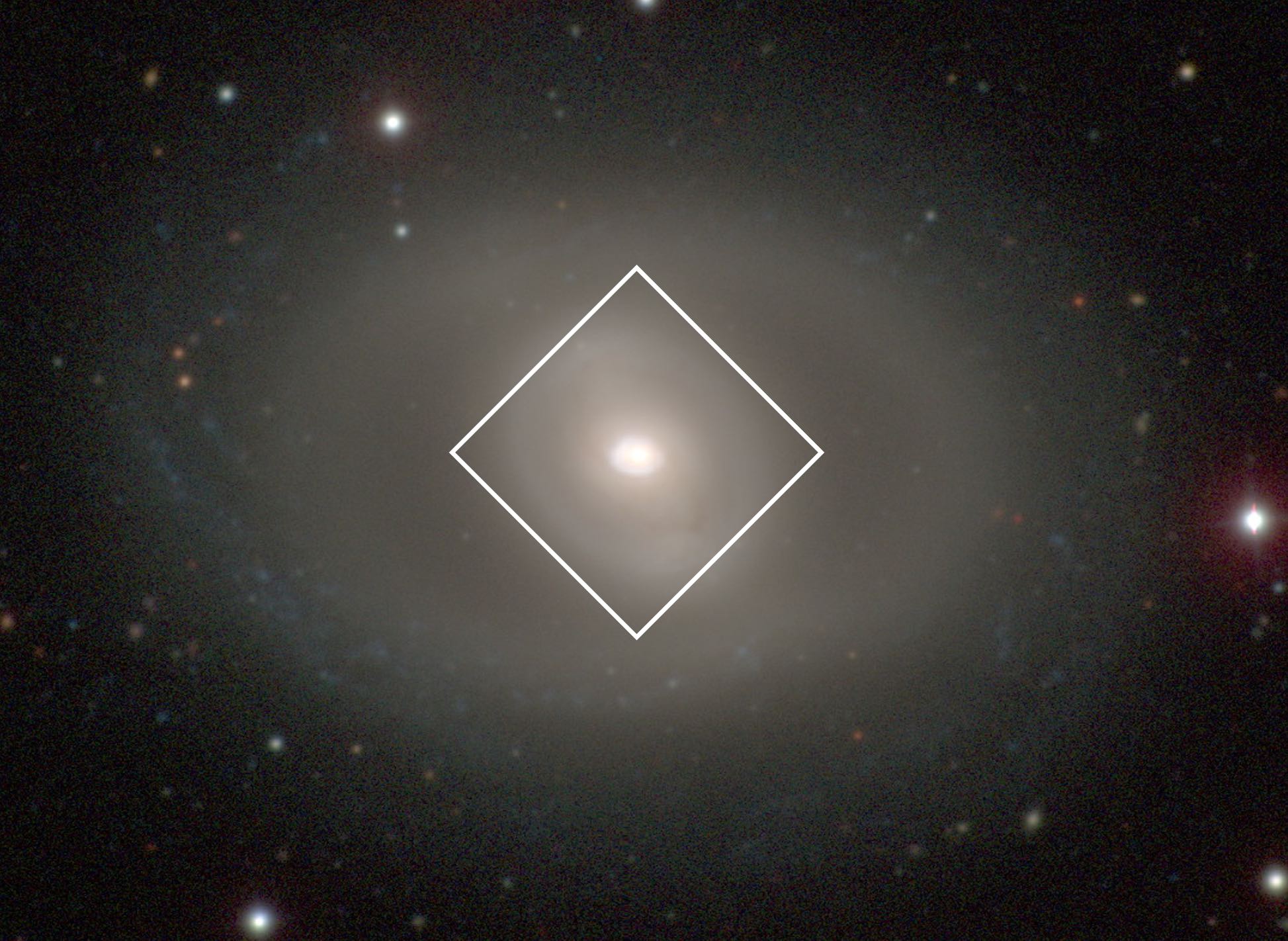}
\caption{NGC\,1326. Left: thumbnail offband and difference images derived from the MUSE data cube. Our PNe candidates are highlighted in blue, while emission-line interlopers are shown in red.   High resolution images can be obtained by clicking on the hyperlink titles. \href{https://cloud.aip.de/index.php/s/xb8pjMG977Egkb4}{\colorbox{yellow}{VIDEO}}.
Right: A broadband image of the galaxy with the location of the MUSE field shown in white
(Credit: CGS). \label{fig:NGC1326_FChart}}
\end{figure*}

\subsection{NGC 1326 (FCC\,029) \label{subsec:NGC1326}}

NGC\,1326, otherwise known as Fornax Cluster Catalog (FCC) galaxy 029 \citep{Ferguson1989}, is a bright lenticular system (Hubble type (R1)SAB(r)0/a) with an outer ring of star formation (roughly 2\arcmin\  from the nucleus) and an inner ring-like structure extending $\sim 9\arcsec$ in diameter.  Although the galaxy has Tully-Fisher measurements in the range of $\sim 13$ to $\sim 19$~Mpc \citep[e.g.,][]{Bottinelli+84,Willick+97,Springob+07}, the system's early Hubble type and moderately face-on ($i \approx 45^\circ$) inclination calls those values into question. Nevertheless, the system's location in the core of the Fornax cluster, $3 \fdg 1$ from the central cD galaxy NGC\,1399, suggests that the Tully-Fisher estimates are reasonable.  There are no Cepheid, TRGB, or SBF distances to the system.


We retrieved from the ESO archive a MUSE-DEEP data cube formed from the combination of two exposures (ESO Archive ID: ADP.2018-04-05T08:26:13.290, PI: M. Carollo, Program ID: 0100.B-0116).  The effective exposure time for these data is 3354~s, and the seeing at 5007~\AA\ is $1\farcs 00$.  The footprint of the IFS is shown in Figure~\ref{fig:NGC1326_FChart}. The region encompasses the system's bright nuclear region, and much of the galactic bar.

\begin{figure}[ht]
\includegraphics[width=0.473\textwidth]{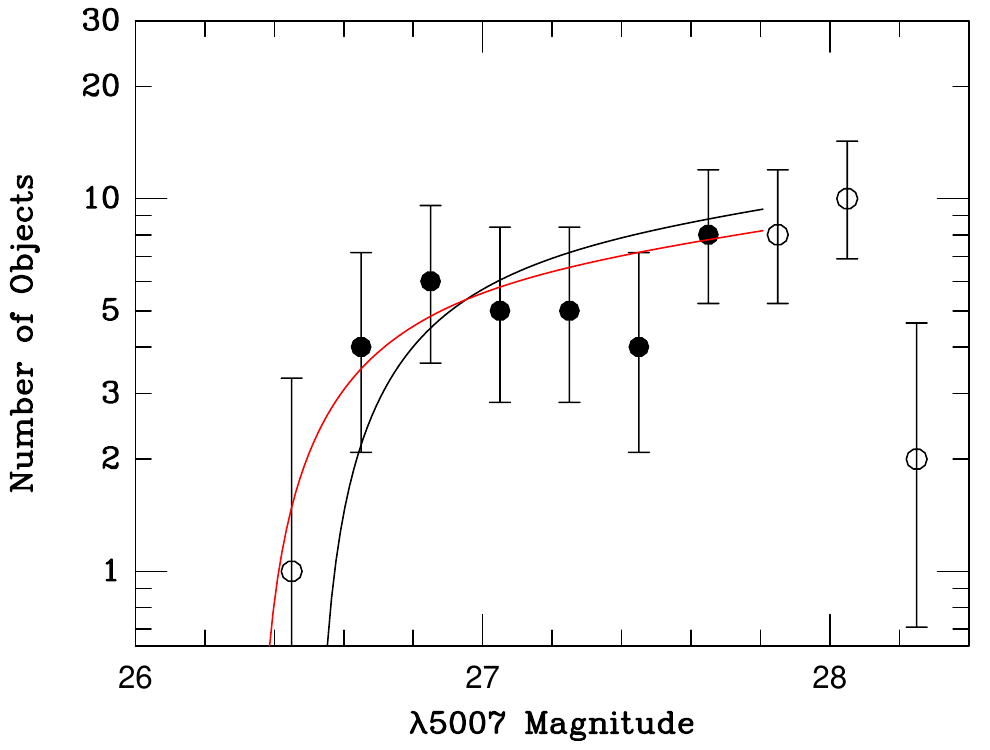}
\caption{The observed PNLF of NGC\,1326 binned into 0.2~mag intervals.  The open circles beyond $m_{5007} \sim 27.8$ represent data beyond the completeness limit; the bright open circle shows PN1.  The error bars represent each bin's $1\sigma$ confidence intervals \citep{Gehrels86}. The red curve illustrates the most-likely fit to equation~(\ref{eq:pnlf}) when PN1 is included in the sample; the black curve shows the fit when PN1 is excluded.  The difference between the solutions is 0.17~mag.
\label{fig:NGC1326_PNLF}}
\end{figure}

\begin{figure}[h!]
\includegraphics[width=0.473\textwidth]{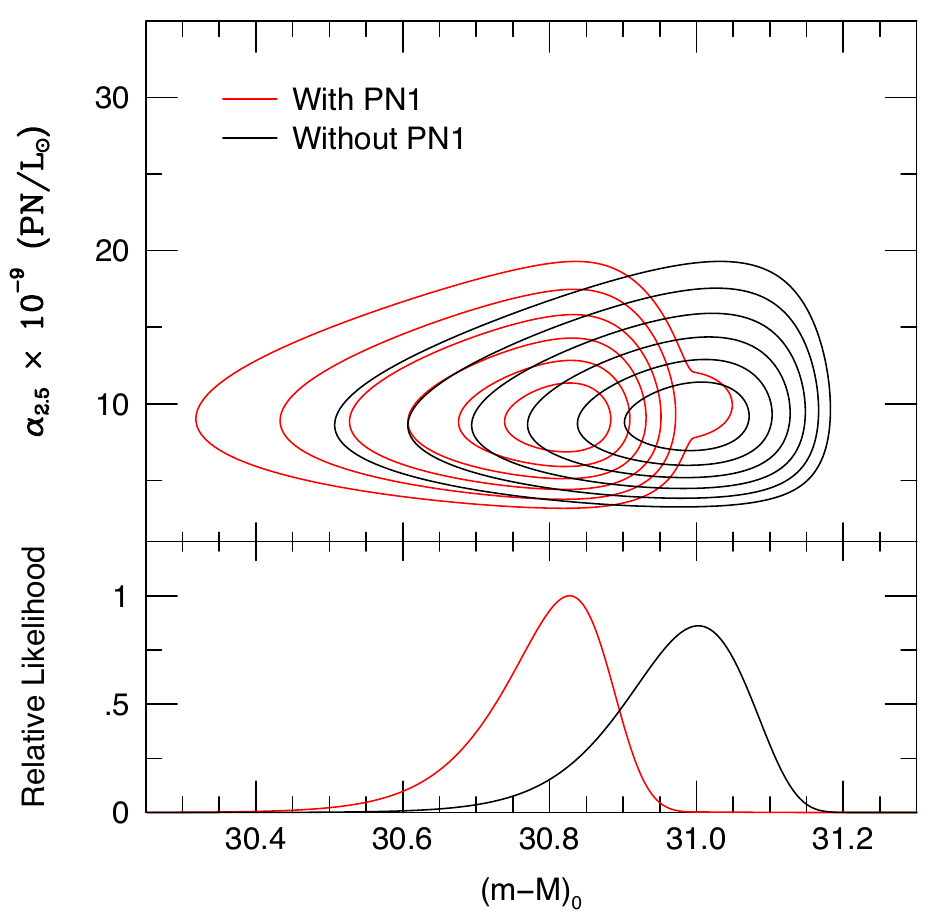}
\caption{The top panel shows the results of the maximum likelihood analysis for NGC\,1326.  The abscissa is the galaxy's true distance modulus, the ordinate is $\alpha_{2.5}$, the number of PNe within 2.5~mag of $M^*$, normalized to the amount of bolometric light sampled. The red contours, shown at $0.5\sigma$ intervals, illustrate the likelihoods when PN1 is included in the sample; the likelihood that PN1 is a superposition of two bright \OIII sources is reflected in the distorted probability contours at the $\sim 3\sigma$ level.  The black contours display the same information when PN1 is excluded. The bottom panel marginalizes these data over $\alpha_{2.5}$.\label{fig:NGC1326_contours}}
\end{figure}

PN detections within the central $12\arcsec$ of NGC 1326's nucleus are difficult, due to the region's high surface brightness and copious amount of diffuse \OIII emission.  Thus we confined our analysis to larger galactocentric radii, where we found 86 point-like emission-line sources.  Many of these objects, especially those close to the nucleus and along the galactic bar (the red circles in Fig.~\ref{fig:NGC1326_FChart}), have the spectral signature of \ion{H}{2} regions or SNRs and are excluded from our list of PN candidates.  The final sample therefore consists of 55 objects; this PNLF is shown in Figure~\ref{fig:NGC1326_PNLF}.

NGC\,1326's PNLF is not very well defined. In part this is because there is one PN candidate that is 0.31~mag more luminous than any other object.  (Fig.~\ref{fig:NGC1326_PNLF} shows the PNLF with the starting edge of the first bin chosen to minimize the discordant appearance of PN1.\footnote{This choice in no way affects our PNLF distance estimates, since our analysis does not use binned data.  The histograms displayed in this paper were created solely for the purpose of visualization.})  However, a more important factor is that, of the 40 PNe brighter than the nominal completeness limit of $m_{5007} = 27.8$, only $\sim 20$ are in the magnitude range that defines the PNLF cutoff ($m \lesssim M^* + 1$).

There are several possible explanations for the overluminous nature of PN1.  The first, and simplest, is that the apparent luminosity of the object (PN1) is simply due to poor photometry.  As can be seen in the ``diff'' image in Fig.~\ref{fig:NGC1326_FChart}, PN1 is located in a region of bright diffuse emission-line gas, adajacent to an interloping object (red circle).  It is quite possible that the source's apparent high luminosity is due to imperfect background subtraction, in which case the object should be excluded from the analysis.  Similarly, if the ``overluminous'' object is something other than a normal PN (such as a high-excitation, metal-poor \ion{H}{2} region), it should also be eliminated from the data set.  There is no evidence for this in the object's spectrum, but the interpretation cannot be completely excluded.  

A third explanation is that there is nothing abnormal about the PN at all, and the 0.31~mag difference between it and the next brightest PNe is simply due to small number statistics.  This seems unlikely, as that would imply that NGC\,1326 is at least $\sim 4$~Mpc closer than the main body of Fornax.  But if the object is a normal PN, then it must be included in the fitting process.  Similarly, it is possible that this bright \OIII source is actually formed from the combined flux of two normal PNe that are superposed upon each other.  To wit: PN1 is located in a region of NGC\,1326 with a $V$-band surface brightness of $\mu_V \sim 19.7$~mag~arcsec$^{-2}$ \citep{Buta+98}. According to equation~(3) of \citet{Chase+23}, this means at the nominal Fornax distance of 19~Mpc, PN1 has a $\sim 20\%$ chance of being a superposition of two objects in the top 2.5~mag of the PN luminosity function, and a $\sim 5\%$ chance that both PNe are in the top 1 mag of the PNLF\null.  If this hypothesis is correct, then, once again, the source must be included in the PNLF analysis.

Finally, it is possible that PN1 is a normal planetary nebula, but that our knowledge of the true shape of the PNLF is incomplete.  Equation~(\ref{eq:pnlf}) was originally defined to fit the magnitude distribution of a statistically complete sample of $\sim 120$~PNe in the bulge and inner disk of M31 \citep{Ciardullo+89}.  In such a finite dataset (with only $\sim 32$~objects in the top $\sim 1$~mag of the luminosity function), extremely rare, short-lived objects may not be present. Indeed, although \OIII PNLFs now exist for dozens of galaxies, the PN samples are rarely large enough to exclude the possible existence of a very low-amplitude, bright-end tail to the luminosity function.  If such a tail does exist, any attempt to fit PN1 with the expression given by equation~(\ref{eq:pnlf}) will result is a systematic underestimate of the galaxy's distance.  We will discuss the true shape of the empirical PNLF in \S\ref{subsec:grand_pnlf}.

We fit the PNLF of NGC\,1326, with and without PN1, using the photometric measurements of \citet{Buta+98} to determine the galaxy surface brightness underlying each PN candidate.  These data imply that the total amount of $V$-band light contained within the MUSE data cube is $V \sim 11.5$.  However, we made no attempt to model the behavior of the stellar velocity dispersion in MUSE field:  since the spectroscopy of \citet{Dalle-Ore+91} and \cite{Gadotti+05} both indicate that the line-of-sight velocity dispersion is much less than the velocity resolution of the MUSE spectrograph, we set $\sigma = 100$~km~s$^{-1}$ throughout the survey region.  Finally, we assume $E(B-V) = 0.016$ as the foreground extinction to the galaxy \citep{Schlafly+11}.

Figure~\ref{fig:NGC1326_contours} shows the results of our fits.  As expected, the galaxy's distance is not well-defined:  23 objects spread out over the top $\sim 1$~mag of the PNLF is insufficient for defining the precise location of the PNLF cutoff.  As a result, whether or not one includes PN1 in the sample does not substantially change the quality of the fit.  If PN1 is treated as a normal planetary nebulae, then the galaxy is forced to a distance that is $\sim 4$~Mpc smaller than the canonical distance to Fornax, i.e., to $14.6_{-0.7}^{+0.4}$~Mpc.  This result is unaffected by the inclusion of superpositions in the analysis:  although it is \textit{possible} that the object is a superposition of two sources, the likelihood that two PNe in sum would  produce an \OIII source $\sim 0.3$~mag brighter than the next most luminous object is quite low and barely registers in the contour plot. 

The exclusion of PN1 from consideration produces a fit that is only slightly better than that which includes the PN\null.  (Without PN1, the Kolmogornov-Smirnov statistic for a comparison of the data with the model is $D_n = 0.162$, for a $p$ value of $p=0.77$; with PN1, $D_n = 0.174$ and $p=0.83$.)  Yet without PN1, the galaxy's most-likely distance modulus is $\Delta \mu = 0.19$~mag greater, $(m-M)_0 = 31.00^{+0.06}_{-0.11}$, or $d = 15.9_{-0.8}^{+0.5}$~Mpc.

\begin{figure*}[ht]
\hspace{20mm}
\href{https://cloud.aip.de/index.php/s/FReEbZpaAjtTtXr}{\bf \colorbox{yellow}{Off-band}}
\hspace{28mm}\href{https://cloud.aip.de/index.php/s/TbADdxfEy2nc2Bj}{\bf \colorbox{yellow}{Diff}} \\
\hspace{10mm}
\includegraphics[width=52mm,bb=0 0  2054 2160,clip]{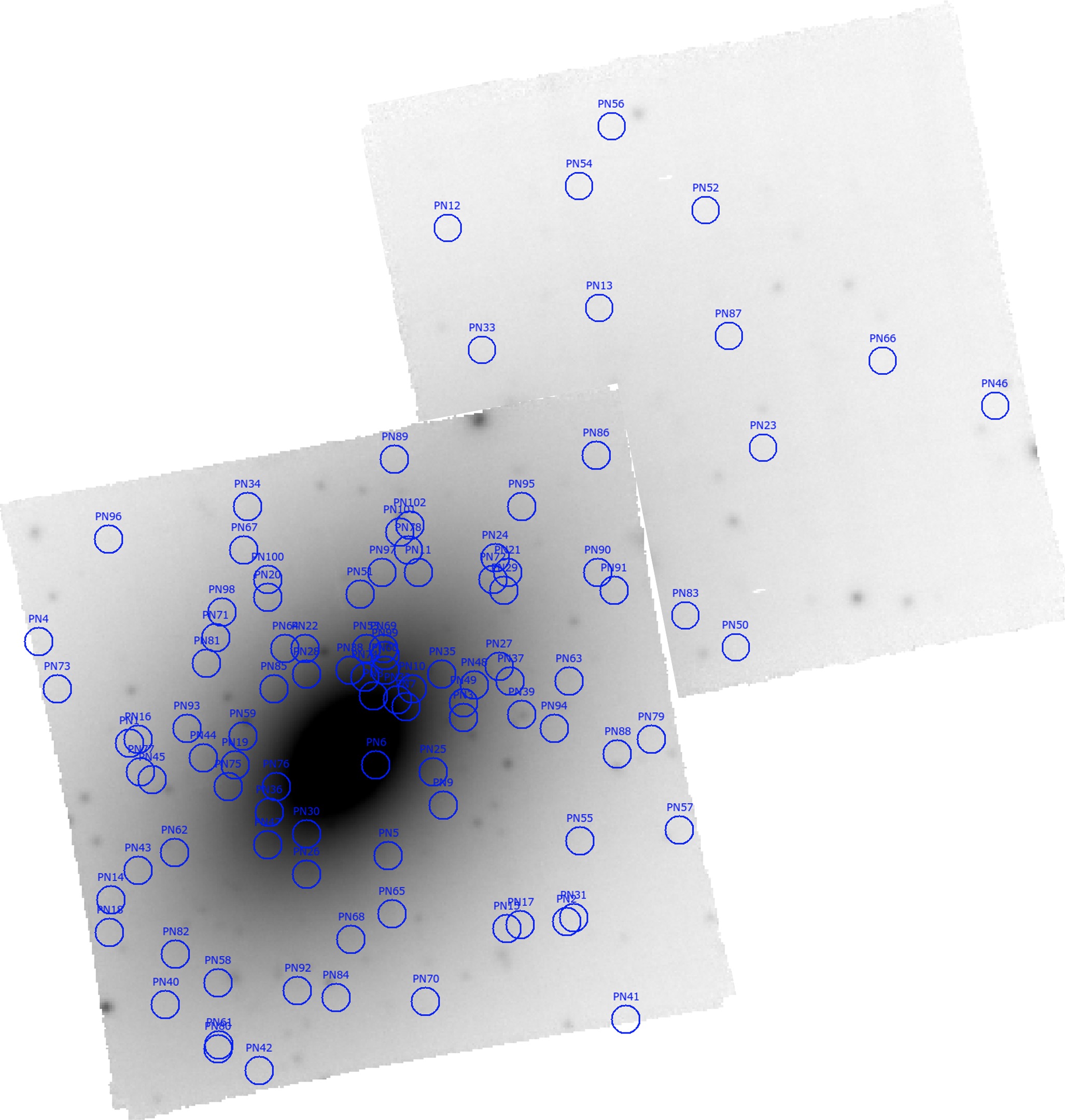}
\includegraphics[width=52mm,bb=0 0  2054 2160,clip]{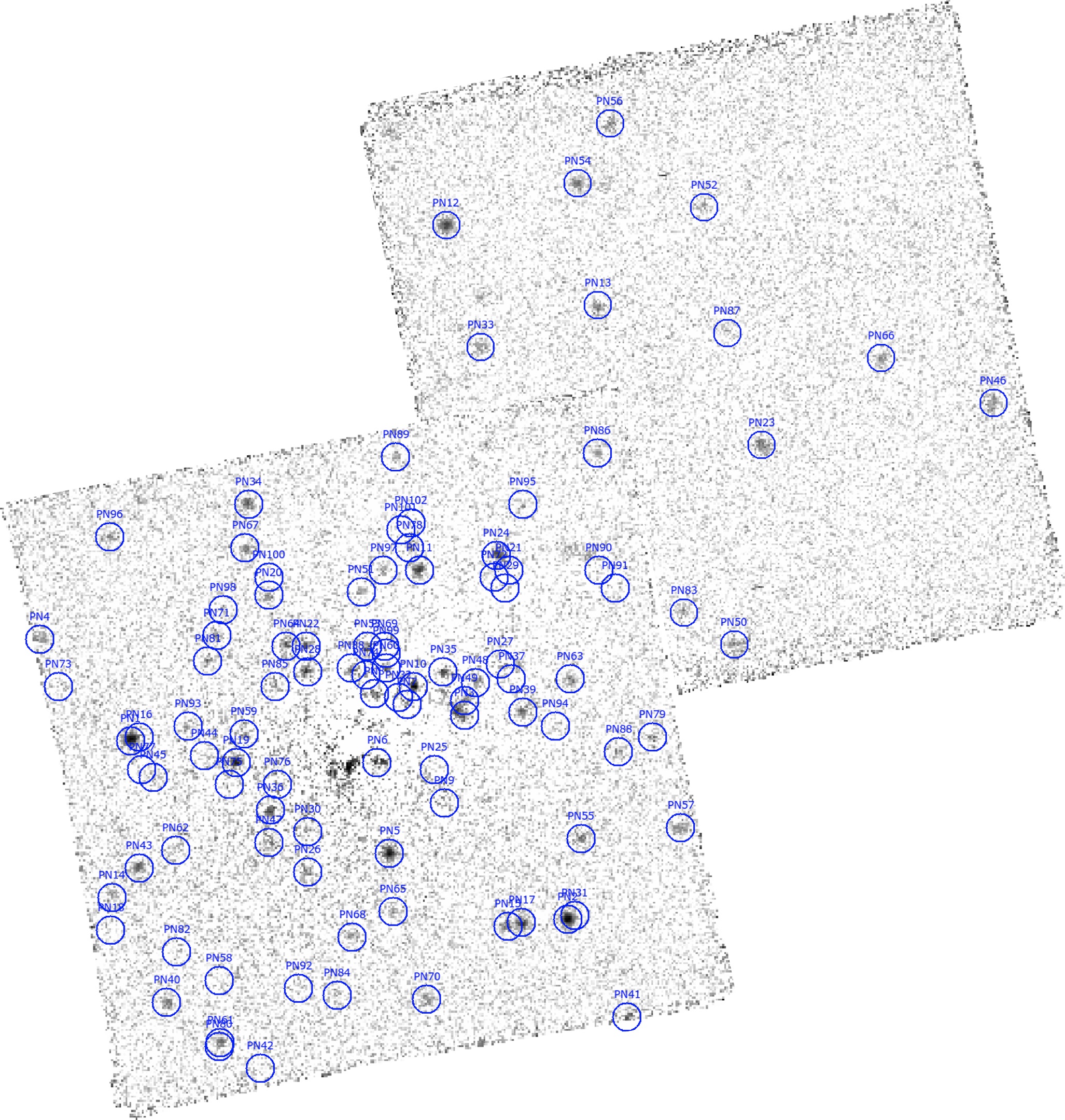}
\hspace{2mm}
\includegraphics[width=70mm,bb=0 20  700 550,clip]{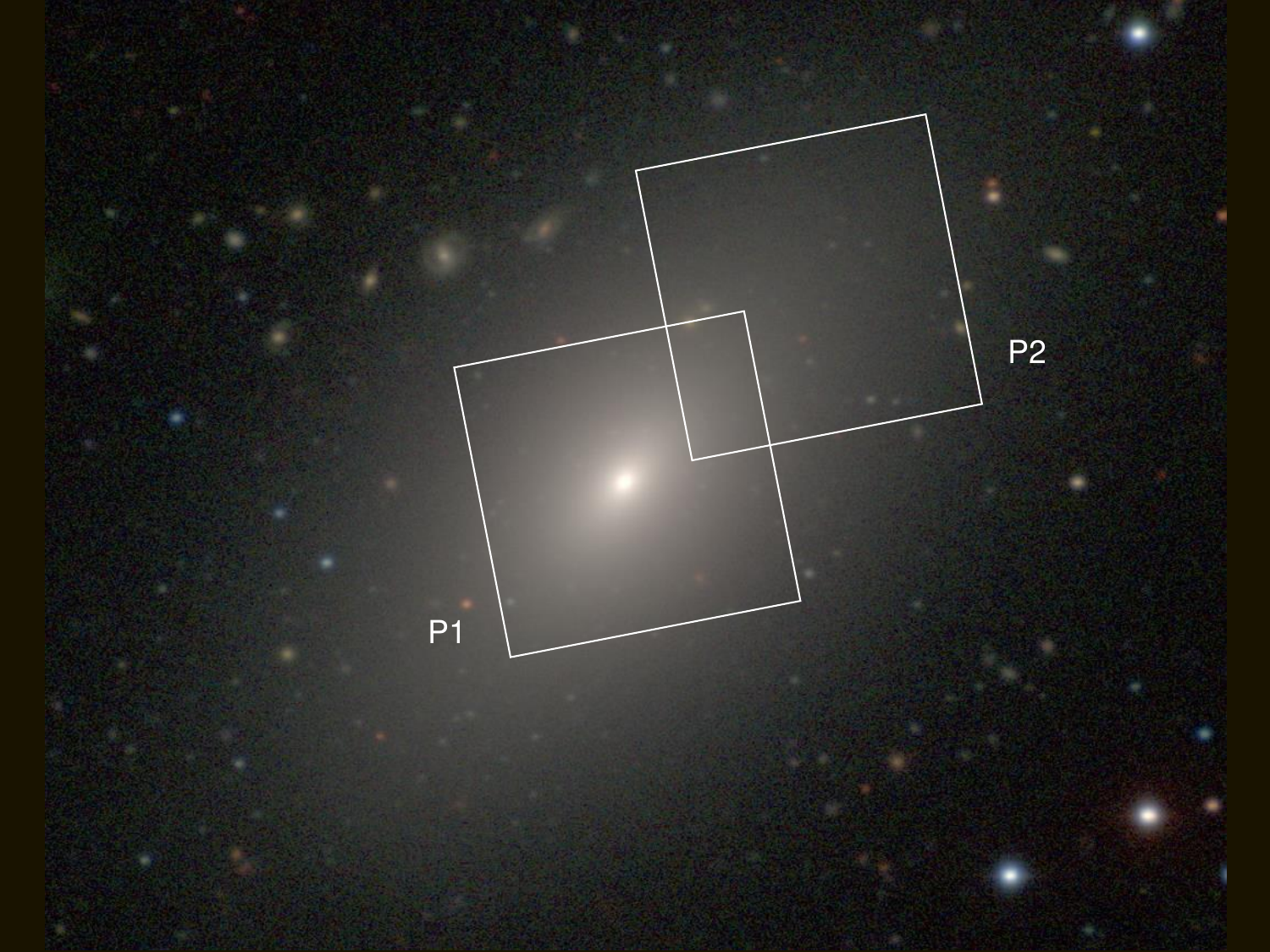}
\caption{NGC\,1351. Left: thumbnail off-band and difference images derived from MUSE data cubes of the galaxy. Our PNe candidates are circled in blue.  High resolution images can be obtained by clicking on the hyperlink titles. \href{https://cloud.aip.de/index.php/s/8ogZ5CHbwKCHY3r}{\colorbox{yellow}{VIDEO}}. Right: a broadband image with the MUSE fields outlined in white. (Credit: CGS). \label{fig:NGC1351_FChart}}
\end{figure*}

The uncertainties quoted above do not include the systematic errors associated with the reddening, MUSE flux calibration, data cube aperture correction, or limitations in the reference PNLF.  In particular, although the aperture correction error obtained from the field's brightest point-like source has a calculated uncertainty of only 0.028 mag, the standard deviation about the mean aperture correction for the three brightest PSF objects is 0.085~mag, implying an 0.05~mag error of the mean.  This uncertainty, along with those for the MUSE flux calibration and foreground reddening, result in a distance modulus of $31.00^{+0.09}_{-0.13}$, or $d = 15.9_{-0.9}^{+0.6}$~Mpc.  This number is still significantly smaller than the distance to Fornax, and suggest that the system may be foreground to the cluster's core.

\subsection{NGC 1351 (FCC\,083) \label{subsec:NGC1351}}

NGC\,1351 (FCC\,083) is another moderately bright Fornax lenticular (classification SAOp) that has two pointings in the ESO archive (ESO Archive IDs: ADP.2017-12-12T15:38:27.863 and ADP.2017-07-19T15:12:54.145, PI: M. Sarzi, Program ID: 296.B-5054).  The first pointing, P1, is  centered on the nucleus and has a nominal seeing at 5007~\AA\ of $0\farcs 88$ and an exposure time of 3379 s; the second, P2, overlaps P1, as it is centered $\sim 35\arcsec$ northwest of the nucleus along the galaxy's major axis.  The latter data cube has a somewhat longer exposure (4680~s) and slightly worse ($1\farcs 11$) image quality.

Our examination of the central pointing of NGC\,1351 found 92 PN candidates, of which 6 were later rejected as interlopers based on their spectra.  P2 had far fewer objects, as it sampled less light with poorer image quality: this pointing only added 16 PNe to the sample.  In total, we detected 102~PNe in NGC\,1351, with more than 30 objects in the top $\sim$1~mag of the luminosity function.  

\begin{figure}
\includegraphics[width=0.473\textwidth]{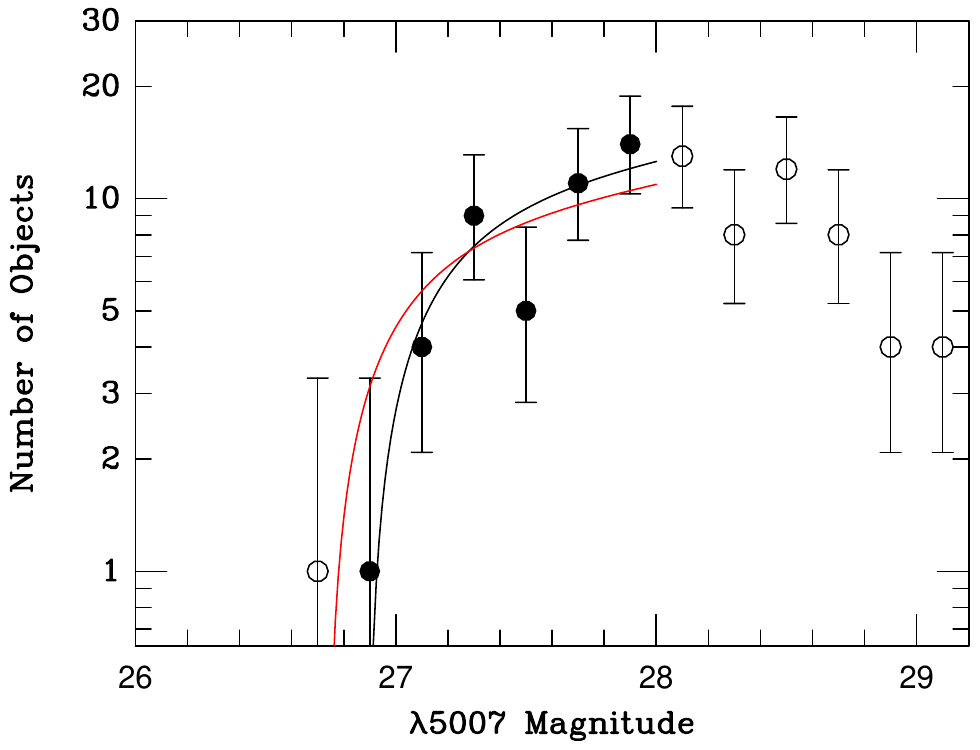}
\caption{The observed PNLF of NGC\,1351 binned into 0.2~mag intervals.  The open circles beyond $m_{5007} \gtrsim 28.0$ show data beyond the completeness limit; the open circle at $m_{5007} \sim 26.8$ represents PN1, an object that is 0.2~mag more luminous than the next brightest source.  The error bars illustrate $1\sigma$ confidence intervals \citep[see][]{Gehrels86}. The red curve illustrates the most-likely fit to equation~(\ref{eq:pnlf}) when PN1 is included in the sample; the black curve shows the fit when PN1 is excluded.  The difference between the values is 0.16~mag.  A K-S test excludes PN1 from the sample with 93\% confidence. 
\label{fig:NGC1351_PNLF}}
\end{figure}

\begin{figure}[h!]
\includegraphics[width=0.45\textwidth]{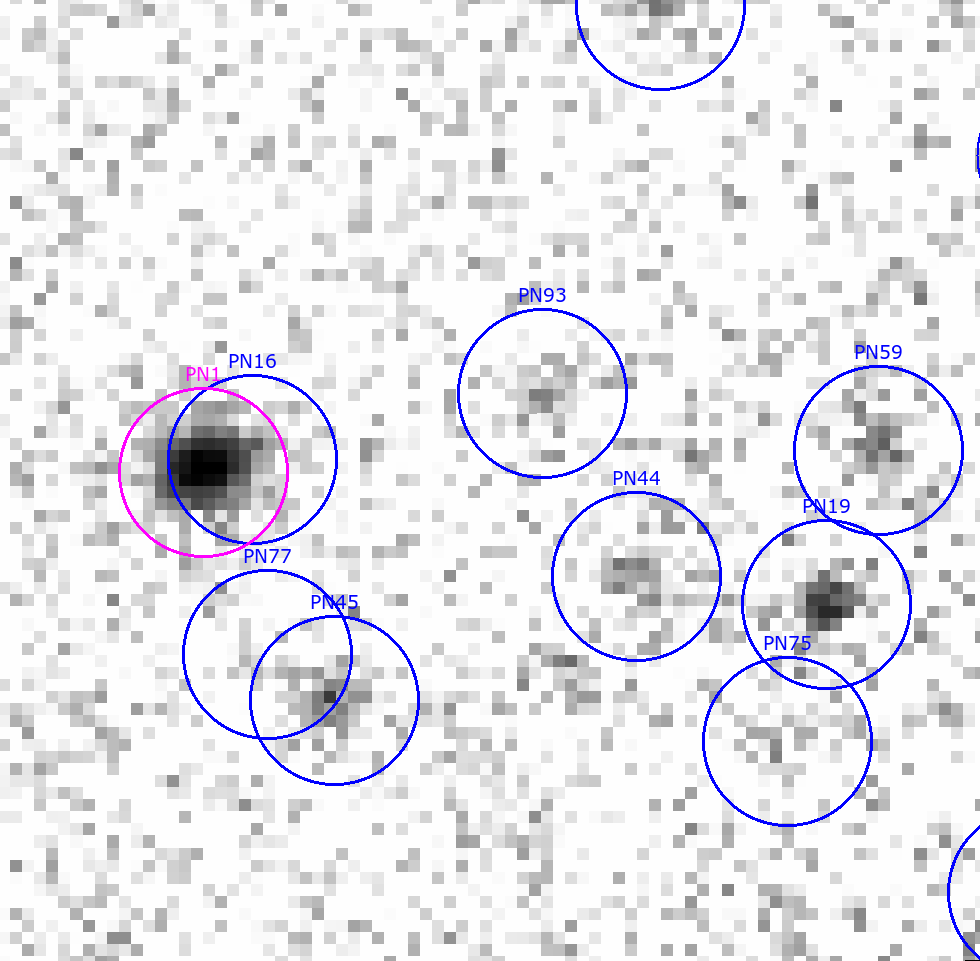}
\caption{A small region of the MUSE data cube for NGC\,1351, showing one wavelength slice centered on \OIII $\lambda 5007$ at the redshift of the galaxy.  Note that PN1 is blended with PN16. The overluminous nature of PN1 may be due to an imperfect deblending of the two sources. 
\label{fig:NGC1351_blend}}
\end{figure}

The observed PNLF is shown in Figure~\ref{fig:NGC1351_PNLF}.  The displayed data points exclude two objects (PN8 and 32) located within $10\arcsec$ of the nucleus, as in this region, the galaxy's bright background severely limits PN detections.  But the remaining objects outline a fairly normal PNLF\null.  In fact, the overall distribution is quite similar to that of NGC\,1326, though with a greater population of PNe in the top $\sim 1$~mag of the luminosity function.  Like NGC\,1326, NGC\,1351's PNLF contains one object whose spectrum is fully consistent with that of normal PNe, but whose luminosity is 0.2~mag brighter than expected from  equation~(\ref{eq:pnlf}).  Again, there are three explanations for the object.  

The first is that the apparent luminosity of PN1 may be due to poor photometry.  PN1's \OIII flux is partially blended with that of PN16, which is projected less than one arcsecond away (see Figure~\ref{fig:NGC1351_blend}).  Since the radial velocities of the two objects are similar, imperfect deblending of their PSFs could result in PN1 appearing overluminous.  The second is that the object is some form of PN (or PN mimic) that does not obey the luminosity function defined by equation~(\ref{eq:pnlf}).  This possibility will be addressed further in \S\ref{subsec:overluminousdiscussion}.  Finally, the object may be an otherwise normal PN that is superposed on another object.  The background upon which PN1 is projected is not particularly bright ($\mu_V \sim 23.4$~mag~arcsec$^{-2}$), but the possibility cannot be entirely discounted.   Thus, we are left with the same ambiguity seen in NGC\,1326, with one object having an outsized influence on the galaxy's derived distance.  The difference between the two galaxies is that the larger number of PNe present in NGC\,1351 makes the anomalous luminosity of PN1 more obvious. Consequently, if one assumes that PN1 is not an unlucky superposition of two sources, a K-S test can exclude its membership in the distribution defined by equation~(\ref{eq:pnlf}) with greater than 93\% confidence.

\begin{figure}[ht]
\includegraphics[width=0.473\textwidth]{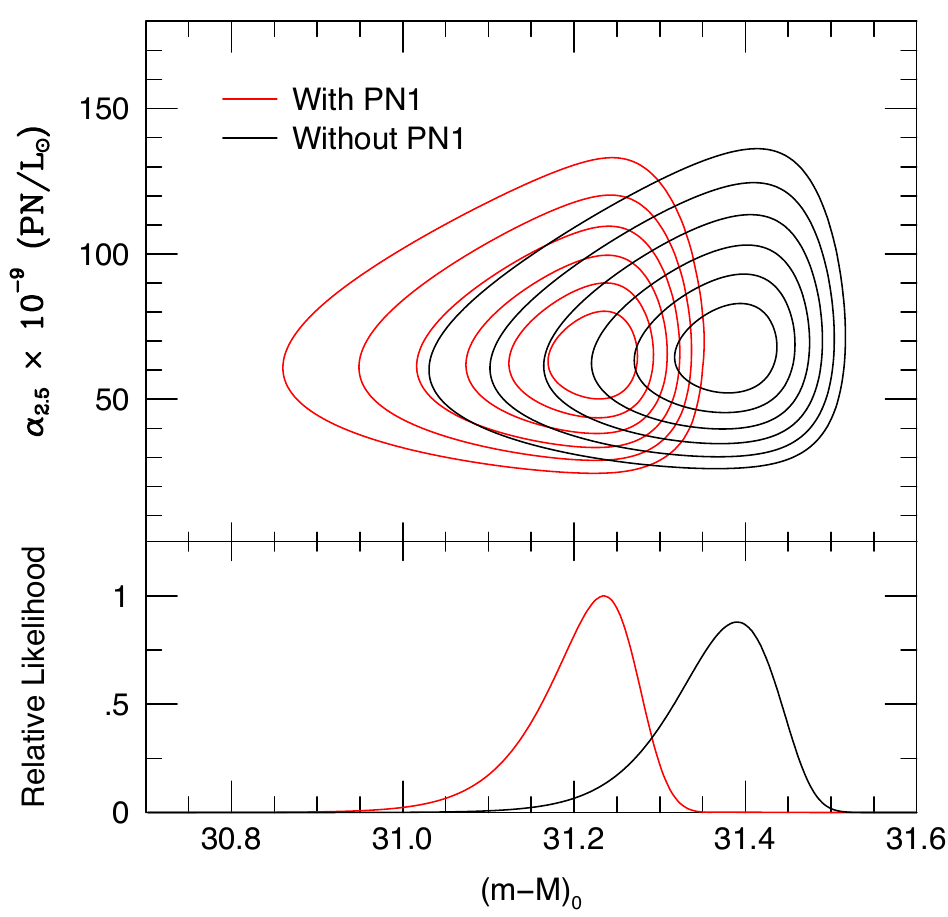}
\caption{
The top panel shows the results of the maximum likelihood analysis for NGC\,1351.  The abscissa is the galaxy's true distance modulus; the ordinate is $\alpha_{2.5}$, the number of PNe within 2.5~mag of $M^*$, normalized to the amount of bolometric light sampled. The red contours, shown at $0.5\sigma$ intervals, illustrate the likelihoods when PN1 is included in the sample; the black contours display the same information when PN1 is excluded. The bottom panel marginalizes these data over $\alpha_{2.5}$.\label{fig:NGC1351_contours}}
\end{figure}

The situation is summarized in Figure~\ref{fig:NGC1351_contours}, which shows the results of fitting equation~(\ref{eq:pnlf}) to the data with and without the brightest PNe.  For the analysis, we calculated the likelihood of unresolved PN superpositions using the $B$-band galaxy surface photometry of \citet{deCarvalho+91}, a mean color of $(B-V) = 0.90$ \citep{Faber+89} and the velocity dispersion data of \citet{DOnofrio+95}.\footnote{The latter required some extrapolation, as the data extend only $\sim 20\arcsec$ from the nucleus and are restricted to the galaxy's major axis. However, in the regions of our survey, the galaxy's stellar motions ($\sigma \lesssim 100$~km~s$^{-1}$) are much less than the $\Delta v = 200$~km~s$^{-1}$ resolvable by MUSE\null.  Thus the exact form of our extrapolation is unimportant.}  From the figure, it is clear that the calculated distance to the galaxy depends on whether PN1's photometry is accurate and whether it is, indeed, a normal PN, rather than some exotic object.  

If PN1 is treated as a normal PN, then the galaxy's inferred distance modulus is $(m-M)_0 = 31.23^{+0.04}_{-0.07}$ ($17.7_{-0.6}^{+0.3}$~Mpc) for $E(B-V) = 0.011$.  This is a formal solution only:  as stated above, a K-S test excludes the null hypothesis that the observed set of PNe is drawn from the distribution defined by equation~(\ref{eq:pnlf}). On the other hand, if the object is excluded from the sample, the galaxy's distance modulus increases to $(m-M)_0 = 31.39^{+0.04}_{-0.08}$ ($19.0_{-0.7}^{+0.4}$~Mpc) and the quality of the fit is much improved.  Consequently, we adopt this larger distance in our analysis.

For NGC\,1351, the aperture correction measurements are quite good, and the total systematic error, due to uncertainties in the MUSE flux calibration, the data cubes' aperture corrections, and foreground extinction, is $\sim 0.06$~mag.  This increases the error bars slightly so that the distance modulus to the galaxy becomes $(m-M)_0 = 31.39^{+0.05}_{-0.10}$, or $19.0_{-0.9}^{+0.7}$~Mpc.  This value is nearly identical to the $z_{850}$-band SBF distance of $31.42\pm0.07$ ($19.2\pm0.6$~Mpc) \citep{Blakeslee+09}. 

\begin{figure*}[t]
\hspace{16mm}
\href{https://cloud.aip.de/index.php/s/6qawyaJ2P6s4Rkk}{\bf \colorbox{yellow}{Off-band}}
\hspace{44mm}\href{https://cloud.aip.de/index.php/s/mznAL9a6XcqkeSq}{\bf \colorbox{yellow}{Diff}} \\
\includegraphics[width=56mm,bb=0 0  800 800,clip]{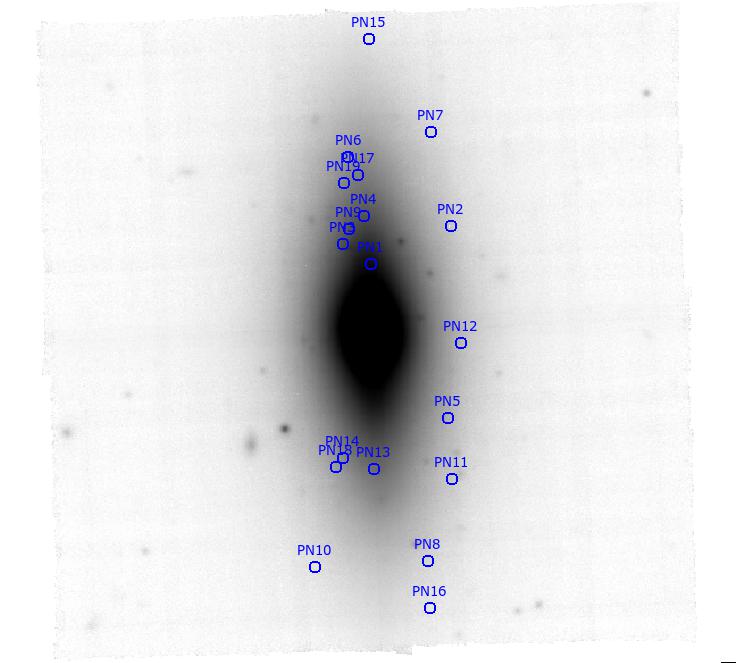}
\includegraphics[width=56mm,bb=0 0  800 800,clip]{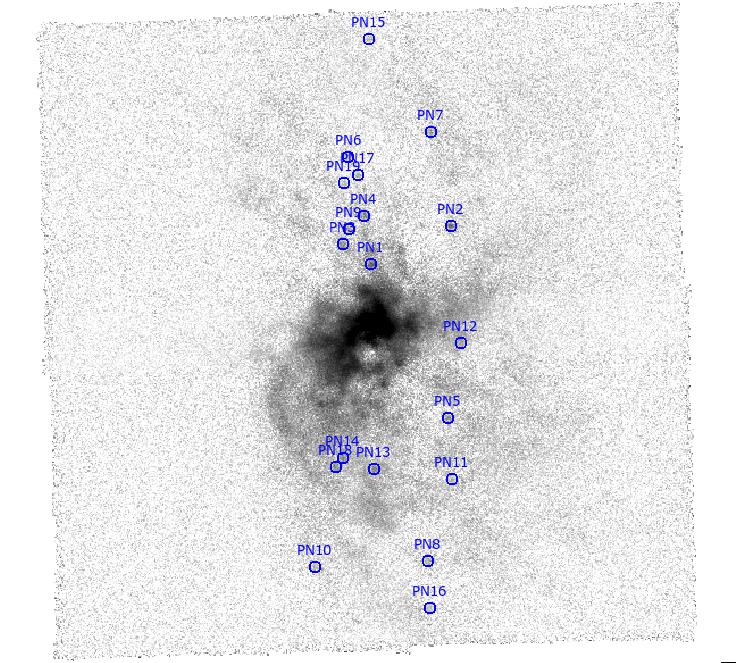}
\includegraphics[width=66mm,bb=0 20  750 550,clip]{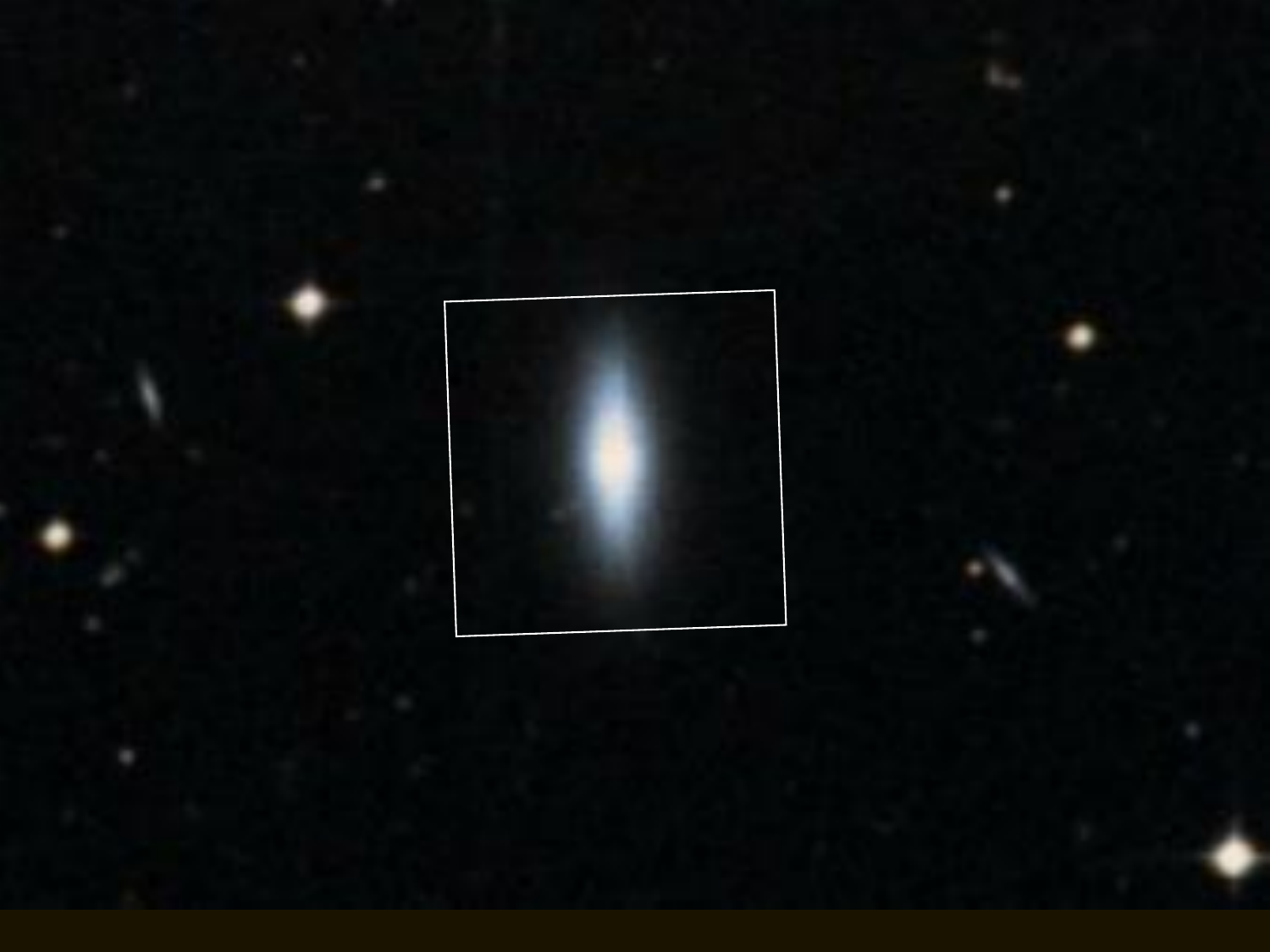}
\caption{NGC\,1366. Left: thumbnail offband and difference images derived from the MUSE data cube. The PN candidates are circled in blue.  High-resolution images can be obtained by clicking on the hyperlink titles. \href{https://cloud.aip.de/index.php/s/6yryXKEZrBKi7AB}{\colorbox{yellow}{VIDEO}}. Right: a broadband image with the MUSE field outlined in white. 
 (Credit: ESO Digital Sky Survey (DSS)).  \label{fig:NGC1366_FChart}}
\end{figure*}

One other feature of Fig.~\ref{fig:NGC1351_PNLF} is worth noting.  The number of PNe observed, normalized to the total amount of galaxy light contained in the MUSE data cubes \citep[found using the surface photometry of ][]{deCarvalho+91}, is higher than that expected for early-type galaxies. This anomaly does not substantially affect the implied distance modulus to the galaxy, since the methodology of \citet{Chase+23} only requires knowledge of the relative amount of light at each position in the MUSE data cube.  But it suggests that there is either a zero-point issue associated with the galactic surface photometry or the stellar population of NGC\,1351 is a bit younger than that of most lenticular systems. 

\subsection{NGC 1366 \label{subsec:NGC1366}}

NGC\,1366 is an edge-on S0 galaxy in the Fornax cluster with an absolute magnitude of $M_I=-19.78$ and evidence for a counter-rotating core \citep{Morelli+17}.  The ESO archive contains a single deep  data cube of the galaxy (ESO Archive ID: ADP.2019-10-10T08:04:58.194, PI: L. Morelli, Program ID: 0103.B-0331) with an effective exposure time of 2446~s and an image quality of $1\farcs 12$ at 5007~\AA. 


Our DELF procedure reveals a large amount of high excitation  (\OIII $\lambda 5007$) line emission in the galaxy's inner $\sim 15\arcsec.$  As can be see in the ``diff'' image in Figure~\ref{fig:NGC1366_FChart}, this light has the appearance of a spiral feature that is oriented \emph{perpendicular} to the plane of the disk. Given the velocity structure of this gas \citep{Morelli+17}, it is hard to imagine that this morphology is the result of an outflow similar to that seen in NGC\,1052 (Section~\ref{subsec:NGC1052}). But the feature does resemble those seen in Illustris TNG100 cosmological simulations of galaxies with counter-rotating cores and infalling gas \citep{Khoperskov+21}. Given that NGC\,1366 is known to have counter-rotation, the models seem to be a plausible explanation for the appearance of the gas. 

\begin{figure}[hb]
\includegraphics[width=0.473\textwidth]{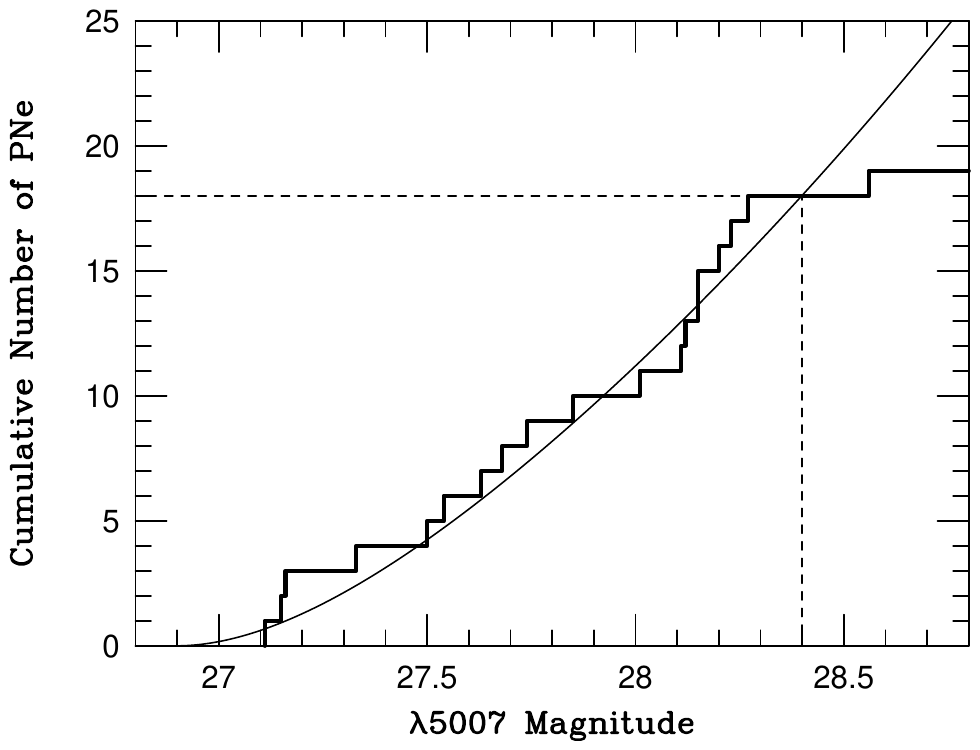}
\caption{The cumulative luminosity function for PNe in NGC\,1366.  The dark line represents the observed data; the curve is equation~(\ref{eq:pnlf}) shifted to the most-likely apparent distance modulus of $(m-M)_0 = 31.39$.  The dashed line shows the location where the incompleteness begins to affect the PN detections.  Data brighter than this are consistent with being drawn from the empirical function.
\label{fig:NGC1366_PNLF}}
\end{figure}

Fig.~\ref{fig:NGC1366_FChart} also identifies a number of point-like \OIII sources superposed on the diffuse emission.  Owing to the modest luminosity of the host galaxy and the data cube's poor image quality, we initially detected a set of only 31 PN candidates.  Upon further inspection of the spectra, this number was pared down by 12, leaving just 19 objects in our PN sample, and only 13 in the brightest $\sim 1$~mag of the luminosity function. 

Figure~\ref{fig:NGC1366_PNLF} shows the cumulative distribution of PN magnitudes in NGC\,1366.  Clearly, the distribution is in good agreement with the empirical law of equation~(\ref{eq:pnlf}).  But it is also clear that with only 18 objects in the statistically complete sample, and just 13 in the critical top magnitude of the luminosity function, the exact location of the PNLF cutoff is poorly defined. 

\begin{figure}
\includegraphics[width=0.473\textwidth]{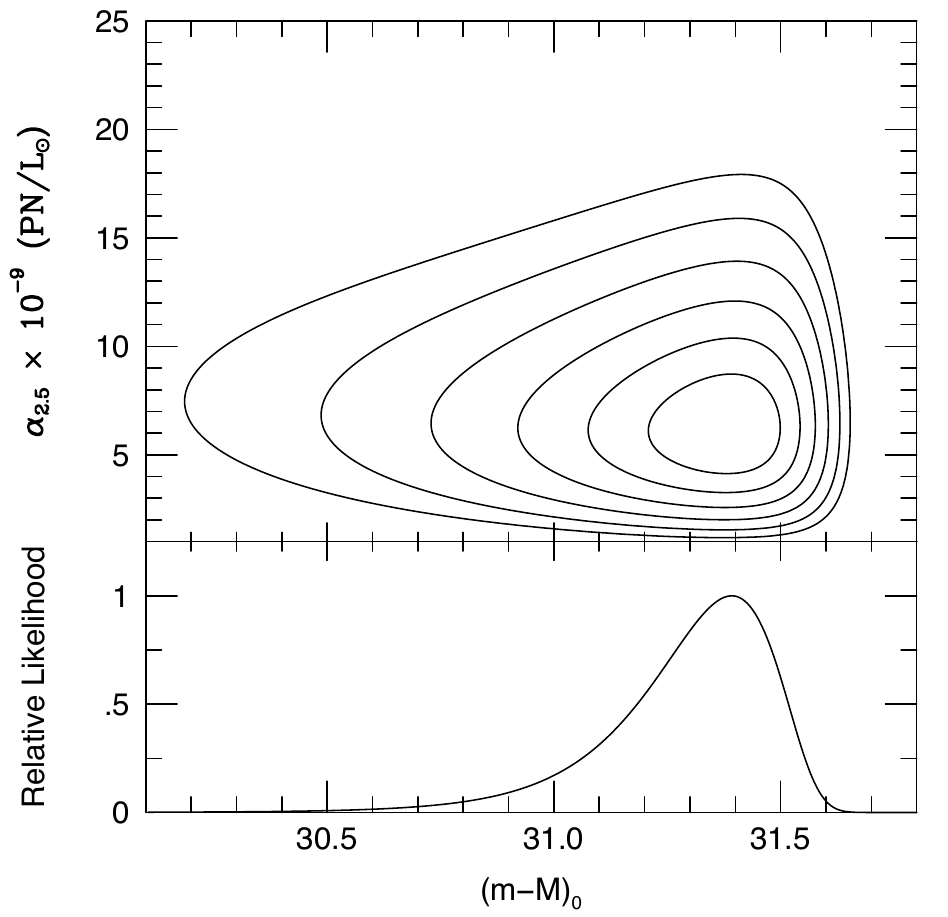}
\caption{The top panel shows the results of the maximum likelihood analysis for NGC\,1366.  The abscissa is the galaxy's true distance modulus, the ordinate is $\alpha_{2.5}$, the number of PNe within 2.5~mag of $M^*$, normalized to the amount of bolometric light sampled.  The contours are drawn at $0.5\sigma$ levels.  The bottom panel marginalizes these data over the PN per luminosity variable.
\label{fig:NGC1366_contours}}
\end{figure}

We analyzed the galaxy's PNLF using the $R$-band surface photometry of \citet{Morelli+08} with an assumed $(V-R)$ color of 0.55 \citep{Prugniel+98}. These data imply that, outside the galaxy's central $10\arcsec$, where no PNe are detected, the MUSE data cube surveys $V \sim 12.5$ of galaxy light.  We also assumed a line-of-sight stellar velocity dispersion of 100~km~s$^{-1}$ throughout the region of PN detections; this is consistent with the spectroscopy of \citet{Morelli+17}.  The results of our analysis are displayed in Figure~\ref{fig:NGC1366_contours}.

With a reddening of $E(B-V) = 0.014$, the formal PNLF fit for NGC\,1366 gives $(m-M)_0 = 31.39^{+0.10}_{-0.22}$; if we assume that flux calibration, aperture correction, and reddening contribute an additional $\sim 0.06$~mag to the error budget, the final distance modulus becomes $(m-M)_0 = 31.39^{+0.11}_{-0.23}$,  or $19.0^{+1.0}_{-1.8}$~Mpc.  This value is consistent with Fornax cluster membership and is in accord with the $I$ band SBF distance modulus of $(m-M)_0 = 31.62\pm0.29$ found by \citet{Tonry+01}.  We note that the value for the PNe per unit luminosity value is slightly low for a galaxy of this type.  However, given the small number of PNe detected and the possibility that we are losing objects due to the bright, high-excitation line-emission in the galaxy's core, the estimate of $\alpha$ is reasonable.


\subsection{NGC 1385 \label{subsec:NGC1385}}

\begin{figure*}[t]
\hspace{14mm}
\href{https://cloud.aip.de/index.php/s/sNyqJ5RrjAGGpjS}{\bf \colorbox{yellow}{Off-band}}
\hspace{43mm}\href{https://cloud.aip.de/index.php/s/Pb7ZDXeYaqrMpeW}{\bf \colorbox{yellow}{Diff}} \\
\includegraphics[width=55mm,bb=0 0  800 1000,clip]{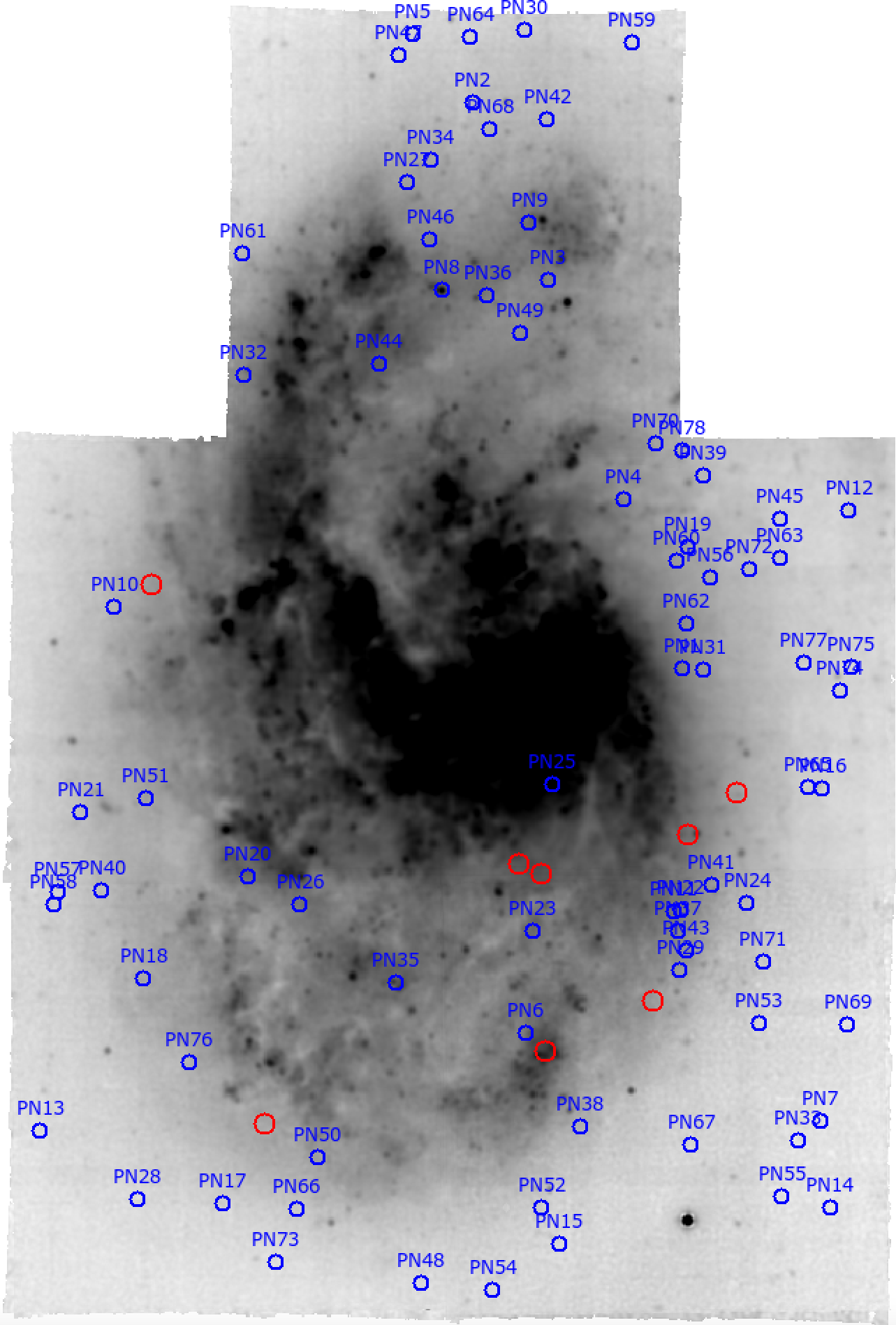}
\includegraphics[width=55mm,bb=0 0  800 1000,clip]{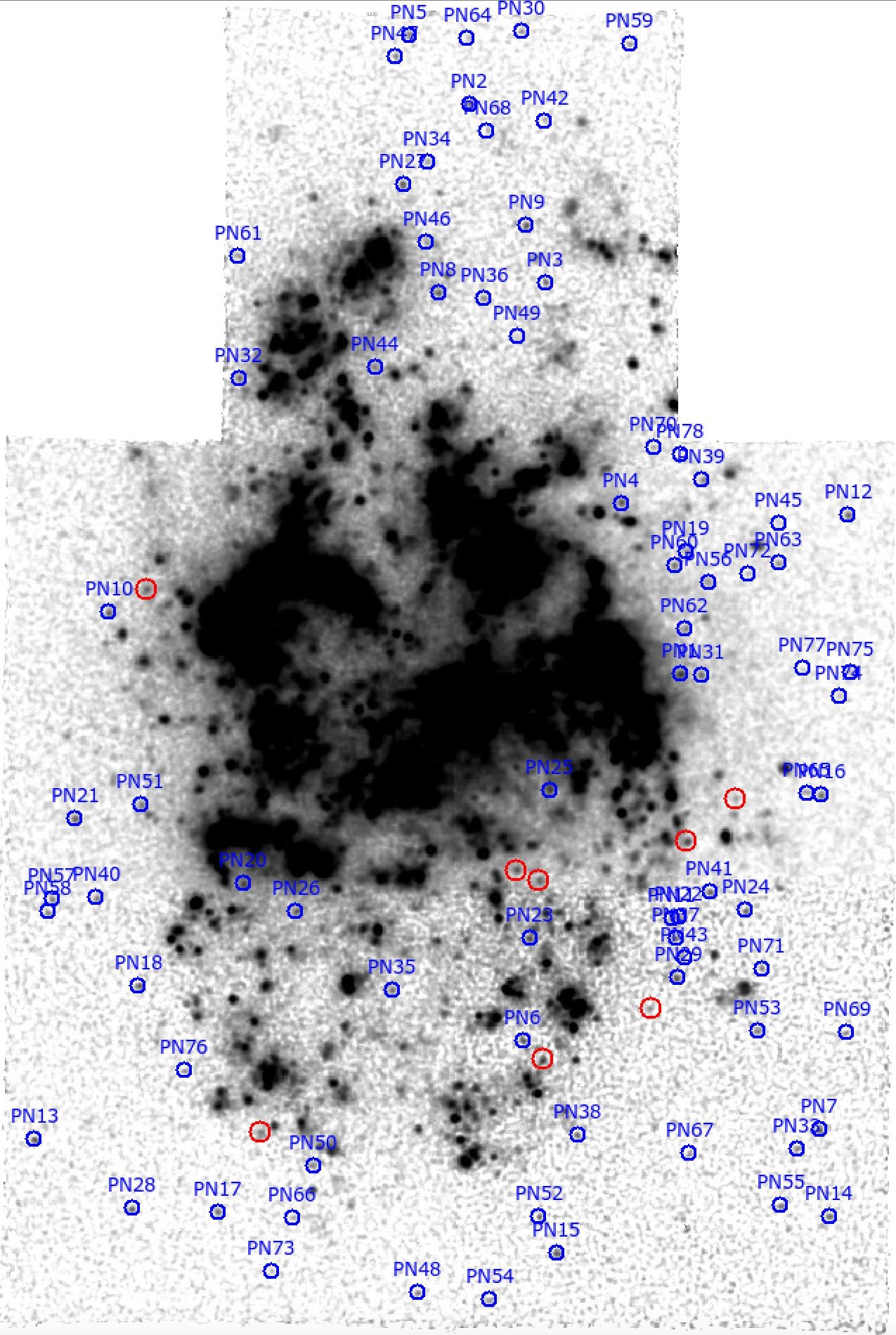}
\hspace{2mm}
\includegraphics[width=65mm,bb=100 0  650 550,clip]{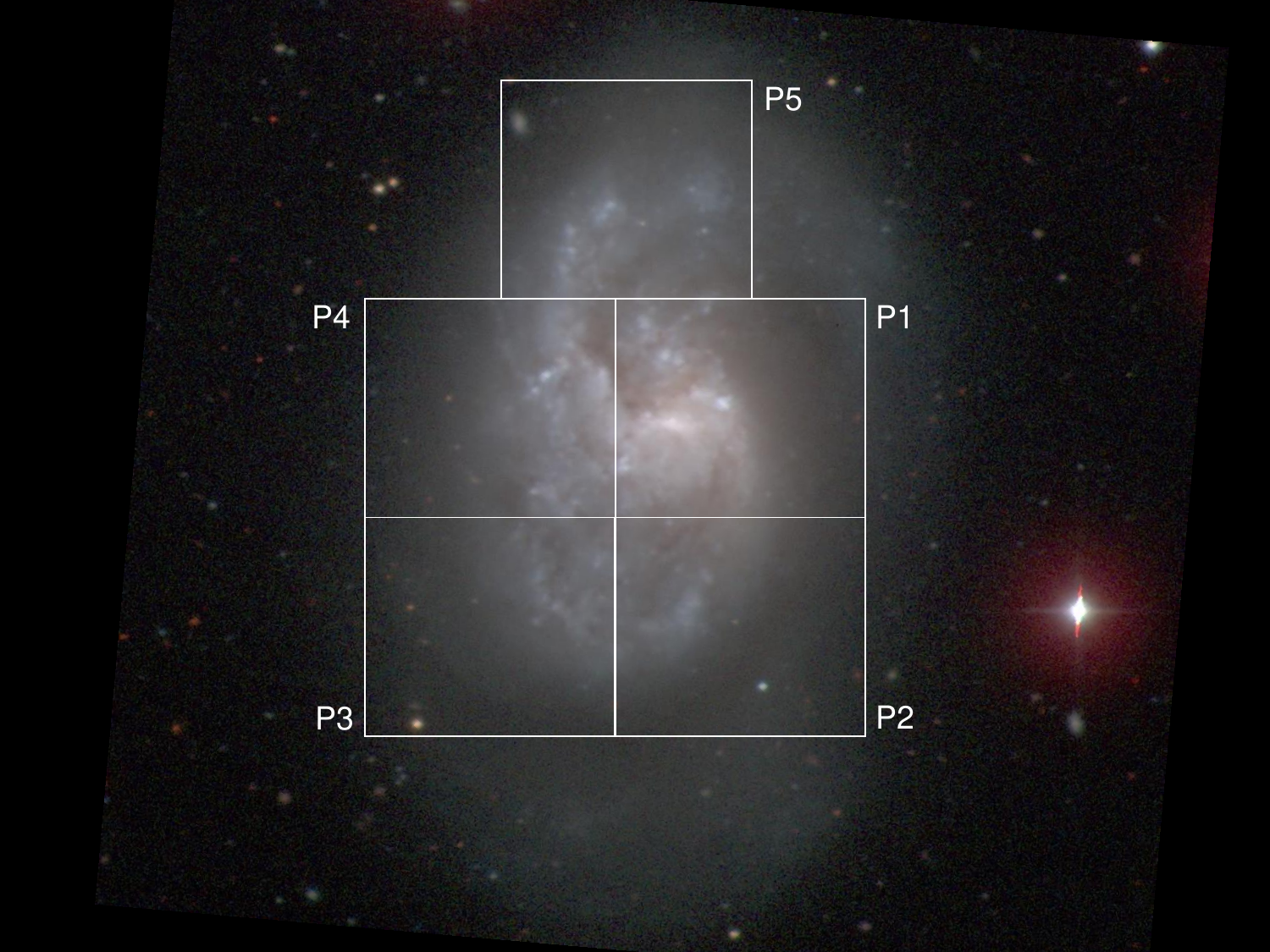}
\caption{NGC\,1385. Left: thumbnail offband and difference images derived from five MUSE data cubes.  Our PNe candidates are highlighted in blue, while emission-line interlopers are shown in red.   High resolution images can be obtained by clicking on the hyperlink titles. \href{https://cloud.aip.de/index.php/s/82eFmJZcH74oBbq}{\colorbox{yellow}{VIDEO}}. Right: a broadband image outlining the MUSE fields. (Credit: CGS). \label{fig:NGC1385_FChart}}
\end{figure*}

NGC\,1385 is an Eridanus cluster 
barred spiral galaxy (Hubble type SB(s)cd) with publicly available data products in the ESO Archive from the PHANGS-MUSE survey \citep[][PI: E. Schinnerer, Program ID: 1100.B-0651]{Emsellem+22}. The galaxy is particularly interesting for a distance scale study since it has no TRGB, SBF, or Cepheid measurements, and its Tully-Fisher distances extend over a very wide range, from $\sim 8$~Mpc \citep[e.g.,][]{Bottinelli+86, Sorce+14} to $\sim 50$~Mpc \citep{Bottinelli+84}.  Figure~\ref{fig:NGC1385_FChart} shows the five pointings in the MUSE archive, which cover most of the galaxy's light.  These pointings, along with their exposure times and seeing at 5007~\AA, are listed in Table~\ref{tab:NGC1385_cubes}.

We note that P1, P4, and P5 were observed twice; however, the combined data cubes are not available in the ESO archive. In addition, the archive also contains two mosaicked data cubes (ESO Archive IDs: ADP.2021-07-16T10:20:56.387, ADP.2021-07-16T10:20:56.381).  The cube that we label as A1 was created by merging all of the individual data sets into one cube, including the fields with two visits.  The second data cube, A2, also combines all the data, but only after convolving each individual cube to a common PSF with a $0\farcs 77$~FWHM\null.  This process is described in \citet{Emsellem+22}. 

To identify the galaxy's PN candidates, we first examined the individual data cubes (P1 -- P5), and, one by one, found and measured 24, 23, 16, 4, and 19 PN candidates, respectively. We then repeated the photometry in mosaic A2, which has greater depth in the regions P1, P4, and P5, due to the fields' second observation. The improved signal-to-noise ratio (S/N) allowed us to exclude 8 of our initial PN candidates as interlopers.  Our final sample of candidates therefore contains 78 objects, with 54 being in the top $\sim 1$~mag of the luminosity function.  

The positions of these PNe are shown in Fig.~\ref{fig:NGC1385_FChart}.  From the figure, it is clear that PN detections within $\sim 30\arcsec$ of the nucleus and along the galaxy's bright northern arm are severely compromised by the galaxy's high surface brightness and bright line emission.  We exclude these regions from our analysis, and thus eliminate one additional object (PN9) from our sample.

\begin{deluxetable}{llcc}
\tablecaption{Data Cubes for NGC 1385
\label{tab:NGC1385_cubes}  }
\tablehead{\colhead{Field} &\colhead{Archive ID} &\colhead{Exp Time}
&\colhead{Seeing} \\ [-0.25cm]
&&\colhead{(sec)} &\colhead{(5007~\AA)}}
\startdata
P1 &ADP.2020-01-11T01:24:00.728 &2420 &$0\farcs 64$ \\
P2 &ADP.2021-01-29T13:19:05.970 &2420 &$0\farcs 77$ \\
P3 &ADP.2020-01-28T16:14:42.084 &2420 &$0\farcs 82$ \\
P4 &ADP.2019-02-13T01:31:18.894 &2420 &$0\farcs 71$ \\
P5 &ADP.2020-01-27T19:05:08.471 &2420 &$0\farcs 64$ \\
\enddata
\end{deluxetable}

\begin{figure}
\includegraphics[width=0.473\textwidth]{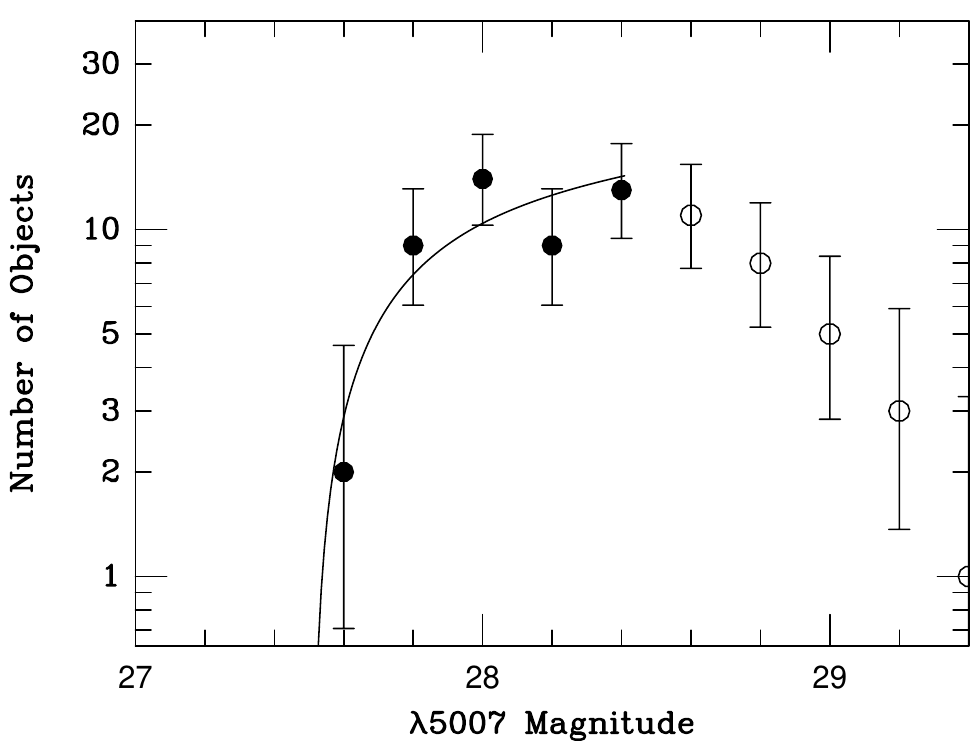}
\caption{The observed PNLF of NGC\,1385 binned into 0.2~mag intervals.  The error bars represent $1\sigma$ confidence intervals \citep[see][]{Gehrels86}; the open circles show data beyond the completeness limit.  The black curve illustrates the most-likely fit to equation~(\ref{eq:pnlf}). 
\label{fig:NGC1385_PNLF}}
\end{figure}

Figure~\ref{fig:NGC1385_PNLF} shows the luminosity function of the remaining PN candidates.  Despite all the active star formation and dust present in the galaxy, the bright-end of NGC\,1385's PNLF is extremely well fit by the empirical law.  This is in agreement with expectations:  because the scale height of PNe should be larger than that of the galactic extinction, the bright-end of the PNLF in large, moderately face-on galaxies should be dominated by objects on the near side of the galaxy, above the dust layer \citep{Feldmeier+97, Rekola+05}.  There is also no evidence for any ``overluminous'' PNe in the system.

Since there are no surface photometry measurements over the face of NGC\,1385, we estimated the amount of light underlying each PN and in the entire MUSE data cube using continuum measurements made directly from the MUSE spectra.  Also, since NGC\,1385 is a disk galaxy seen relatively face-on (inclination angle of $\sim 44^\circ$), the line-of-sight velocity dispersion of its stars should be much less than the galaxy's $\sim 136$~km~s$^{-1}$ rotation speed \citep{Lang+20}.  Thus, we can expect superposed PNe to be unresolved, and  adopt $\sim 50$~km~s$^{-1}$ as a typical value for the system's line-of-sight velocity dispersion.

\begin{figure}[ht]
\includegraphics[width=0.473\textwidth]{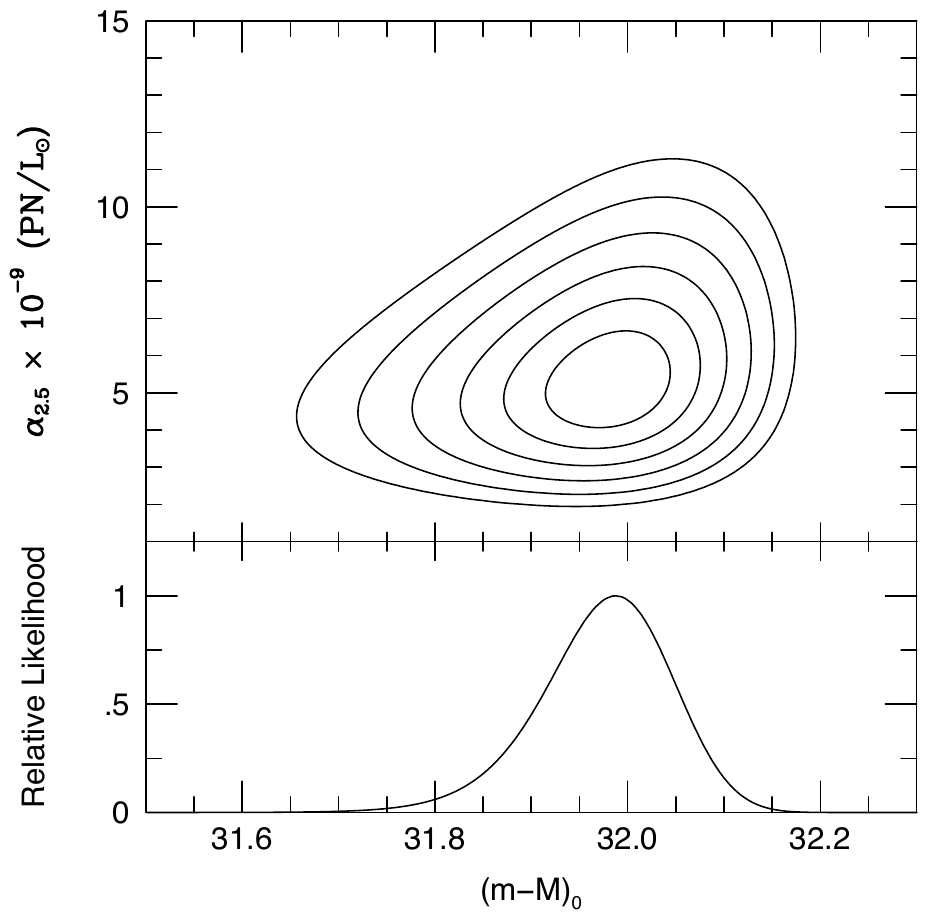}
\caption{The top panel shows the results of the maximum likelihood analysis for NGC\,1385. The contours are drawn at $0.5\sigma$ intervals.  The abscissa is the galaxy's true distance modulus; the ordinate is the number of PNe within 2.5~mag of $M^*$, normalized to the amount of bolometric light sampled.   The bottom panel marginalizes over the latter variable.
\label{fig:NGC1385_contours}}
\end{figure}

Figure~\ref{fig:NGC1385_contours} shows the result of fitting the observed PNLF to equation~(\ref{eq:pnlf}).  For a Milky Way foreground extinction of $E(B-V) = 0.017$ \citep{Schlafly+11}, the fitted distance to NGC\,1385 is $(m-M)_0 = 31.99^{+0.06}_{-0.08}$.  

Unfortunately, obtaining the true error bars for the NGC\,1385's distance is difficult.  Pointings P2 and P5 have reasonably bright point sources in the field, and estimates for their aperture corrections are consistent to within 0.03 and 0.04~mag, respectively.  However, in fields P1, P2, and P5, the PSF star photometry displayed large uncertainties with formal errors on the aperture corrections as large as 0.25~mag.  It is highly unlikely that the true errors are as large as this, since the effect would blur out the PNLF cutoff and create a luminosity function that is a poor match for equation~(\ref{eq:pnlf}) (see \S\ref{subsec:NGC1433}).  Conservatively, we adopt $\sim 0.1$ as the total systematic error associated with our distance measurement.  NGC\,1385's distance is then  $(m-M)_0 = 31.99^{+0.11}_{-0.12}$, or $25.0_{-1.5}^{+1.4}$~Mpc.  This places the galaxy well beyond the core of the Fornax cluster, and in the mid-range of the Tully-Fisher estimates.



We note that NGC\,1385 was included in the recent PNLF study by \citet{Scheuermann+22} that is based on the same dataset that we have used. Their analysis resulted in a significantly shorter distance, with $(m-M)_0 = 29.96^{+0.14}_{-0.32}$, or only $9.8^{+0.6}_{-1.5}$~Mpc. Interestingly, the authors caution that their derived distance modulus was subject to a large uncertainty and should be considered an \textit{upper} limit to the galaxy's true distance.  Specifically, they reported only 11 PNe brighter than their completeness limit of $m_{\rm lim} = 28$, with the most luminous PN being brighter than $m_{5007} = 26$. This is not supported by our measurements, as our brightest PN has $m_{5007} = 27.55$. Also the number of PNe contributing to our fit is almost 5 times larger than theirs. 

Since the agreement of our PNLF distances for other galaxies in the \citet{Scheuermann+22} sample is quite good, we have looked into the possible causes of the discrepancy. A detailed account of the comparison is given in Appendix~\ref{sec:notesN1385}\null. In summary, we note the following:

1) Our sample of 77 \OIII sources with spectra consistent with that of a PN extends $\sim 0.4$ mag deeper than the \citet{Scheuermann+22} dataset. This is readily understood from the superior capabilities of the DELF extraction, which typically yields a factor of $\sim 2$ improvement in S/N over other techniques (see Paper I).  From their list of 11 candidates, only two are confirmed by our analysis.

2) The majority of the \citet{Scheuermann+22} PN candidates are located within giant \ion{H}{2} region complexes.  This is problematic due to issues associated with background subtraction.  With the exception of the two PNe mentioned above, we classify all these objects as \ion{H}{2} regions or SNRs.  This explains why two-thirds of their objects have measured magnitudes brighter than our PNLF cutoff.  Due to the difficulty of distinguishing \ion{H}{2} regions from planetary nebulae, our survey largely avoided those regions of the galaxy with the highest star formation rates.

3) The short distance of 9.8~Mpc is not compatible with the SBF distances of other Fornax/Eridanus group galaxies (see Appendix~\ref{sec:notesN1385}, Table~\ref{tab:Eridanusdistances}). In fact, the system's radial velocity and our PNLF distance place NGC\,1385 near the center of the Eridanus cloud \citep{Willmer+89}.

\begin{figure*}[t]
\hspace{18mm}
\href{https://cloud.aip.de/index.php/s/5DidbsPJBiCwjRa}{\bf \colorbox{yellow}{Off-band}}
\hspace{45mm}\href{https://cloud.aip.de/index.php/s/FSEkXQ8a82kzrxs}{\bf \colorbox{yellow}{Diff}} \\
\includegraphics[width=56mm,bb=0 0  1000 1100,clip]{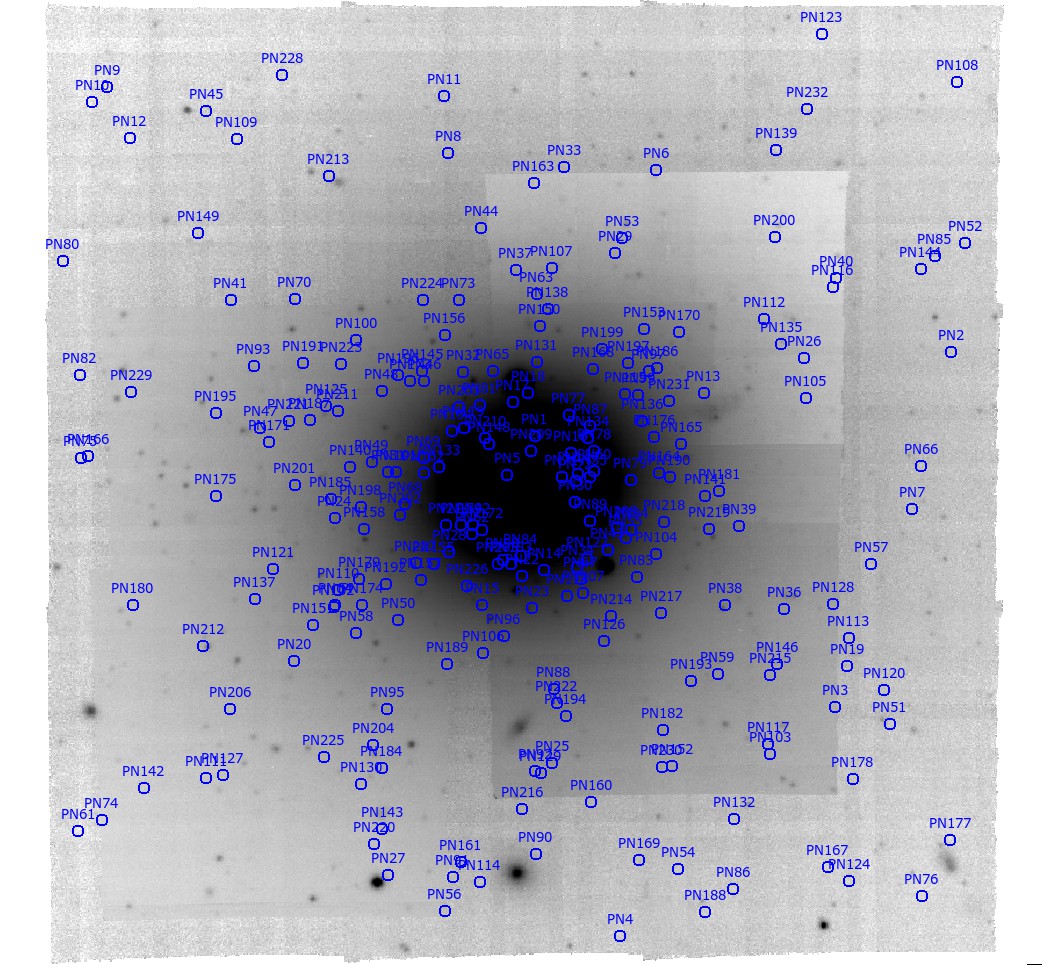}
\includegraphics[width=56mm,bb=0 0  1000 1100,clip]{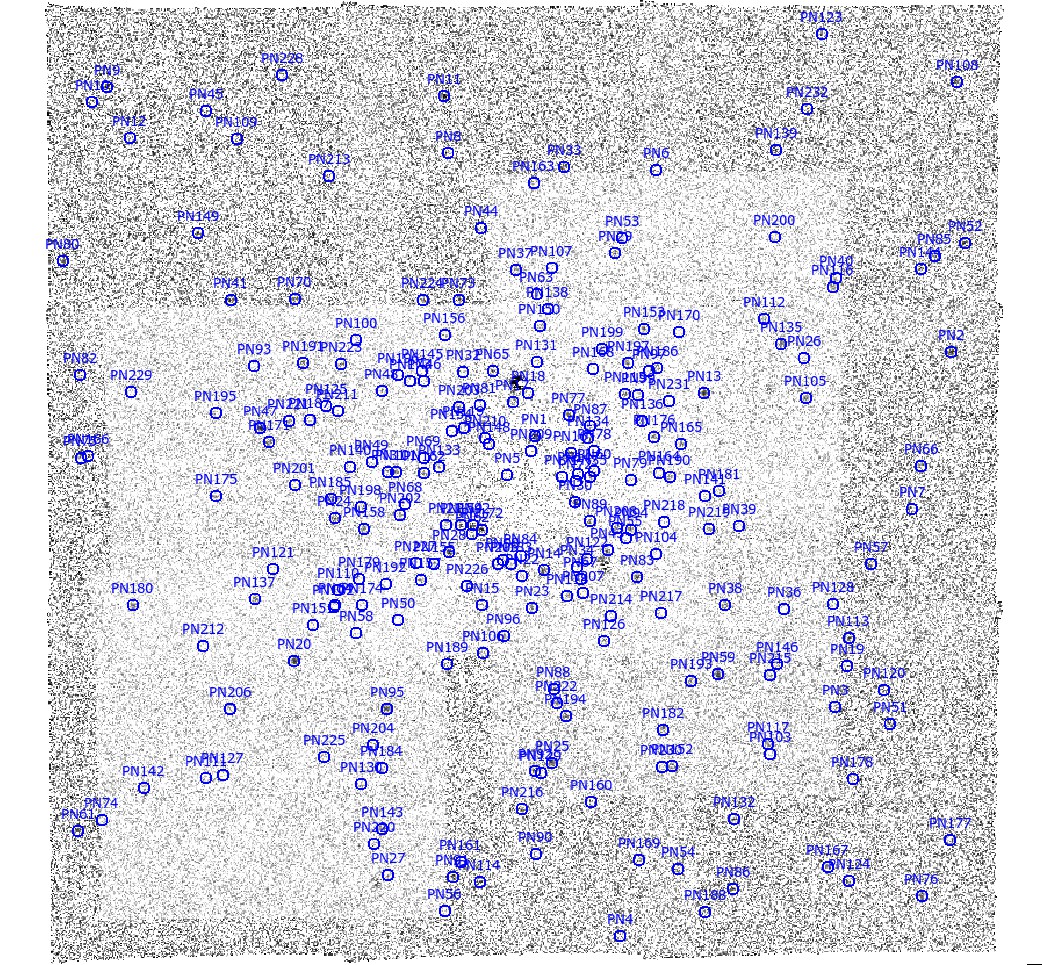}
\includegraphics[width=60mm,bb=100 0  700 600,clip]{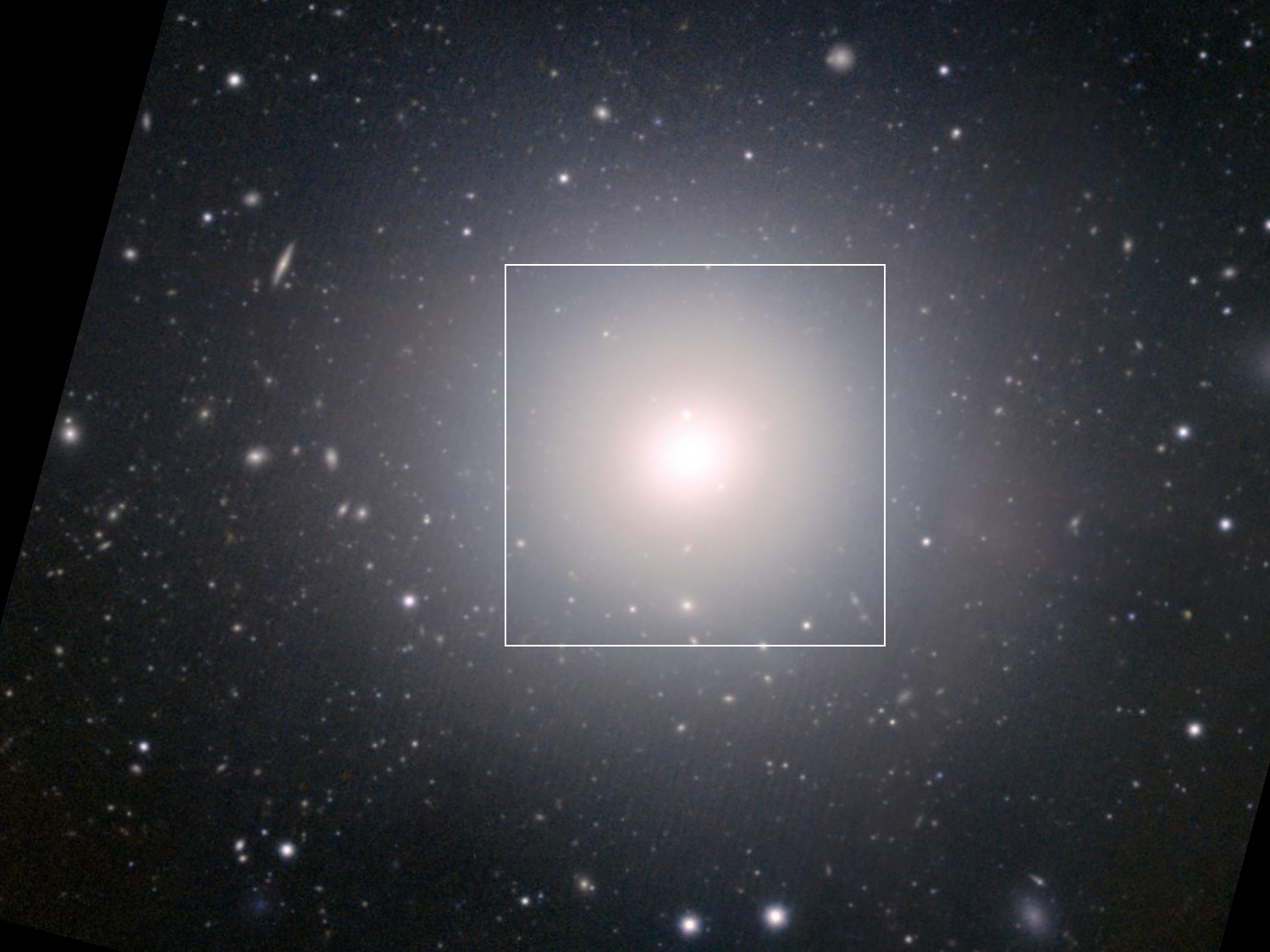}
\caption{NGC\,1399. Left: thumbnail difference and off-band images derived from the MUSE data cube. Our PN candidates are circled in blue.  High resolution images can be obtained by clicking on the hyperlink titles. \href{https://cloud.aip.de/index.php/s/2RrWr9Z2Dc4oX5p}{\colorbox{yellow}{VIDEO}}. Right: a broadband image with the region of the data cube shown in white. (Credit: ESO). \label{fig:NGC1399_FChart}}
\end{figure*}

4) The brightest object in the \citet{Scheuermann+22} sample is 1.83 mag more luminous than the brightest PN in our sample. Following the analysis of \citet{Soemitro+23}, who examined the PNLF for NGC\,300, another disk galaxy with active star formation, if the galaxy is at the distance of Eridanus, then their PN1 would need to be excited by a central star with a luminosity of at least $\log L/L_{\odot} = 4.48$.  This is much brighter than the post-AGB evolutionary track of any PN central star \citep{Gesicki+18}.  

\subsection{NGC 1399 (FCC\,213) \label{subsec:NGC1399}}

NGC\,1399, the central cD/giant elliptical galaxy of the Fornax cluster, has been surveyed for planetary nebulae as far back as the early 1990s \citep{McMillan+93, Arnaboldi+94}.  Two different MUSE programs have observed the galaxy (ESO Archive ID: ADP.2017-03-27T13:16:27.827, PIs: J. Walcher and S. Zieleniewski, Program IDs: 094.B-0903 and 094.B-0298) and these data have been combined into one large ($3\farcm 92$~FoV) data cube with a quoted effective integration time of 954~s and an image quality of $0\farcs 81$.  In fact, according to the ESO archive provenance chain, this is a full-sized mosaic with 295~s exposure times everywhere, plus five 900s exposures on the central regions that provide field overlap.  This co-addition was not seamless: as the processed ``off'' and ``diff'' images of Figure~\ref{fig:NGC1399_FChart} show the data suffer from imperfect flatfield corrections and different levels of noise. Nevertheless, our DELF technique allowed us to identify 232~PN candidates over the body of the galaxy.

Surface photometry of NGC\,1399 exists from a number of studies \citep[e.g.,][]{Franx+89, Caon+94, Iodice+16}, and we used these data to determine the amount of $V$-band galactic light at every position in the MUSE data cube.  For estimating the line-of-sight velocity dispersion at the position of each PN, we used the spectroscopy of \citet{Saglia+00}, who obtained major and minor axis kinematic data over all but the outermost regions of the MUSE survey.

A comparison of the distribution of MUSE PNe and that of the galaxy's light reveals that we are likely missing PNe within $15\arcsec$ of the galaxy's nucleus.  This is not unexpected as the exposure times for this galaxy are relatively short.  Consequently, the region's steep surface brightness gradient causes the limiting magnitude for PN detections to change rapidly with position.  While we could perform artificial star experiments to model and incorporate the effect into our analysis, it is simpler to just exclude the entire region from our study.  This reduces the PNe sample by less than 10\% and still leaves over 100 PNe, covering $V \sim 10.3$~mag of galaxy light, in the brightest $\sim 1$~mag of the luminosity function.   

\begin{figure}[ht]
\includegraphics[width=0.473\textwidth]{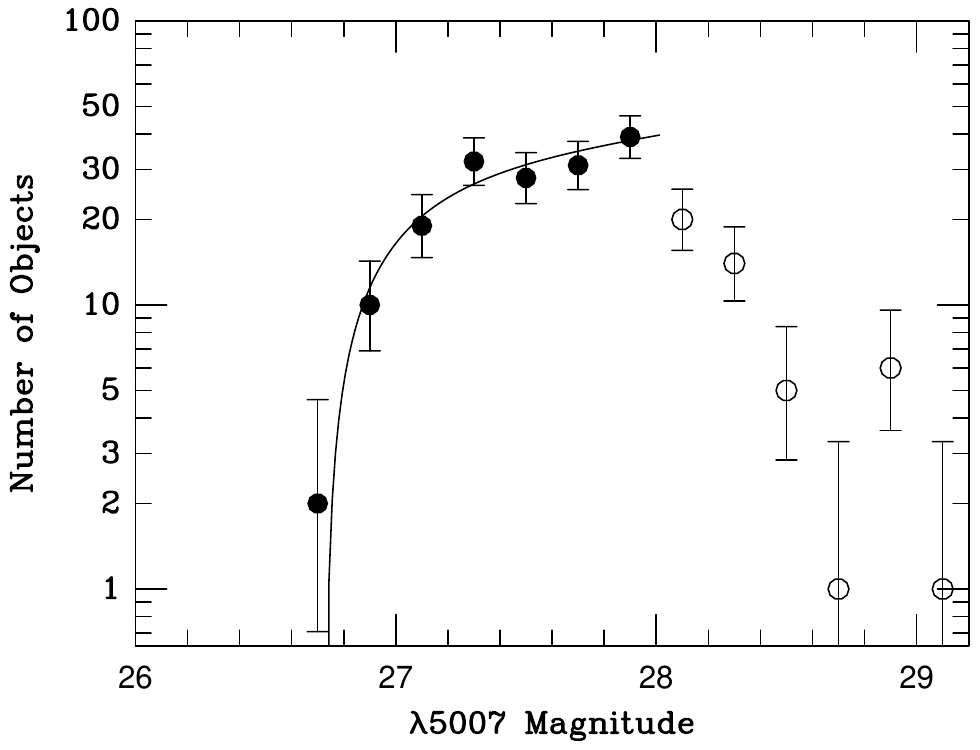}
\caption{The observed PNLF of NGC\,1399 binned into 0.2~mag intervals.  The open circles show data beyond the completeness limit; the error bars represent $1\sigma$ confidence intervals \citep[see][]{Gehrels86}.  The black curve illustrates the most-likely fit to equation~(\ref{eq:pnlf}). \label{fig:NGC1399_PNLF}}
\end{figure}

Figure~\ref{fig:NGC1399_PNLF} shows the observed PNLF of the galaxy.  Since the bright end of the function is so well-populated, the shape of the PNLF is exquisitely defined.  Moreover, despite the large number of PNe, there is no evidence for any ``overluminous'' objects.  The lack of such objects can be partially explained by the galaxy's mass.  NGC\,1399 is a central cD galaxy whose line-of-sight velocity dispersion ranges from $\sim 370$~km~s$^{-1}$ near the nucleus to $\sim 250$~km~s$^{-1}$ in the outer regions surveyed by MUSE\null.  Consequently, even when two PNe fall onto the same spatial element, there is a good chance that their fluxes can be disentangled via the objects' discordant radial velocities.  Overluminous objects caused by PN superpositions should therefore be less common in such an environment, and agreement between the observed PN distribution and equation~(\ref{eq:pnlf}) certainly supports this.

\begin{figure}[ht]
\includegraphics[width=0.473\textwidth]{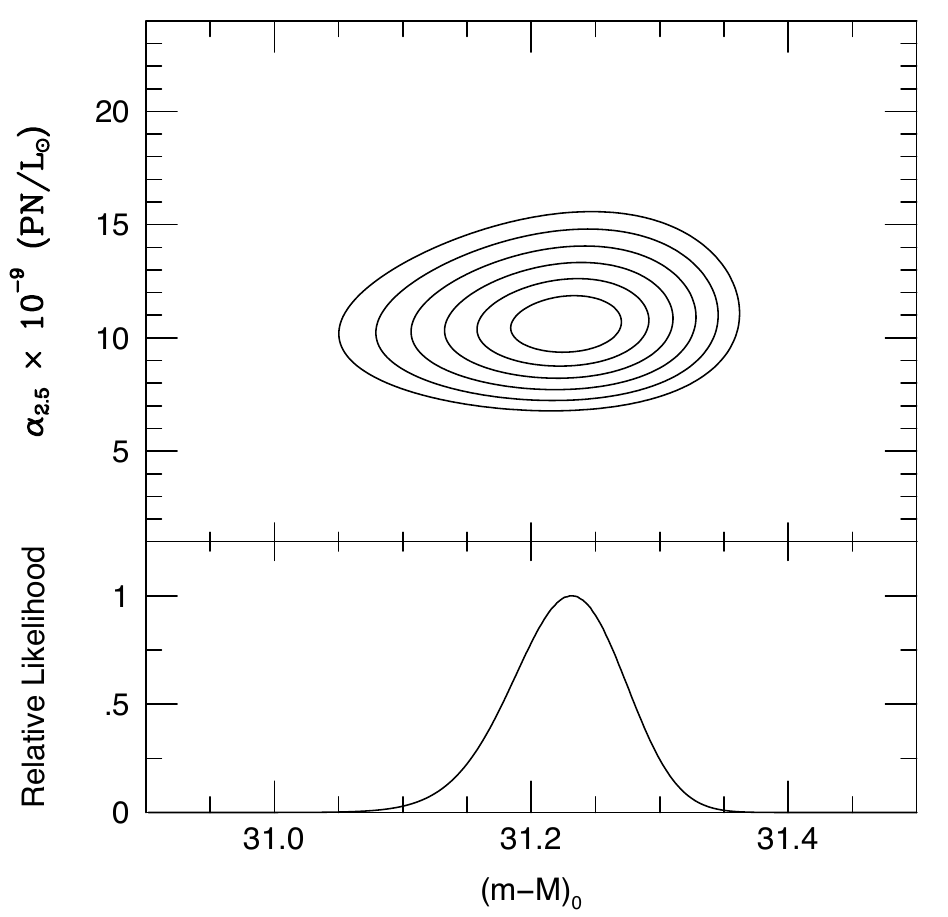}
\caption{The top panel shows the results of the maximum likelihood solution for NGC\,1399's distance modulus. The abscissa is the galaxy's true distance modulus, the ordinate is $\alpha_{2.5}$, the number of PNe within 2.5~mag of $M^*$, normalized to the amount of bolometric light sampled.  The contours are drawn at $0.5\sigma$ intervals.  The bottom panel marginalizes over $\alpha_{2.5}$.
 \label{fig:NGC1399_contours}}
\end{figure}

Figure~\ref{fig:NGC1399_contours} shows the results of our maximum likelihood analysis.  If we assume a foreground reddening of $E(B-V) = 0.011$ \citep{Schlafly+11}, then a fit of equation~(\ref{eq:pnlf}) to NGC\,1399's observed PNLF results in a distance modulus of $(m-M)_0 =  31.23_{-0.05}^{+0.04}$.  When folded in with the $\sim 0.05$~mag zero-point uncertainty from the MUSE flux calibration, our determination of the data cube's aperture correction, and the uncertainty in the reddening, the inferred distance becomes $(m-M)_0 =  31.23_{-0.07}^{+0.06}$ or  $17.6_{-0.6}^{+0.5}$~Mpc.

\begin{figure*}[th!]
\hspace{15mm}
\href{https://cloud.aip.de/index.php/s/NJNQwwaK6Hx2eXD}{\bf \colorbox{yellow}{Off-band}}
\hspace{28mm}\href{https://cloud.aip.de/index.php/s/kc44XmNkTzD3dXF}{\bf \colorbox{yellow}{Diff}} \\
\includegraphics[width=41mm,bb=0 0  1362 2210,clip]{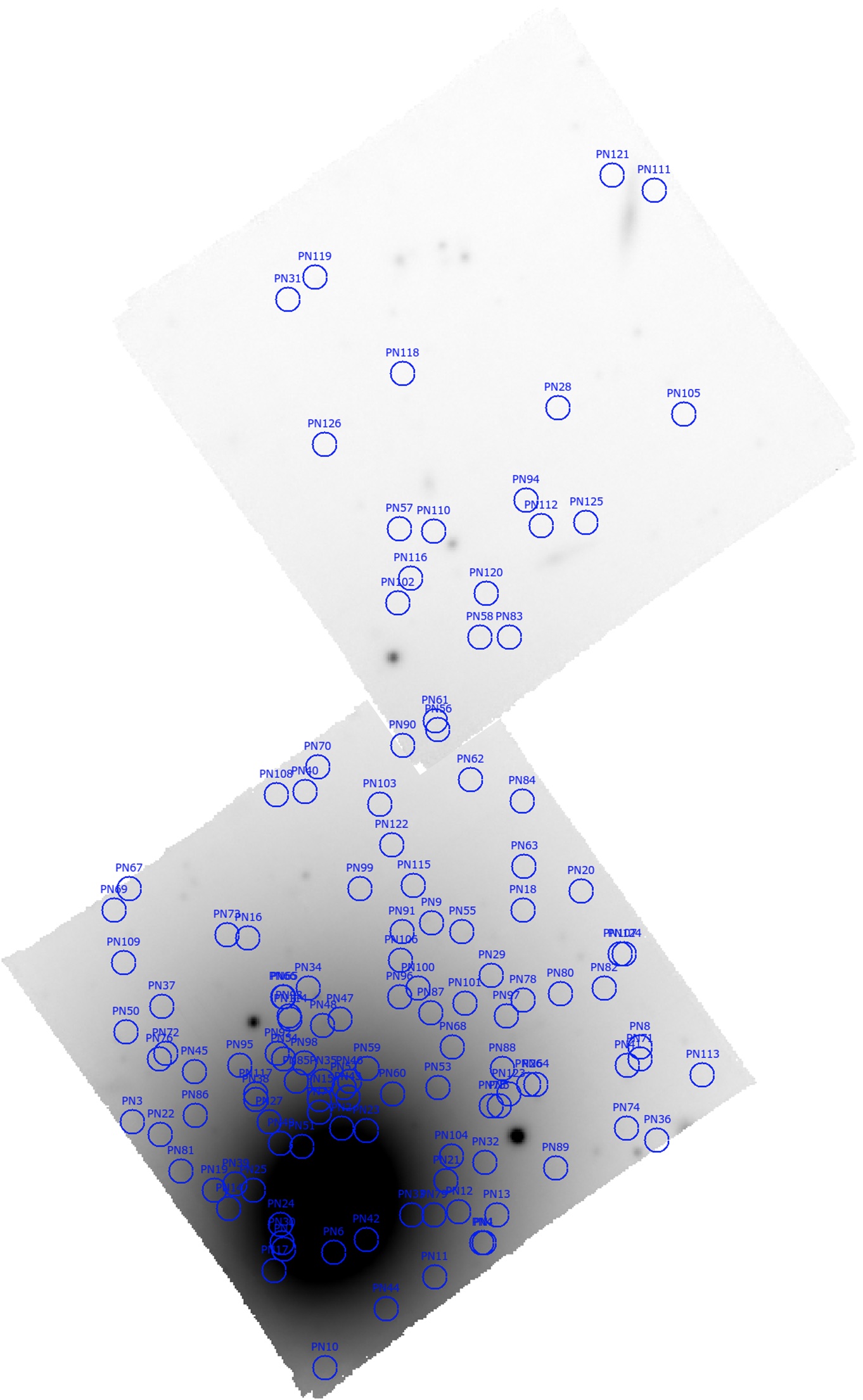}
\includegraphics[width=41mm,bb=0 0  1362 2210,clip]{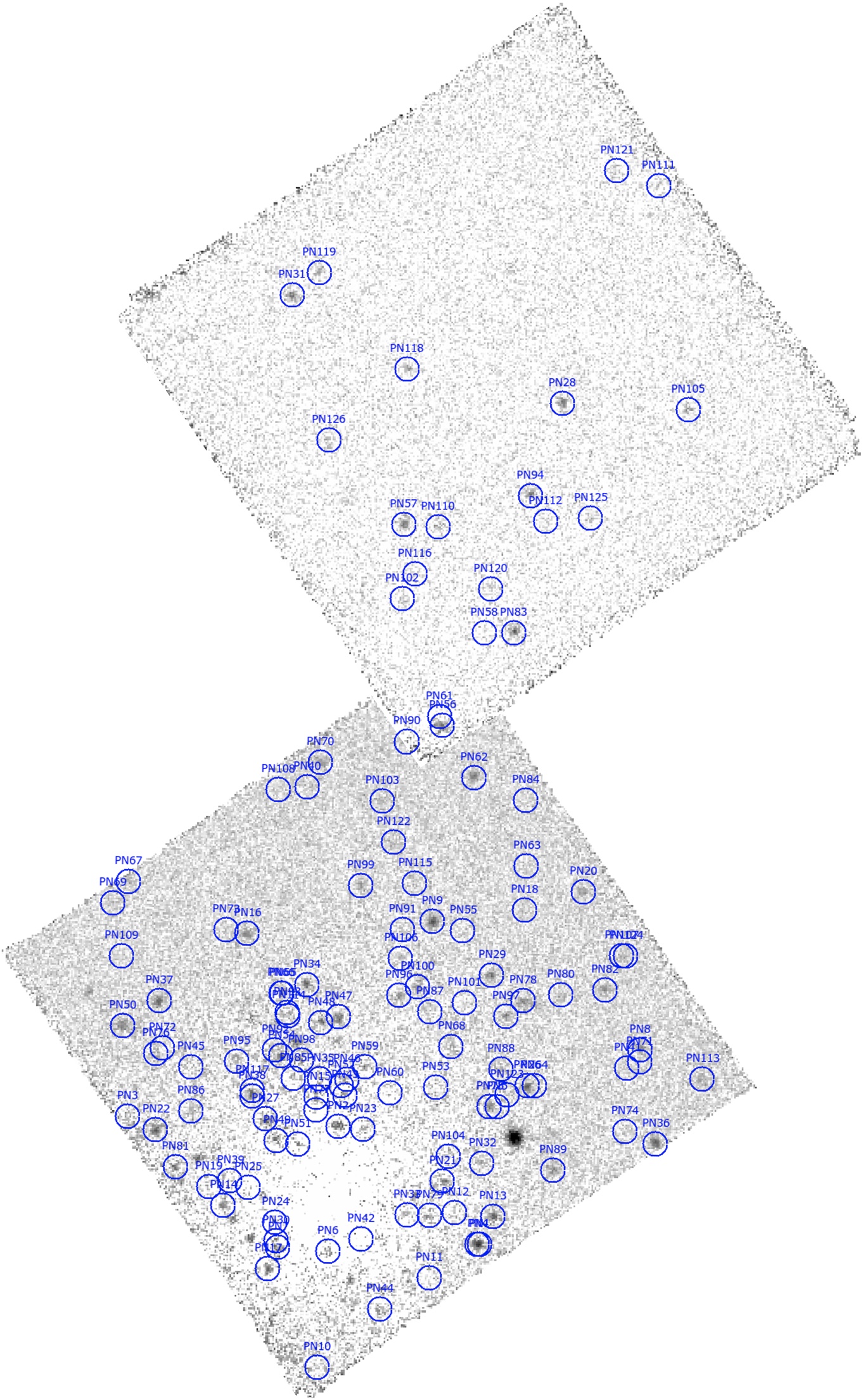}
\hspace{25pt}
\includegraphics[width=85mm,bb=150 150  650 600,clip]{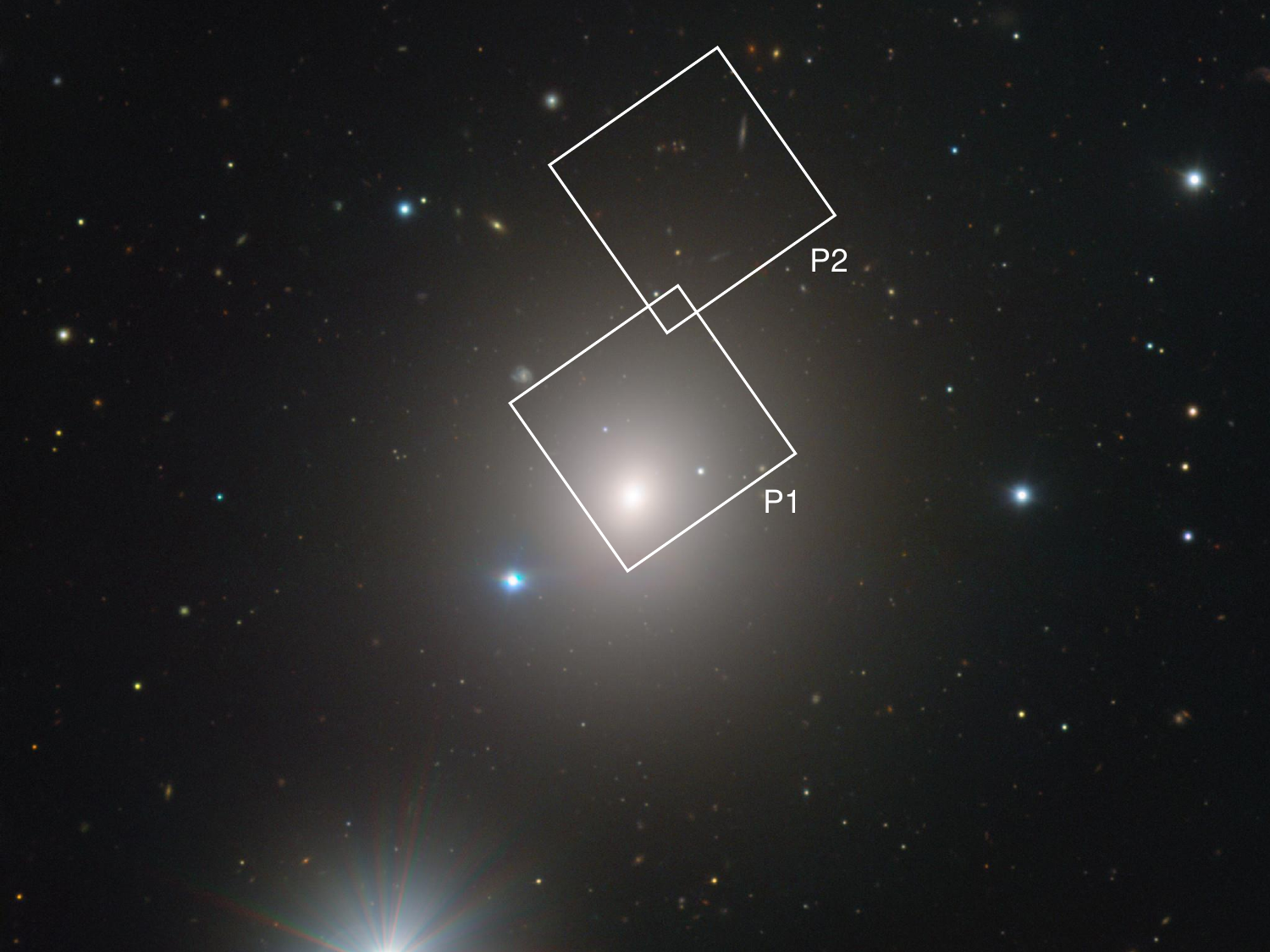}
\caption{NGC\,1404. Left: thumbnail offband and difference images derived from the MUSE data cubes. The PN candidates are circled in blue.  High-resolution images can be obtained by clicking on the hyperlink titles. \href{https://cloud.aip.de/index.php/s/wQ4CCLwHswmp3sP}{\colorbox{yellow}{VIDEO}}. Right: a broadband image with the MUSE survey regions outlined in white. (Credit: ESO).\label{fig:NGC1404_FChart}}
\end{figure*}

Both the distance and PN/luminosity values are almost identical to those measured by \citet{McMillan+93} using interference filter observations with the Blanco 4-m telescope through significantly worse ($1\farcs 3$) seeing.  However, our value is almost 20\% shorter than the latest \textit{HST}-based SBF distance of $21.1\pm 0.7$~Mpc from \citet{Blakeslee+10}.  It is also slightly closer than recent Cepheid and TRGB distances to the Fornax cluster, though not to NGC\,1399 itself \citep[e.g.,][]{Jang+18, Riess+16, Hoyt+21}.  However, the uncertainties in these measurements do largely overlap.

\subsection{NGC 1404 (FCC\,219) \label{subsec:NGC1404}}

NGC\,1404 is another Fornax galaxy, located near the very center of the cluster. For the long-term objective of our study, this E1 galaxy is particularly interesting, as it has hosted two Type Ia supernovae, SN\,2007on and SN\,2011iv, whose distance estimates differ by as much as 14\% \citep{Gall+18}.  The galaxy is also compelling for technical reasons since a bright foreground star is projected onto the body of the galaxy.  This star provides an excellent measure of the field's PSF\null.  

There are four useful data cubes of NGC\,1404 in the MUSE archive. Field P1, which includes the galaxy's nucleus, is a 3287~s exposure with $0\farcs 88$ seeing at 5007~\AA\ (ESO Archive ID: ADP.2017-12-13T01:47:07.213, PI: M. Sarzi, Program ID: 296.B-5054).  The three other exposures are of a halo field (P2), which is located north-northwest of the nucleus along the system's major axis.  The exposure times for these three cubes are 1680~s, and their net seeing is $1\farcs 01$ at 5007~\AA\ (IDs: ADP.2017-12-05T15:14:58.786, ADP.2017-12-01T13:57:32.491, ADP.2017-12-01T13:57:32.480; same PI and ID as above). As the archive did not provide a combined data cube for P2, we chose to reduce the entire dataset anew.  The locations of both fields are shown in Figure~\ref{fig:NGC1404_FChart}.

Our DELF analysis of NGC\,1404's data cubes initially found 179 PN candidates, with the vast majority located in the central field.  Further inspection of the candidates' spectra eliminated 53 objects from consideration, either on the basis of classification as interlopers, or as spurious detections that were too faint to meet the detection criteria.  This left us with a sample of 126 PNe candidates: 107 in P1, 18 in P2, and one object common to both data cubes.  (The two measurements of this object,   $27.92\pm0.05$ on P1 and $27.88\pm0.07$ P2 are in excellent agreement.)  Most importantly, $\sim 64$ of the PNe populate the top $\sim 1$~mag of the luminosity function. 

Like NGC\,1399, NGC\,1404 has been well studied photometrically \citep[e.g.,][]{Franx+89, Sparks+91, Munoz+09} and kinematically \citep{DOnofrio+95, Iodice+19}, allowing us to estimate the amount of light and line-of-sight velocity dispersion at every location in the data cube.  Also like NGC\,1399, a comparison of the distribution of light and PNe in the galaxy reveals that our survey is incomplete within $\sim 10\arcsec$ of the nucleus. The exclusion of this region removes 7 PNe from the sample and leaves $V \sim 10.8$~mag of galaxy light in the two data cubes. 

\begin{figure}
\includegraphics[width=0.473\textwidth]{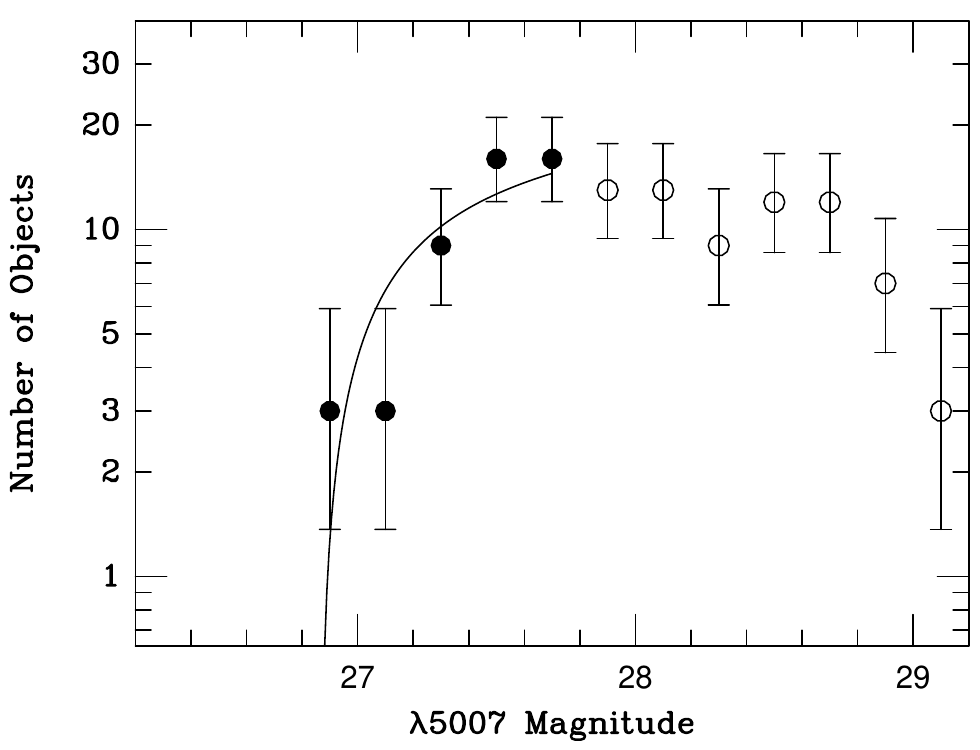}
\caption{The observed PNLF of NGC\,1404 binned into 0.2~mag intervals.  The open circles denote data beyond the completeness limit; the error bars represent $1\sigma$ confidence intervals \citep[see][]{Gehrels86}.  The black curve shows the most-likely fit to equation~(\ref{eq:pnlf}).  Note that the process of binning makes the brightest PNe appear exceptional; in fact, it is only 0.04~mag more luminous than the second brightest object. \label{fig:NGC1404_PNLF}} 
\end{figure}

\begin{figure}[h!]
\includegraphics[width=0.473\textwidth]{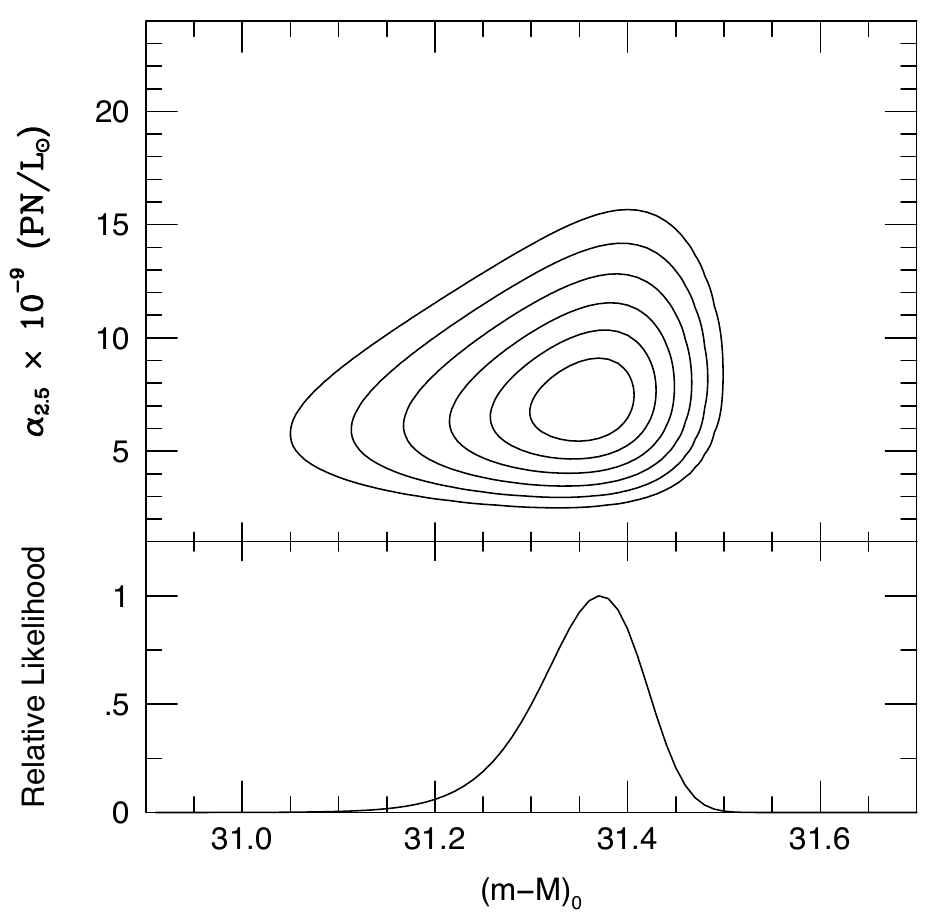}
\caption{The upper panel shows the results of the maximum likelihood solution for NGC\,1404. The abscissa is the galaxy's true distance modulus, the ordinate is the number of PNe within 2.5~mag of $M^*$, normalized to the amount of bolometric light sampled, and the contours are drawn at $0.5\sigma$ intervals. The lower panel marginalizes over the PN/light variable.  \label{fig:NGC1404_contour}}
\end{figure}

\begin{figure}[h]
\includegraphics[width=0.473\textwidth]{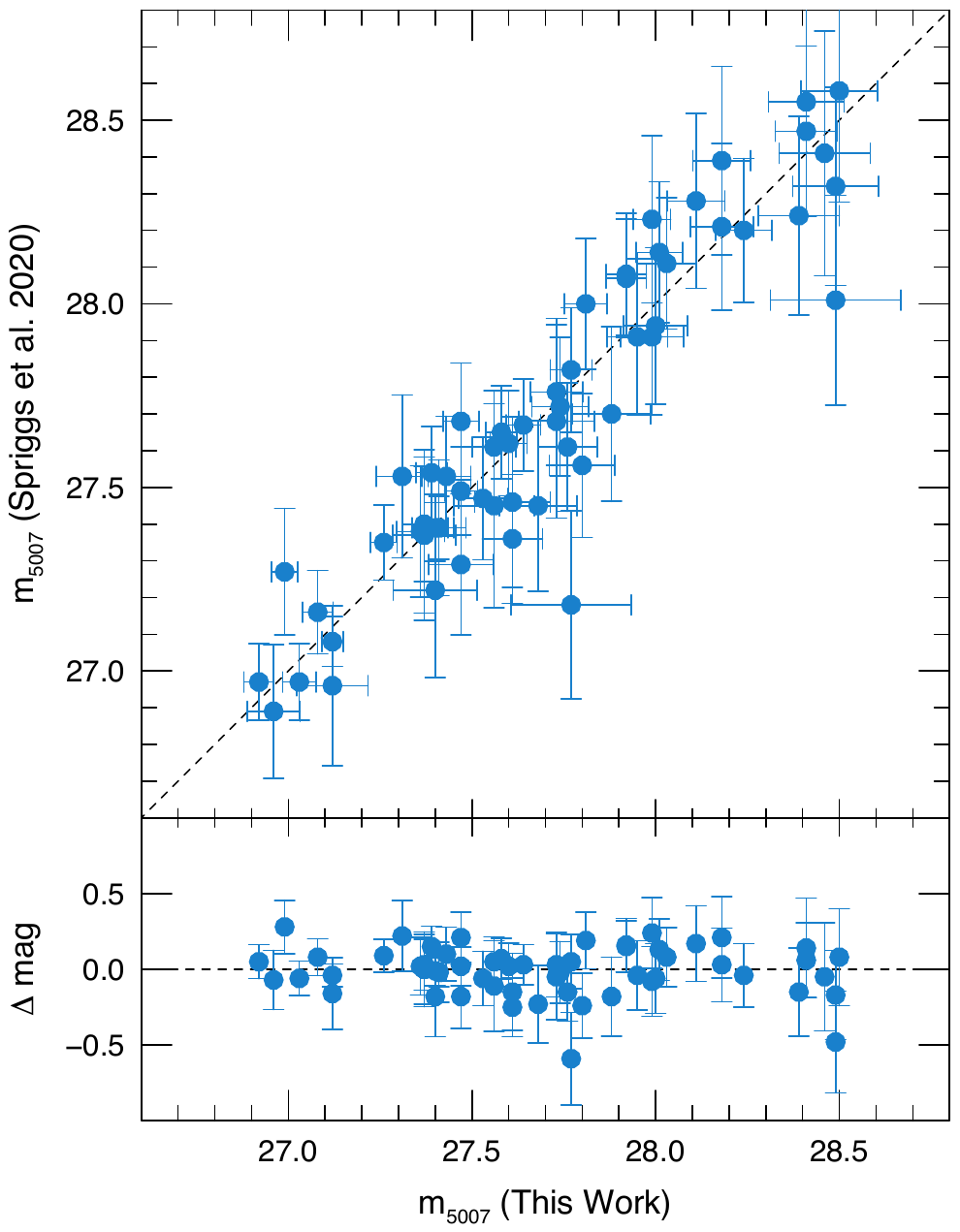}
\caption{A comparison between our PN photometry for NGC\,1404 and that of  \citet{Spriggs+20}; the error bars for the latter are estimated from the inverse of the quoted peak amplitude to residual-noise level ratio (A/rN).  The standard deviation of the distribution is 0.13 mag, and the mean offset between the two datasets (excluding the two outliers) is 0.006 mag.
 \label{fig:comp_Spriggs2020}}
\end{figure}

\begin{figure*}[t!]
\hspace{15mm}
\href{https://cloud.aip.de/index.php/s/9rrSS3n2SxygNWt}{\bf \colorbox{yellow}{Off-band}}
\hspace{45mm}\href{https://cloud.aip.de/index.php/s/pQrXCKM6GgCnt9t}{\bf \colorbox{yellow}{Diff}} \\
\includegraphics[width=57mm,bb=0 0  800 800,clip]{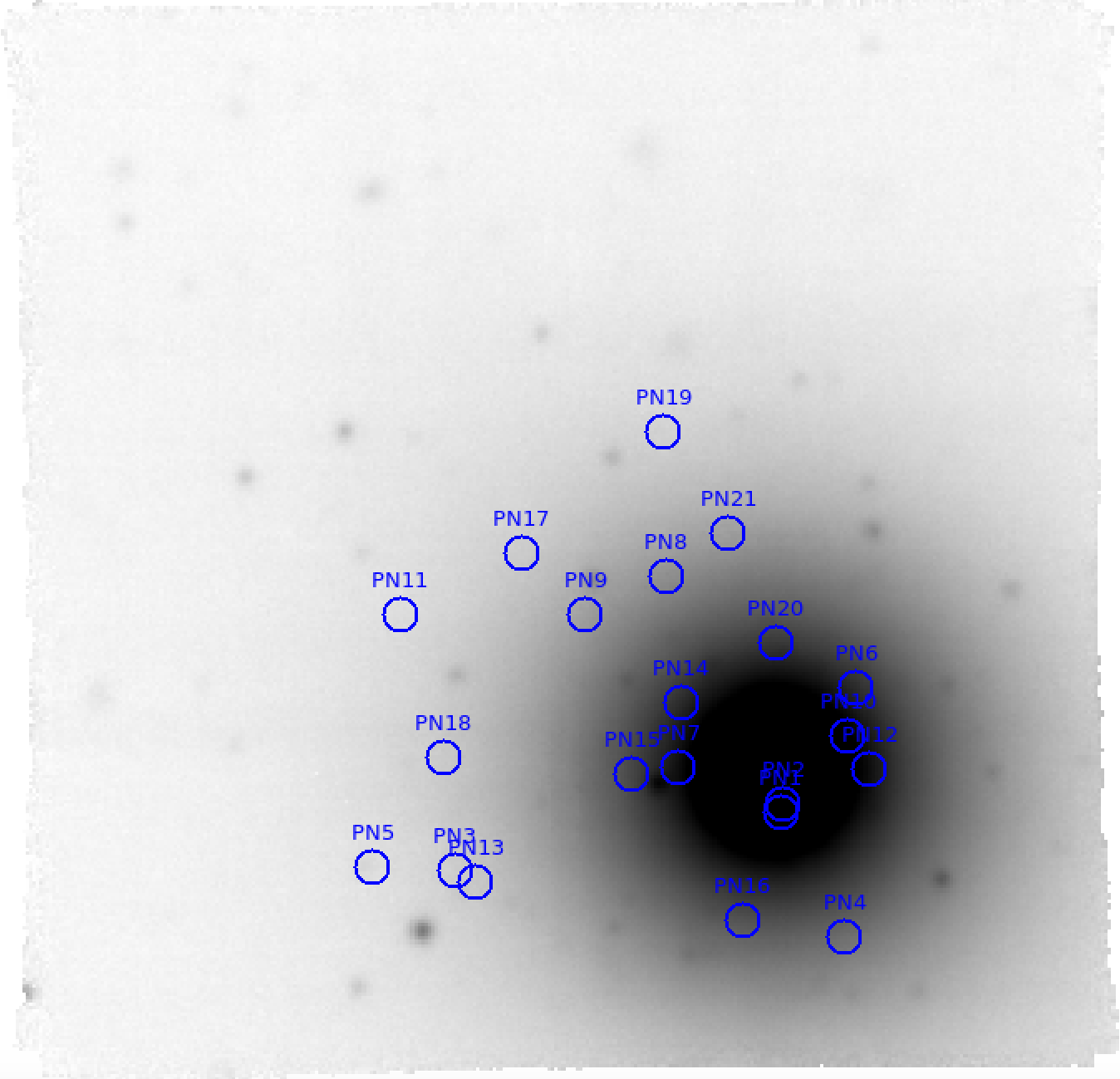}
\includegraphics[width=57mm,bb=0 0  800 800,clip]{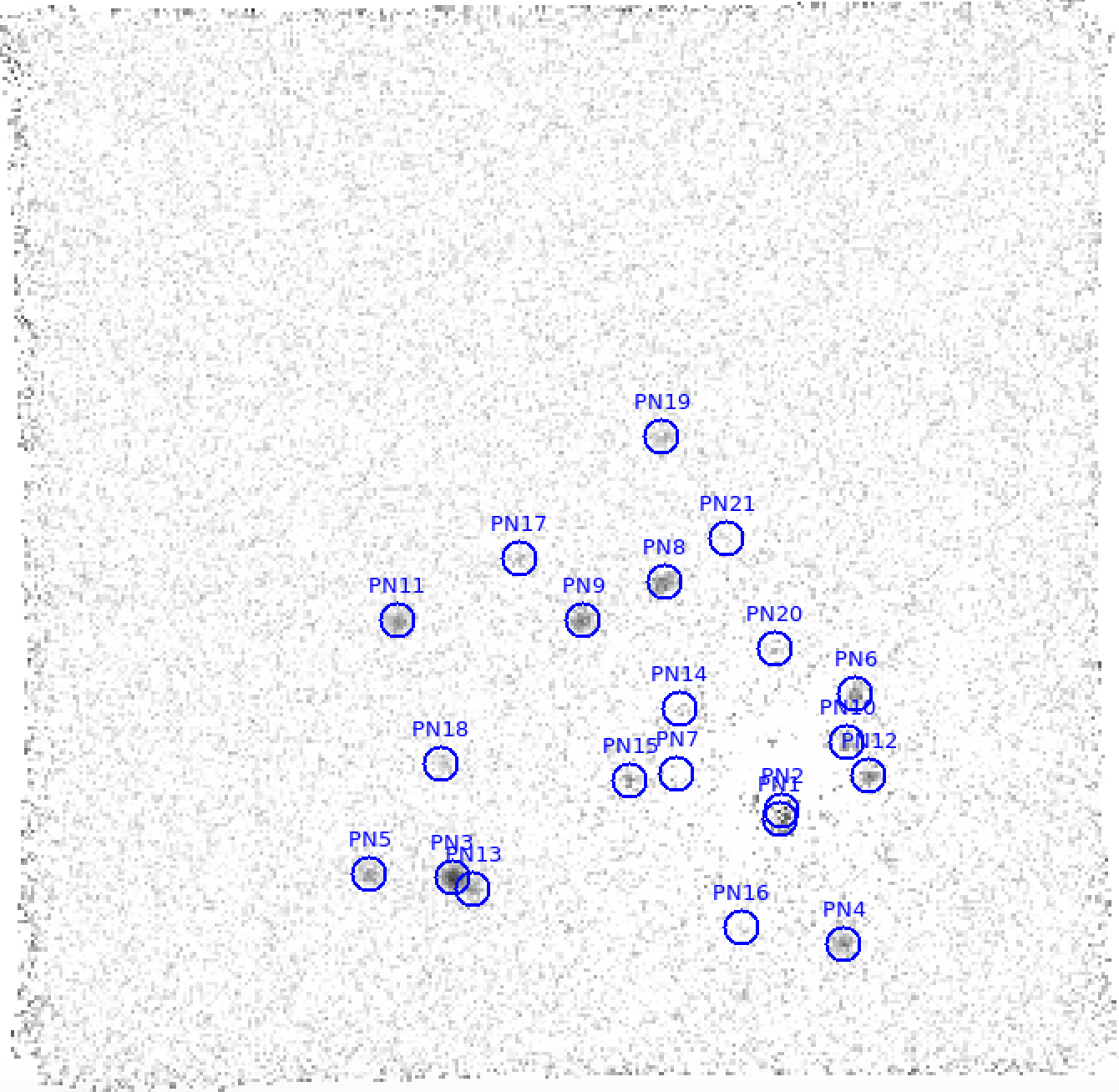}
\includegraphics[width=65mm,bb=120 160  450 400,clip]{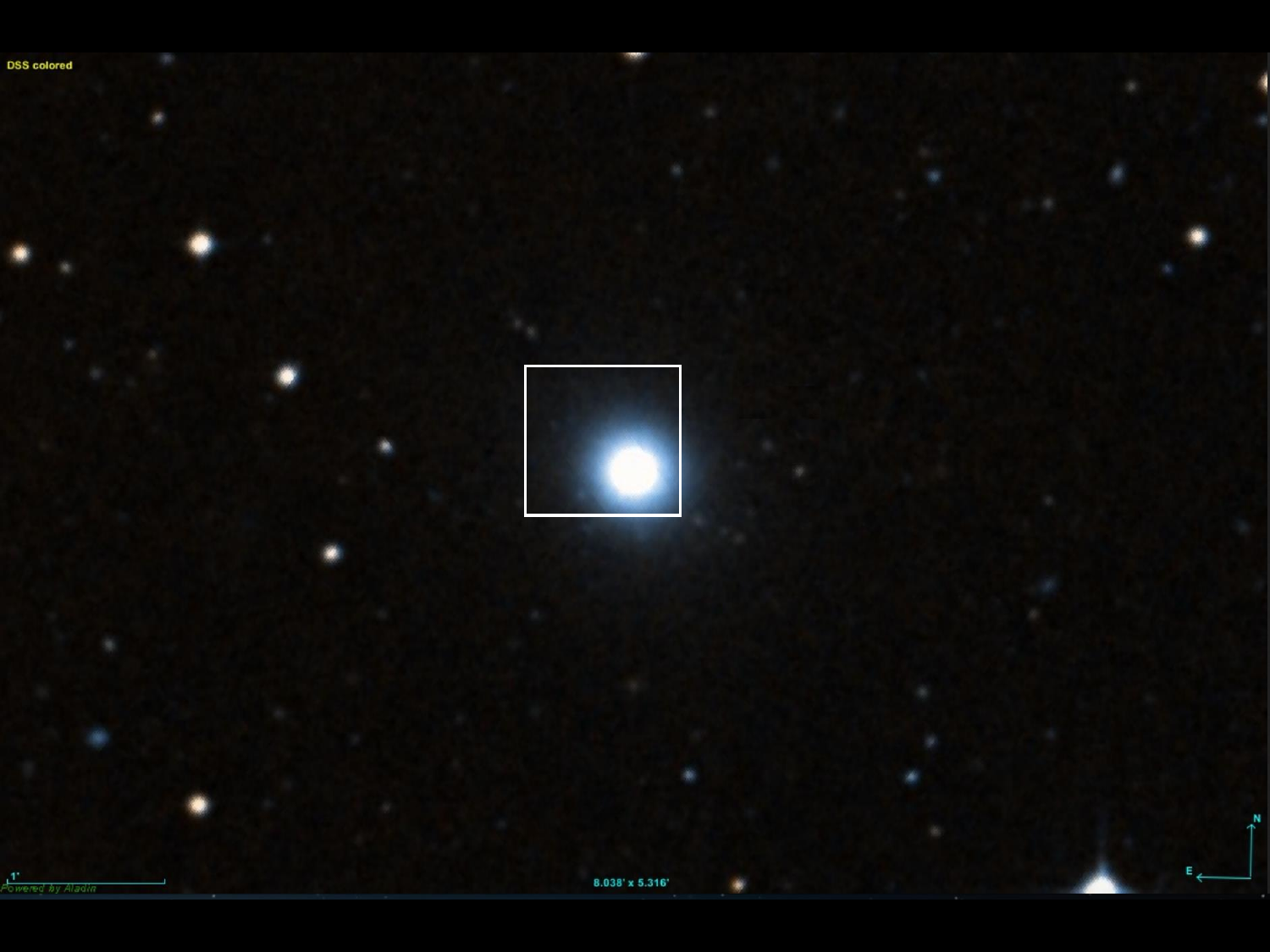}
\caption{NGC\,1419. Left: thumbnail off-band and difference images derived from the MUSE data cube. High resolution images can be obtained by clicking on the hyperlink titles. \href{https://cloud.aip.de/index.php/s/ZKACFzjjeZ6XJYH}{\colorbox{yellow}{VIDEO}}. Right: a broadband image with the location of the MUSE data cube outlined in white. (Credit: DSS). \label{fig:NGC1419_FChart}}
\end{figure*}

\begin{figure}[h]
\includegraphics[width=0.473\textwidth]{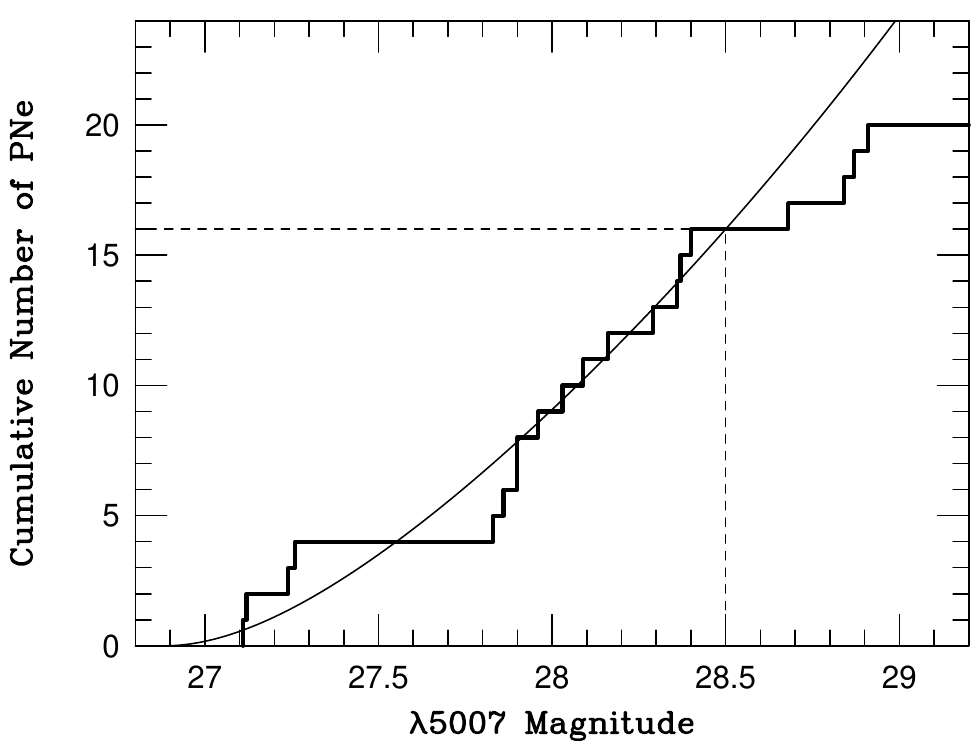}
\caption{The cumulative luminosity function for PNe in NGC\,1419.  The dark line represents the observed data; the curve is equation~(\ref{eq:pnlf}) shifted to the most-likely apparent distance modulus of $(m-M)_0 = 31.42$.  The dashed line shows where incompleteness begins to affect the detections.  Data brighter than this are consistent with being drawn from the empirical function.\label{fig:NGC1419_PNLF}}
\end{figure}

\begin{figure}[h]    
\includegraphics[width=0.473\textwidth]{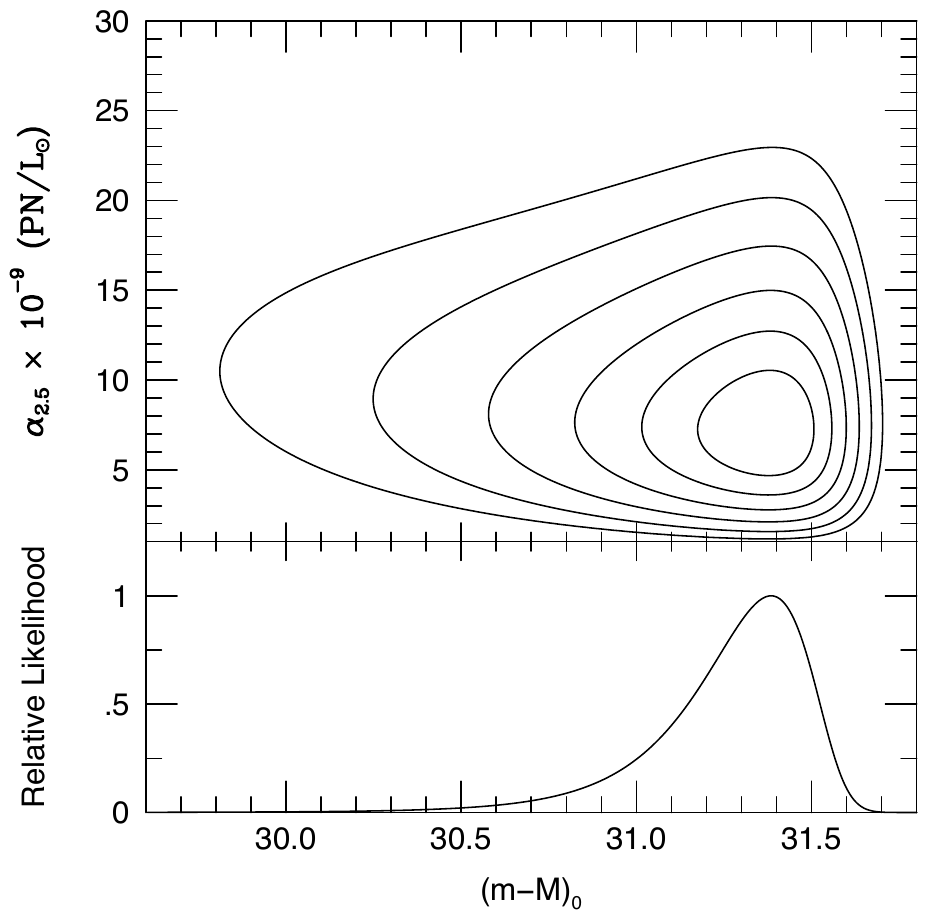}
\caption{The upper panel shows the results of the maximum likelihood solution for NGC\,1419. The abscissa is the galaxy's true distance modulus, the ordinate is the number of PNe within 2.5~mag of $M^*$, normalized to the amount of bolometric light sampled, and the contours are drawn at $0.5\sigma$ intervals. The lower panel marginalizes over the PN per unit light variable.  
\label{fig:NGC1419_contours}\label{NGC1419_contours}}
\end{figure}

Figure~\ref{fig:NGC1404_PNLF} shows NGC\,1404's PNLF, along with the best-fitting PN luminosity function.  The figure illustrates that our PN detections in NGC\,1404 do not go as deep as in some other galaxies of Fornax, and faint-end incompleteness sets in only $\sim 0.8$~mag below $M^*$.  Nevertheless, the PNLF cutoff is very well defined, and, for a foreground reddening of $E(B-V) = 0.010$, the analysis displayed in Figure~\ref{fig:NGC1404_contour} yields a distance modulus of $(m-M)_0 = 31.37^{+0.04}_{-0.07}$.  When combined with the $\sim 3\%$ error on the MUSE flux calibration \citep{Weilbacher+20}, and the very small uncertainties associated with the aperture correction (see Appendix~\ref{sec:notesapcor}) and reddening, the final distance modulus becomes $(m-M)_0 = 31.37^{+0.05}_{-0.08}$, or $18.8^{+0.4}_{-0.6}$~Mpc.  

Unsurprisingly, as Figure~\ref{fig:comp_Spriggs2020} illustrates, our distance is in excellent agreement with the $(m-M)_0 = 31.42 \pm 0.1$ value derived by \citet{Spriggs+20}, as the measurements use the same MUSE data cubes and are not systematically different (although the error bars on the DELF photometry are smaller; see Paper~I\null).  Our value is also essentially identical to the galaxy's two TRGB distance moduli, $(m-M)_0 = 31.36 \pm 0.04$ (stat) $\pm 0.05$ (sys) measured by \citet{Hoyt+21}, and $31.29 \pm 0.07$ derived by \citet{Anand2022}.   It is, however, significantly more distant than the poorer-seeing PNLF estimate of $(m-M)_0 = 31.15^{+0.07}_{-0.10}$ found by \citet{McMillan+93}.

\subsection{NGC 1419 \label{subsec:NGC1419}}

NGC\,1419 is a small elliptical galaxy in the Fornax cluster, with SBF distances generally ranging from 19 to 22~Mpc \citep{Tonry+01, Blakeslee+01, Blakeslee+09}.  The ESO archive contains a single MUSE pointing of the galaxy, with a footprint as shown in Figure~\ref{fig:NGC1419_FChart} (ESO Archive ID: ADP.2018-03-26T15:02:26.469, PI: M. Sarzi, Program ID: 296.B-5054). This MUSE-DEEP data cube was assembled from two observations, with an effective observation time of 4921~s, and a seeing at 5007~\AA\ of $0\farcs 87$.  Owing to the low luminosity of the host galaxy (the MUSE data cube encompasses only $V \sim 12.9$ of galaxy light), just 21 PNe were detected, with only $\sim 12$ in the top $\sim\,$1 mag of the luminosity function where the data are likely complete.  This limits our ability to constrain the systemn's distance.

Figure~\ref{fig:NGC1419_PNLF} shows the cumulative PNLF of the galaxy.  As with NGC\,1366, the shape of the bright end of the distribution is in good agreement with the empirical law defined by equation~(\ref{eq:pnlf}).  However, the exact magnitude of the PNLF cutoff is not well-defined due to the limited PN dataset.

We measured the amount of galaxy light underlying the position of each PNe and over the MUSE data cube as a whole using the $B$-band surface photometry of \citet{deCarvalho+91}; these data were then converted to $V$ by assuming a color of $(B-V) = 0.89$ \citep{Prugniel+98}.   To estimate the line-of-sight velocity dispersion, we used the measurements of \citet{Graham+98} for objects in the inner $\sim 20\arcsec$ of the galaxy and adopted 50~km~s$^{-1}$ for PNe at larger galactocentric radii.

\begin{figure*}[th!]
\hspace{20mm}
\href{https://cloud.aip.de/index.php/s/NZNBcY7ezTwJFgL}{\bf \colorbox{yellow}{Off-band}}
\hspace{45mm}\href{https://cloud.aip.de/index.php/s/T7bF2nNjacjeoks}{\bf \colorbox{yellow}{Diff}} \\
\includegraphics[width=55mm,bb=0 0  800 900,clip]{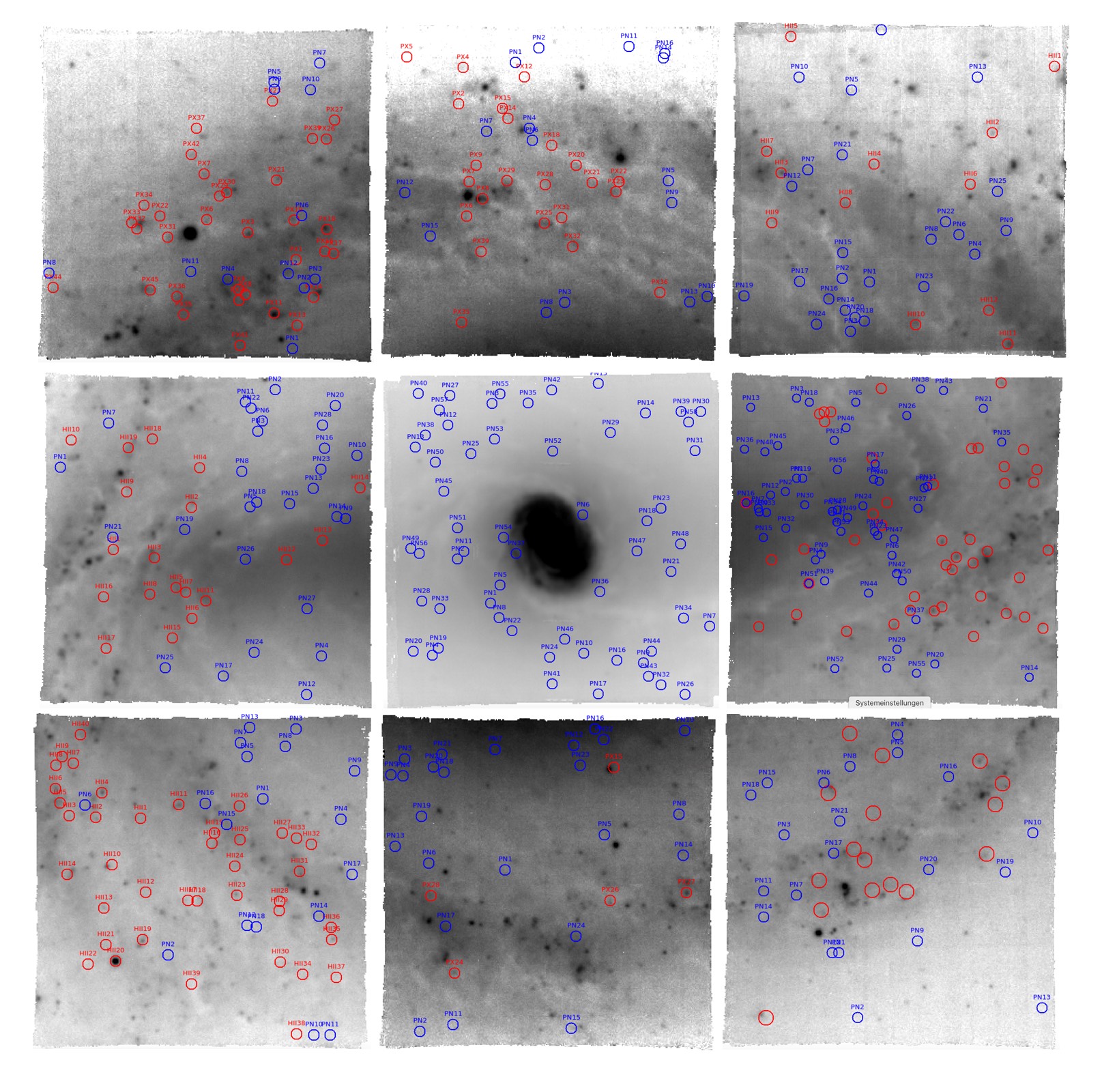}
\includegraphics[width=55mm,bb=0 0  800 900,clip]{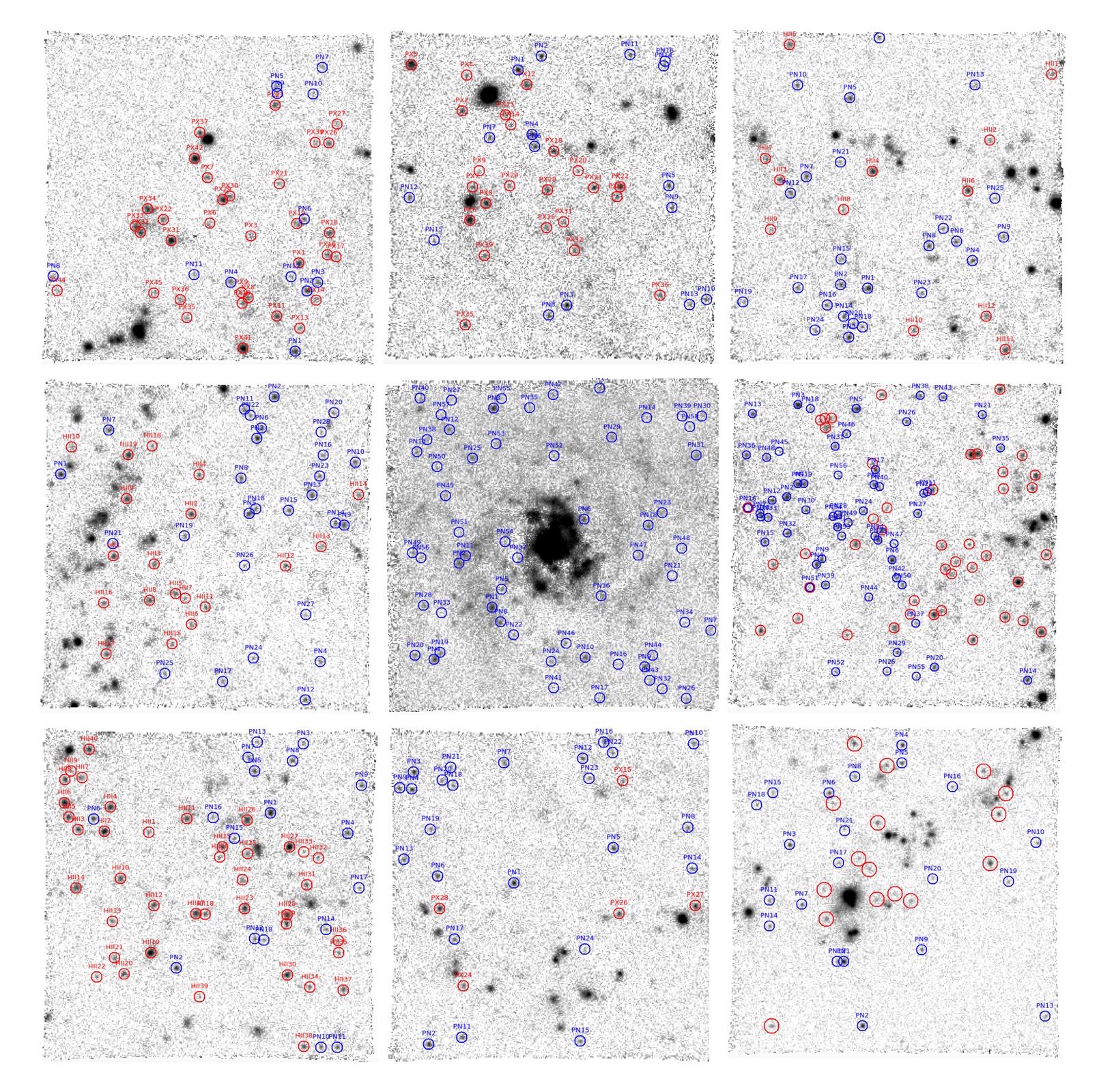}
\includegraphics[width=65mm,bb=50 0  700 600,clip]{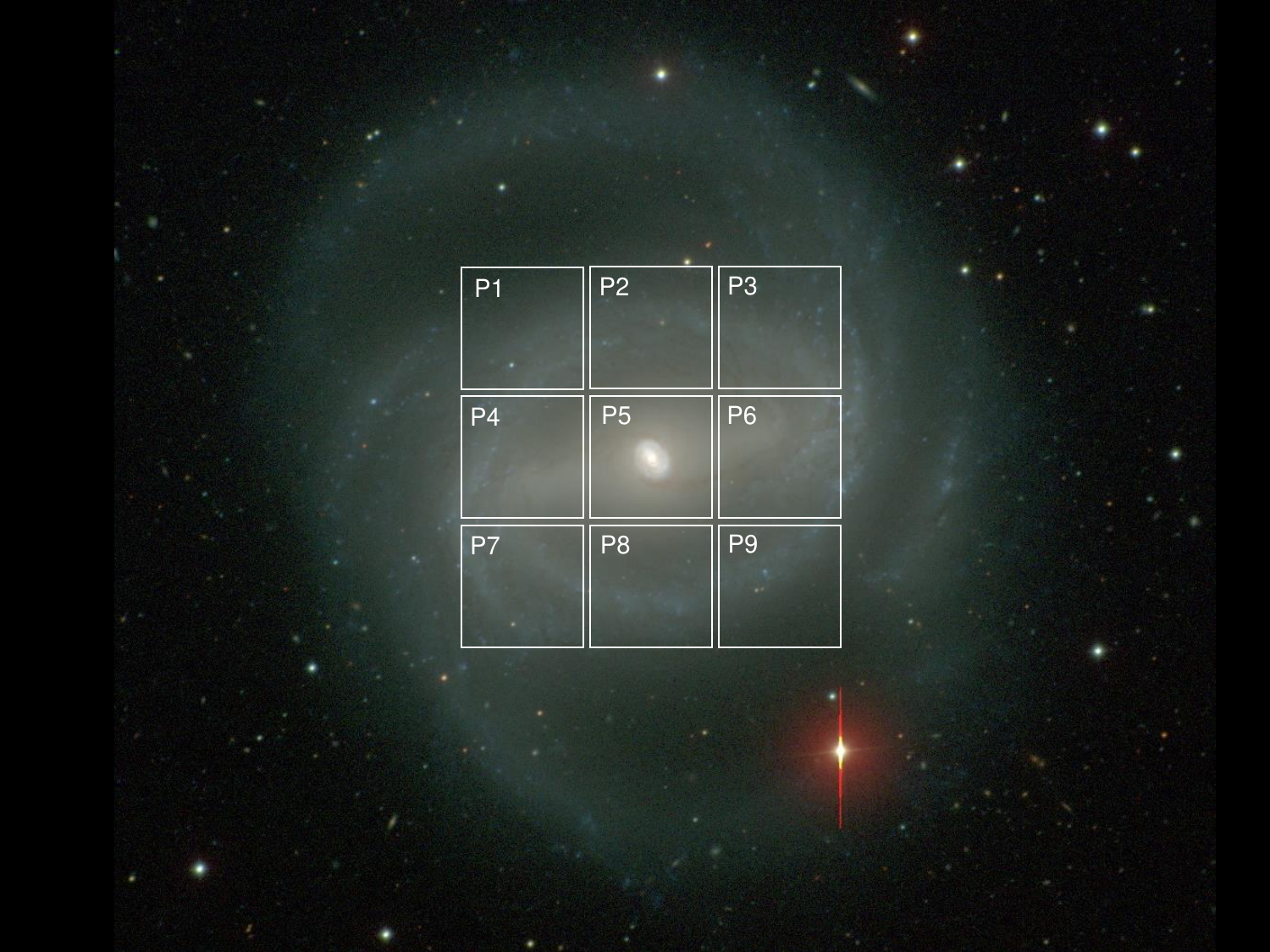}
\caption{NGC\,1433. Left: thumbnail off-band and difference images derived from the MUSE data cubes.  Our PNe candidates are highlighted in blue, while emission-line interlopers are shown in red. Note that the numbering of markers is assigned to the individual pointings P1$\ldots$P9, not to the final sample of the galaxy as a whole. High-resolution images can be obtained by clicking on the hyperlink titles. Right: a broadband image of the galaxy with the MUSE fields outlined in white. (Credit: CGS).  \label{fig:NGC1433_FChart}}
\end{figure*}

The results of our analysis are shown in Figure~\ref{fig:NGC1419_contours}.  With a reddening of $E(B-V) = 0.011$, our fit to the PNLF gives $(m-M)_0 = 31.39^{+0.10}_{-0.26}$; when we include a $\sim 0.06$~mag error due to the flux calibration, aperture correction, and extinction our final value is $(m-M)_0 = 31.39^{+0.12}_{-0.27}$, or $18.9^{+1.1}_{-2.5}$~Mpc.  This value is consistent with membership in the Fornax cluster and is essentially identical to our distance to NGC\,1404, but on the low side of the distribution of SBF distances and smaller than the most recent SBF distance of $d=22.9\pm0.9$ \citep{Blakeslee+09}.  Given the known zero-point offset between the PNLF and SBF distance scales, the difference is a bit larger than expected.


\subsection{NGC 1433 \label{subsec:NGC1433}}

NGC\,1433 is a barred spiral galaxy (Hubble type (R')SB(r)ab) which shows a double ring structure, an active galactic nucleus, and a bright central region of intense star formation. The galaxy has been observed extensively as part of the PHANGS program (PI: E. Schinnerer, Program ID: 1100.B-0651), from which we have used their 9 MUSE pointings. The IDs of these data cubes, along with their exposures times and seeing at 5007~\AA\ are listed in Table~\ref{tab:NGC1433_cubes}; their positions are shown in Figure~\ref{fig:NGC1433_FChart}.

The identification and measurement of PNe in NGC\,1433's were made difficult by two problems: diffuse emission from the system's interstellar gas, and the lack of point sources (i.e., PSF stars) in the field.  The former issue increases the random photometric errors of the PN measurements, as the complex morphology and excitation of the emission-line gas results in uncertain background subtractions around the \OIII $\lambda 5007$ and H$\alpha$ lines.  The latter problem introduces a systematic error into the photometry of each pointing, as without bright PSF stars, it is impossible to accurately determine the aperture correction applicable to our faint-object photometry.  When the data from all nine fields are combined, this systematic error then manifests itself as a pseudo-random error, which smooths out and distorts the shape of the PNLF\null.  The net effect of this smoothing is to bias the PNLF results towards smaller distances.

\begin{deluxetable}{llcc}[bh!]
\tablecaption{Data Cubes for NGC 1433
\label{tab:NGC1433_cubes}  }
\tablehead{\colhead{Field} &\colhead{Archive ID} &\colhead{Exp Time}
&\colhead{Seeing} \\ [-0.25cm]
&&\colhead{(sec)} &\colhead{(5007~\AA)}}
\startdata
P1 &ADP.2019-12-22T01:43:14.831 &2580 &$0\farcs 88$ \\
P2 &ADP.2019-12-22T01:22:27.745 &2580 &$0\farcs 90$ \\
P3 &ADP.2019-12-17T22:49:20.671 &2580 &$0\farcs 94$ \\
P4 &ADP.2019-11-27T01:02:00.695 &2580 &$0\farcs 93$ \\
P5 &ADP.2017-06-14T09:12:09.412 &3840 &$0\farcs 89$ \\
P6 &ADP.2019-11-27T02:44:54.831 &2580 &$0\farcs 73$ \\
P7 &ADP.2020-01-09T22:41:58.291 &2580 &$0\farcs 74$ \\
P8 &ADP.2020-01-11T02:39:27.620 &2580 &$0\farcs 76$ \\
P9 &ADP.2020-01-08T07:08:40.807 &2580 &$0\farcs 59$ \\
\enddata
\end{deluxetable}

\begin{figure}
\includegraphics[width=0.473\textwidth]{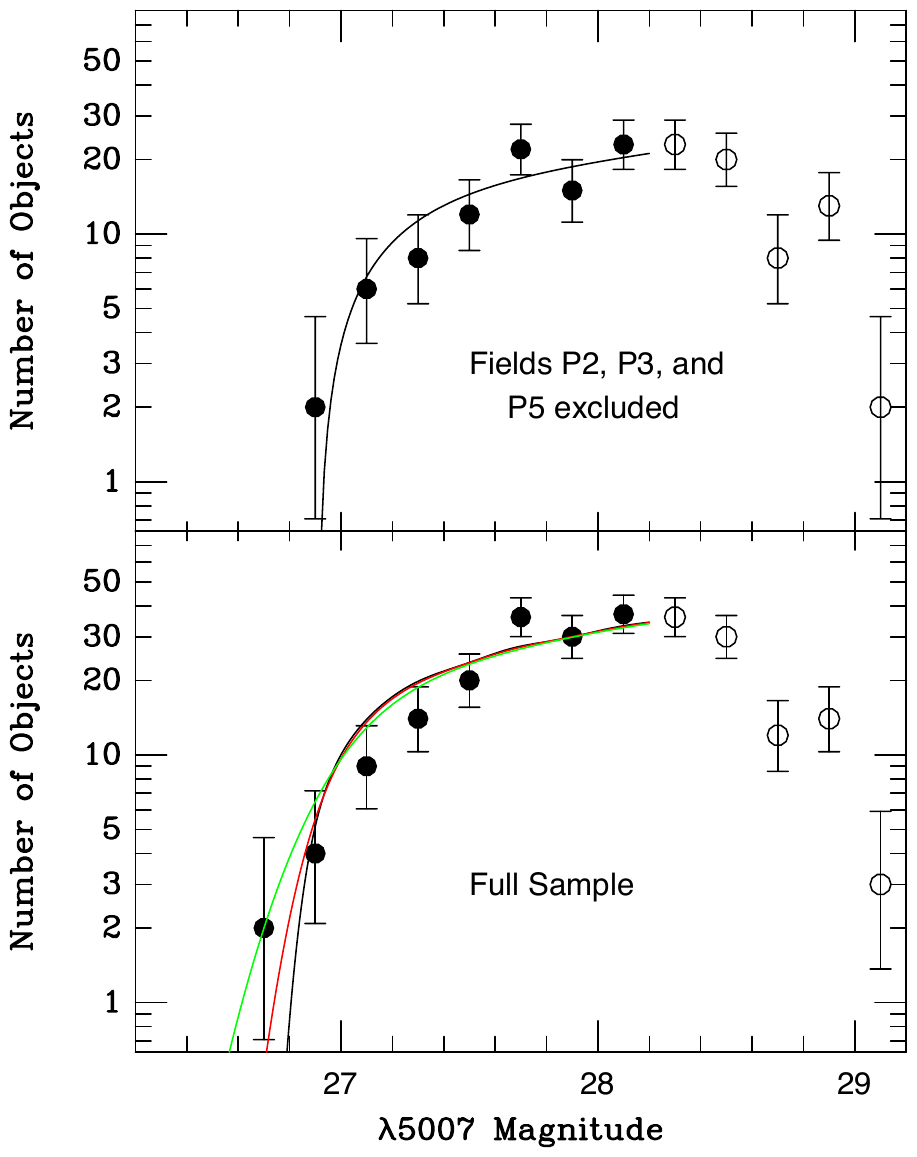}
\caption{The observed PNLF for NGC\,1433 binned into 0.2~mag intervals.  Open circles denote data beyond the completeness limit; the error bars are from small number counting statistics \citep{Gehrels86}. The bottom panel displays the full dataset, along with three curves which all assume an apparent distance of $(m-M) = 31.36$, but with different amounts of an additional photometric error added in quadrature (0.0~mag for the black curve, 0.1~mag for red, and 0.2~mag for green).  The top panel shows the data when the three fields with poorly-known aperture corrections are excluded.  The magnitude distribution is a much better match to equation~(\ref{eq:pnlf}) and implies a most-likely apparent distance modulus of $(m-M) = 31.45$.  
 \label{fig:NGC1433_PNLF}}
\end{figure}

Using the DELF technique we identified almost 500 point-like \OIII $\lambda 5007$ sources across NGC\,1433's disk; after examining the spectra of these objects, almost half were excluded as interlopers.  This left us with 258 PN candidates, with over 120 in the top $\sim 1$~mag of the luminosity function. As the lower panel of Figure~\ref{fig:NGC1433_PNLF} illustrates, the luminosity function of these data is very well-defined and complete to at least $m_{5007} = 28.2$.  However, the bright-end cutoff of the PNLF has a shape that is closer to that of a power-law than an exponential. Part of this behavior is due to the existence of two sources that are $\sim 0.3$~mag more luminous than the next brightest planetary. A careful inspection of the DELF-extracted data cube surrounding these sources reveals that their \OIII measurements are likely contaminated by light from the regions' bright and irregularly distributed emission-line gas.  However, even if these two bright objects are excluded, the shape of the PNLF's cutoff is still somewhat less abrupt than expected. 

The behavior of NGC\,1433's PNLF is consistent with the hypothesis that the PN photometry of the galaxy is compromised by the zero-point issues described above. The system's nine non-overlapping data cubes were taken under a range of seeing conditions (from $0\farcs 64$ to $1\farcs 05$ seeing), which made placing all the PN candidates on a unified flux scale difficult.  This is evidenced by the distribution of PN magnitudes within the fields. Pointing P5 contains no good PSF stars, and the zero-point error associated with the uncertain aperture correction may be as large as 0.1~mag. Not coincidentally the field contains the three brightest PNe in our sample. Pointing P3, which also has a highly uncertain ($\sigma \sim 0.1$~mag) zero-point due to its lack of PSF stars, contains the fourth and fifth brightest PNe. Conversely, the most luminous PNe in P2, a third field with no bright point sources, is 0.3~mag fainter than the brightest PN in P6, a pointing that does have a reliable aperture correction. 
 
Thus, it is quite likely that the bright tail of NGC\,1433's PNLF is entirely due to zero-point errors in the PN photometry.  If so, there are three ways to handle this type of error.  The first is simply to discard the PNe in fields where the aperture correction and/or flux calibration is untrustworthy. This removes a third of the sample, but, as shown in the upper panel of Fig.~\ref{fig:NGC1433_PNLF}, it produces a PNLF more in line with the expectations of equation~(\ref{eq:pnlf}).  

A second possibility is to artificially shift the PNLF's of those MUSE pointing with poor photometric zero points to match the data acquired from data cubes with robust aperture correction measurements.  If the PN samples within the data cubes were larger, this might be a viable methodology.  However, since each cube only contains 5 or 6 PNe in the top magnitude of the luminosity function, it would be impossible to determine the appropriate shifts with the accuracy needed for co-addition.

The third way to handle variations in each data cube's zero-point is to increase the random photometric errors assigned to each individual PN\null.  This, of course, is an approximation: the aperture correction errors of each data cube act in a systematic fashion, shifting all the PN magnitudes in the same manner.  However, if one assumes that a) each MUSE pointing contains a similar number of PNe and b) the distribution of zero-point errors over the 9 data cubes is roughly Gaussian, then expected distortion in the PNLF's shape can be easily modeled and fit to the data in the manner outlined by \citet{Chase+23}.  For example, the curves shown in the bottom panel of Fig.~\ref{fig:NGC1433_PNLF} all assume an apparent distance modulus of $(m-M) = 31.35$ and the PNLF of equation~(\ref{eq:pnlf}).  However, while the black curve shows the expected PNLF in the absence of additional errors, the red and blue curves show what the curve would look like with an additional 0.1 and 0.2~mag of randomly distributed errors, respectively.  These errors, which could either be due to poor PN background subtraction or the effect of imperfect aperture corrections in each of the nine fields, produce a function that adequately represents the observed data.

\begin{figure}
\includegraphics[width=0.42\textwidth]{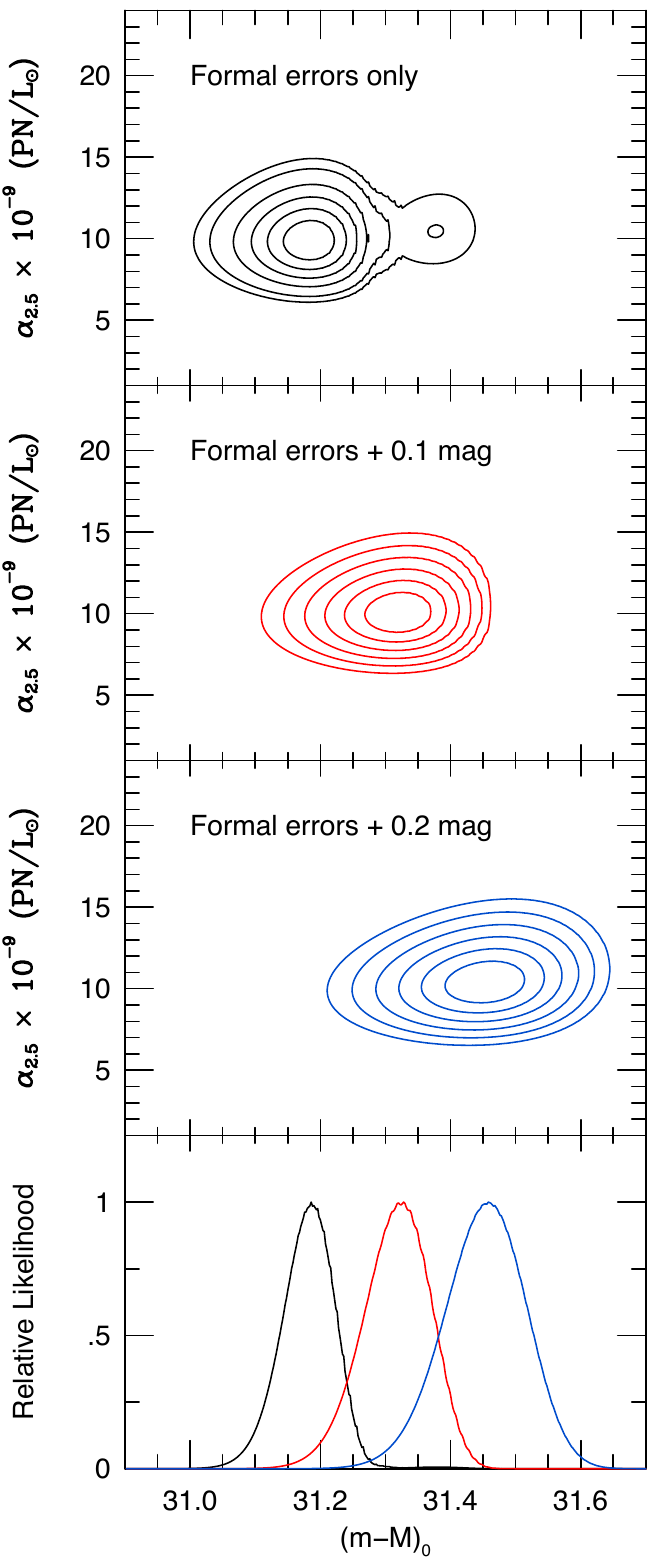}
\caption{The top three panels show our maximum likelihood solution for NGC\,1433. The abscissa is the galaxy's true distance modulus, while the ordinate for the top three panels is $\alpha_{2.5}$, the number of PNe within 2.5~mag of $M^*$, normalized to the amount of bolometric light sampled.  The contours are drawn at $0.5\sigma$ intervals. Three different solutions are shown:  one where the errors are solely those reported by the PN photometry, and two others where we assume the existence of additional random uncertainties due to poor background subtraction and our inability to accurately measure each cube's aperture correction.  The bottom panel marginalizes these solutions over $\alpha_{2.5}$.  Note how the galaxy's inferred distance depends on the photometric precision. \label{fig:NGC1433_contours}}
\end{figure}

To derive the best-fit curves shown in Fig.~\ref{fig:NGC1433_PNLF} we started with the galaxy continuum measurements from the MUSE data cube. While \citet{Buta86} and \citet{Buta+01} do provide a plot of NGC\,1433's azimuthally averaged surface brightness profile, the galaxy's complex two-dimensional morphology, with its bulge, bar, and ring-like structures precludes the use of any simple photometric model. Consequently, we estimated the amount of light underlying each PN directly from stellar continuum measurements on the MUSE data cubes.  Excluding the central $\sim 3\arcsec$ of the galaxy, where PN detections are difficult, the total amount of $V$-band light contained in the 9 MUSE pointings is $V \sim 10.5$.  We also adopted a single number for the galaxy's line-of-sight stellar velocity dispersion.  \citet{Buta+01} have shown that in the inner $\sim 30\arcsec$ of the galaxy, this number is $\sim 75$~km~s$^{-1}$. a value significantly smaller than the velocity difference needed to deblend the emission-lines of two superposed PNe.  Consequently, we simply use this number throughout the galaxy.

Figure~\ref{fig:NGC1433_contours} uses the entire 9-field data sample to illustrate how the inferred distance to NGC\,1433 depends on the accuracy of the photometry.  If the formal errors of the photometry are accurate, then the galaxy's most likely apparent distance modulus is $31.21_{-0.04}^{+0.04}$, though there is some likelihood that superpositions are biasing the result.  However, if we assume that there are additional sources of photometric error, such as those produced by incorrect estimates of the PSF aperture corrections or poor sky subtraction of the galaxy's diffuse emission, then the best-fit distances increase to $(m-M) = 31.35_{-0.06}^{+0.04}$ (for $\sigma_{\rm add} = 0.1$~mag) and $(m-M) = 31.49_{-0.07}^{+0.06}$ (for $\sigma_{\rm add} = 0.2$~mag). 

To understand the trend seen in Fig.~\ref{fig:NGC1433_contours}, consider the basis behind the PNLF method. PNLF distances are defined by measuring the shape of the function's rapid bright-end cutoff.  In defining this shape, a single PNe near $M^*$ carries more weight than a similar PN at fainter magnitudes -- the brighter the PN, the greater its affect on the measured value of $M^*$.  Consequently, random photometric errors will turn into a systematic fitting error, and an underestimate of galaxy distance.  The only way to avoid this systematic is to carefully model the effect of errors on the PNLF, either through the application of an \citet{Eddington1913} correction to the binned data, or by fitting the PN magnitudes, not to the expression given by equation~(\ref{eq:pnlf}), but to the result of a convolution of the fitted function with a kernel that represents the expected amplitude of the errors \citep{Ciardullo+89}.  

In the case of NGC\,1433, the aperture corrections for pointings P2, P3, and P5 are poorly known, as are their errors.  In the absence of this information, magnitude excursions to the bright side of the true value push the resultant distance to smaller values more than those excursions which work the other way.  Thus, the PNLF's best-fit distance modulus is underestimated.




A confirmation of this effect comes from the PNLF solution when objects from the three fields with poorly-determined aperture corrections, P2, P3, and P5, are removed from the analysis.  For these higher-quality data, the distance modulus for an assumed $E(B-V) = 0.008$ is $(m-M)_0 = 31.42_{-0.06}^{+0.04}$.  If we then assume a nominal uncertainty of 0.06~mag for the flux calibration and aperture correction, then distance to the galaxy becomes $(m-M)_0 = 31.42_{-0.08}^{+0.07}$, or $19.2 \pm 0.7$~Mpc.

\begin{figure*}[ht]
\hspace{20mm}
\href{https://cloud.aip.de/index.php/s/pcMxgfFxXJ6iAY6}{\bf \colorbox{yellow}{Off-band}}
\hspace{40mm}\href{https://cloud.aip.de/index.php/s/y4Mgf5sfJ2fXcfX}{\bf \colorbox{yellow}{Diff}} \\
\includegraphics[width=53mm,bb=0 0  1000 1100,clip]{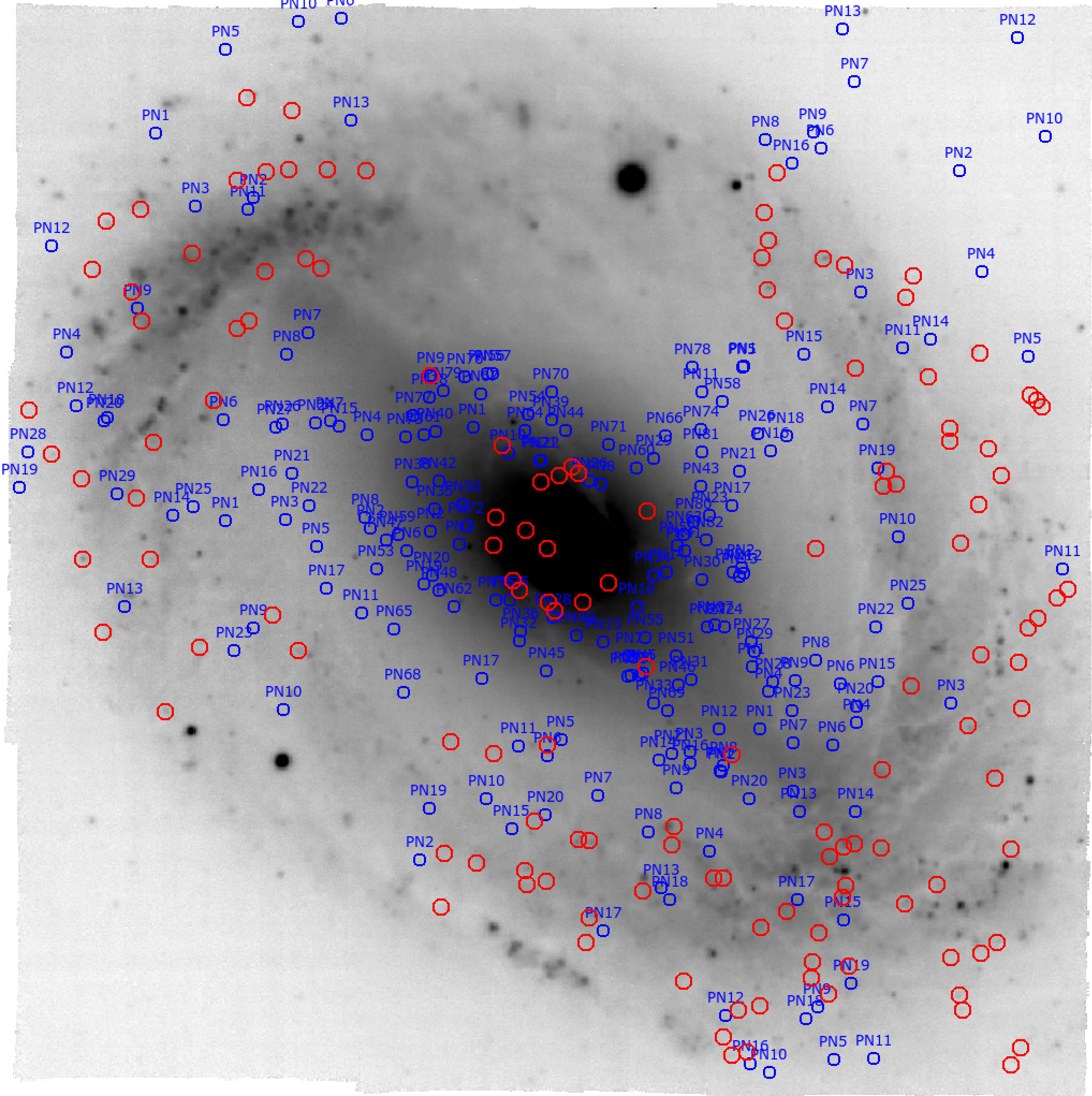}
\includegraphics[width=53mm,bb=0 0  1000 1100,clip]{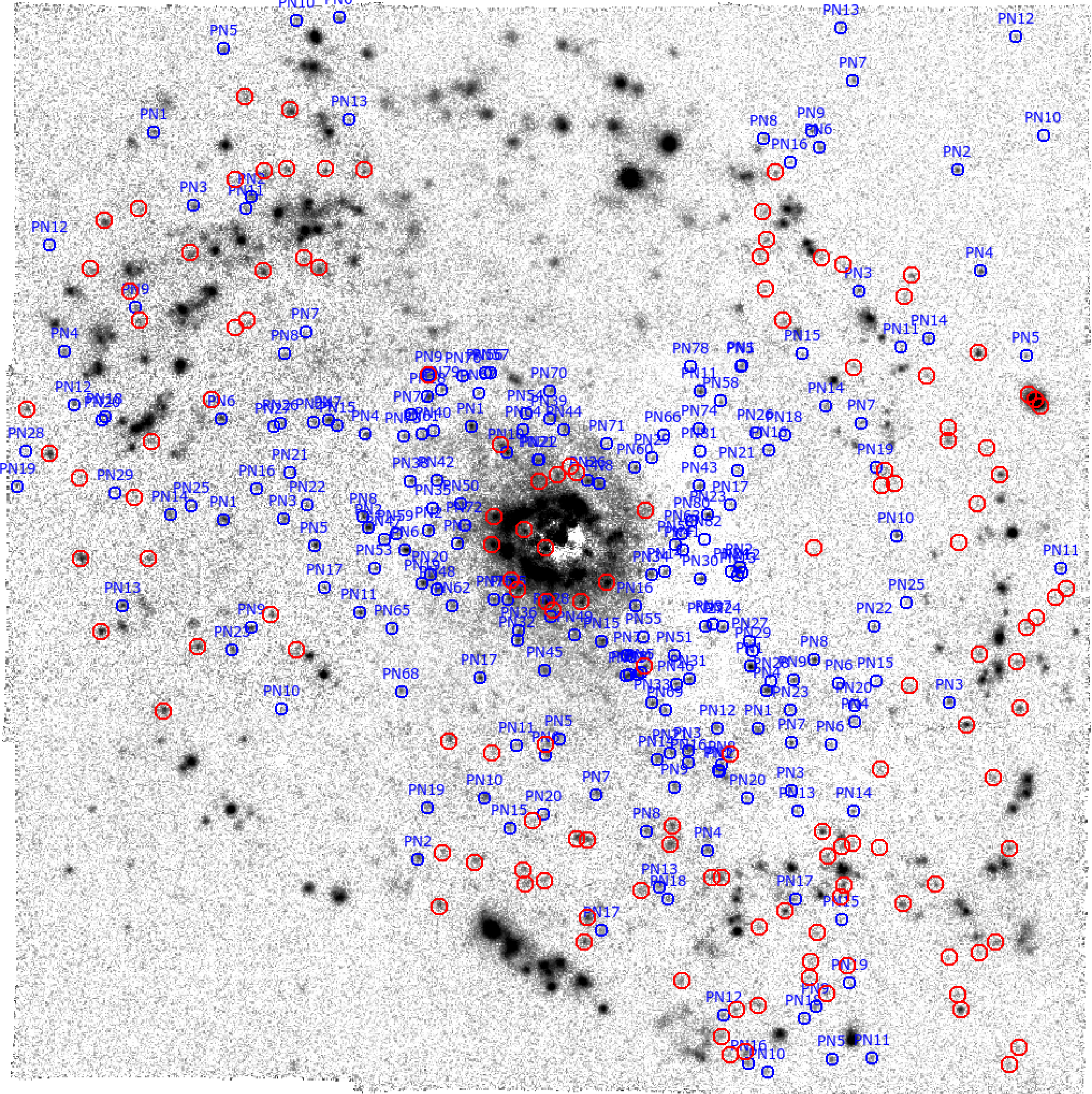}
\hspace{2mm}
\includegraphics[width=65mm,bb=100 50  650 500,clip]{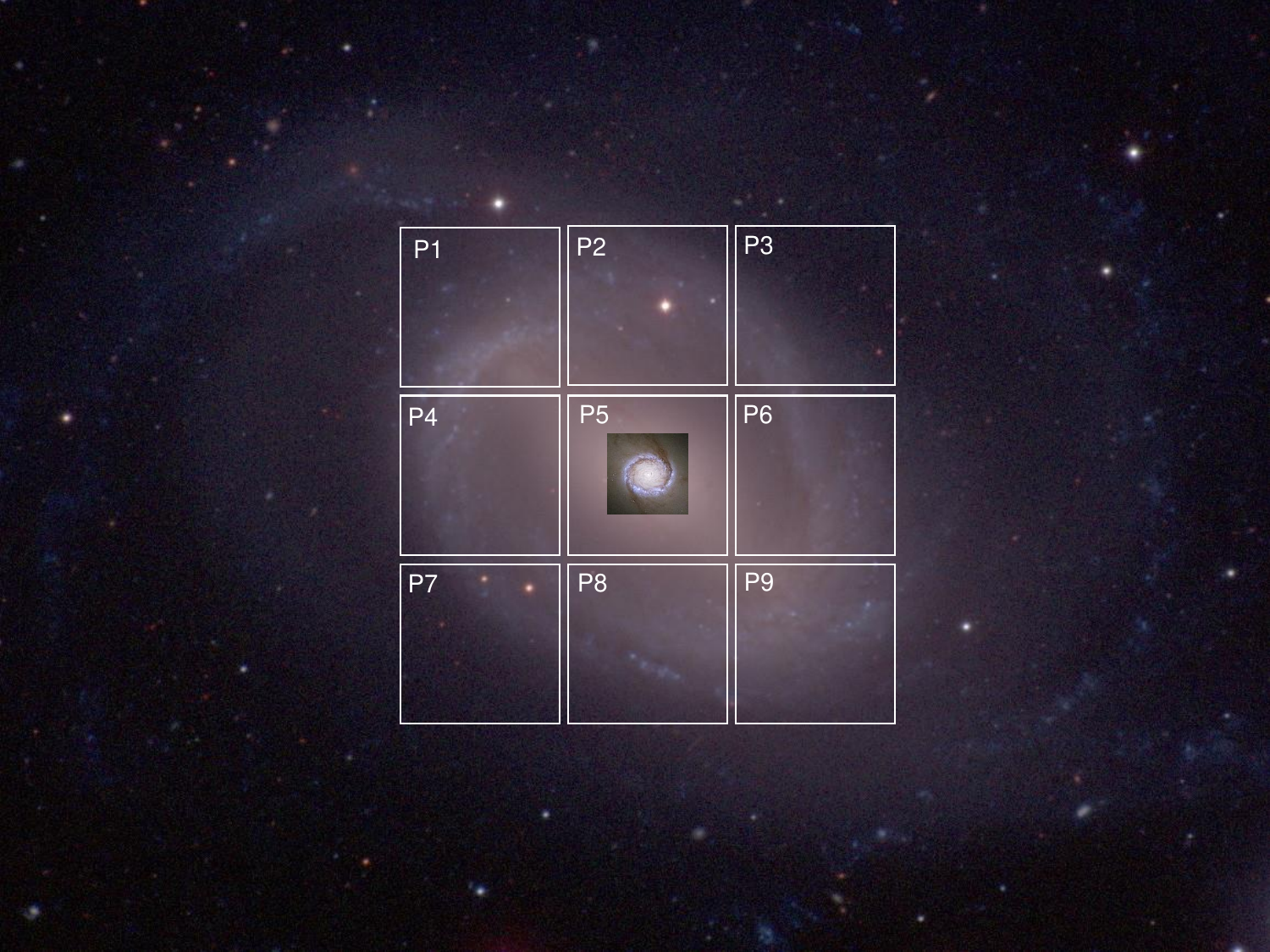}
\caption{NGC\,1512. Left: thumbnail off-band and difference images derived from the MUSE data cubes. 
The PN candidates are highlighted in blue, while the emission-line interlopers are shown in red. The numbering of markers is assigned to the individual pointings P1 through P9. High resolution images can be obtained by clicking on the hyperlink titles.  \href{https://cloud.aip.de/index.php/s/7kknz8x7y8HxJ5Q}{\colorbox{yellow}{VIDEO}}. Right: broadband image of the galaxy, including a high resolution insert of nucleus from the Hubble Space Telescope.  The MUSE data cubes are outlined in white. (Credit: CGS and NASA/ESA/Hubble/LEGUS.)  \label{fig:NGC1512_FChart}}
\end{figure*}

Our result for NGC\,1433 is essentially identical to that of \citet{Scheuermann+22}, who use the same MUSE data cube to identify 90~PNe in the galaxy and derive a distance modulus of $(m-M)_0 = 31.39^{+0.04}_{-0.07}$ ($18.94^{+0.39}_{-0.46}$)~Mpc.    We support their explanation of the conflict between our PNLF distance and the TRGB result of $\sim 9$~Mpc \citep{Sabbi+18}:  confusion between RGB and AGB stars.  The PNLF of Fig.~\ref{fig:NGC1433_PNLF} rules out a distance modulus that would be smaller by as much as 1.6 mag: PNe that bright are simply not seen in this galaxy. 

\subsection{NGC 1512 \label{subsec:NGC1512}}

NGC\,1512 is a barred spiral galaxy, classified SB(r)a, with both a traditional outer ring and an inner disk that encompasses a UV-bright nucleus with intense star formation.  As the galaxy is relatively inclined ($i \sim 72^\circ$), it has been the target of a myriad of Tully-Fisher investigations, all of which place the system between $\sim 11$ and 15~Mpc away \citep[see][]{Tully+09}. A TRGB analysis derived from multiwavelength \textit{HST} photometry agrees with this assessment, assigning a distance at the near end this range \citep[11.6~Mpc;][]{Sabbi+18}.  A grid of nine MUSE pointings obtained by the PHANGS program (PI: E. Schinnerer, C.M. Carollo, IDs: 1100.B-0651, 099.B-0242) covers the galaxy and is available in the ESO archive.  These pointings are listed in Table~\ref{tab:NGC1512_cubes} and shown in Figure~\ref{fig:NGC1512_FChart}.  

Since there is little to no surface photometry available for NGC\,1512, we used continuum measurements from the MUSE IFUs to estimate the amount of galaxy light falling onto the nine data cubes and at the position of each PN\null.  Also, since NGC\,1512 is a disk galaxy, we did not attempt to model the system's line-of-sight velocity dispersion, as we expect this number to be much less than the minimum velocity separation that can be deblended by the MUSE spectrograph.  In our analysis, we adopted 50~km~s$^{-1}$ for this value.

\begin{deluxetable}{llcc}[b!]
\tablecaption{Data Cubes for NGC 1512
\label{tab:NGC1512_cubes}  }
\tablehead{\colhead{Field} &\colhead{Archive ID} &\colhead{Exp Time}
&\colhead{Seeing} \\ [-0.25cm]
&&\colhead{(sec)} &\colhead{(5007~\AA)}}
\startdata
P1 &ADP.2019-02-05T06:15:31.835 &2580 &$0\farcs 82$ \\
P2 &ADP.2018-03-02T17:54:29.654 &3225 &$1\farcs 75$ \\
P3 &ADP.2018-03-08T17:26:28.765 &2580 &$0\farcs 73$ \\
P4 &ADP.2019-02-05T06:15:31.827 &2580 &$0\farcs 81$ \\
P5 &ADP.2017-12-12T10:52:12.342 &3600 &$0\farcs 82$ \\
P6 &ADP.2018-03-08T18:22:48.347 &2580 &$0\farcs 76$ \\
P7 &ADP.2019-02-12T01:24:29.983 &2580 &$1\farcs 27$ \\
P8 &ADP.2019-02-12T01:24:30.018 &2580 &$1\farcs 10$ \\
P9 &ADP.2019-02-12T01:24:29.997 &2580 &$0\farcs 89$ \\
\enddata
\end{deluxetable}
 
\begin{figure}[h]
\includegraphics[width=0.473\textwidth]{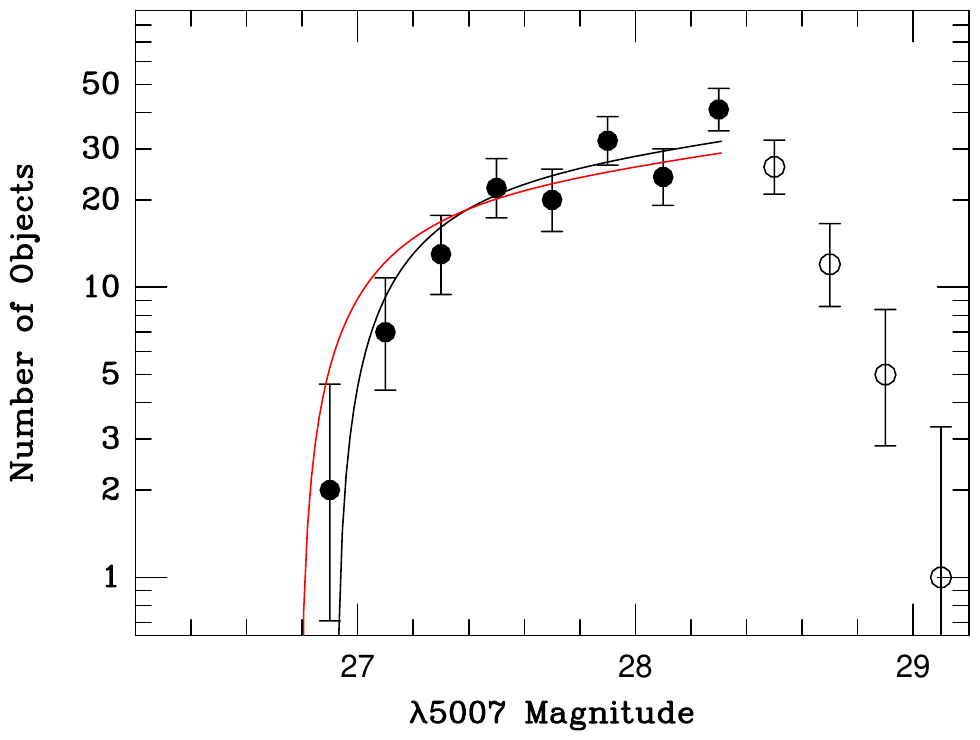}
\caption{The observed PNLF for NGC\,1512 binned into 0.2~mag intervals.  Open circles denote data beyond the completeness limit; the error bars are from small number counting statistics \citep{Gehrels86}.  The red curve shows the most likely fit to equation~(\ref{eq:pnlf}) when PN1 is included in the analysis; the black curve shows the solution when PN1 is assumed to be made up of a superposition of two sources (or is excluded from the analysis all together). 
\label{fig:NGC1512_PNLF}}
\end{figure}

\begin{figure}[ht]   \includegraphics[width=0.473\textwidth]{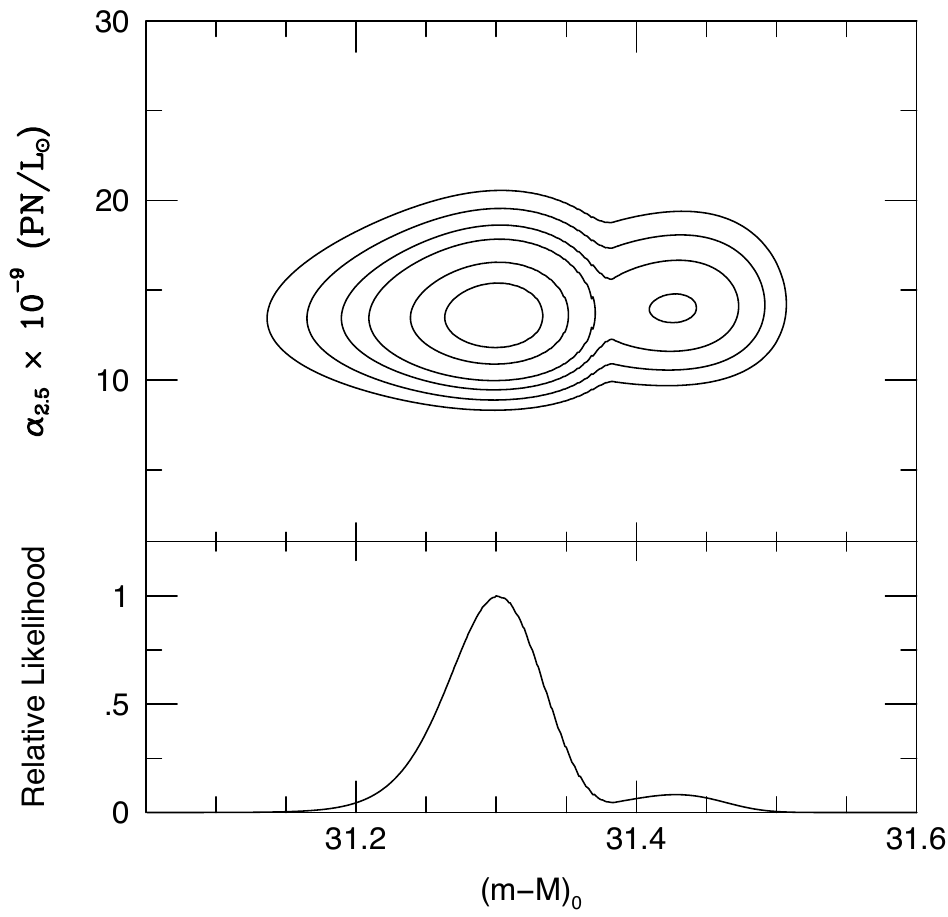}
\caption{The top panel shows the results of the maximum likelihood analysis for NGC\,1512. The abscissa is the galaxy's true distance modulus, the ordinate is $\alpha_{2.5}$, the number of PNe within 2.5~mag of $M^*$, normalized to the amount of bolometric light sampled.  The contours are drawn at $0.5\sigma$ intervals.  The bottom panel marginalizes over $\alpha_{2.5}$.  Note the bimodal solution: there is a non-negligible probability that PN1 is a superposition of two objects.\label{fig:NGC1512_contours}}
\end{figure}

PN detections within an isophotal radius of $\sim 13\arcsec$ of NGC\,1512's nucleus are extremely difficult, due to the region's high surface brightness and bright diffuse emission, so we eliminated the region from our analysis.  Also, poor seeing precluded faint point source detections in fields P2 and P7, so these areas were also excluded from consideration.  That left us with $V \sim 11$~mag of galactic light hosting 210 PN candidates found by our DELF analysis.  Over 60 of these objects are in the top $\sim 1$~mag of the luminosity function.  Figure~\ref{fig:NGC1512_PNLF} displays the object's PNLF.

As Figure~\ref{fig:NGC1512_PNLF} illustrates, the bright-end of NGC\,1512's PNLF is very well defined, with a cutoff magnitude of $m_{5007} \sim 27$.  However, there is one slight ambiguity: the shape of the PNLF, as defined by the PNe with $m_{5007} > 27$, suggests a PNLF cutoff that is very close to $m^* \sim 26.9$.  Yet the brightest PN in the galaxy (PN1) is $\sim 0.1$~mag brighter than this threshold.  In other words, the object appears \textit{slightly} overluminous compared to the apparent PNLF cutoff.  One potential explanation for this brightness is a PN superposition.  

\begin{figure*}[!th]
\hspace{20mm}
\href{https://cloud.aip.de/index.php/s/R6TwwELK9dPEc5A}{\bf \colorbox{yellow}{Off-band}}
\hspace{50mm}\href{https://cloud.aip.de/index.php/s/xf9dm6TjHHgZzRr}{\bf \colorbox{yellow}{Diff}} \\
\includegraphics[width=60mm,clip]{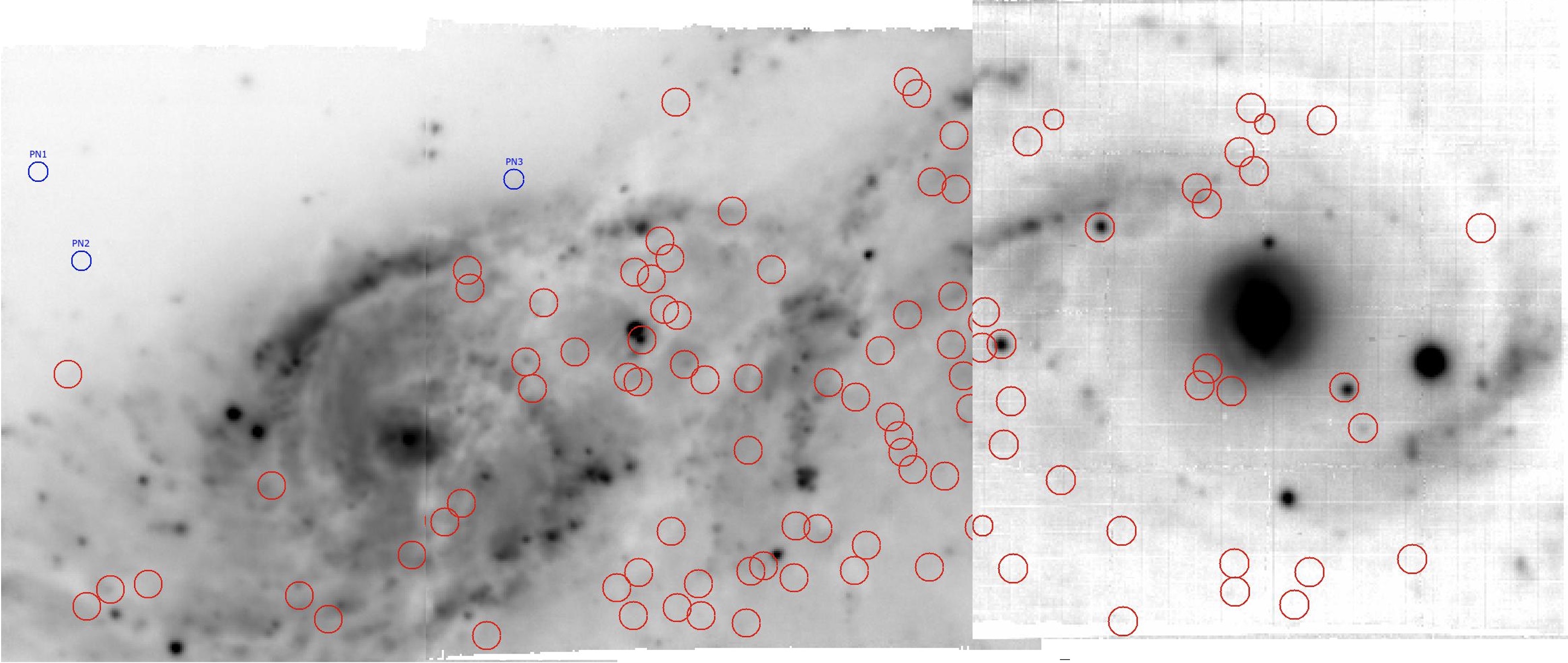}
\includegraphics[width=60mm,clip]{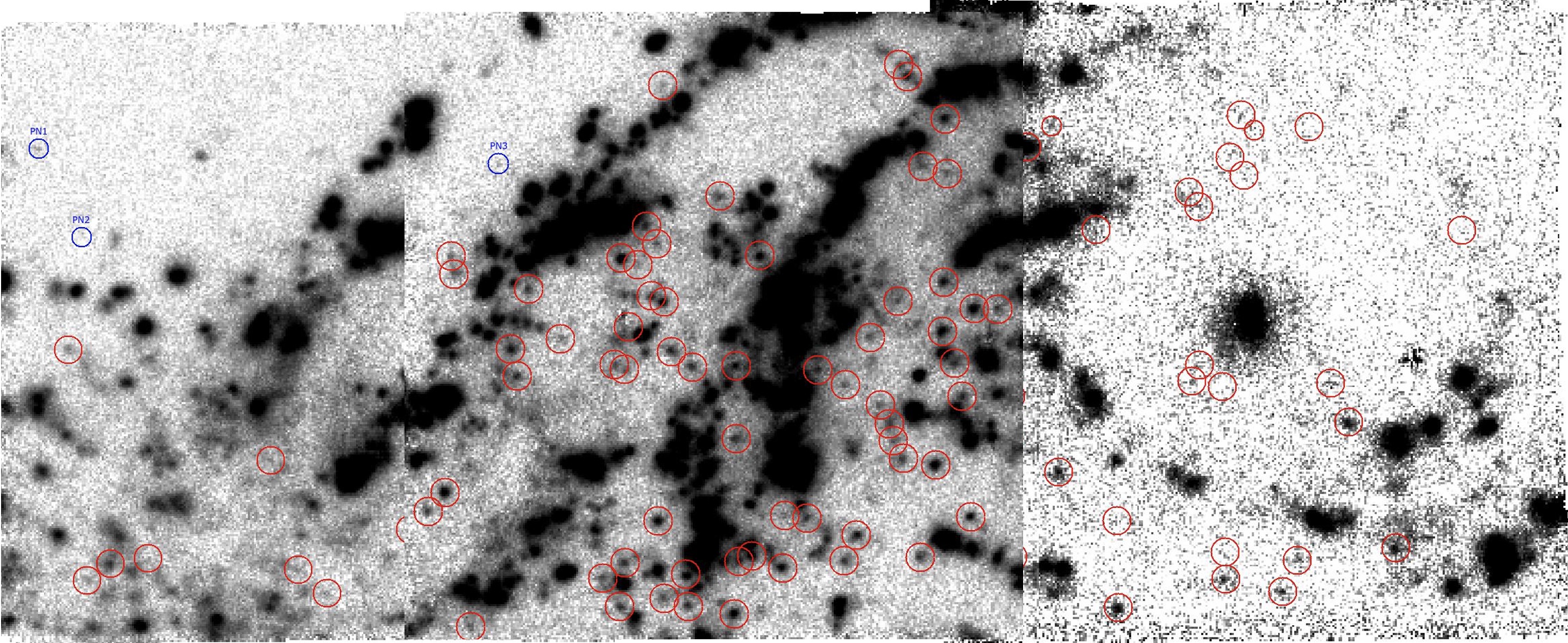}
\hspace{2mm}
\includegraphics[width=55mm,bb=100 500  2888 1800,clip]{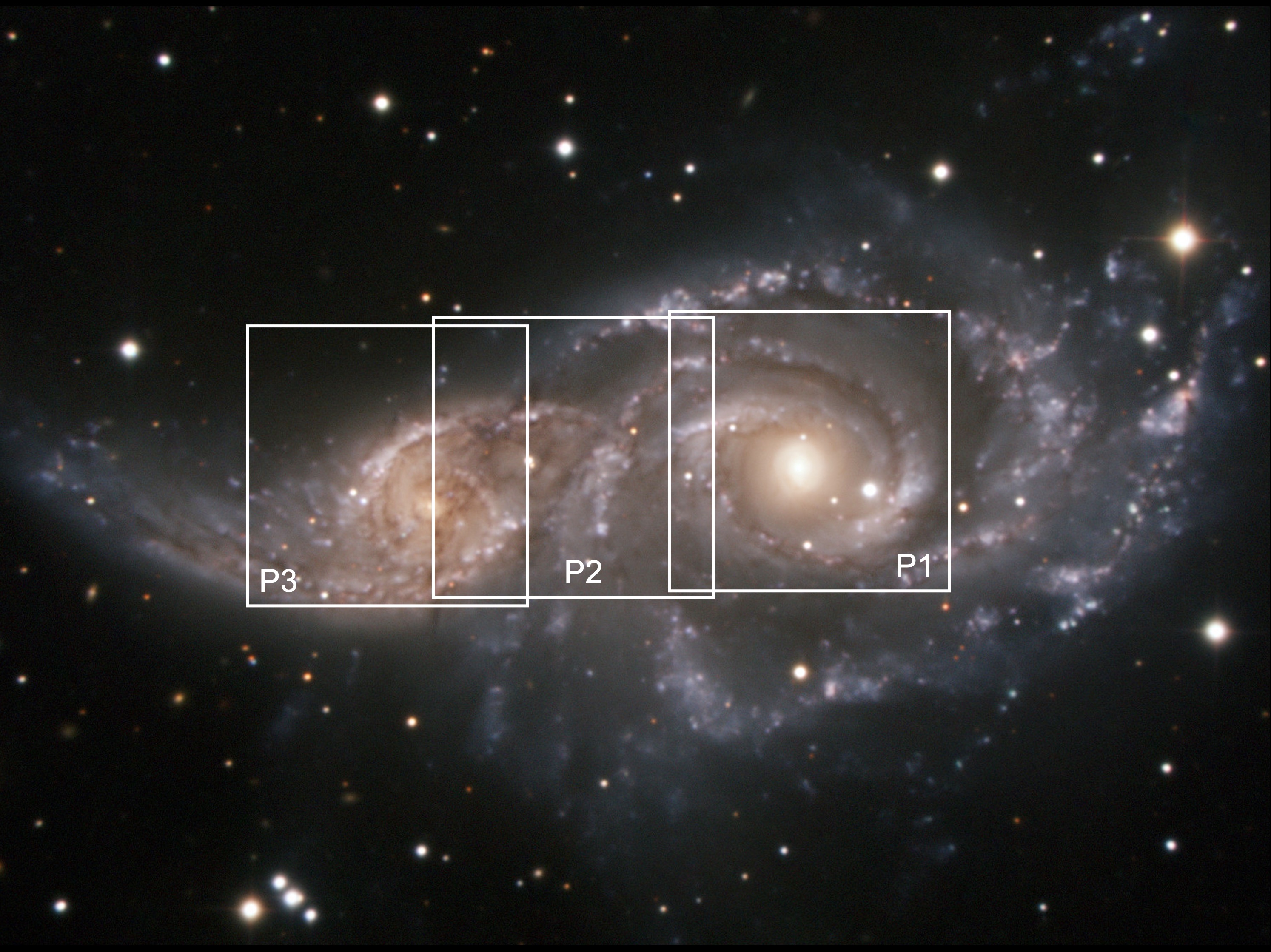}
\caption{NGC\,2207/IC\,2163. Left: thumbnail off-band and difference images derived from the MUSE data cubes. The majority of point-like \OIII sources are interloping objects (red markers); there are only three PN candidates (blue markers) present in the field.  High resolution images can be obtained by clicking on the hyperlink titles. Right: a broadband image with the MUSE pointings outlined in white.  (Credit: ESO).  \label{fig:NGC2207_FChart}}
\end{figure*}

This possibility is summarized in Figure~\ref{fig:NGC1512_contours}.  Because the region upon which PN1 is superposed is relatively bright ($\mu_V \sim 21$), the likelihood that the object is composed of two superposed sources is moderately high.  But still, the hypothesis is disfavored:  the best-fit solution considers PN1 to be a single object and yields a distance modulus of $(m-M)_0 = 31.30_{-0.04}^{+0.04}$, or $18.2 \pm 0.3$~Mpc (for an $E(B-V) = 0.01$).  Yet if PN1 is a single, ordinary PN, the quality of the fit is rather poor, and a Kolmogorov-Smirnov test can rule out the null hypothesis with 86\% confidence.  Conversely, if PN1 is a superposition of two objects, the preferred solution is $(m-M)_0 = 31.43 \pm 0.04$; not coincidentally, this is identical to the distance of $31.43_{-0.04}^{+0.03}$ ($19.3_{-0.4}^{+0.3}$~Mpc) which is derived when PN1 is excluded entirely from the sample.  In this case, the fit to equation~(\ref{eq:pnlf}) is excellent and the Kolmogorov-Smirnov statistic is a miniscule $D_n = 0.062$.  We therefore prefer this value for the distance.

Like NGC\,1433, the PNLF of NGC\,1512 is compromised by uncertainties in the zero-point errors of some of the data cubes.  Specifically, pointings P3, P6, and P8 show an unusual wave-like pattern in the aperture correction versus wavelength relation and/or have noise estimates which may point to an issue in the data reduction process.  Fortunately, unlike NGC\,1433, the PNe of these data cubes do not dominate the bright-end of the luminosity function: the 2nd and 6th brightest PNe are in P6, while P3 and P8 contribute very little to the definition of the PNLF's cutoff.  Since we estimate that the flux calibration and aperture correction errors for the remaining fields are no greater than 0.05~mag, we adopt this number as the systematic component to our error budget.  Thus, if we assume a foreground reddening of $E(B-V) = 0.01$, our final distance to the galaxy is $(m-M)_0 = 31.43_{-0.06}^{+0.06}$ ($19.3_{-0.6}^{+0.5}$~Mpc).

Our result is slightly larger than the distance modulus of $(m-M)_0 = 31.27^{+0.07}_{-0.11}$ found by \citet{Scheuermann+22} from the same data cube, albeit with a much smaller sample (43) of PNe.  But it is significantly larger than the value of $\sim 11.7$~Mpc found from the analysis of UV and optical \textit{HST} images taken as part of the Legacy ExtraGalactic UV Survey \citep{Sabbi+18}. As noted for NGC\,1433, the TRGB distance is much smaller than both PNLF distance estimates, and probably for the same reason discussed above, i.e., confusion between the tip of the RGB and that of the AGB.

\vspace{50pt}
\subsection{NGC 2207/IC~2163 \label{subsec:NGC2207}}

NGC\,2207 and IC\,2163 are a pair of interacting spiral galaxies in the early stages of a merger. The larger system, NGC\,2207, is classified SAB(rs)bc pec and has a total $B$ magnitude of $10.8$; IC\,2163 is a barred spiral of type SB(rs)c pec with $B=11.4$ \citep{RC3-1991}.  Both galaxies exhibit robust star formation, and are subject to a relatively large amount of foreground extinction \citep[$A_V=0.238$;][]{Schlafly+11}. 

There are no TRGB or Cepheid distances to the system, and estimates from the Tully-Fisher relation \citep[13 -- 17~Mpc;][]{Bottinelli+84, Bottinelli+86, Russell2002, Theureau+07} are discordant with that determined from SN~Ia 1975A \citep[33 -- 50~Mpc;][]{Arnett1982, Davis+21}.  Thus, a PNLF distance to the galaxies would be interesting.  However, if the SN-based distances are correct, then it would be a real challenge to measure a PNLF solely from the MUSE archival data.

\begin{deluxetable}{llcc}[b!]
\tablecaption{Data Cubes for NGC 2207/IC 2163
\label{tab:NGC2207_cubes}  }
\tablehead{\colhead{Field} &\colhead{Archive ID} &\colhead{Exp Time}
&\colhead{Seeing} \\ [-0.25cm]
&&\colhead{(sec)} &\colhead{(5007~\AA)}}
\startdata
P1 &ADP.2019-03-06T02:54:12.126 &2238 &$0\farcs 88$ \\
P2 &ADP.2018-04-05T08:26:13.265 &3338 &$0\farcs 66$ \\
P3 &ADP.2018-04-05T08:26:13.281 &2892 &$0\farcs 84$ \\
\enddata
\end{deluxetable}

Three MUSE data cubes of the NGC\,2207/IC\,2163 system exist in the ESO archive; these data are summarized in Table~\ref{tab:NGC2207_cubes}.  Field P1, which is centered on the nucleus of NGC\,2207 was surveyed as part of Program  0102.D-0095 (PI: J. Anderson); data for field P3 (which covers most of IC\,2163) and P2 (which connects P1 and P3) were obtained through Program ID: 0100.B-0116 (PI: C.M. Carollo).  As can be appreciated from the difference image in Figure~\ref{fig:NGC2207_FChart}, the system's copious diffuse emission, coupled with the dataset's relatively short exposure times and only moderate image quality, made PN detections challenging.  Nevertheless, the two galaxies are luminous enough so that, if the system were at the distance implied by the Tully-Fisher measurements, then a large number of PN candidates would be observable.

\begin{figure*}[th!]
\hspace{10mm}
\href{https://cloud.aip.de/index.php/s/55bgP2pc5wzYLxS}{\bf \colorbox{yellow}{Off-band}}
\hspace{50mm}\href{https://cloud.aip.de/index.php/s/QASfz4BAoHWkaEL}{\bf \colorbox{yellow}{Diff}}\\
\includegraphics[width=60mm,bb=0 0  700 800,clip]{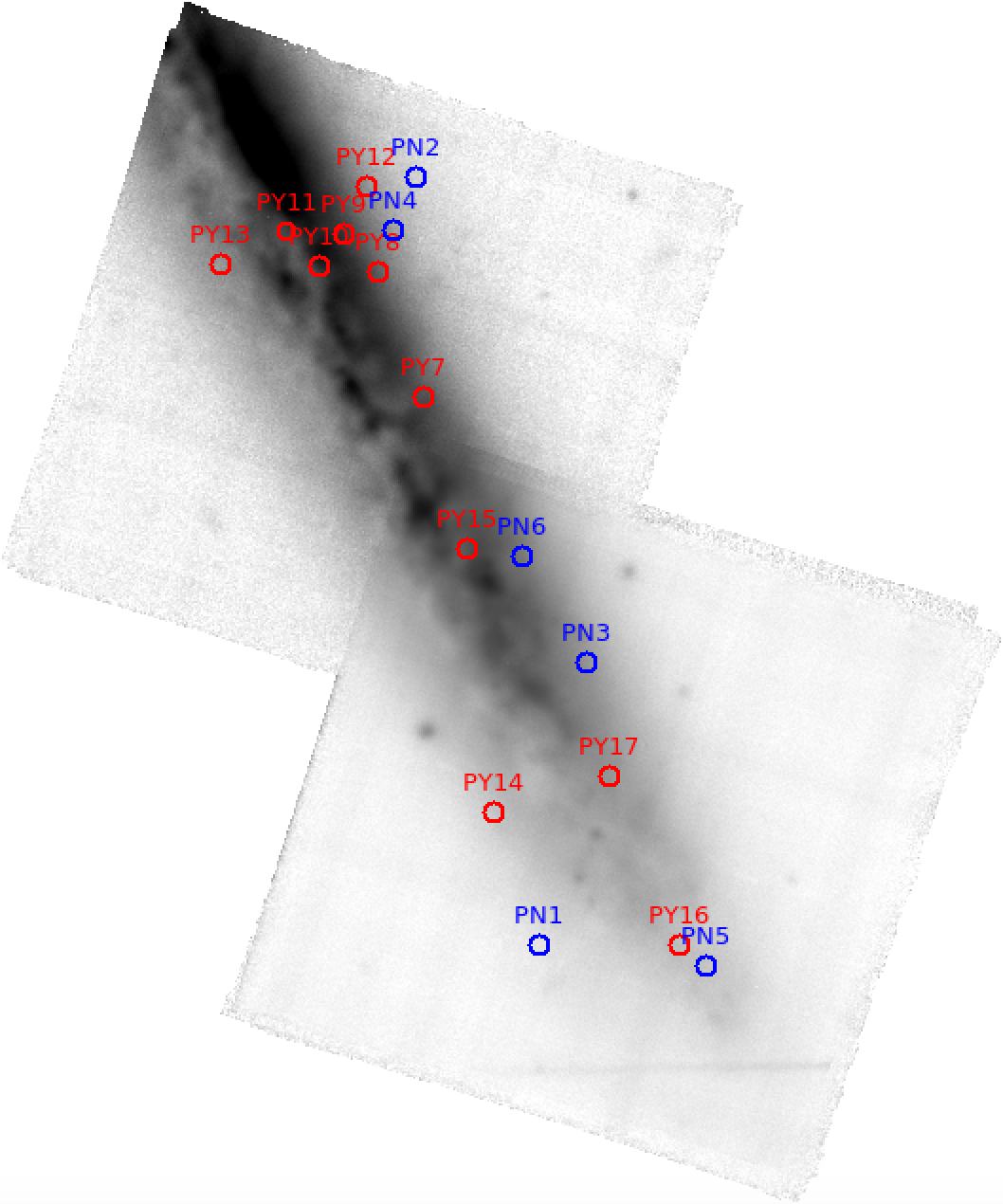}
\includegraphics[width=60mm,bb=0 0  700 800,clip]{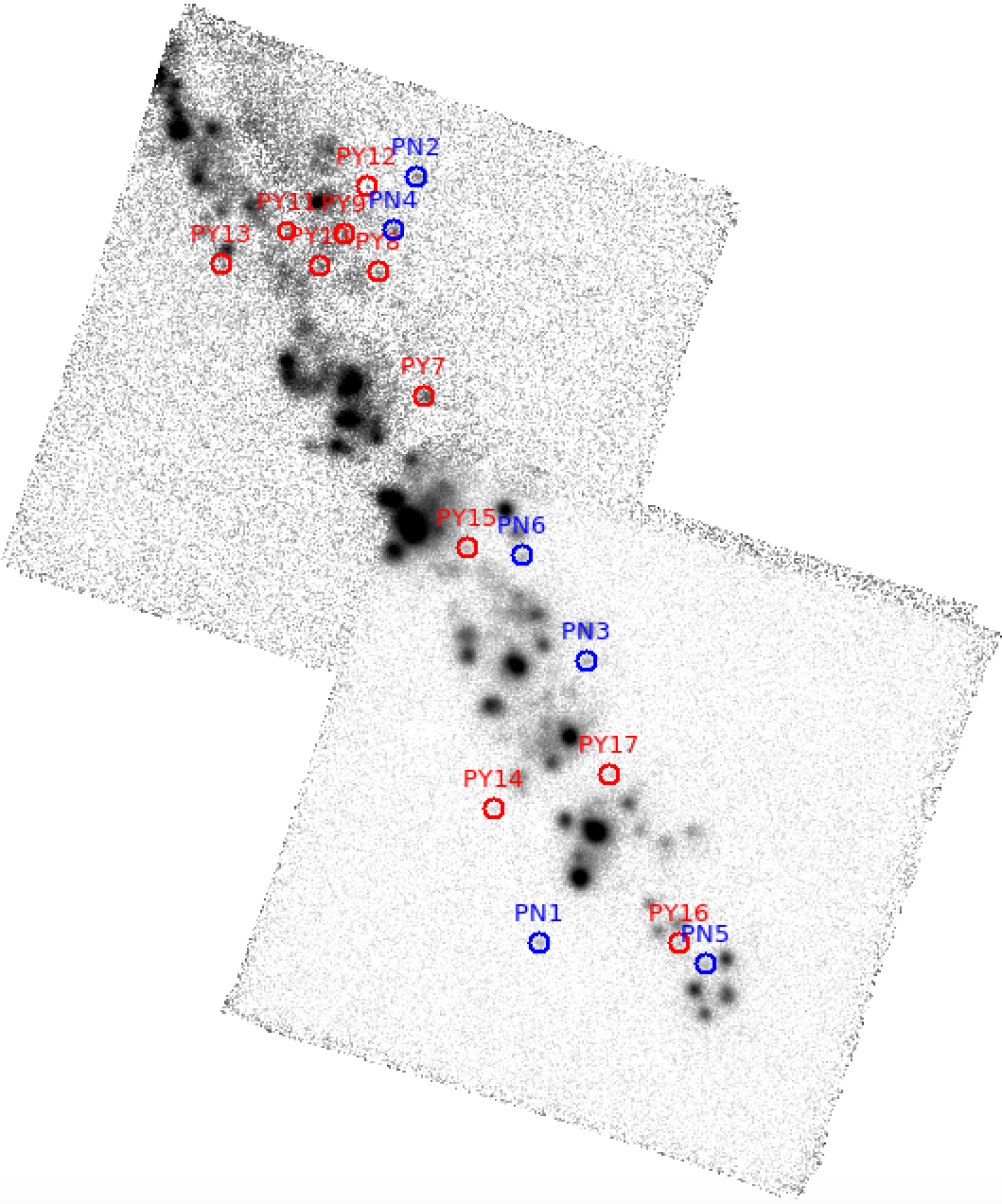}
\includegraphics[width=55mm,bb=50 0  600 600,clip]{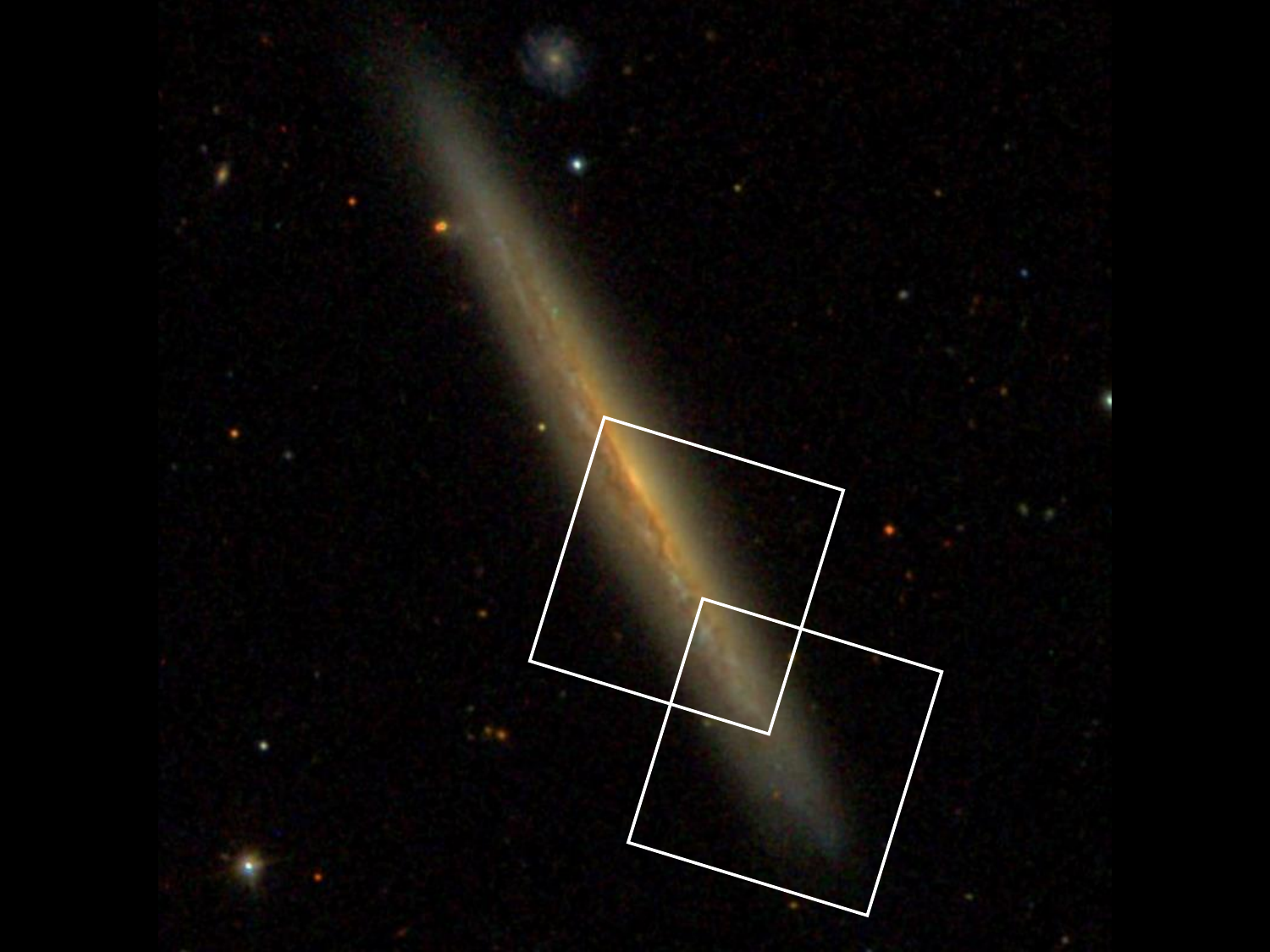}
\caption{NGC\,3501. Left: thumbnail off-band and difference images derived from the MUSE data cubes. High resolution images can be obtained by clicking on the hyperlink titles. \href{https://cloud.aip.de/index.php/s/jkQZ3PZ26NRtHfq}{\colorbox{yellow}{VIDEO}}. Right: a broadband image with the locations of the MUSE fields outlined in white. (Credit: SDSS DR14). \label{fig:NGC3501_FChart}}
\end{figure*}

Our initial examination of the MUSE data cubes identified 112 point-like \OIII sources in the two galaxies.  However, after screening the objects' spectra for evidence of strong H$\alpha$ (typical of an \ion{H}{2} region) and emission from [\ion{S}{2}] and [\ion{N}{2}] (the signature of a supernova remnant), only 3 of these PN candidates survived.  These potentially true PNe are all near the detection limit of the survey and have \OIII\ magnitudes of $m_{5007} = 28.77 \pm 0.13$, $29.70 \pm 0.28$, and $29.91 \pm 0.33$.  

If the brightest of the objects is indeed a PN with an absolute magnitude near $M^*$, then the implied distance to the galaxy is close to $\sim 40$~Mpc, in agreement with the analysis of SN 1975A. Scaling from Paper~I's survey of NGC\,474, an elliptical galaxy at a similar distance, a proper PNLF study of the system would likely require at least $\sim 5$~hours of MUSE exposure time with $\sim 0\farcs 6$ seeing.  With the present data, all we can say is that the PN observations are inconsistent with the galaxies' Tully-Fisher distances.

\subsection{NGC 3501 \label{subsec:NGC3501}}

NGC\,3501 is an edge-on spiral, with an uncertain Hubble type of Scd.   The galaxy has more than a dozen Tully-Fisher distance estimates in the literature, all within the range of 19.5 to 26.5~Mpc. However, the viewing angle presents an obvious challenge to standard candles such as Cepheids and the PNLF, and no TRGB distances have been published.  Nevertheless, it has been shown that \OIII observations in the halos of edge-on systems can overcome the issues associated with galaxy orientation and produce reliable PNLF distances \citep[e.g.,][]{Ciardullo+91, Jacoby+96}.  So MUSE has the potential to obtain a PNLF distance to the galaxy.

As depicted in Figure~\ref{fig:NGC3501_FChart}, the one NGC\,3501 MUSE data cube in the ESO archive (ID: ADP.2017-10-16T11:12:01.527, PI: F. Pinna, Program ID: 098.B-0662) consists of the combination of two pointings. The effective exposure time for these data is quoted as 9600~s, but the image quality of the observation was quite poor, $1\farcs 51$.  Nevertheless, the DELF technique did allow us to detect 6 of the system's brightest PNe, from an initial list of 17 candidates.  Such a sparse sample is not suitable for a PNLF analysis, though it can be used to place limits on a galaxy's distance.  

\begin{figure}[h!]
\centerline{
\includegraphics[width=0.4\textwidth,bb=0 0  600 500,clip]{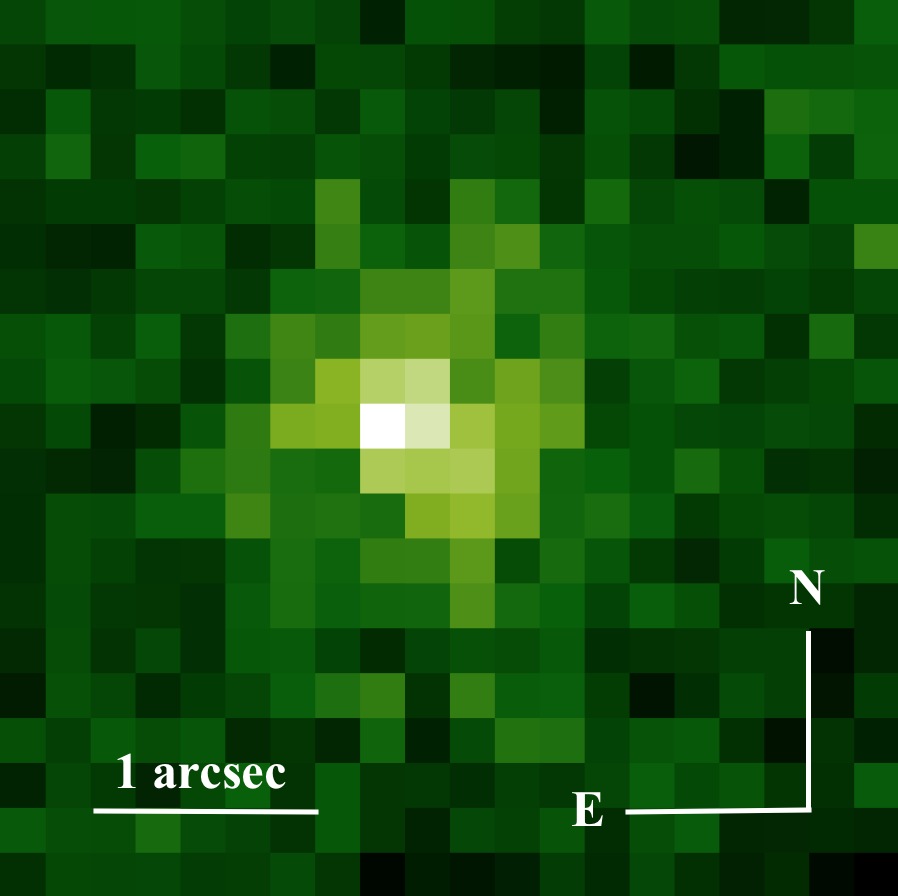}}
\caption{A possible overluminous PN in NGC\,3501. The PN may be a superposition of two \OIII emission-line sources with a separation of $0\farcs4$, but the poor ($1\farcs 51$) image quality precludes a definitive statement.
\label{fig:NGC3501_PN1}}
\end{figure}

It is worth noting that, although the magnitudes of the detected PNe range from $27.69 \leq m_{5007} \leq 28.47$, the difference between the brightest and second brightest PN is 0.54~mag. This gap may simply be due to the sparsity of the sample.  However, a careful examination of the \OIII line of the brightest PN hints at the possibility that the object is actually composed of a marginally resolved pair of objects; if so, the two objects are of comparable brightness and have a separation of $\sim 0\farcs4$ with a position angle of $125^\circ$ (Figure~\ref{fig:NGC3501_PN1}). Unfortunately, the poor image quality does not allow us to draw any definite conclusions about the object.

As with NGC\,2207, we can estimate a crude upper limit to NGC\,3501's distance using the magnitudes of the brightest PNe.  If we believe that the brightest \OIII source is the superposition of two objects and   adopt the second brightest object as our indicator, then, for a foreground reddening of $E(B-V)=0.02$, the upper limit on the galaxy's distance is $\sim 38$~Mpc.  Alternatively, if we believe the \OIII flux from PN1 comes from a single PN, then the upper limit drops to $\sim 33$~Mpc.  Either way, these estimates are well beyond the system's Tully-Fisher values. We emphasize that this estimate is for an upper limit on the distance, so it is still consistent with other measurements.  The only firm conclusion we can draw is that, under more favorable observing conditions, MUSE should be able to yield an accurate PNLF distance to the galaxy.


\begin{figure*}[!th]
\hspace{20mm}
\href{https://cloud.aip.de/index.php/s/dxnzGbzNfK9WC29}{\bf \colorbox{yellow}{Off-band}}
\hspace{48mm}\href{https://cloud.aip.de/index.php/s/pxcPK3aR5HPQJnz}{\bf \colorbox{yellow}{Diff}}\\
\includegraphics[width=58mm,bb=0 0  500 550,clip]{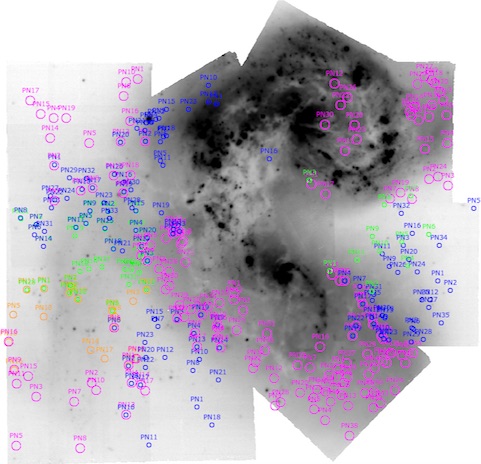}
\includegraphics[width=58mm,bb=0 0  500 550,clip]{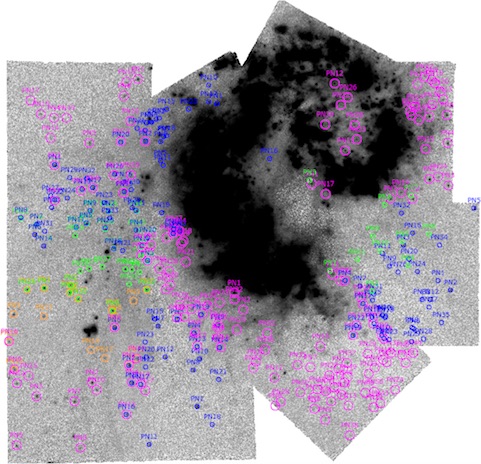}
\includegraphics[width=60mm,bb=150 100  600 500,clip]{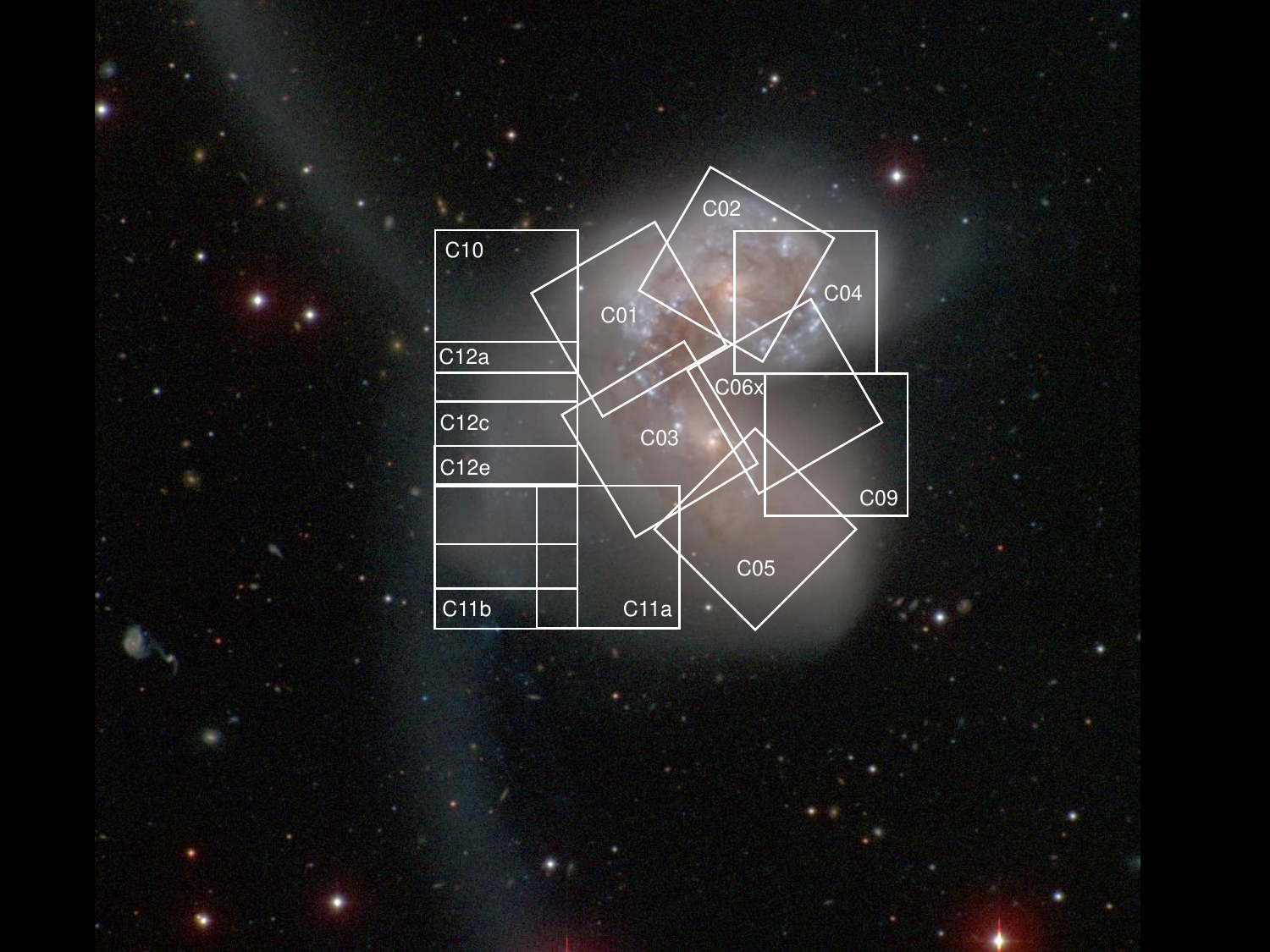}
\caption{NGC\,4038/39. Left: thumbnail off-band and difference images based on the 13 MUSE data cubes. The PN candidates are circled, with the colors the individual pointing (C01 through C12e).  High resolution images can be obtained by clicking on the hyperlink titles. \href{https://cloud.aip.de/index.php/s/ggQKjzSC8Kj8Ggr}{\colorbox{yellow}{VIDEO}}. Right: a broadband image with the positions of the data cubes outlined in white. 
 (Credit: CGS).  \label{fig:NGC4038_FChart}}
\end{figure*}

\subsection{NGC 4038/39 \label{subsec:NGC4038}}

The Antennae galaxy (NGC\,4038/NGC\,4039) is the closest, and therefore the prototypical, example of a major merger, and is well-known for its spectacular antennae-like tidal tails (hence its name) and vigorous starburst activity. A PNLF study of the Antennae is particularly interesting, both because of an earlier controversial distance determination for SN2007sr \citep{Schweizer+08}, and because it has been the focus of both TRGB and Cepheid observations. In principle, NGC~4038/39 can serve as a check point for four different distance indicators, although the PNLF and SN~Ia methods have ties to the Cepheid and/or TRGB scales and thus, are not fully independent.
\begin{deluxetable}{llcc}[h!]
\tablecaption{Data Cubes for NGC 4038/39
\label{tab:Antennae_cubes}  }
\tablehead{\colhead{Field} &\colhead{Archive ID} &\colhead{Exp Time}
&\colhead{Seeing} \\ [-0.25cm]
&&\colhead{(sec)} &\colhead{(5007~\AA)}}
\startdata
C01  & ADP.2017-03-28T13:08:20.713   & 4951 & $0\farcs 74$ \\
C02  & ADP.2017-03-28T13:08:20.697   & 5013 & $0\farcs 59$ \\
C03  & ADP.2017-03-28T13:08:20.689   & 4811 & $0\farcs 62$ \\
C04  & ADP.2017-03-28T13:08:20.681   & 4985 & $0\farcs 59$ \\
C05  & ADP.2017-03-28T13:08:20.705   & 5117 & $0\farcs 84$ \\
C06x & ADP.2016-06-15T08:55:13.262   & 2578 & $1\farcs 35$ \\
C09  & ADP.2017-05-24T12:39:01.277   & 2584 & $0\farcs 81$ \\
C10  & ADP.2016-09-29T05:21:54.086   & 2592 & $0\farcs 86$ \\
C11a & ADP.2017-05-24T11:10:28.457   & 2498 & $0\farcs 98$ \\
C11b & N/A\tablenotemark{\footnotesize a}   & 2523 & $0\farcs 78$ \\
C12a & N/A\tablenotemark{\footnotesize a}   & 2700 & $0\farcs 88$ \\
C12c & N/A\tablenotemark{\footnotesize a}   & 2700 & $0\farcs 75$ \\
C12e & N/A\tablenotemark{\footnotesize a}   & 2700 & $0\farcs 60$ \\
\enddata
\tablenotetext{a}{Not available in public archive, own data reduction (see text).}
\end{deluxetable}

NGC\,4038/39 has been observed extensively with MUSE to study the ionization mechanism of the diffuse ionized gas \citep{Weilbacher+18}.  As a result, there are numerous data cubes available in the ESO Archive (PI: P. Weilbacher, Program IDs: 093.B-0023, 095.B-0042). However, we chose to optimally re-reduce all the cubes with the most recent version of the MUSE pipeline \citep{Weilbacher+20}; this process removed many of the artifacts (such as H$\alpha$ saturation and high sky-line residuals) present in the prior reduction. The re-reduction also allowed us to produce a set of 13 independent data cubes without the complication of changing aperture corrections in the regions of field overlap.  These fields are listed in Table~\ref{tab:Antennae_cubes} and illustrated in Figure~\ref{fig:NGC4038_FChart}.

\begin{figure}[h!]
\includegraphics[width=0.473\textwidth]{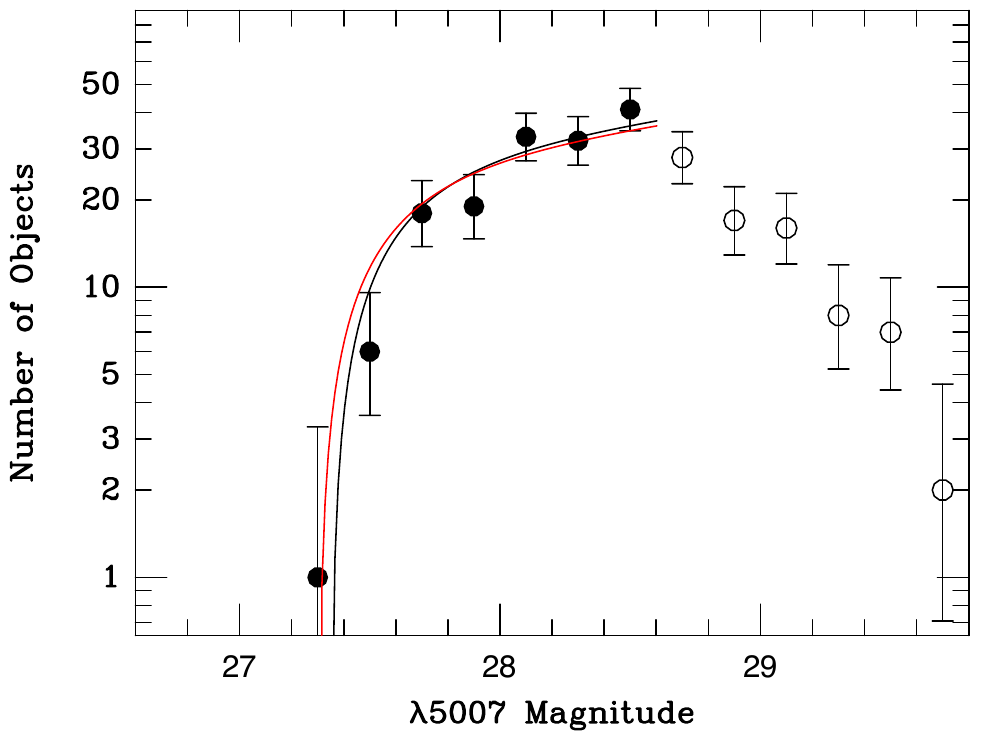}
\caption{The observed PNLF for the Antennae galaxy binned into 0.2~mag intervals.  Open circles denote data beyond the completeness limit; the error bars are from small number counting statistics \citep{Gehrels86}.  The red curve shows the best-fit to equation~(\ref{eq:pnlf}) when PN1 is included in the analysis; the black curve shows the best fit when this one object is excluded.  The two distance moduli differ by 0.048~mag.
 \label{fig:NGC4038_PNLF}}
\end{figure}

Our DELF analysis found 234 PN candidates scattered over the 13 fields, with almost a hundred in the top $\sim 1$~mag of the luminosity function.  In general, the spatial distribution of these PNe do not follow that of the light:  in the regions of intense star formation and high obscuration, PN detections are difficult, and little effort was made to perform a comprehensive search in those areas.  However, the large number of PNe found between and beyond the regions of star formation create a very well defined PNLF\null.   This function is shown in Figure~\ref{fig:NGC4038_PNLF}.  

As is illustrated in the figure, the distribution of observed PN magnitudes in the Antennae galaxy is generally well-fit by the empirical luminosity function described by equation~(\ref{eq:pnlf}).  But there is one exception: PN1, which is 0.14~mag more luminous than the second and third brightest objects, appears \textit{slightly} too bright for the best-fit curve.  Moreover, it is difficult to argue that the PN's \OIII flux is due to a superposition of two sources, as the location upon which the object is projected is not especially bright.  Specifically, in terms of underlying surface brightness, the position of PN1 barely falls in the top $\sim 1/3$ of the sample.  But its existence forces the best-fit curve towards smaller distances.  

\begin{figure}[ht]
\includegraphics[width=0.473\textwidth]{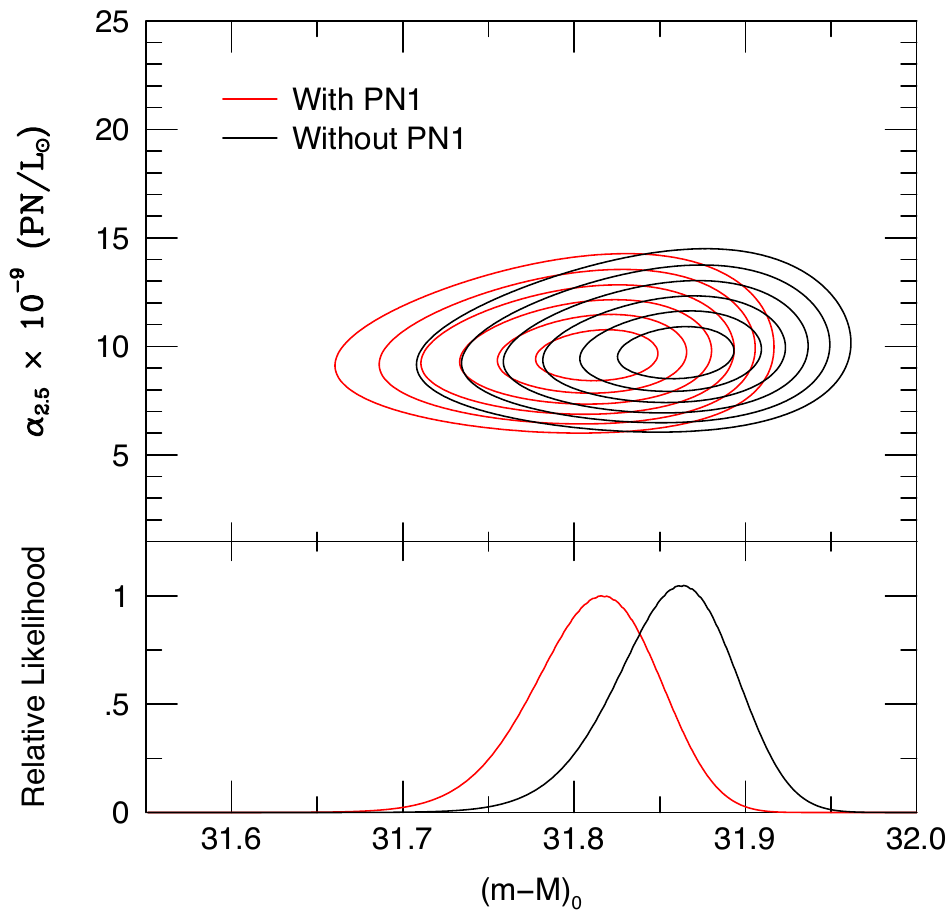}
\caption{The top panel shows the results of the maximum likelihood solution for the Antennae Galaxy. The abscissa is the galaxy's true distance modulus, the ordinate is $\alpha_{2.5}$, the number of PNe within 2.5~mag of $M^*$, normalized to the amount of bolometric light sampled.  The contours are drawn at $0.5\sigma$ intervals. The red contours are for the analysis that includes PN1, while the black contours show the likelihoods without the object. As the marginalization in the bottom panels show, the difference in the solutions is 
$\sim 0.05$~mag.
\label{fig:NGC4038_contours}}
\end{figure}

The effect of PN1 is quantified in Figure~\ref{fig:NGC4038_contours}, where we fit equation~(\ref{eq:pnlf}) to the observed distribution of PN magnitudes with and without PN1. For the analysis, we assumed that the line-of-sight velocity dispersion of the galaxy's stars is $\sim 150$~km~s$^{-1}$ throughout the MUSE survey region.  Given the complexity of the system and the lack of any spatially-resolved stellar radial velocity measurements, this was the simplest assumption we could make.  Similarly, due to the irregular isophotes of the galaxy, our estimates for the amount of galaxy luminosity underlying the position of each PN were made from the MUSE spectra themselves, rather than any published photometry.  These data imply that, after excluding those regions of the galaxy where PN detections were impossible, the 13 pointings shown in Fig.~\ref{fig:NGC4038_FChart} contain $V \sim 10.8$~mag of galaxy light.

Figure~\ref{fig:NGC4038_contours} confirms that, due to the large number of PNe in the top $\sim 1$~mag of the luminosity function, the formal errors of the PNLF fits are quite small.  If PN1 is excluded, the distance modulus of the system (for an $E(B-V) = 0.04$) is $(m-M)_0 = 31.86_{-0.04}^{+0.03}$, or $23.6 \pm 0.4$~Mpc.  In other words, the statistical error for the distance is less than 2\%.  However, if PN1 is included, then the Antennae galaxy's most-likely distance modulus decreases by 0.048~mag, a value that is greater than the statistical error of the fit. The fit with PN1 is slightly worse than that without the object, but not so much that it can be deemed inconsistent with the shape of equation~(\ref{eq:pnlf}).  Since both solutions are easily allowed by the Kolmogorov-Smirov statistic, we prefer keeping PN1 in the dataset.  The distance is then $(m-M)_0 = 31.82_{-0.04}^{+0.03}$, or $23.1 \pm 0.4$~Mpc, where the uncertainties represent only the internal errors of the fits.

\begin{figure*}[t]
\hspace{18mm}
\href{https://cloud.aip.de/index.php/s/dWpQLYmKLqBM77w}{\bf \colorbox{yellow}{Off-band}}
\hspace{47mm}\href{https://cloud.aip.de/index.php/s/DWQxtirFbJ5HwP6}{\bf \colorbox{yellow}{Diff}}\\
\includegraphics[width=60mm,bb=0 0  800 850,clip]{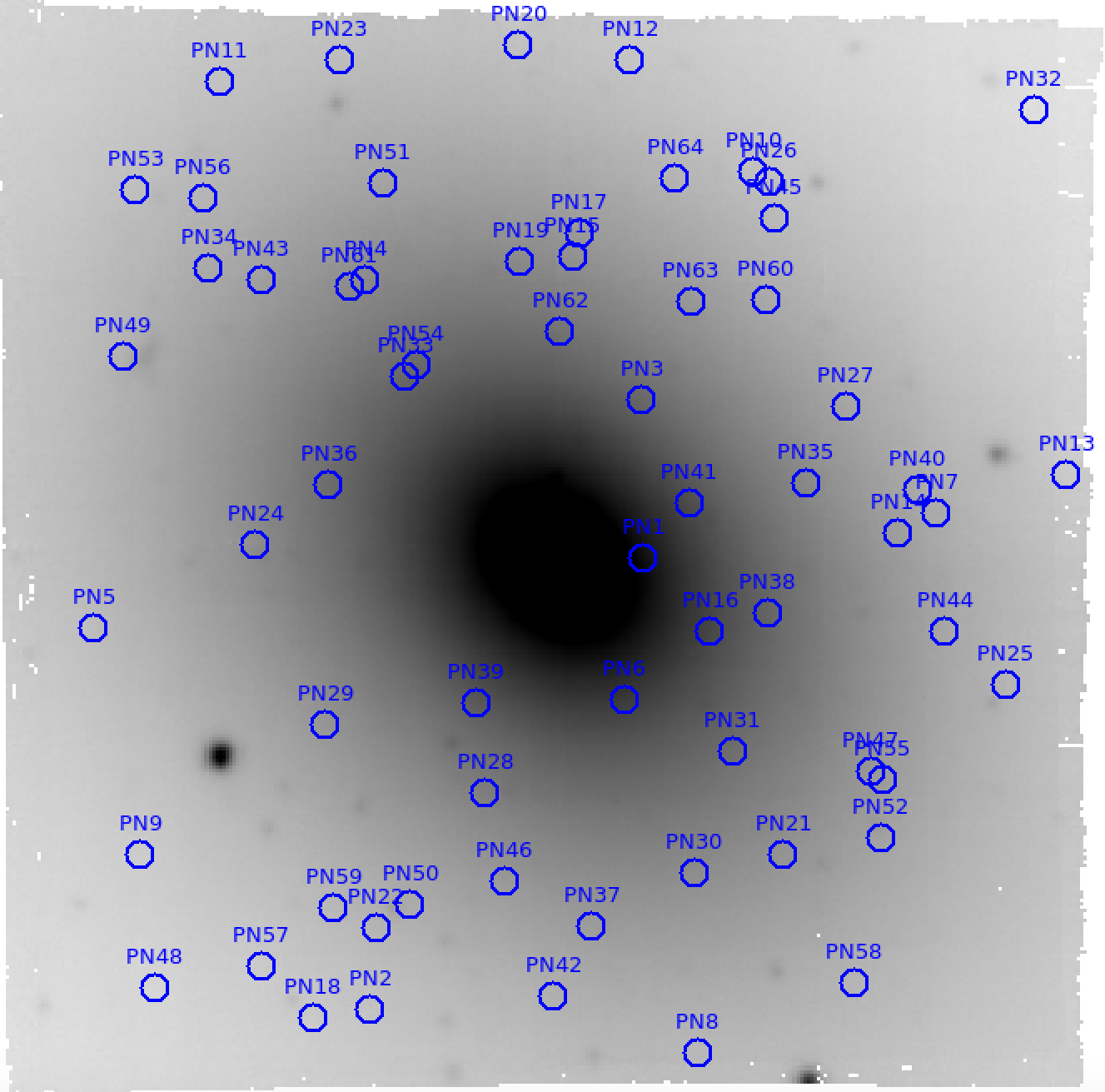}
\includegraphics[width=60mm,bb=0 0  800 850,clip]{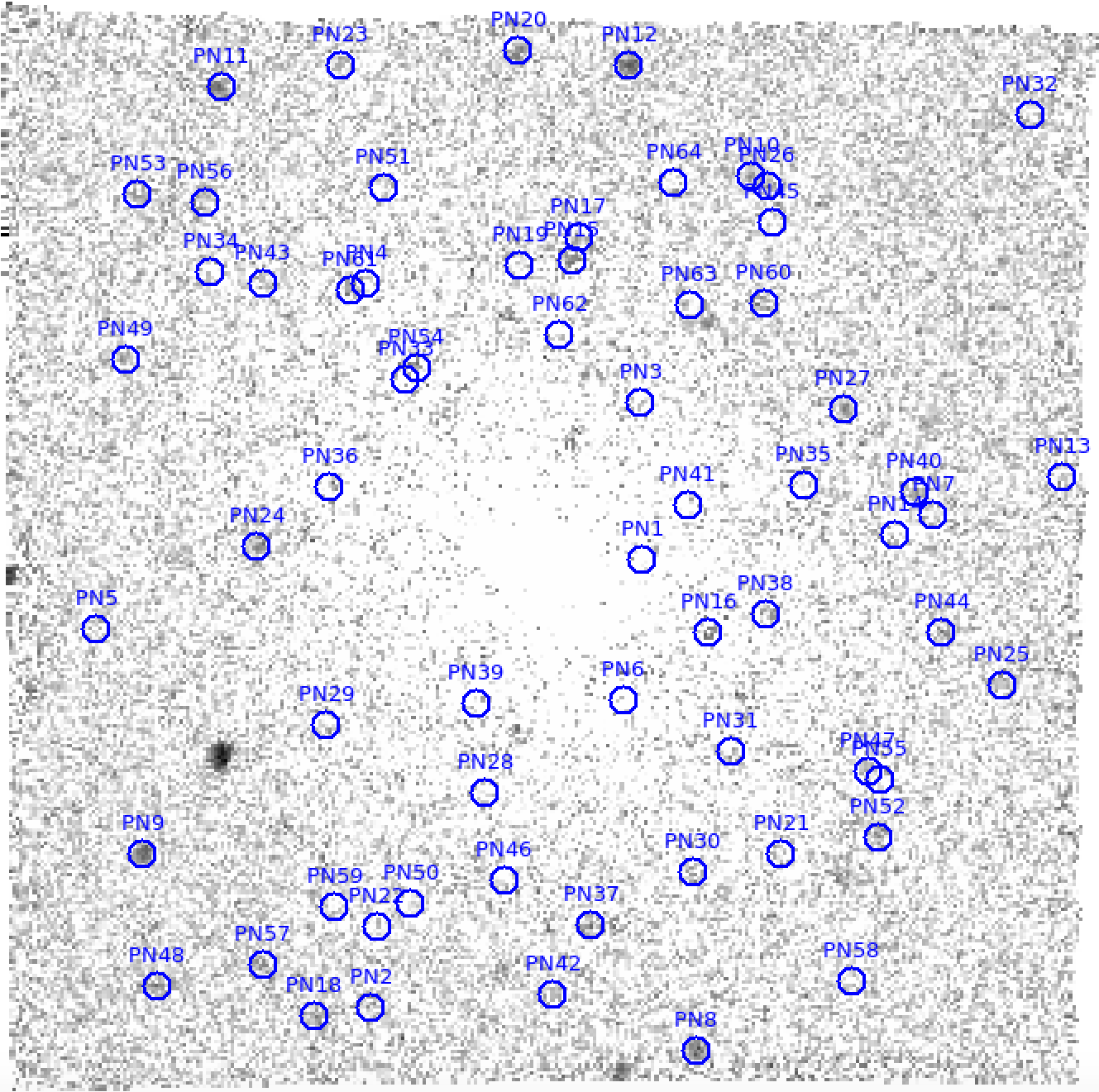}
\includegraphics[width=60mm,bb=100 50  600 450,clip]{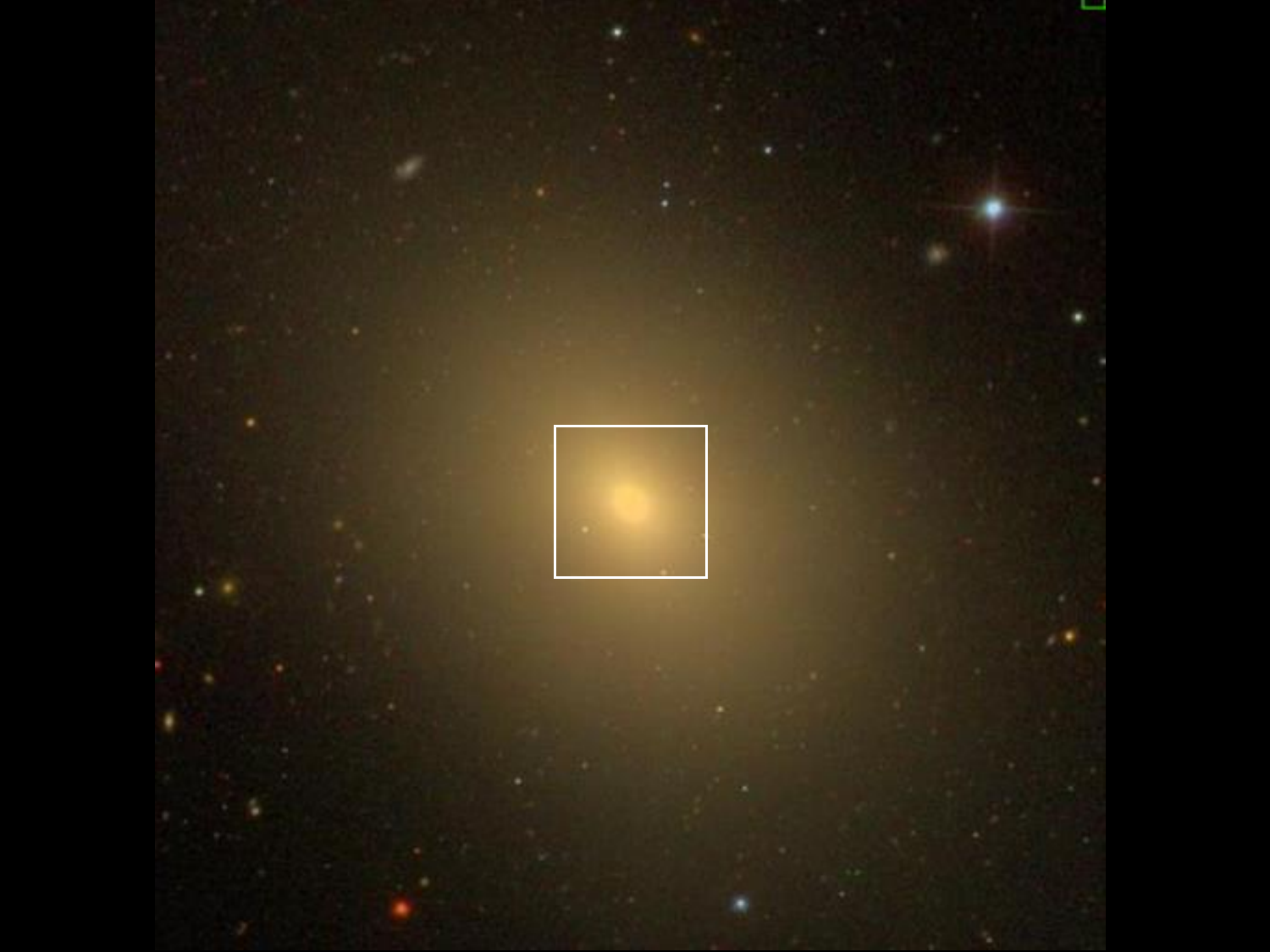}
\caption{NGC\,4365. Left: thumbnail off-band and difference images derived from the MUSE data cube. The blue circles show the positions of the PN candidates.  High-resolution images can be obtained by clicking on the hyperlink titles. \href{https://cloud.aip.de/index.php/s/FGnqWcZ3cHaDGCJ}{\colorbox{yellow}{VIDEO}}. Right: a broadband image with the MUSE field outlined in white. (Credit: SDSS DR14).  \label{fig:NGC4365_FChart}}
\end{figure*}

Since the Antennae system was observed using 13, sometimes overlapping MUSE pointings, the analysis of the systematic component of the distance determination's error budget is complicated.  Most of the PNe that occupy the brightest 0.25~mag of the PNLF lie in fields C10, C5, C9, and C1. None of these fields have any anomalies in the aperture correction estimates: formally, the uncertainties in their aperture corrections are 0.029, 0.010, 0.045, 0.001~mag, respectively.  Conversely, the C4 pointing contained no suitable point-like object for an aperture correction measurement; the best we could do is stack the field's brightest PNe and determine the aperture correction for \OIII $\lambda 5007$ as explained in Paper~I\null. This procedure likely has an uncertainty of $\sim 0.05$~mag. For pointing C11b, which also contained no useful PSF stars, we took a different tact:  since the seeing of the data cube was extremely similar to that pointing C1, we simply used the latter's aperture correction in our analysis.  Fortunately, the PNe contained on the pointings with the poorest aperture correction measurements do not contribute significantly to the PNLF's bright-end cutoff.  We, therefore, adopt $\sim 0.06$~mag as the combined zero-point uncertainty due to the MUSE data cube's flux calibration and aperture correction.

Our final distance to NGC\,4038/39 is $(m-M)_0 = 31.82_{-0.07}^{+0.07}$ (or $23.1_{-0.8}^{+0.7}$~Mpc) for an $E(B-V) = 0.04$.  This distance is statistically identical to the SN~Ia-based value of $(m-M)_0 = 31.74 \pm 0.27$ \citep{Schweizer+08}, and is only slightly larger (but still consistent with) the system's TRGB distances, which fall near $31.67 \pm 0.05$ \citep{Jang+15, Freedman+19}.  Compared to the system's Cepheid distances, our value is much larger than the discordant estimate of $31.29\pm0.11$ by \citet{Riess+16}, but is in reasonable agreement with the measurements of $(m-M)_0 = 31.55\pm0.06$ \citep{Fiorentino+13} and $31.615\pm0.117$ \citep{Riess+22}.  These results suggest that PN1 does belong in the sample, and the 0.14~mag offset between PN1 and PN2 is merely a statistical fluke, possibly due to the zero-point differences between the different MUSE paintings. 

One final issue to keep in mind is that some of the dust associated with the NGC\,4038/39's starburst regions may have been forced outward over the body of the galaxy by the tidal forces associated with the interaction.  In such a scenario, this dust could extinct some of the PNe and cause an overestimate of the galaxy's distance.  The fact that our PNLF distance to the system is slightly larger than those produced by the TRGB and Cepheid techniques lends support for this internal extinction hypothesis.


\subsection{NGC 4365 \label{subsec:NGC4365}}

The giant E3 elliptical galaxy NGC\,4365 is the central member of Virgo's W$^\prime$ group, roughly 6~Mpc behind the system's central core \citep{Blom+14}. The galaxy is a popular target for studies of extragalactic globular clusters, and it has been analyzed kinematically with the aid of the Planetary Nebula Spectrograph \citep{Pulsoni+18}.  The published SBF distances to the galaxy span the range $16.9 < D < 24.4$~Mpc, with the latest SBF measurement placing the galaxy at 23.1~Mpc \citep{Blakeslee+09}. 

There is a single MUSE data cube of NGC\,4365 in the ESO archive (ID: ADP.2016-06-07T11:11:26.095, PI: L. Coccato, Program ID: 094.B-0225).  The exposure, which is centered on the galaxy's nucleus (see Figure~\ref{fig:NGC4365_FChart}) has an exposure time of 2343~s, and an image quality of $0\farcs 76$ at \OIII $\lambda 5007$.

Our DELF analysis initially found 115 point-like \OIII sources in the MUSE data cube; however, upon closer inspection of their spectra, 51 were excluded as either spurious objects (i.e., too faint to meet the detection criteria) or some type of interloper.  That left us with 64 PN candidates, with $\sim 50$ in the top $\sim 1$~mag of the PNLF\null.

As illustrated in Fig.~\ref{fig:NGC4365_FChart}, only five of these detections were made in the galaxy's central $\sim 12\arcsec$, an area that contains $\sim 35\%$ of the light falling on the MUSE frame.  This is not unexpected, as faint-object detections are severely incomplete against a bright, rapidly varying background. We therefore eliminate the region and its PNe from the analysis.  

Interestingly, one of the excluded objects is PN1, a source that is 0.37~mag brighter than any other PN candidate in the galaxy.  PN1 is located just $5\farcs 1$ from NGC\,4365's nucleus, on an isophote with an $R$-band surface brightness of 17.2~mag~arcsec$^{-2}$ \citep{Lauer1985}.  If we assume a color of $(V-R) = 0.60$ \citep{Buta+95} and a distance of 24~Mpc \citep{Blakeslee+09}, then equations 3, 4, and 5 of \citet{Chase+23} imply that there is a $\sim 30\%$ chance that the observed \OIII flux is formed from the emission of multiple sources. In fact, a careful inspection of the narrow-band images made from the MUSE data cube reveals that the position of PN1 shifts slightly (by $0\farcs 34$) between the wavelengths of 5021.5~\AA\ and 5024.0~\AA\null.  This supports the hypothesis that the \OIII flux of PN1 is produced by multiple superposed sources.   Thus, the extreme luminosity of the source is not surprising and further justifies its rejection in the PNLF analysis.

\begin{figure}[t!]
\includegraphics[width=0.473\textwidth]{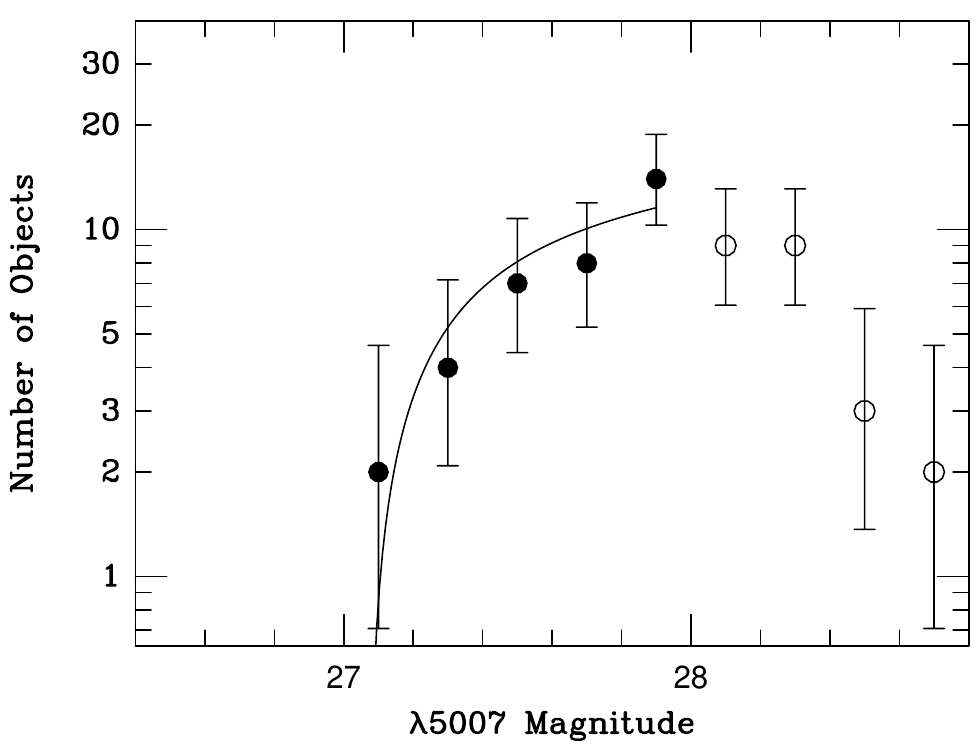}
\caption{The observed PNLF for NGC\,4365 binned into 0.2~mag intervals.  Open circles denote data beyond the completeness limit; the error bars reflect the small number counting statistics \citep{Gehrels86}. The black curve shows equation~(\ref{eq:pnlf}) shifted by the best-fit apparent distance modulus of $(m-M) = 31.61$. \label{fig:NGC4365_PNLF}}
\end{figure}

The luminosity function of PNe outside the central $12\arcsec$ of NGC\,4365 is shown in Figure~\ref{fig:NGC4365_PNLF}.  The function is very well defined, and shows no obvious deviation from the empirical law of equation~(\ref{eq:pnlf}): the PN number counts decline rapidly to zero at magnitudes brighter than $m_{5007} \sim 27.2$, and there is no evidence for the existence of any ``overluminous'' objects.  Although the data are only complete over the brightest $\sim 0.8$~mag of the luminosity function, the data produce a very precise measure of distance.

\begin{figure}[t!]    \includegraphics[width=0.473\textwidth]{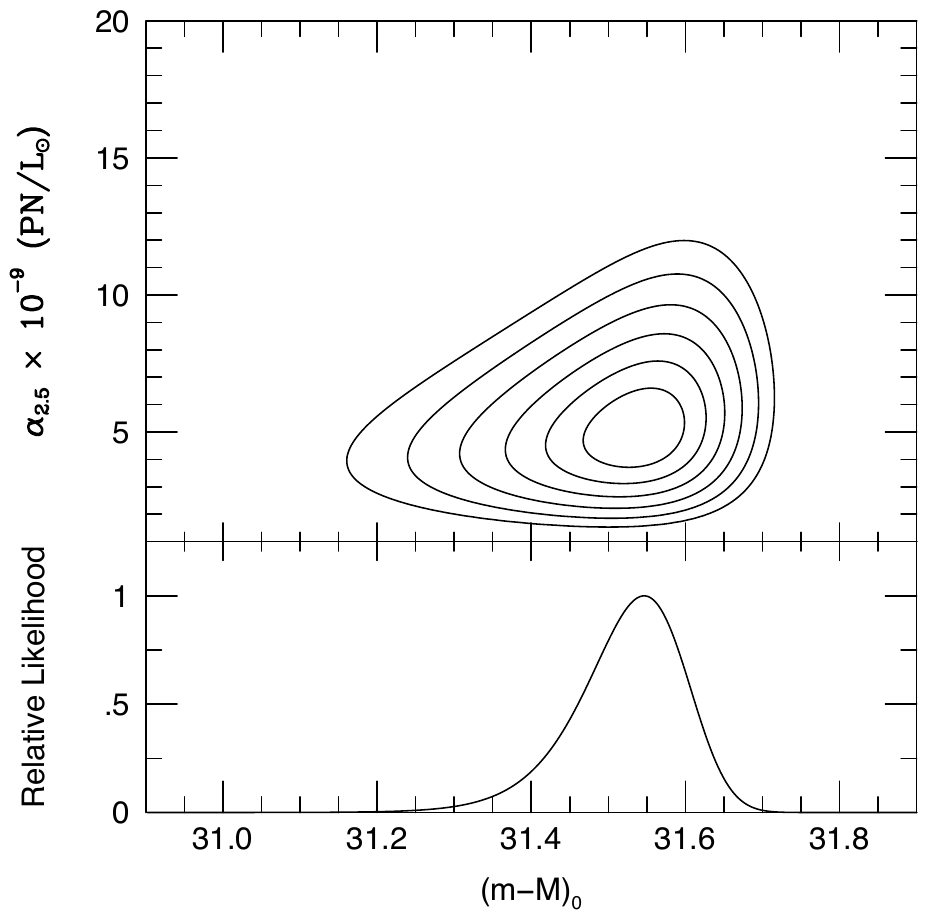}
\caption{The upper panel shows the results of the maximum likelihood solution for NGC\,4365. The abscissa is the galaxy's true distance modulus, the ordinate is the number of PNe within 2.5~mag of $M^*$, normalized to the amount of bolometric light sampled, and the contours are drawn at $0.5\sigma$ intervals. The lower panel marginalizes over the PN/light variable.  
\label{fig:NGC4365_contours}}
\end{figure}

To fit the PN magnitudes of Fig.~\ref{fig:NGC4365_PNLF}, we used the surface photometry of \citet{Lauer1985} and \citet{Michard+88}, and the kinematic data of \citet{Foster+16} to estimate the amount of galaxy light and the line-of-sight velocity dispersion at every location in the data cube.  The former measurements imply that the MUSE survey contains $V \sim 10.8$ of galaxy light; the latter quantities range from $\sim 225$ to 207~km~s$^{-1}$.  (Both values exclude the central $12\arcsec$ of the galaxy.)  We then applied the algorithms of \citet{Chase+23} to derive the likelihood contours of Figure~\ref{fig:NGC4365_contours}.  Assuming a foreground reddening of $E(B-V) = 0.02$, the resultant distance modulus to the system is $(m-M)_0 = 31.55_{-0.08}^{+0.05}$.  

\begin{figure*}[t!]
\hspace{16mm}
\href{https://cloud.aip.de/index.php/s/cLLCEBFTs5gxCNd}{\bf \colorbox{yellow}{Off-band}}
\hspace{50mm}\href{https://cloud.aip.de/index.php/s/DWQxtirFbJ5HwP6}{\bf \colorbox{yellow}{Diff}}\\
\includegraphics[width=60mm,bb=0 0  800 850,clip]{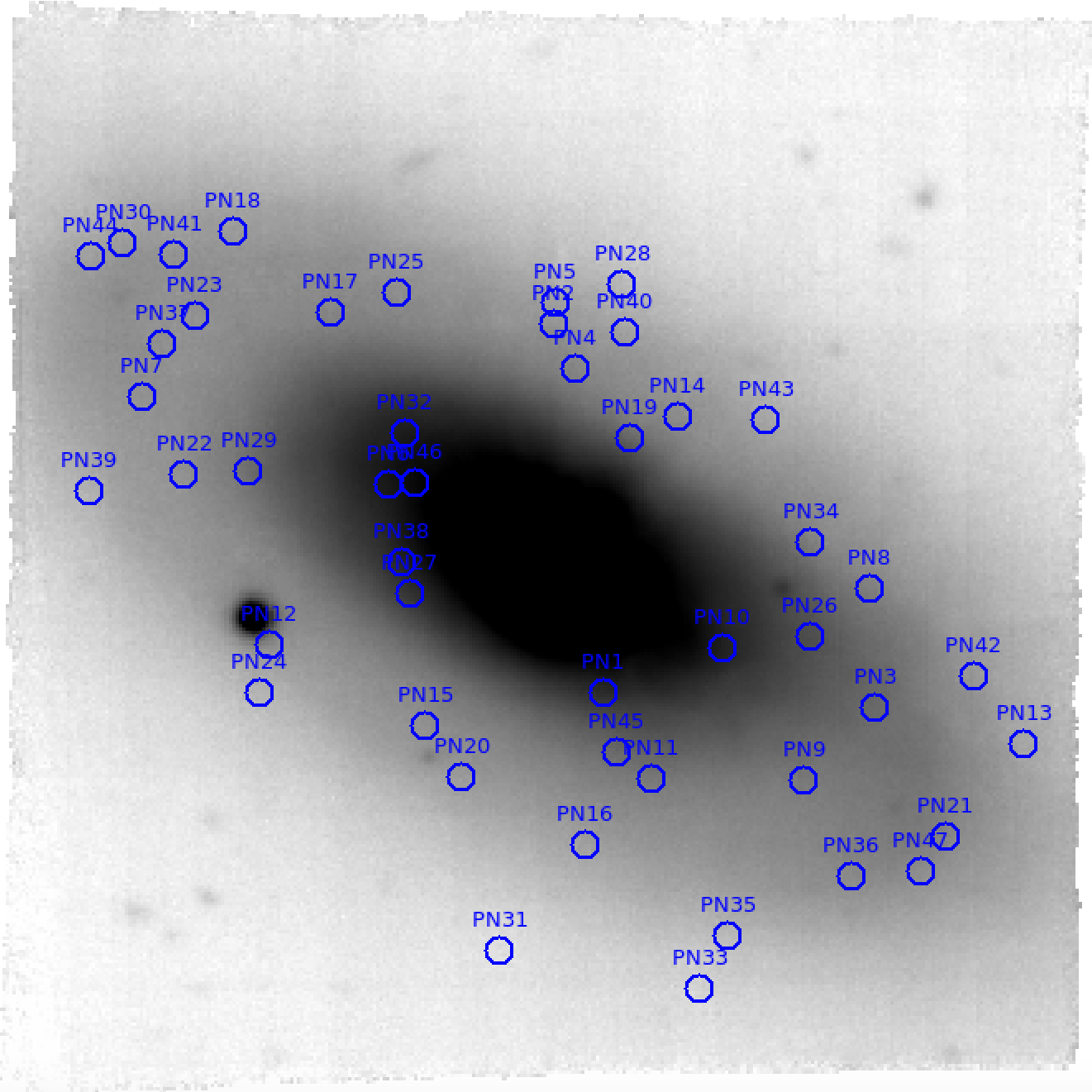}
\includegraphics[width=60mm,bb=0 0  800 850,clip]{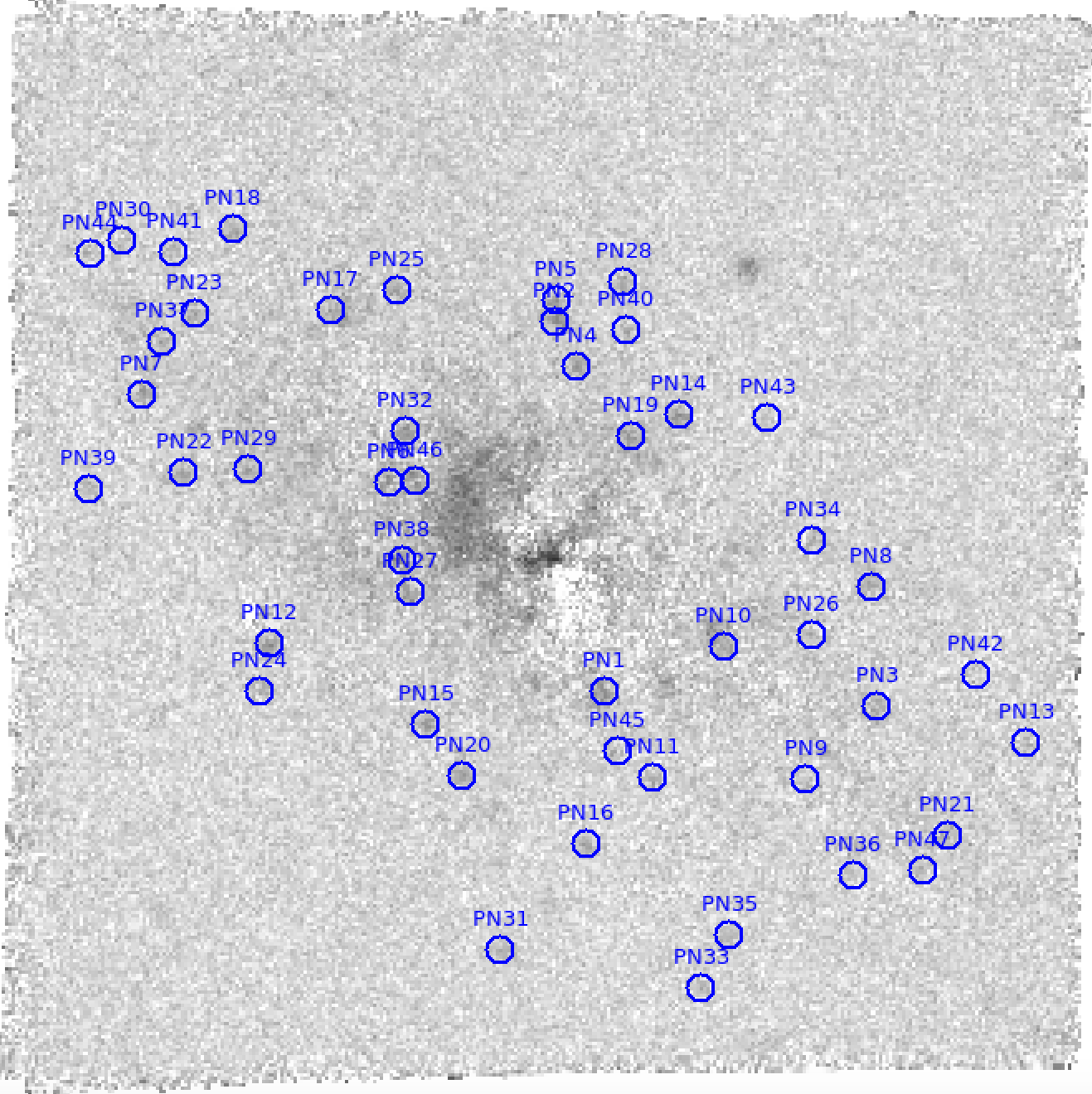}
\includegraphics[width=58mm,bb=220 150  500 390,clip]{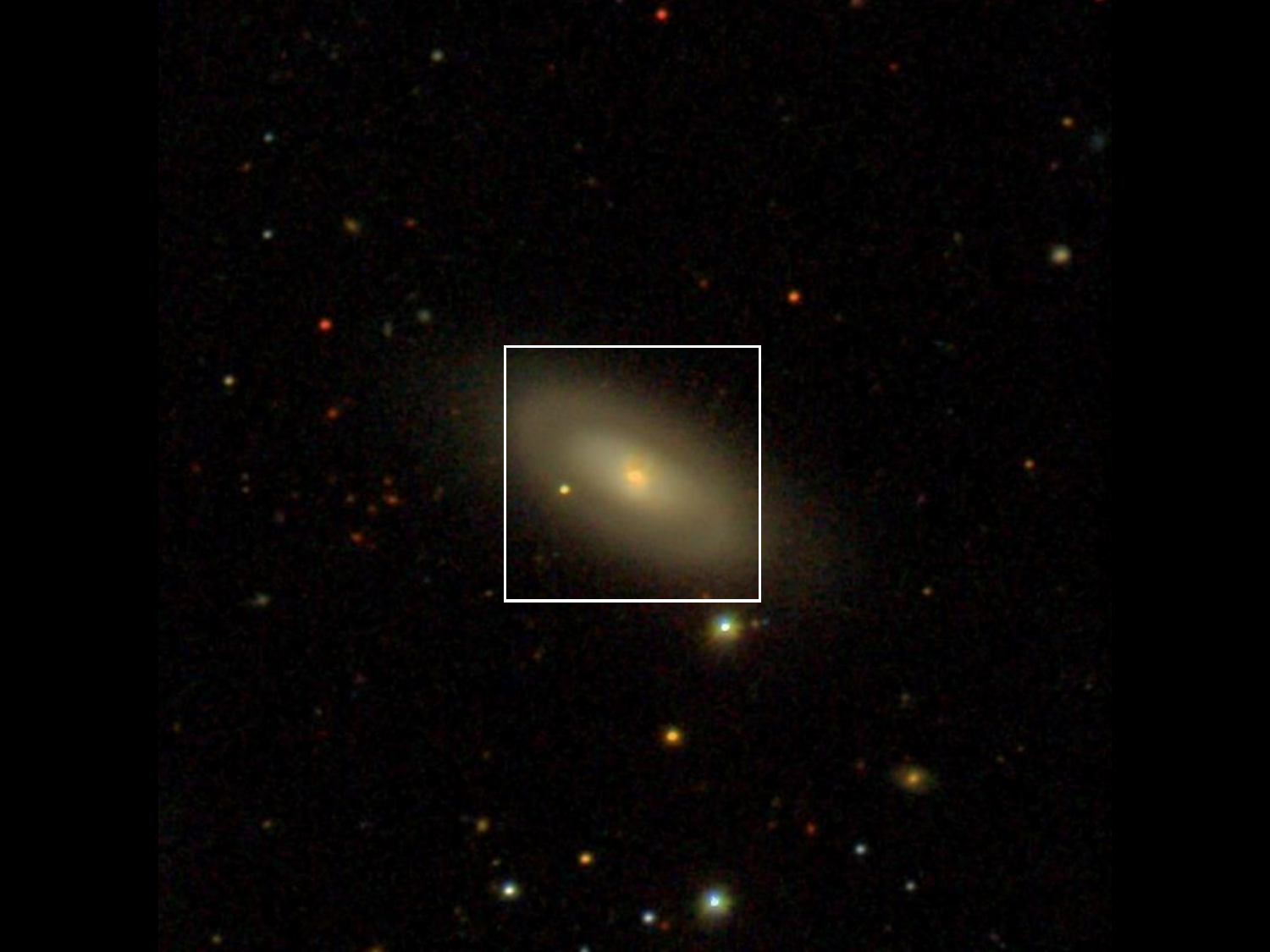}
\caption{NGC\,4418. Left: thumbnail off-band and difference images derived from the MUSE data cube. The blue circles show the positions of our PN candidates.  High resolution images can be obtained by clicking on the hyperlink titles. \href{https://cloud.aip.de/index.php/s/3JLi5ZLa9KaSPAa}{\colorbox{yellow}{VIDEO}}. Right: a broadband image with the location of the MUSE data cube outlined in white.  (Credit: SDSS DR14). 
\label{fig:NGC4418_FChart}}
\end{figure*}

The systematic component to NGC\,4365's error budget is quite low, as the data cube contains a bright and well-behaved PSF star.  (The formal error on the aperture correction is just 0.002~mag.)  If we assume a nominal 3\% error due to the MUSE data cube's flux calibration, then the distance modulus to the galaxy becomes $(m-M)_0 = 31.55_{-0.09}^{+0.06}$ or $20.4_{-0.8}^{+0.6}$~Mpc.  This value is on the low end of the range of SBF results, but consistent with the known zero-point offset between the two distance scales, which can be as high as 15\% \citep{Ciardullo+02}.

\vspace{10mm}


\begin{figure}[h!]
\includegraphics[width=0.473\textwidth]{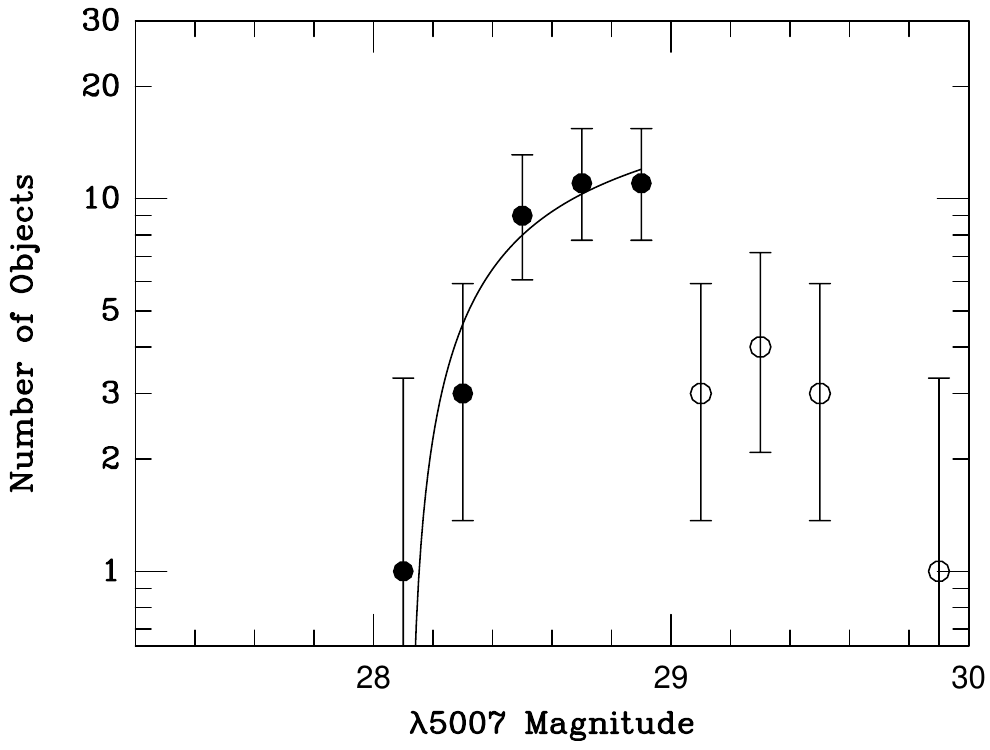}
\caption{The observed PNLF for NGC\,4418 binned into 0.2~mag intervals.  Open circles denote data beyond the completeness limit; the error bars show the small-number counting statistics \citep{Gehrels86}, and the curve shows equation~(\ref{eq:pnlf}) shifted by the best-fit apparent distance modulus of $(m-M) = 32.66$.  
\label{fig:NGC4418_PNLF}} 
\end{figure}

\subsection{NGC 4418 \label{subsec:NGC4418}}

NGC\,4418 (also known as NGC\,4355 and IRAS\,12243-0036) is a lenticular galaxy (Hubble type S0/a), with unusually luminous infrared emission and a mid-IR silicate absorption features that indicates a bright, but deeply obscured nucleus. The nature of the nuclear region (starburst or AGN) is unclear, but evidence from radio data suggests the presence of a compact super star cluster with intense star formation \citep{Varenius+14}.  The only distances available to the system come from a single Tully-Fisher analysis, which places the galaxy $\sim 22$~Mpc away \citep{Theureau+07}.   

\begin{figure}[h!]
\includegraphics[width=0.473\textwidth]{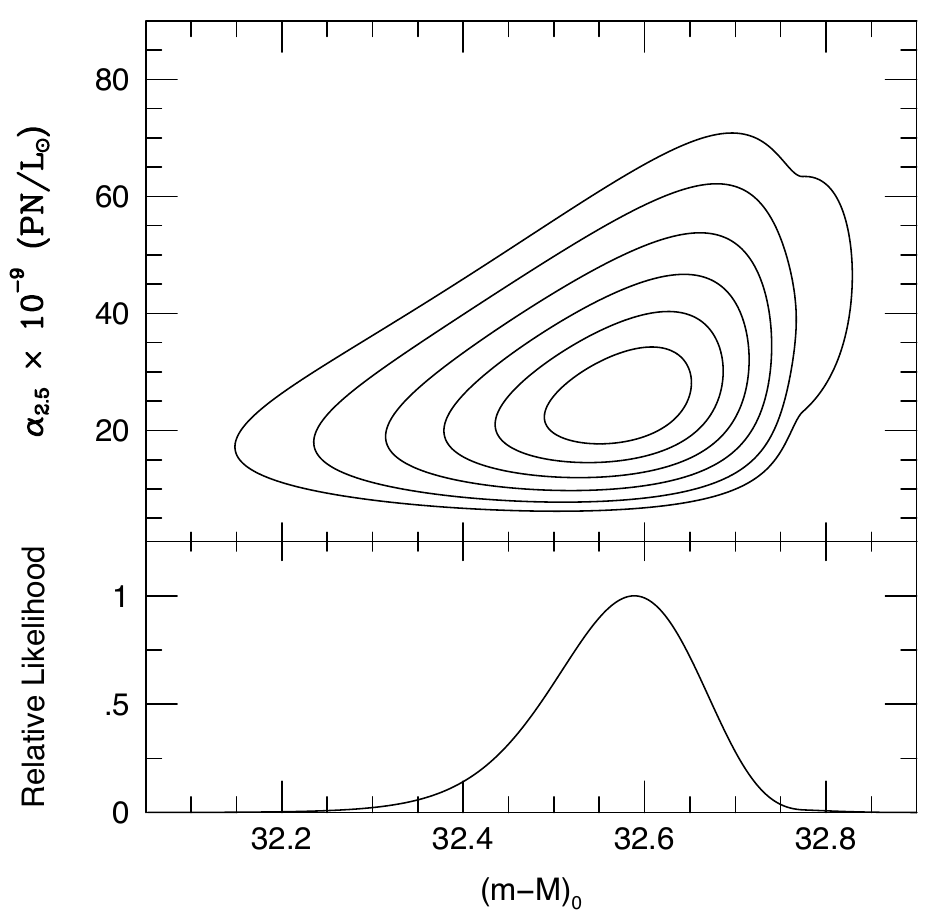}
\caption{The top panel shows the results of the maximum likelihood solution for NGC\,4418. The abscissa is the galaxy's true distance modulus, the ordinate is $\alpha_{2.5}$, the number of PNe within 2.5~mag of $M^*$, normalized to the amount of bolometric light sampled.  The contours are drawn at $0.5\sigma$ intervals.  There is a slight possibility that the galaxy's brightest PN candidate is the superposition of two sources; this is reflected in the ``bump'' in the outermost likelihood contour.
\label{fig:NGC4418_contours}} 
\end{figure}

One promising MUSE-DEEP data cube exists in the ESO archive (archive ID: ADP.2020-02-13T10:53:27.677, PI: F. Stanley, Program ID: 0104.B-0668); these data have an exposure time of 5999~s, an image quality of $0\farcs 89$ at 5007~\AA, and a position centered on the galaxy's nucleus (see Figure~\ref{fig:NGC4418_FChart}).  Our examination of the DELF images created from the data cube resulted in a sample of 56 PN candidates, of which 47 survived spectral examination.  

The locations of NGC\,4418's PN candidates are shown in Figure~\ref{fig:NGC4418_FChart}.  Within the $11\arcsec$ semi-major axis isophote, there are no PN candidates; this reflects the effect that the region's high surface brightness has on PN detections.  The PNLF of the objects outside this radius are shown in Figure~\ref{fig:NGC4418_PNLF}.

NGC\,4418's PNLF is not as well populated as some of the other galaxies in this study.  The data start to become incomplete $\sim 0.7$~mag down the PNLF, and there are only $\sim 24$~PN detections brighter than this limit.  Consequently, stochasticity is an issue.  For example, PN1 is 0.27~mag brighter than any other planetary.  In galaxies with a well-population PNLF, such a gap might be indicative of an anomaly.  Here, it is perfectly consistent with expectations.  In fact, a Kolmogorov-Smirnov test prefers a fit that includes PN1.

NGC\,4418 has no published surface photometry, nor are there any measurements of the disk's line-of-sight velocity dispersion.  We compute the former using continuum measurements from the MUSE spectra; after excluding the galaxy's inner $\sim 11\arcsec$, we estimate that $V \sim 13.6$ of light is contained in the survey region.  For the latter we adopt 75~km~s$^{-1}$ throughout the galaxy;  this number is much less than the minimum velocity separation needed to deblend two sources and is consistent with expectations for the velocity ellipsoid of a disk galaxy. 

Figure~\ref{fig:NGC4418_contours} displays the likelihood contours for NGC\,4418. The distortion of the lowest-level contours reflects the possibility that PN1 may be a superposition:  the galaxy continuum underlying the PN is relatively bright and the object is the most luminous PN in the sample.  However, as the marginalized probability curve in the lower panel of the figure indicates, the possibility has almost no effect on the most-likely distance to the galaxy: for a foreground reddening $E(B-V) = 0.02$ \citep{Schlafly+11} we obtain $(m-M)_0 = 32.59_{-0.10}^{+0.07}$.  When we include the small systematic component to the error budget (as shown in Fig.~\ref{fig:NGC4418_FChart}, the error on the aperture correction error is minimal due to the presence of a bright field star), the distance modulus becomes $(m-M)_0 = 32.59_{-0.10}^{+0.07}$, or $32.9_{-1.5}^{+1.2}$~Mpc.  This places the galaxy amongst the most distant objects ever studied by the PNLF technique, and well beyond the Tully-Fisher estimate.  The observation demonstrates what is now possible with the MUSE instrument.

We do note that, since the galaxy is a bright infrared source, it is possible that some of the emitting dust could be distributed across the face of the galaxy.  So it is possible that the PN measurements are attenuated by this component.  In that case, our PNLF distance would be an overestimate.  Assuming that this is not the case, the galaxy provides an opportunity for a direct measurement of the Hubble Constant, as the system is distant, relatively isolated, and has projected onto its body and an unusually bright field star for a precise measure of the aperture correction (see \S\ref{subsec:H0discussion}).



\subsection{NGC 4472 (M49) \label{subsec:NGC4472}}

\begin{figure*}[t]
\hspace{16mm}
\href{https://cloud.aip.de/index.php/s/ZAe6nPBXtgbM5T8}{\bf \colorbox{yellow}{Off-band}}
\hspace{44mm}\href{https://cloud.aip.de/index.php/s/nCjyKbQDeyQjr6r}{\bf \colorbox{yellow}{Diff}}\\
\includegraphics[width=56mm]{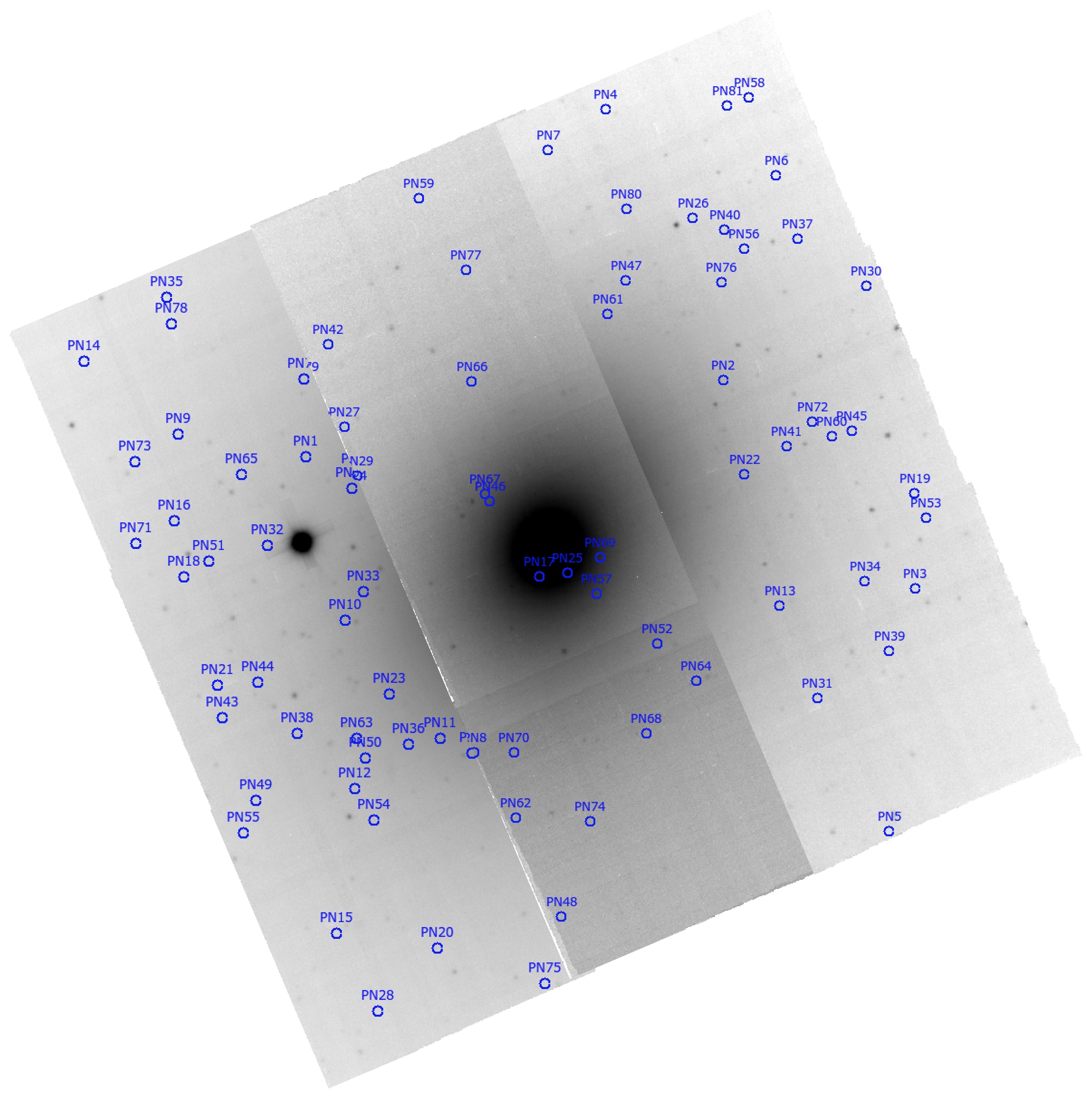}
\includegraphics[width=56mm]{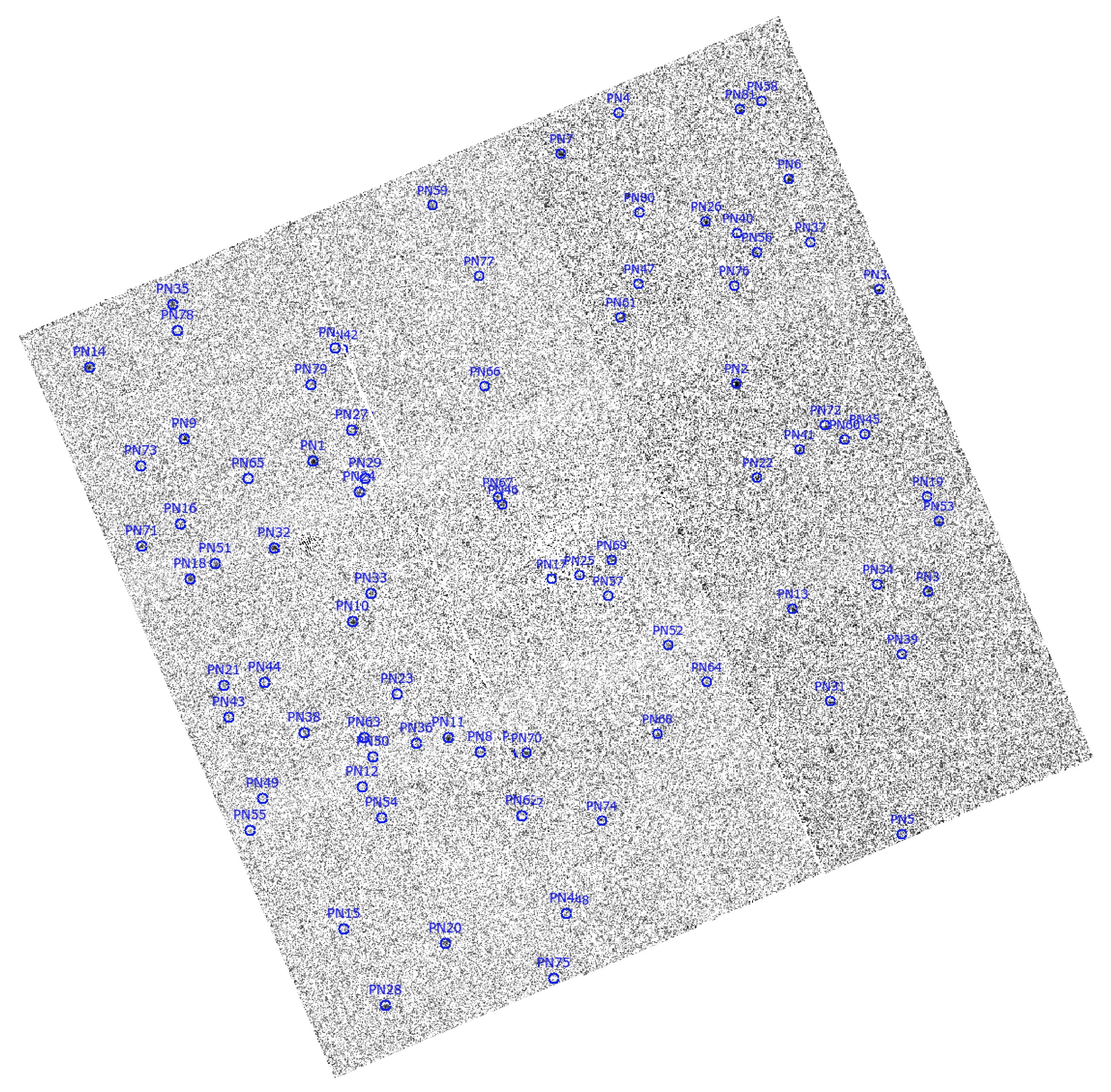}
\hspace{5mm}
\includegraphics[width=55mm]{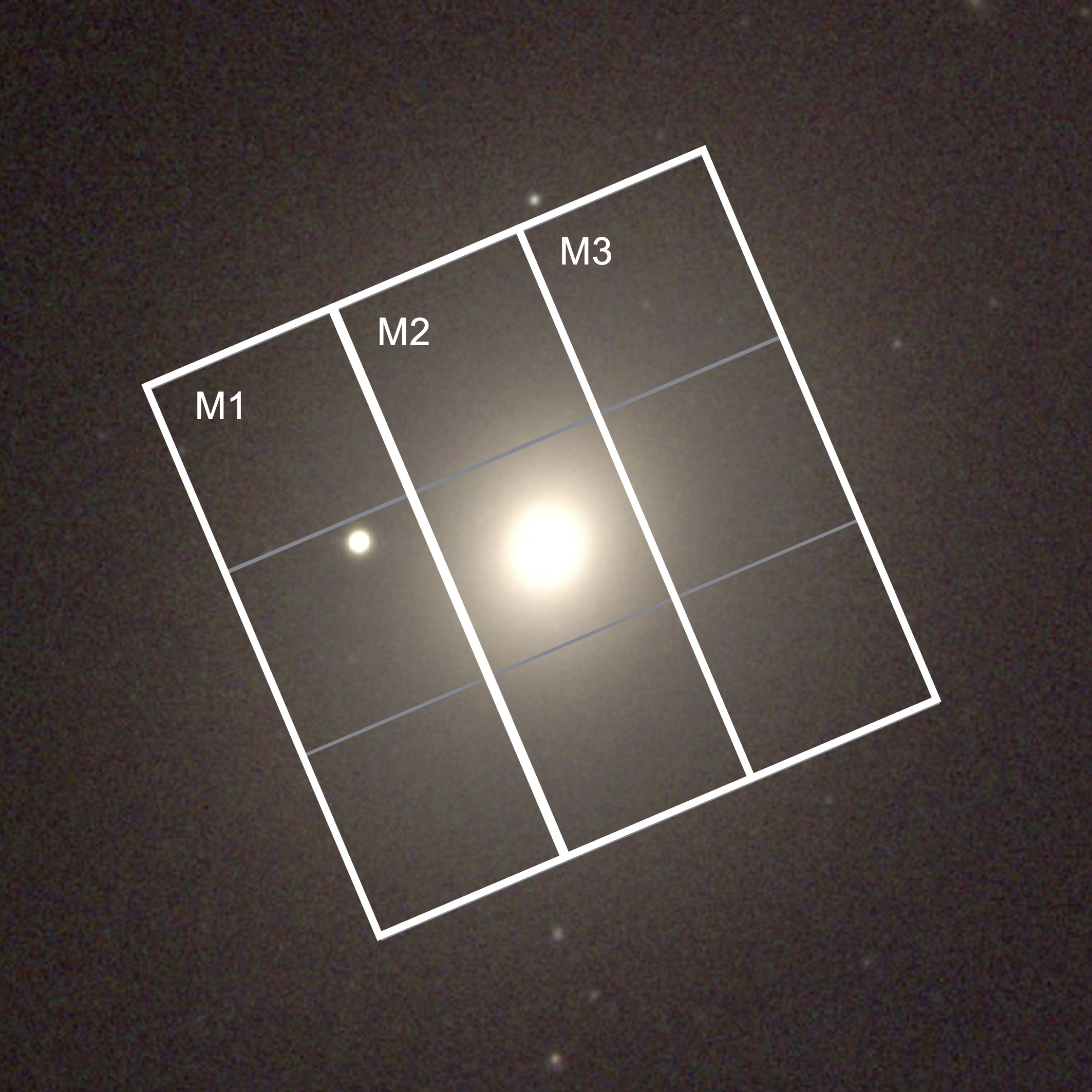}
\caption{NGC\,4472. Left: thumbnail off-band and difference images derived from the MUSE data cubes. The blue circles denote the PN candidates.  High-resolution images can be obtained by clicking on the hyperlink titles. \href{https://cloud.aip.de/index.php/s/6KipRkACwkmmP8X}{\colorbox{yellow}{VIDEO}}. Right: a broadband image with the MUSE data cubes outlined in white. (Credit: NOIRlab/NSF/AURA).  \label{fig:NGC4472_FChart}}
\end{figure*}

The E2 giant elliptical NGC\,4472 is the brightest galaxy in the Virgo cluster and is at the center of the Virgo B subclump. Previous PNLF measurements have placed this  galaxy at a distance of 13.9~Mpc \citep{Jacoby+90}, 14.4~Mpc with a new calibration from the \textit{HST} key project \citep{Ferrarese+00}, and 18.1~Mpc \citep{Hartke+17}. NED lists 22 SBF distances for the system, ranging from 14.3 to 17.8~Mpc. 

\begin{deluxetable}{llcc}
\tablecaption{Data Cubes for NGC 4472
\label{tab:NGC4472_cubes}  }
\tablehead{\colhead{Field} &\colhead{Archive ID} &\colhead{Exp Time}
&\colhead{Seeing} \\ [-0.25cm]
&&\colhead{(sec)} &\colhead{(5007~\AA)}}
\startdata
M1 &ADP.2016-06-21T00:50:22.928 &626 &$0\farcs 72 $ \\
M2 &ADP.2016-07-11T13:38:27.884 &623 &$0\farcs 82 $ \\
M3 &ADP.2016-06-21T00:31:05.332 &621 &$0\farcs 93 $ \\
\enddata
\end{deluxetable}

There are a total of nine relatively shallow MUSE data cubes of NGC\,4472 in the ESO archive (Program ID: 095.B-0295 PI: J. Walcher) that are provided as three mosaics; each combines three MUSE pointings in one cube.  We list these data in Table~\ref{tab:NGC4472_cubes} and show their locations in Figure~\ref{fig:NGC4472_FChart}.   Despite the short exposure times, we managed to identify and measure 81 of the galaxy's PN candidates with the DELF technique.

\begin{figure}[h!]
\includegraphics[width=0.473\textwidth]{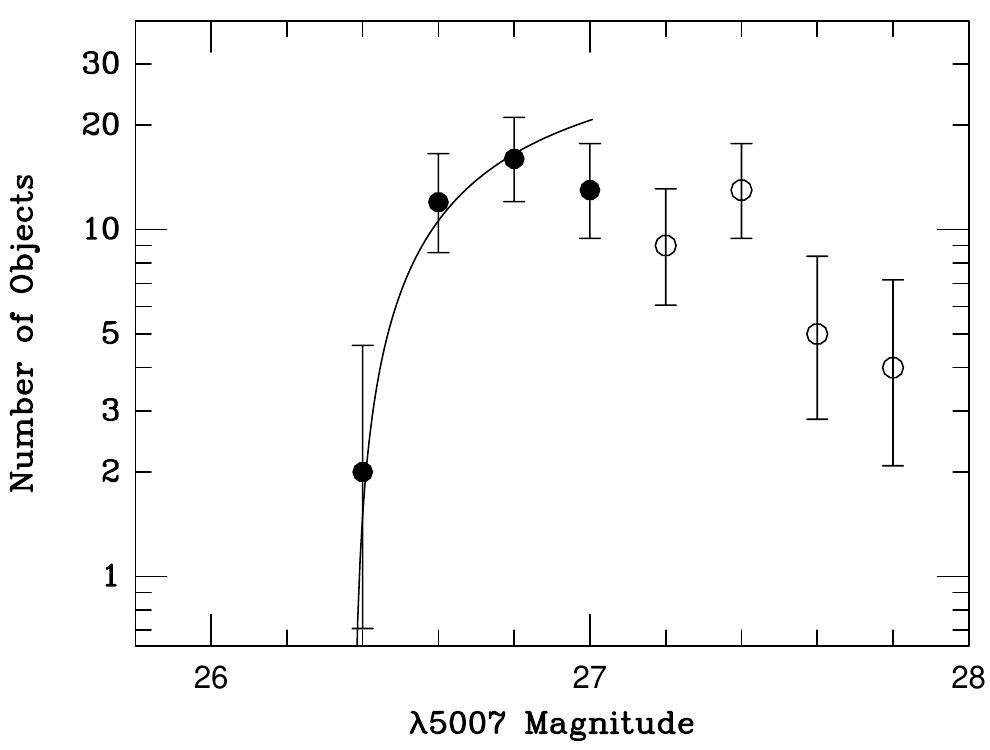}
\caption{The observed PNLF for NGC\,4472 binned into 0.2~mag intervals. Open circles denote data beyond the completeness limit; the error bars illustrate the uncertainties of small-number counting statistics \citep{Gehrels86}.  The curve shows equation~(\ref{eq:pnlf}) shifted by the best-fit apparent distance modulus of $(m-M) = 30.91$. \label{fig:NGC4472_PNLF}} 
\end{figure}

Figure~\ref{fig:NGC4472_PNLF} displays the PNLF of the objects.  In the figure, we have excluded the PNe in the central $\sim 40\arcsec$ of the galaxy, since, as Figure~\ref{fig:NGC4472_FChart} illustrates, the limited depth of the data made PN detections in this high surface-brightness region difficult.  Even outside this radius, the PN luminosity function is not especially deep, as the data only extend $\sim 0.7$~mag fainter than PNLF's bright-end cutoff.  This is just deep enough to determine a robust distance to the galaxy.

\begin{figure}[h!]
\includegraphics[width=0.473\textwidth]{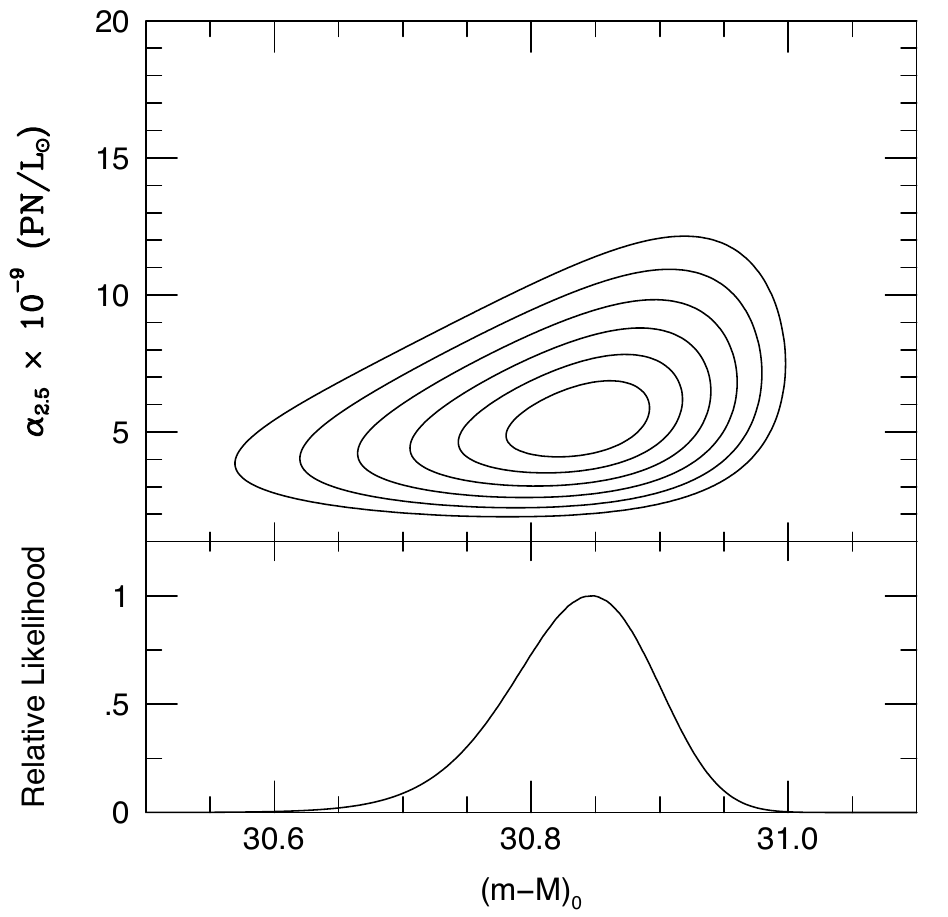}
\caption{The top panel shows the results of the maximum likelihood solution for NGC\,4472. The abscissa is the galaxy's true distance modulus, the ordinate is $\alpha_{2.5}$, the number of PNe within 2.5~mag of $M^*$, normalized to the amount of bolometric light sampled.  The contours are drawn at $0.5\sigma$ intervals.  The bottom panel marginalizes this distribution over the PN/light variable. 
\label{fig:NGC4472_contours}}
\end{figure}

To determine the amount of light underlying the location of each PN, we used the surface photometry of \citet{Cohen1986}; outside the $40\arcsec$ radius, the total amount of galaxy light falling onto the MUSE data cubes is $V \sim 9.5$~mag.  The stellar velocity dispersion, as a function of galactic radius, was taken from \citet{Veale+17}.  For the PNe being analyzed here, this quantity ranges from $\sim 230$ to $\sim 255$~km~s$^{-1}$.  Effectively, this means that even if two PNe are spatially co-located on the same location on the sky, there is better than a 50\% chance that the flux from the two objects can be disentangled via their differing radial velocities. This is consistent with a visual inspection of Fig.~\ref{fig:NGC4472_PNLF}: our dataset displays no evidence for any superposed or overluminous objects.

The aperture correction error for this dataset is particularly difficult to assess.  The $3\arcmin \times 3\arcmin$ region surveyed by MUSE is composed of three mosaics, and each mosaic consists of data from three MUSE pointings. Consequently, assumptions must be made as to whether the aperture correction derived from any given star is applicable to the faint sources found in other regions of the same mosaic.  Moreover, the short exposure time means that the data are rather shallow, leading to low signal-to-noise measurements. Mosaic M1 does contain one very bright star that yields a formal aperture correction error of 0.001 mag, and the photometry of two other stars in the field yields corrections that are consistent with that from the bright star.  At least in the M1 mosaic, the zero-point error appears to be dominated by the uncertainty of the MUSE flux calibration, rather than our ability to define the point-source aperture correction.

In contrast, M2 covers the high surface brightness region around NGC\,4472's nucleus, where faint point-like sources are difficult to measure. In this mosaic, the formal aperture correction errors may be as large as 0.17~mag.  M3 lies somewhere in the middle:  our value for the aperture correction seems to be reasonably accurate, although the measurement largely depends on one well-measured object.  

Of the brightest 50 PNe in our sample, only 3 are contained in M2, and none are amongst the brightest 16 objects.  This suggests that our estimate of the aperture correction may be underestimated.  However, as pointed out in \S\ref{subsec:NGC1433}, such an error will likely not affect our derived distance to the galaxy.  We therefore conservatively assign 0.06~mag as the zero-point error of our PNLF measurements.  

The results of our PNLF fit to NGC\,4472 are shown in Figure~\ref{fig:NGC4472_contours}.  Despite the limited depth of the survey, which restricted the number of PNe in our statistically complete sample to 43, the data are very well fit to the curve of equation~(\ref{eq:pnlf}). Our distance modulus to NGC\,4472 for $E(B-V) = 0.02$ is therefore $(m-M)_{0} = 30.85^{+0.08}_{-0.09}$, or $14.8_{-0.6}^{+0.5}$~Mpc. 

For comparison, the almost two dozen SBF distances derived for the galaxy give slightly larger values ($30.78 \leq (m-M)_0 \leq 31.25$) with the most recent value of $(m-M)_0 = 31.12$ falling on the high side of the interval \citep{Blakeslee+09}.  This offset is consistent with the historical difference between the two distance methods.  Somewhat curiously, there are no TRGB measurements to the galaxy.



Our MUSE-based measurement of $14.8_{-0.4}^{+0.3}$~Mpc is slightly more distant, but still consistent, with the first PNLF distance to the galaxy \citep{Jacoby+90}: when scaled to today's zero point, those authors inferred a value of $14.3 \pm 0.4$~Mpc.  However, our new value is significantly smaller than the results of \citet{Hartke+17}, who used the PN observations from Subaru's Suprime Cam to derive a distance $18.1\pm0.6$~Mpc.  These on-band/off-band data extend much deeper than the MUSE spectra (to $m_{5007}=28.8$), contain many more (624) PNe, and extend over a much wider field-of-view than the MUSE observations. We hypothesize that about half the discrepancy is due to a metallicity gradient, with the \citet{Hartke+17} observations principally sampling the bluer outer envelope of the galaxy, while MUSE (and the \citealp{Jacoby+90} study) are drawn from the system's inner regions.  In a metal-poor environment, a PNLF shift of $\sim 0.2$~mag towards fainter values is possible \citep[See, for example][]{Dopita+92}.

\subsection{NGC 5248 \label{subsec:NGC5248}}

\begin{figure*}[ht]
\hspace{20mm}
\href{https://cloud.aip.de/index.php/s/4Dy8N5dSsJqyPWa}{\bf \colorbox{yellow}{Off-band}}
\hspace{45mm}\href{https://cloud.aip.de/index.php/s/HdKZc3eo588GH5o}{\bf \colorbox{yellow}{Diff}} \\
\includegraphics[width=56mm,bb=0 0  944 932,clip]{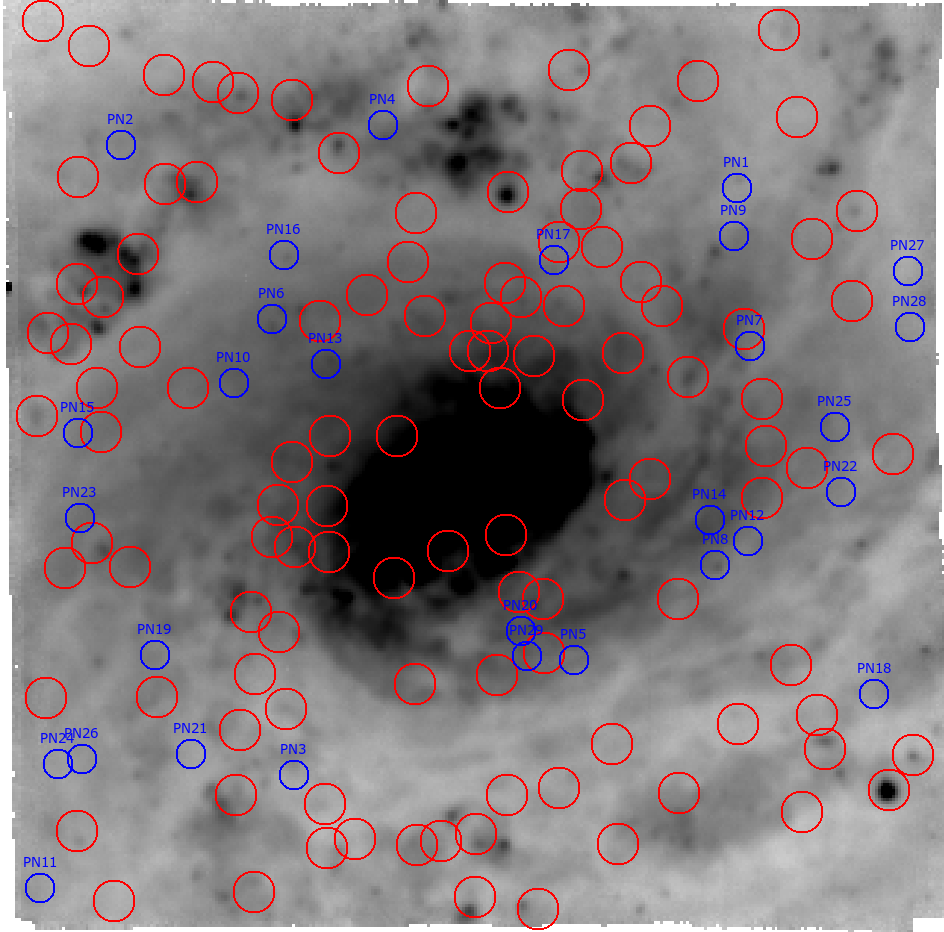}
\includegraphics[width=56mm,bb=0 0  944 932,clip]{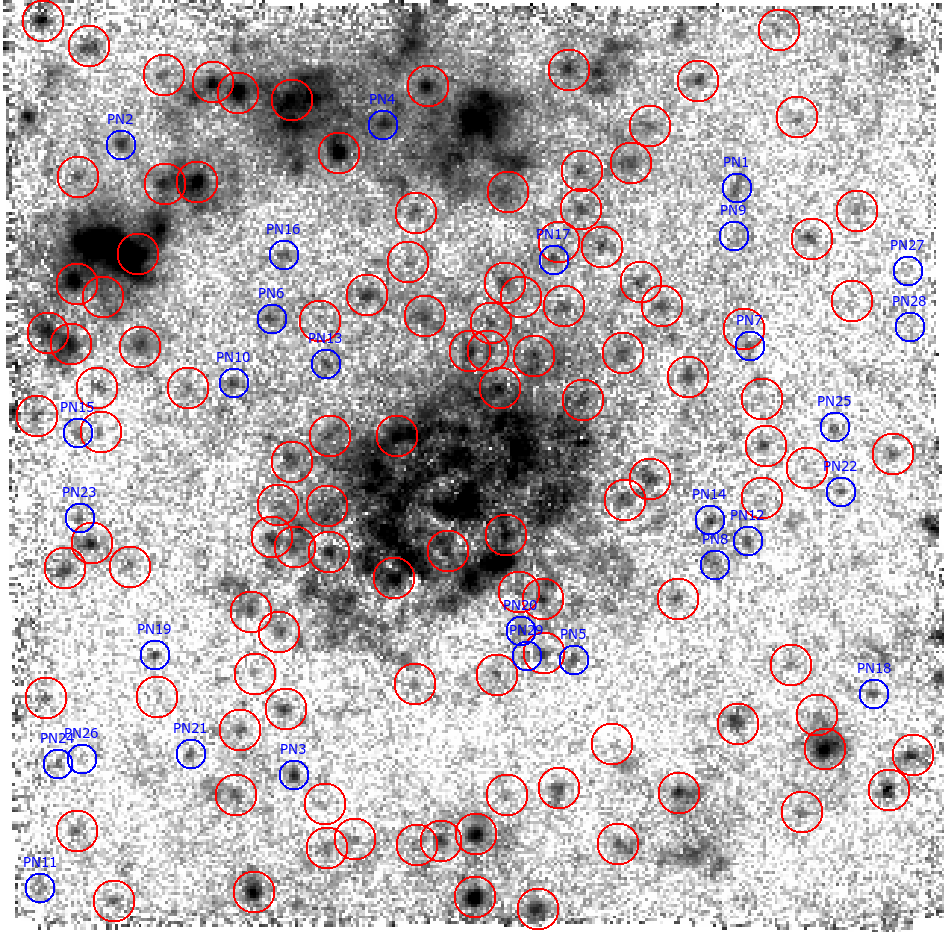}
\includegraphics[width=61mm,bb=100 50  600 500,clip]{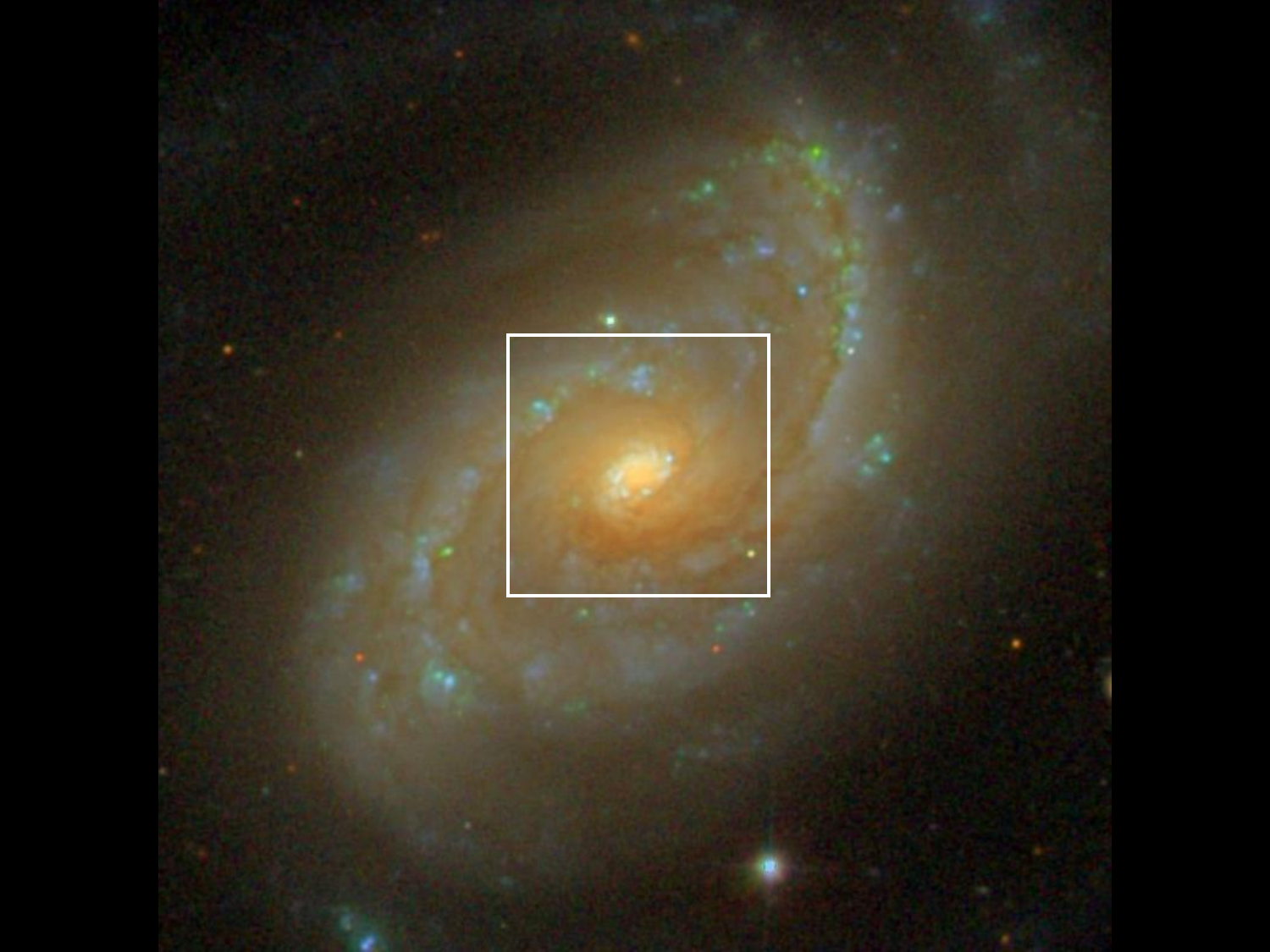}
\caption{NGC\,5248. Left: thumbnail off-band and difference images derived from the MUSE data cube.  The PN candidates are circled in blue; likely \ion{H}{2} and other interlopers are shown in red.  High resolution images can be obtained by clicking on the hyperlink titles. \href{https://cloud.aip.de/index.php/s/NSS6jZZwEBGswwy}{\colorbox{yellow}{VIDEO}}. Right: a broadband image with the location of the MUSE field outlined in white. (Credit: SDSS DR14).  \label{fig:NGC5248_FChart}}
\end{figure*}

NGC\,5248 is a barred SAB(rs)bc galaxy projected within the Virgo\,III group, east of the Virgo cluster core.  The galaxy's distance is uncertain: while the system has been measured using the Tully-Fisher relation, the results span a wide range of values, from $\sim 10$~Mpc \citep[e.g.][]{Bottinelli+84, Sorce+14} to $\sim 22$~Mpc \citep{Ekholm+00}.  Given the complexity of the region, none is preferred.

Due to its prominent bar and nuclear ring, the galaxy has been targeted by MUSE as part of the TIMER survey \citep{Gadotti+19}.   \citet{Neumann+20} have used these data to create H$\alpha$ maps of the galaxy's central ring/bar region with the goal of exploring the area's star formation rate history. These authors consider NGC\,5248 to be peculiar with respect to other galaxies of its type, with strong star formation in a very large nuclear disc and attached spiral-like features. 

We retrieved the archival MUSE-DEEP data cube (ESO Archive ID: ADP.2017-06-14T09:12:09.276, PI: D. Gadotti, Program ID: 097.B-0640) obtained from two observations with an effective exposure time of 3411~s, and $0\farcs 76$ seeing at 5007~\AA\null. This pointing is centered on the galaxy's nucleus and includes the system's nuclear ring, two spiral arms, and a number of prominent dust lanes (see Figure~\ref{fig:NGC5248_FChart}). Our difference image confirms the presence of strong emission lines throughout the field, and, out of an initial sample of 143 candidates, 114 had to be rejected as \ion{H}{2} regions or SNRs (the red circles in the figure). The final sample of PN candidates therefore contains only 29 objects, with 15 being in the top magnitude of the luminosity function.

\begin{figure}[h!]
\includegraphics[width=0.473\textwidth]{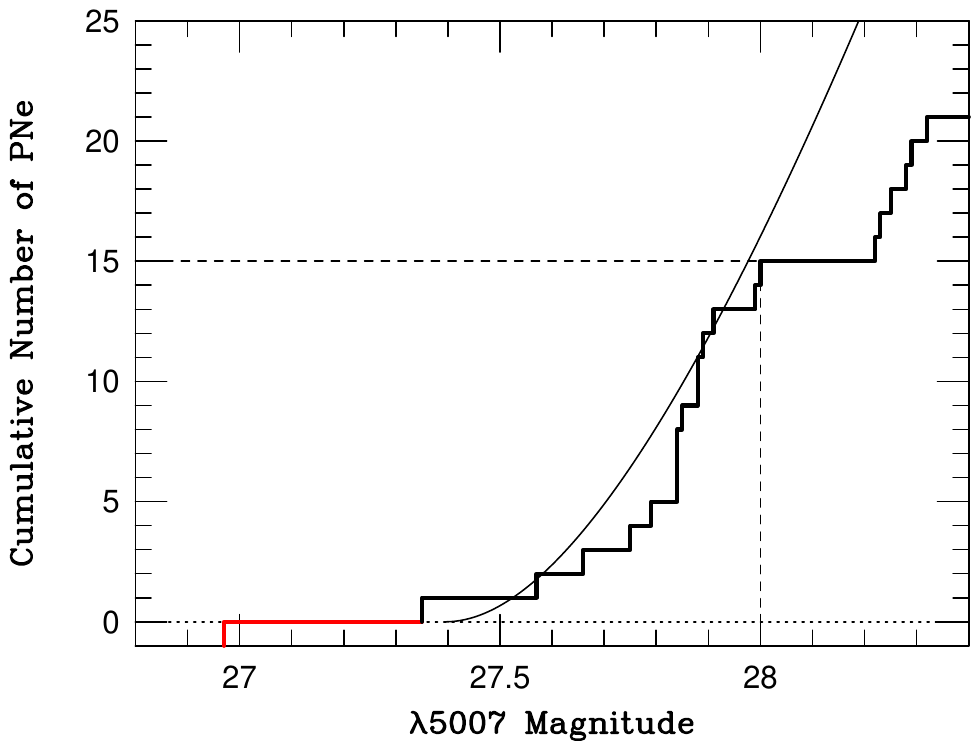}
\caption{The cumulative luminosity function for PNe in NGC\,5248.   The dark line shows the observed data and the curve is equation~(\ref{eq:pnlf}) shifted to the most-likely apparent distance modulus of $(m-M)_0 = 31.93$.  The dark red line shows PN1 and the dashed line denotes where incompleteness begins to affect the detections.  Data brighter than this are consistent with being drawn from the empirical function. A K-S test excludes the inclusion of PN1 at the 99\% confidence level. \label{fig:NGC5248_PNLF}}
\end{figure}

\begin{figure}[h!]
\includegraphics[width=0.473\textwidth]{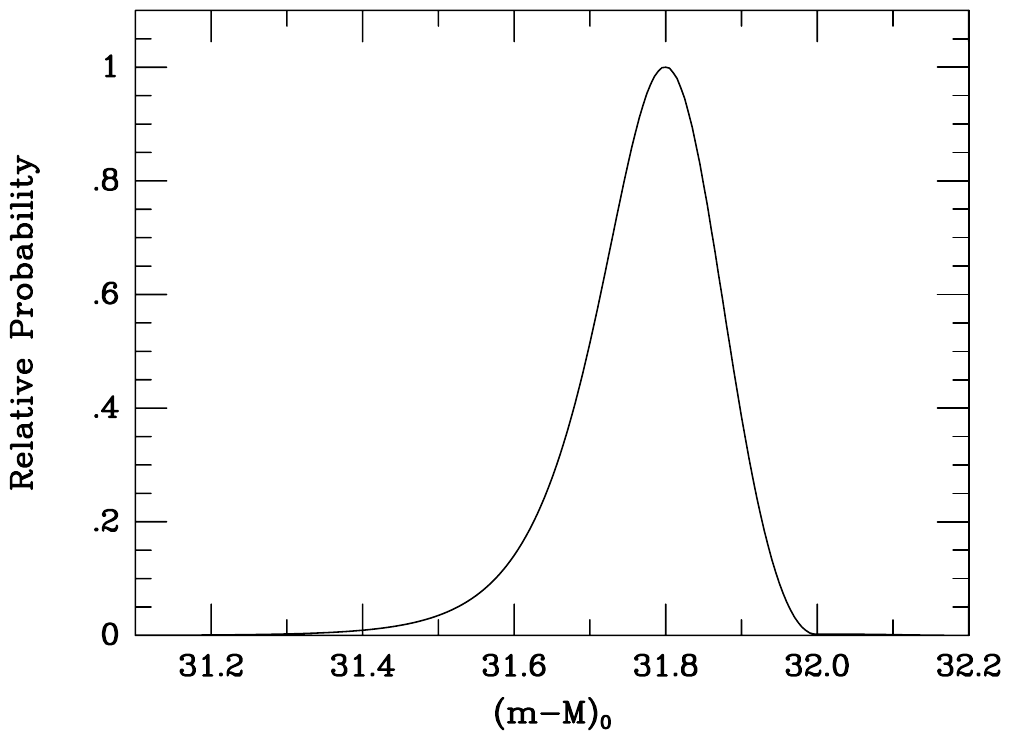}
\caption{The results of the maximum likelihood analysis for NGC\,5248.  The abscissa is the galaxy's true distance modulus, the ordinate the relative likelihood of the solution. PN1 has been excluded from the analysis.  This distance is consistent with the Tully-Fisher-based estimate of \citet{Ekholm+00}.
\label{fig:NGC5248_prob}}
\end{figure}

NGC\,5248's cumulative PNLF is plotted in Figure~\ref{fig:NGC5248_PNLF}. There are two immediate features to note.  The first is the luminosity of PN1, which is 0.42~mag brighter than the next brightest object.  For galaxies with larger numbers of PNe, such a difference would be highly improbable.  Considering NGC\,5248's sparsely populated PNLF, though, the possibility that PN1 is normal cannot immediately be discounted.

The second important property of the distribution is the depth of the survey.  If PN1 is overluminous, and not well-described by the luminosity function defined by equation~(\ref{eq:pnlf}), then the statistically complete sample of PNe only extends $\sim 0.5$~mag down the PNLF\null.  While this lack of depth does not preclude a PNLF distance from being obtained, it does lessen the reliability of the result.

Given the complex morphology of NGC\,5248 inner regions, we did not attempt to derive the galaxy's brightness distribution from the literature; instead, we used the information contained within the MUSE data cube to estimate the continuum brightness at the location of each PN\null.  Similarly, since the line-of-sight velocity dispersion in the inner regions of the galaxy are much smaller than MUSE's spectral resolution \citep[e.g.,][]{Fabricius+12, Rosado-Belza+20}, we simply assume $\sigma = 75$~km~s$^{-1}$ throughout the region. 

As stated above, the only previous distances estimates for this galaxy come from the Tully-Fisher relation. Only one of these determinations places the galaxy beyond 17~Mpc \citep{Ekholm+00}, and most prefer distances on the near-side of Virgo \citep[e.g.,][]{Bottinelli+84, Theureau+07, Sorce+14}.  Our PNLF values are clearly inconsistent with such a location; they agree more with the larger distances. 

\begin{figure*}[t!]
\hspace{20mm}
\href{https://cloud.aip.de/index.php/s/MN86Q9BFjbEkXSx}{\bf \colorbox{yellow}{Off-band}}
\hspace{45mm}\href{https://cloud.aip.de/index.php/s/SQdwpHtqqLnbxF8} {\bf \colorbox{yellow}{Diff}}\\
\includegraphics[width=57mm,bb=0 0  700 700,clip]{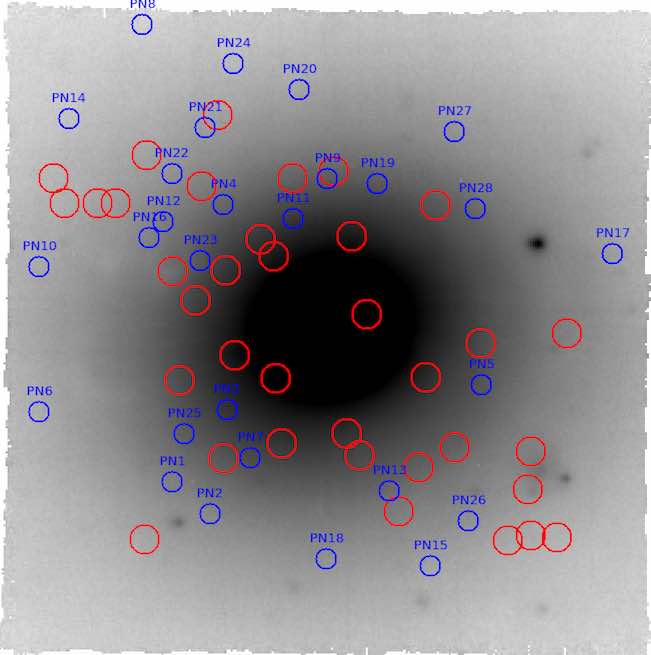}
\includegraphics[width=57mm,bb=0 0  700 700,clip]{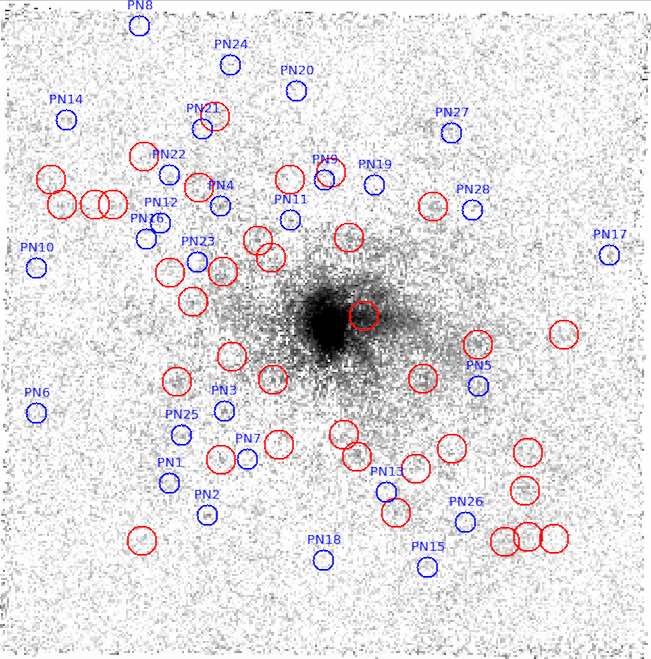}
\includegraphics[width=59mm,bb=100 50  600 500,clip]{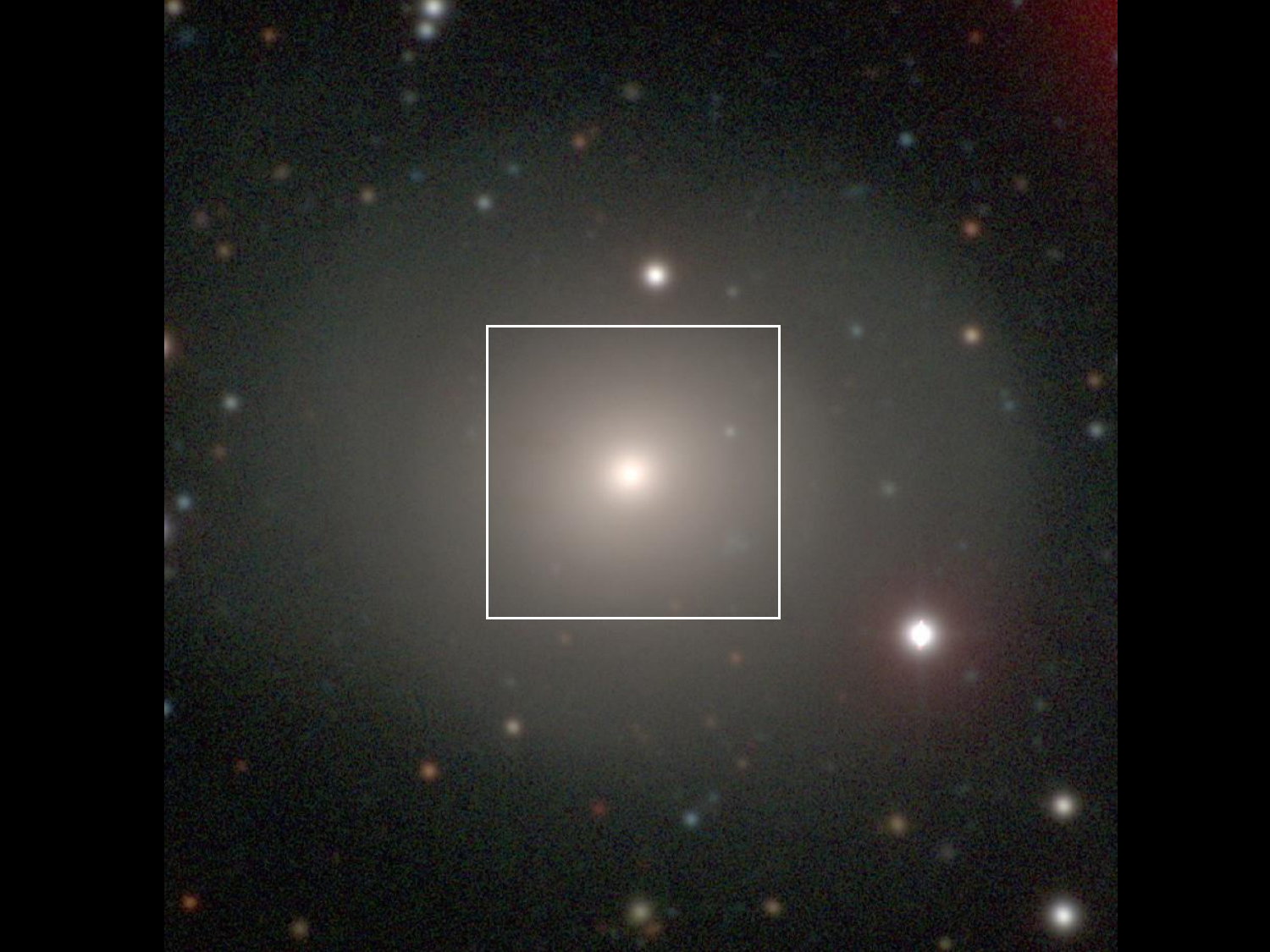}
\caption{NGC\,6958. Left: thumbnail off-band and difference images derived from the MUSE data cube. Blue circles show the positions of the PN candidates; the red circles \ion{H}{2} regions and other interlopers.  High-resolution images can be obtained by clicking on the hyperlink titles. 
\href{https://cloud.aip.de/index.php/s/z5rJar9Tfn54QKE}{\colorbox{yellow}{VIDEO}}. Right: a broadband image with the location of the MUSE pointing outlined in white.  (Credit: CGS). \label{fig:NGC6958_FChart}}
\end{figure*}

If we include PN1 in our PNLF analysis, then for an $E(B-V) = 0.02$, our maximum-likelihood analysis yields a distance modulus of $(m-M)_{0} = 31.37^{+0.06}_{-0.13}$, while assigning less than a 0.4\% probability that the overluminous PN is a blend of two sources.  However, if we accept that the object is not the result of a superposition and adopt the galaxy's most-likely distance modulus, then a Kolmogorov-Smirnov test excludes the possibility that the observed PNe are drawn from the distribution defined by equation~(\ref{eq:pnlf}) at the 99\% confidence level.  Alternatively, if we remove PN1 from the sample, then the distance to NGC\,5248 increases to $(m-M)_{0} = 31.80^{+0.07}_{-0.10}$ and the hypothesis that the PNe are drawn from the empirical function cannot be refuted. On that basis, we adopt the larger distance to the galaxy.  The distribution of likelihoods for this assumption is shown in Figure~\ref{fig:NGC5248_prob}. 

Since the MUSE data cube of the galaxy contains a bright star, the error associated with the photometric aperture correction is very small, formally 0.005~mag.  Thus, the systematic component of the error budget is dominated by the $\sim 0.03$~mag uncertainty in the MUSE flux calibration.  Our derived distance to the galaxy is then $(m-M)_0 = 31.80^{+0.08}_{-0.11}$, or $22.9_{-1.1}^{+0.8}$~Mpc.

\subsection{NGC 6958 \label{subsec:NGC6958}}

\begin{figure}[h!]
\includegraphics[width=0.473\textwidth]{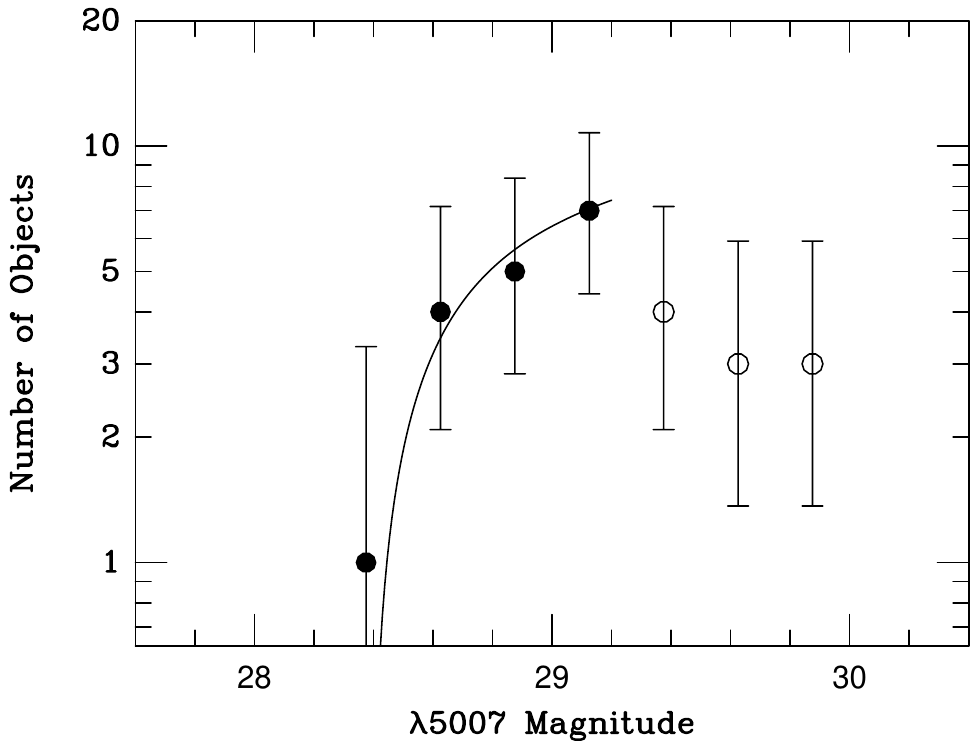}
\caption{The observed PNLF for NGC\,6958 binned into 0.25~mag intervals.  Although incompleteness likely affects our entire survey, we only consider PNe brighter than $m_{5007} = 29.2$ in our fit. Fainter objects are denoted by open circles.  The error bars illustrate the uncertainties of small-number counting statistics \citep{Gehrels86}.  The curve shows equation~(\ref{eq:pnlf}) shifted by the best-fit apparent distance modulus of $(m-M)_0 = 32.80$.
\label{fig:NGC6958_PNLF}}
\end{figure}

\begin{figure}[h!]
\includegraphics[width=0.473\textwidth]{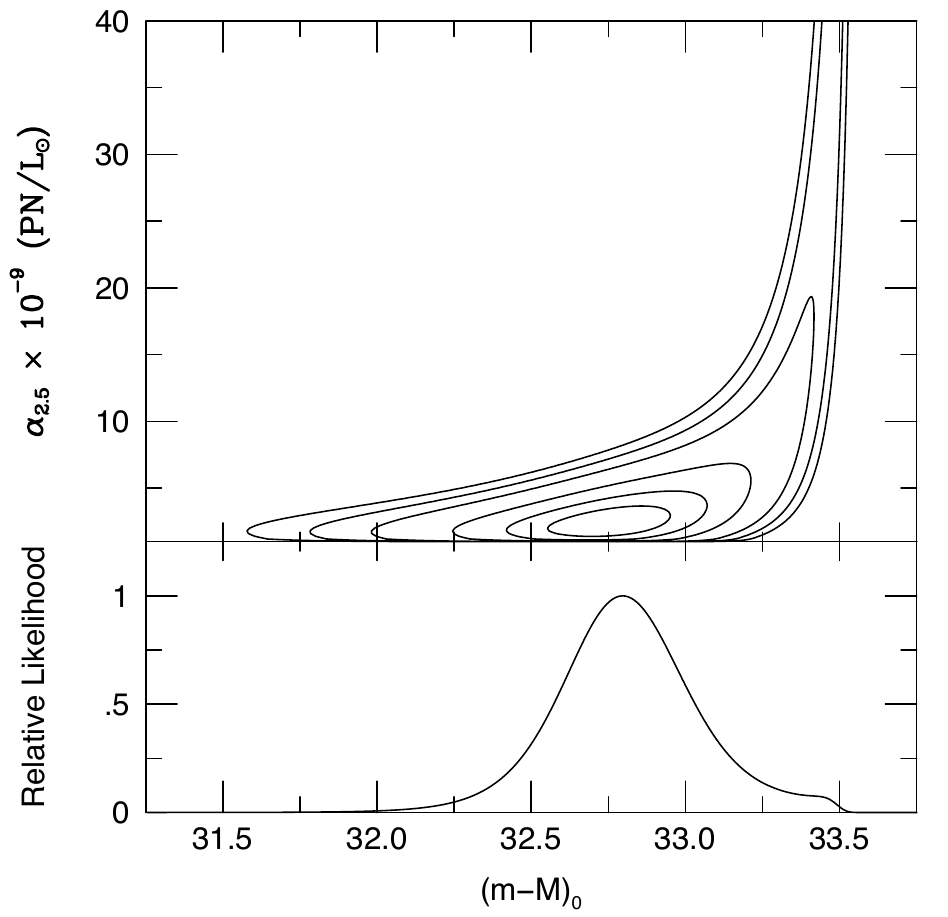}
\caption{The top panel shows the results of the maximum likelihood solution for NGC\,6958. The abscissa is the galaxy's true distance modulus, the ordinate is $\alpha_{2.5}$, the number of PNe within 2.5~mag of $M^*$, normalized to the amount of bolometric light sampled.  The contours are drawn at $0.5\sigma$ intervals.  The bottom panel marginalizes this distribution over the PN/light variable. The low value of $\alpha_{2.5}$ is most likely due to incompleteness in our survey.
\label{fig:NGC6958_contours}}
\end{figure}

NGC\,6958 is an isolated early-type galaxy, classified as S0 according to \citet{Sandage1994} and \citet{Laurikainen2010}, but misclassified as a cD by \citet{RC3-1991}. \citet{Thater+22} have used MUSE data of the galaxy to measure the mass of its supermassive black hole via stellar kinematics.  Their study also used the dynamical scaling relations for early-type galaxies to place the system at a distance of $35\pm8$~Mpc.  If this is correct, then a PNLF measurement would be challenging.  On the other hand, it would also offer us an opportunity to apply the technique outside the local supercluster in a region where the Hubble flow is relatively unperturbed  (see \S\ref{subsec:H0discussion}).

We retrieved the data cube of a science verification observation from the ESO Archive (ID: ADP.2017-11-09T16:25:30.895, PI: D. Krajnovic, Program ID: 60.A-9193).  The metadata for the observation lists an exposure time of 1906~s and a seeing of $0\farcs 96$.  However, we have sometimes found discrepancies between the recorded metadata and our own PSF analysis, and, indeed, \citet{Thater+22} measured the data cube's PSF FWHM top be $0\farcs 75$ at 5000~\AA, and $\leq 0\farcs 6$ at wavelengths longward of 7400~\AA\null.  Unfortunately, this adaptive optics supported observation was secured under poor weather conditions, so the data were not optimal. 

Despite this shortcoming, we were able to detect 64 PN candidates with our DELF technique.  Further inspection confirmed that 28 had spectra consistent with that of a planetary nebula. The fact that more than half of the initial candidates appear to be either \ion{H}{2} regions or SNRs (red circles in Figure~\ref{fig:NGC6958_FChart}) supports the classification of the galaxy as a lenticular, rather than a cD system. Furthermore, the difference image reveals rotation or outflow of the diffuse emission component.

Figure~\ref{fig:NGC6958_PNLF} shows the PNLF of NGC\,6958.  Clearly, the data are limited: despite the bin size being increased to 0.25~mag, no interval contains more than 7 objects.  Similarly, as the data only extend $\sim 0.5$~mag down the luminosity function, the shape of the exponential cutoff is ill-defined, which means the PNLF technique may be susceptible to systematic errors.  Most importantly, even for the brightest objects, the photometric uncertainties associated with the PN measurements are greater than 0.15~mag.  In other words, the signal-to-noise of even the brightest detections is only $\sim 7$.  Thus, incompleteness may be non-negligible, even at the brightest magnitudes.  

\begin{figure*}[t!]
\hspace{17mm}
\href{https://cloud.aip.de/index.php/s/LWkXQazmkF2jRKZ}{\bf \colorbox{yellow}{Off-band}}
\hspace{42mm}\href{https://cloud.aip.de/index.php/s/EApdB46NKqpmgAM}{\bf \colorbox{yellow}{Diff}}\\
\includegraphics[width=55mm,bb=0 0  750 800,clip]{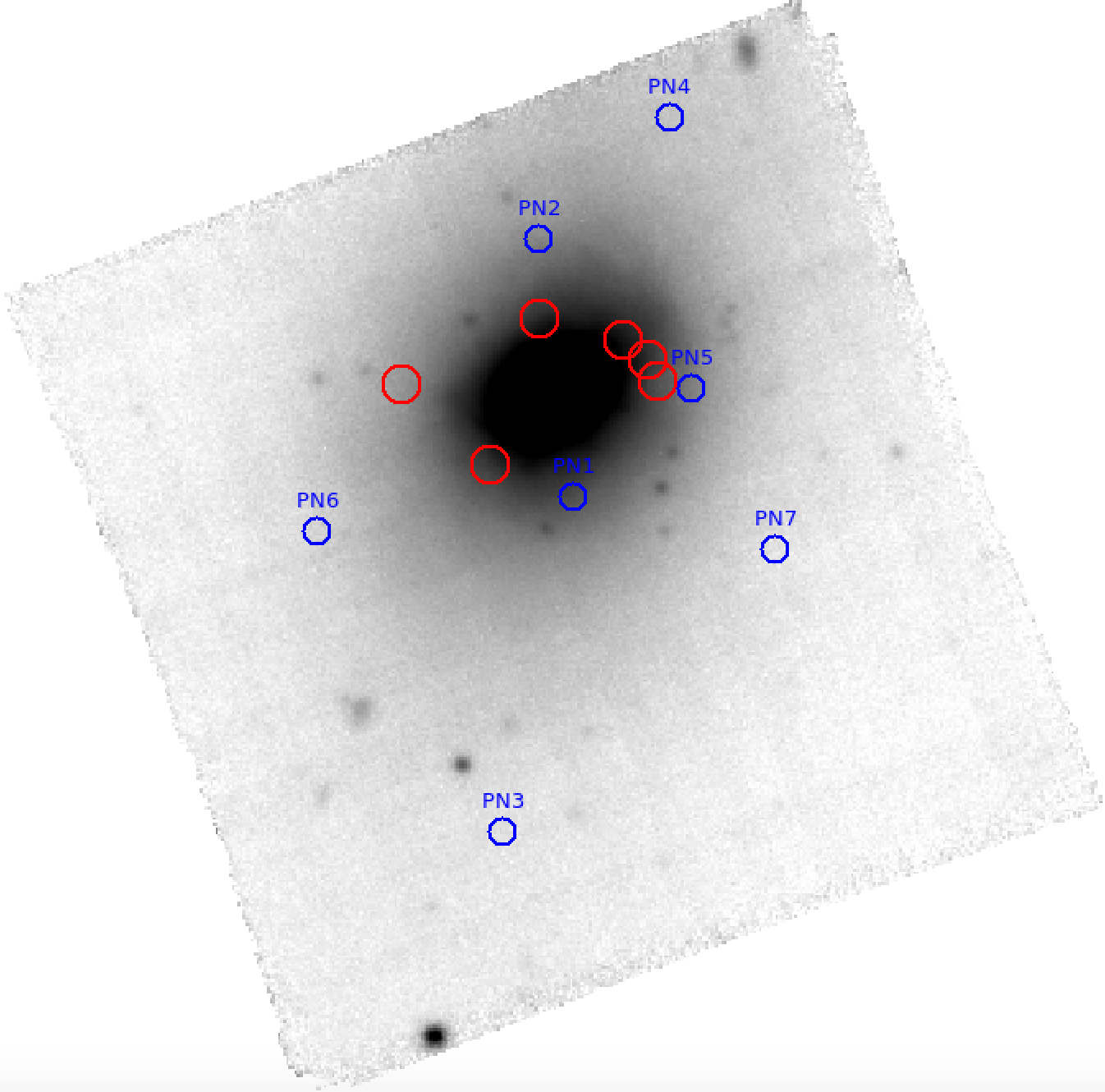}
\includegraphics[width=55mm,bb=0 0  750 800,clip]{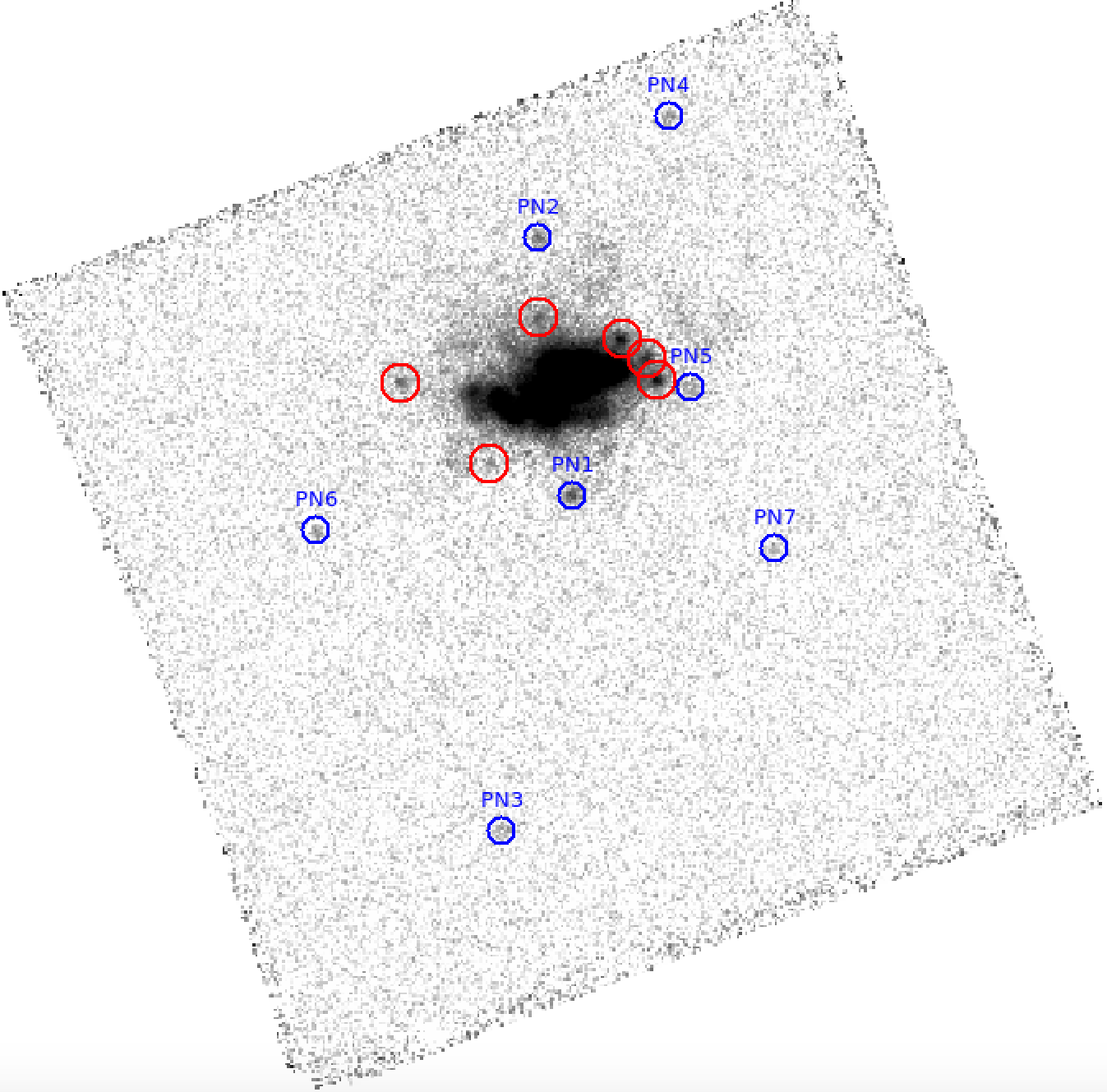}
\includegraphics[width=67mm,bb=150 100  550 380,clip]{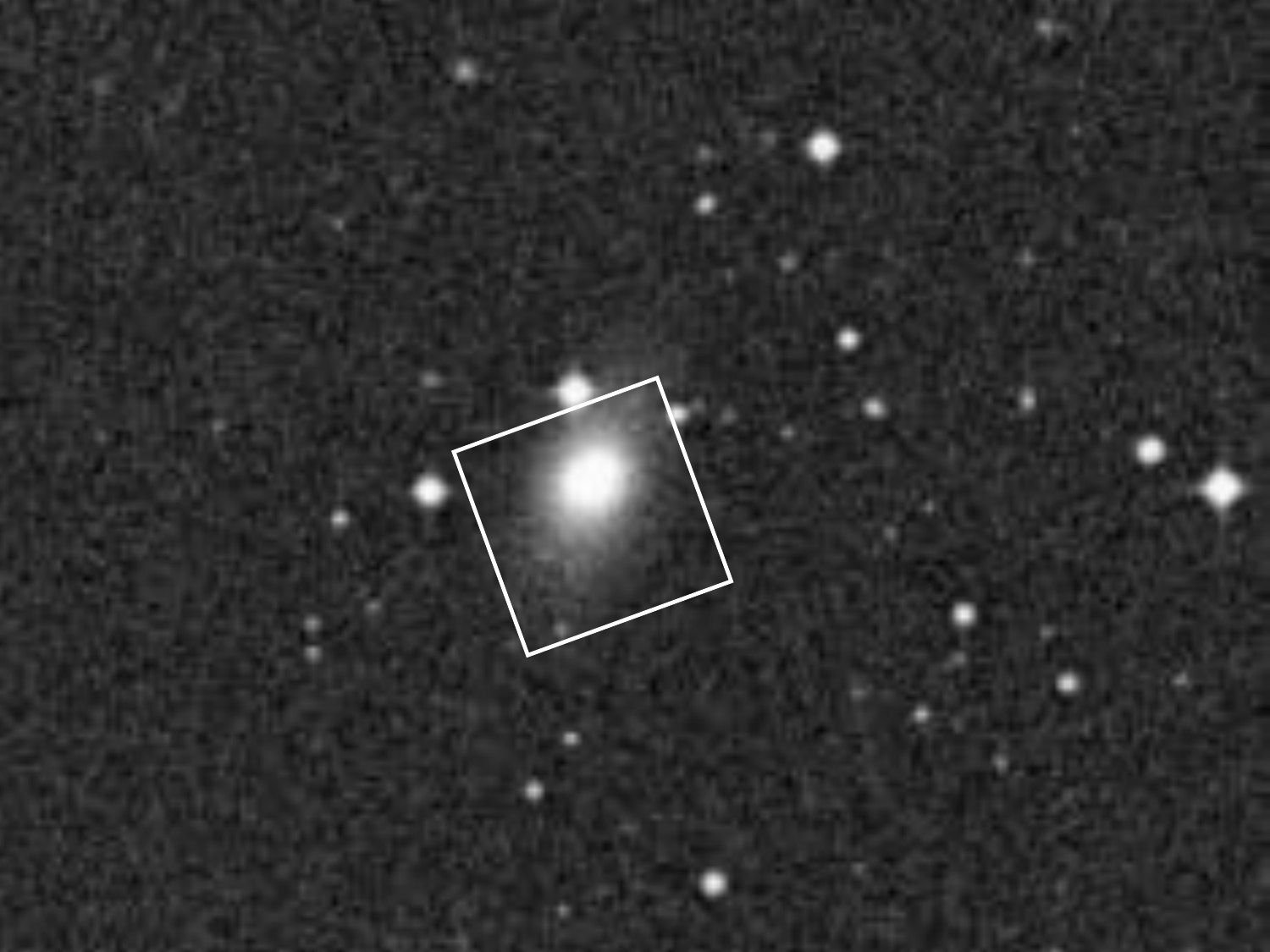}
\caption{MCG-06-08-024. Left: thumbnail off-band and difference images derived from the MUSE data cube.  The blue circles show PN candidates; the red markers show interloping objects. High resolution images can be obtained by clicking on the hyperlink titles. 
\href{https://cloud.aip.de/index.php/s/g84qPLNxF7W6ZDx}{\colorbox{yellow}{VIDEO}}. Right: a broadband image outlining the location of the MUSE field. (credit: ESO DSS1).  \label{fig:FCC090_FChart}}
\end{figure*}

Nevertheless, we fit the data to equation~(\ref{eq:pnlf}), as per the previous galaxies.  Since NGC\,6958 has no published surface photometry or kinematic measurements, we used the MUSE continuum spectroscopy for the former and adopted 75~km~s$^{-1}$ for the line-of-sight stellar velocity dispersion.  As a lenticular system with an inclination of $\sim 45^\circ$, the latter value is a reasonable estimate for the stellar kinematics.  We also excluded all data from the central $\sim 11\arcsec$ of the galaxy, where the surface brightness was too bright for PN detections.  This left us with $V \sim 12.2$ of galaxy light to study.

Figure~\ref{fig:NGC6958_contours} shows the results of our analysis.  Two issues stand out.  The first is that, for the best-fit solutions, $\alpha_{2.5}$, the PN per unit light ratio, is lower than any other galaxy considered in this study.  This suggests that we are missing PNe, even at the bright end of the luminosity function. In such a situation, the PNLF tends to overestimate distance. The second issue is that our distance determination is not very precise, with the $3\sigma$ error contours that extend almost two magnitudes in distance modulus.  Again, this is due to the small number of PN detections and the limited depth of the survey. 

Formally, our distance to the system is $(m-M)_0 = 32.80_{-0.21}^{+0.24}$ ($36.2_{-3.3}^{+4.3}$~Mpc) for a foreground reddening of $E(B-V) = 0.04$.  Perhaps fortuitously, this value is consistent with the Tully-Fisher distance of 36.4~Mpc published by \citet{Theureau+07}. Nevertheless, it is fair to conclude that with better image quality and longer exposure times, this object would be well within reach of the PNLF technique. Although our distance exhibits larger errors than usual and may be a slight overestimate due to the underpopulated bright end of the PNLF, it is one of the few targets in this study with a very good aperture correction star. Thus it is good candidate for a future program for a PNLF measurement of the Hubble Constant, see Section \ref{subsec:H0discussion}.

\subsection{MCG-06-08-024 (FCC\,090) \label{subsec:FCC090}}

MCG-06-08-024, also listed as FCC\,090, is an early-type dwarf galaxy in the Fornax Cluster with a Hubble classification of SA0 (Figure~\ref{fig:FCC090_FChart}). It is known from ALMA observations to harbor a disturbed gas reservoir with a molecular gas tail \citep{Zabel+20}.  Distance from the globular cluster luminosity function and surface brightness fluctuations range from 17 to 20.5~Mpc, consistent with the system being a Fornax cluster member. We note that FCC\,090 is not listed in the sample of 21 Fornax early-type galaxies whose PNLF was studied by \citet{Spriggs+21}, hence it could be an interesting test case to probe the superior sensitivity of our technique.  However, as a dwarf galaxy, we would not expect the galaxy to contain a large population of PNe.

We examined a MUSE-DEEP data cube (ESO Archive ID: ADP.2018-03-26T15:02:26.516, PI: M. Sarzi, Program ID: 296.B-5054)  created from two exposures with a total effective exposure time of 4821~s and $0\farcs 67$ seeing. From an original list of 13 point-like \OIII sources, 6 were eliminated based on their spectra.  The two brightest objects, with $m_{5007}=27.96$ and 28.54, have magnitudes that are roughly 1~mag fainter than the PNLF cutoff of other Fornax cluster galaxies.  This precludes us deriving a useful distance estimate but it does allow us to place an upper limit on the system's distance of $<30.8$ Mpc.

\vspace{10pt}

\section{Discussion}
\label{sec:discuss}

\subsection{Distance Comparisons}

With few exceptions, our PNLF distances agree very well with those of other PNLF results. We already have commented on one major disagreement with the \citet{Scheuermann+22} distance to NGC\,1385 and a more mild disagreement with \citet{Hartke+17} measurement of NGC\,4472.

Unfortunately, there is limited overlap between our heterogeneous set of galaxies culled from the MUSE archive and systems with reliable distances found by other techniques. NGC\,4038/39 is the only galaxy in our sample with a Cepheid distance, and our PNLF analysis yields a distance that is 6 to 26\% larger, depending on which Cepheid value is selected \citep{Riess+11, Riess+16, Fiorentino+13}. That is, the difference between the PNLF and Cepheid distances is similar to that of the Cepheid distances against themselves. Consequently, it is difficult to make any conclusions from the discrepancy.

A distance comparison to the six galaxies with TRGB measurements is equally problematic.  The PNLF distance to the edge-on spiral NGC\,253 is $\sim 54\%$ greater than that found from the system's red giants; however, the sparseness of the luminosity function (less than a 10 objects within 1 magnitude of $M^*$) and the possible effects of internal extinction, weaken confidence in the PNLF result.  Conversely, for the more face-on spirals of NGC\,1433 and 1512, the factor of $\sim 2$ discrepancies between the PNLF and TRGB distances \citep{Sabbi+18} are likely due to the galaxies' large population of AGB stars and the misidentification of the RGB tip.  As for the 3 remaining galaxies, NGC~1404 \citep{Hoyt+21}, NGC~4038/9 \citep{Jang+15}, and Paper~I's measurement of NGC~628 \citep{Sabbi+18} the two distance methods agree within +0.5\%, +4\%, and +8\% respectively, where the PNLF distances are always larger. This encouraging result is consistent with previous TRGB--PNLF comparisons \citep[e.g.,][]{Ciardullo2012, Ciardullo2022}.  

The richest area for a PNLF comparison is with SBF method where we have ten galaxies in common, eight in this paper and two in Paper I\null.  On average, the PNLF distances are smaller by $-6\%$, a systematic that has been known for decades and has usually been attributed to the effect of internal extinction in the calibrating galaxies \citep{Ciardullo+93, Ciardullo+02}. Briefly stated, the PNLF and SBF respond to undetected attenuation in different ways:  distances from the PNLF are overestimated, while those from the SBF are underestimated (due to the color term associated with the fluctuation magnitude).  Since both methods are calibrated in the bulges of nearby spirals, but then applied to elliptical and lenticular systems, any internal attenuation in the calibration galaxies that is not present in the program systems will lead to a systematic offset in the distance scales.  Notably, the $-6\%$ offset is less than half that found in the older studies, lowering the requisite amount in attenuation in spiral bulges to less than $E(B-V) = 0.01$~mag.   

\citet{Scheuermann+22} presented another approach to comparing PNLF distances to other methods in their Figure~9. In comparison to Cepheids and TRGB, their PNLF distances tended to be too large by $\sim 7\%$, but with significant scatter, due in large part, to limited repeatability within each method.

\begin{deluxetable*}{lrccrl}
\tabletypesize{\scriptsize}
\tablecaption{PNLF Galaxy Distances}
\label{tab:galaxydistances}
\tablehead{
\colhead{Galaxy}  &\colhead{Number\tablenotemark{\footnotesize a}} &\colhead{($m-M)_0$} & \colhead{PN Distance} & \colhead{Other Distance} & \colhead{Notes}  \\
& \colhead{of PNe} &     &   \colhead{Mpc}   &  \colhead{Mpc}   & 
}
\startdata
\hline
NGC  253   &  34/10  & $28.66^{+0.12}_{-0.28}$ & $05.4^{+0.3}_{-0.6}$  & $ 3.5\pm0.1$ (TRGB) & possible dust within galaxy \\
NGC 1052   &  86/50  & $31.26^{+0.07}_{-0.08}$ & $17.9^{+0.3}_{-0.6}$  & $18.0\pm2.4$ (SBF) & \\
NGC 1326   &  55/20  & $31.00^{+0.09}_{-0.13}$ & $15.9^{+0.6}_{-0.9}$  &                    & overluminous PN rejected \\
NGC 1351   & 102/45  & $31.39^{+0.04}_{-0.08}$ & $19.0^{+0.4}_{-0.7}$  & $19.2\pm0.6$ (SBF)  & overluminous PN rejected \\
NGC 1366   &  22/13  & $31.39^{+0.11}_{-0.23}$ & $19.0^{+1.0}_{-1.8}$  & $21.1\pm3.0$ (SBF) & \\ 
NGC 1385   &  78/54  & $31.99^{+0.11}_{-0.12}$ & $25.0^{+1.4}_{-1.5}$  &   &  \\
NGC 1399   & 232/164 & $31.23^{+0.06}_{-0.07}$ & $17.6^{+0.5}_{-0.6}$  & $21.1\pm0.7$ (SBF) & \\
NGC 1404   & 124/64  & $31.37^{+0.05}_{-0.08}$ & $18.8^{+0.4}_{-0.6}$  & $18.7\pm0.6$ (TRGB), $20.4\pm0.6$, $20.2\pm0.6$ (SBF) & \\
NGC 1419   &  21/12  & $31.39^{+0.12}_{-0.27}$ & $18.9^{+1.1}_{-2.5}$  & $22.9\pm0.9$ (SBF)  &  \\
NGC 1433   & 258/160 & $31.42_{+0.07}^{-0.08}$ & $19.2^{+0.7}_{-0.7}$  &   & 2 overluminous PNe included \\
NGC 1512   &  210/144 & $31.30^{+0.04}_{-0.04}$ & $18.2^{+0.3}_{-0.3}$  & $11.7\pm1.1$ (TRGB)  & overluminous PN rejected \\
NGC 2207   &   3/0   &                         & $<40$                 &                      & \\
NGC 3501   &   6/0  &                         & $<38$                 &                      & \\
NGC 4038/9 & 228/154 & $31.82^{+0.03}_{-0.04}$ & $23.6^{+0.3}_{-0.4}$  & $21.7\pm0.5$ (TRGB) & ``Antennae Galaxies'' \\
           &         &                         &                       & $18.1\pm0.9$, $20.4\pm0.6$, $21.4\pm0.8$ (CEPH)             &  \\
NGC 4365   &  64/30  & $31.55^{+0.05}_{-0.08}$ & $20.4^{+0.5}_{-0.8}$  & $23.1\pm0.8$ (SBF)  & \\
NGC 4418   &  47/24  & $32.59^{+0.07}_{-0.10}$ & $33.0^{+1.1}_{-1.5}$  &   & overluminous PN included \\
NGC 4472   &  67/43  & $30.85^{+0.05}_{-0.07}$ & $14.8^{+0.3}_{-0.4}$  & $15.9\pm1.0$ (SBF)  & SBF is average of 22 values in NED \\
NGC 5248   &  29/15  & $31.80^{+0.07}_{-0.10}$ & $22.9^{+0.8}_{-1.1}$  &   & overluminous PN rejected \\
NGC 6958   &  28/15  & $32.80^{+0.24}_{-0.21}$ & $36.2^{+4.3}_{-3.3}$  &   &  \\
MCG-06-08-024 &  7/0 &                         &                       &   &  \\
\enddata
\tablenotetext{a}{Given as Total number of PNe / Approximate number of PNe contributing to distance}
\end{deluxetable*}

While it is premature to declare that the PNLF compares as well to other premium distance techniques as they do to each other, the results are quite promising. For the PNLF to be validated further, it is essential to obtain MUSE data cubes taken with absolute photometry of faint point sources in mind.  This means a) choosing fields with foreground stars bright enough to define the observation's PSF and aperture correction, b) observing when the atmospheric seeing is good enough to support faint PN identifications, c) ensuring that each data cube is taken under the best photometric conditions, and d) including enough galaxy luminosity such that, with the proper exposure time, at least $\gtrsim 30$~PNe will be detected in the top 1 mag of the luminosity function.

In addition, the PNLF technique can benefit from additional methodological improvements that we summarize in \ref{conclusions}.





\subsection{Contamination of the PNLF
\label{subsec:overluminousdiscussion}}

In Paper~I, we described several classes of objects that can be confused with PNe and the ways in which MUSE spectra can be used to reject them from our PN samples.  The possible interlopers which were discussed included SNRs, \ion{H}{2} regions, and background galaxies.  Not mentioned in the discussion was another possible source of ``overluminous PNe'' that can cause confusion.  As first noted by \citet{Jacoby+96}, Wolf-Rayet (WR) nebulae \citep{Naze+03} can possibly contaminate the PNLF in star-forming galaxies. Because some of these high-mass nebulae are excited by very luminous O-stars with temperatures of $\sim100,000$ K (similar to the central stars of PNe, their spectra can be indistinguishable from PNe \citep{Chu2016}.  Yet the \OIII luminosities of these PN mimics can be nearly 10 times brighter than $M^*$.  Figures 1b and 4b of \citet{Esteban+94} show spectra of two WR nebulae in M33. At large distances, these objects can appear point-like and can contaminate the PNLF.  

Fortunately, in many cases, this source of confusion should not be a problem.  Specifically,

$\bullet$ In classic E/S0 galaxies, Pop~I objects such as WR nebulae are unlikely to exist.

$\bullet$ In star-forming galaxies closer than $\sim 20$~Mpc, the continuum from the nebulae's exciting O-star ($M_V$ as luminous as $-7$) can be bright enough to detect \citep{Naze+03, Kehrig+11}.  Additionally, in good seeing, the nebulae will generally be large enough ($r > 20$~pc) to be spatially resolved. 

$\bullet$ In systems where WR nebulae are relatively rare, the objects will likely either be much more luminous than the PNLF cutoff, and be identifiable as outliers, or fainter than $M^*$ and have a negligible effect on the PNLF distance.  It is only when WR nebulae are marginally brighter than $M^*$ in star-forming galaxies beyond Virgo and Fornax that they can influence the PNLF maximum likelihood fit. In such galaxies, the line diagnostics of WR nebulae will place them on the border between \ion{H}{2} regions and PNe, and thus make them hard to distinguish from true PNe.  Among the galaxies in this study, PN1 in NGC\,5248 is a possible candidate for a WR contaminant.   

There may also be unusual emission-line sources, such as the ultraluminous x-ray source in Holmberg-II \citep{Lehmann+05} and the black hole in the NGC 4472 globular cluster RZ 2109 \citep{Zepf+08}that can be mistaken for planetary nebulae.  At the distance of NGC\,1385 ($d \sim 25$ Mpc), these objects would have magnitudes, as bright or brighter than $M^*$, i.e., $m_{5007} \sim 24-25$.  Fortunately, such objects are rare, and can have spectral features which discriminate them from PN\null.  Thus they are unlikely to be a problem. 

\subsection{The Shape of the PNLF}
\label{subsec:grand_pnlf}

PNLF distances rely on fitting the apparent magnitude distribution of PNe within a galaxy to some assumed form for the distribution of absolute PN magnitudes. This paper uses the expression first proposed by \citet{Ciardullo+89} as a template, but other forms of the function are possible \citep[e.g.,][]{Longobardi+13, Valenzuela+19}.  The better we can define the true shape of the PNLF, the better the results of the fit.

To that end, it is instructive to sum the absolute \OIII magnitudes of all the galaxies in this survey to produce a single ``grand'' PNLF\null.  In theory, such an analysis could lead to an improved expression for the shape of the PNLF cutoff:  by reducing the errors introduced by counting statistics, subtle deviations from equation~(\ref{eq:pnlf}) could become apparent. On the other hand, any error associated with the conversion of apparent to absolute magnitude will propagate into the summed PNLF; such errors include the formal uncertainties of the galaxy distance determinations, the field-to-field zero-point errors for systems with more than one MUSE data cube, and any population-dependent change in the PNLF's shape.  These errors will all work to soften the abruptness of the function's bright-end cutoff, (slightly) flatten the PNLF's power-law slope, and smooth over possible high-frequency features in the distribution. 

\begin{deluxetable}{llcc}
\tablecaption{Combined PNLF
\label{tab:Grand_PNLF}  }
\tablehead{ \colhead{Galaxy} &\colhead{$M_{\rm lim}$} &\colhead{Total PNe}
&\colhead{PN with $M < M_{\rm lim}$} }
\startdata
NGC 628  &$-3.19$    &202   &90 \\  
NGC 1052 &$-3.74$    &87    &39 \\  
NGC 1326 &$-3.26$    &55    &39 \\  
NGC 1351 &$-3.43$    &102   &45 \\  
NGC 1380 &$-3.22$    &112   &78 \\  
NGC 1385 &$-3.65$    &78    &40 \\  
NGC 1399 &$-3.27$    &232   &163 \\  
NGC 1404 &$-3.70$    &126   &38 \\  
NGC 1433\tablenotemark{\footnotesize a} &$-3.25$    &258   &95 \\  
NGC 1512 &$-3.16$    &210   &144 \\  
NGC 4038/9 &$-3.40$  &228   &152 \\  
NGC 4365 &$-3.71$    &64    &30  \\  
NGC 4418 &$-3.86$    &47    &24  \\
NGC 4472 &$-3.91$    &81    &39  \\
\enddata
\tablenotetext{a}{PN from Pointings P2, P3, and P5 have been excluded, due to their poorly determined aperture corrections.}
\end{deluxetable}

\begin{figure}[h]
\includegraphics[width=0.473\textwidth]{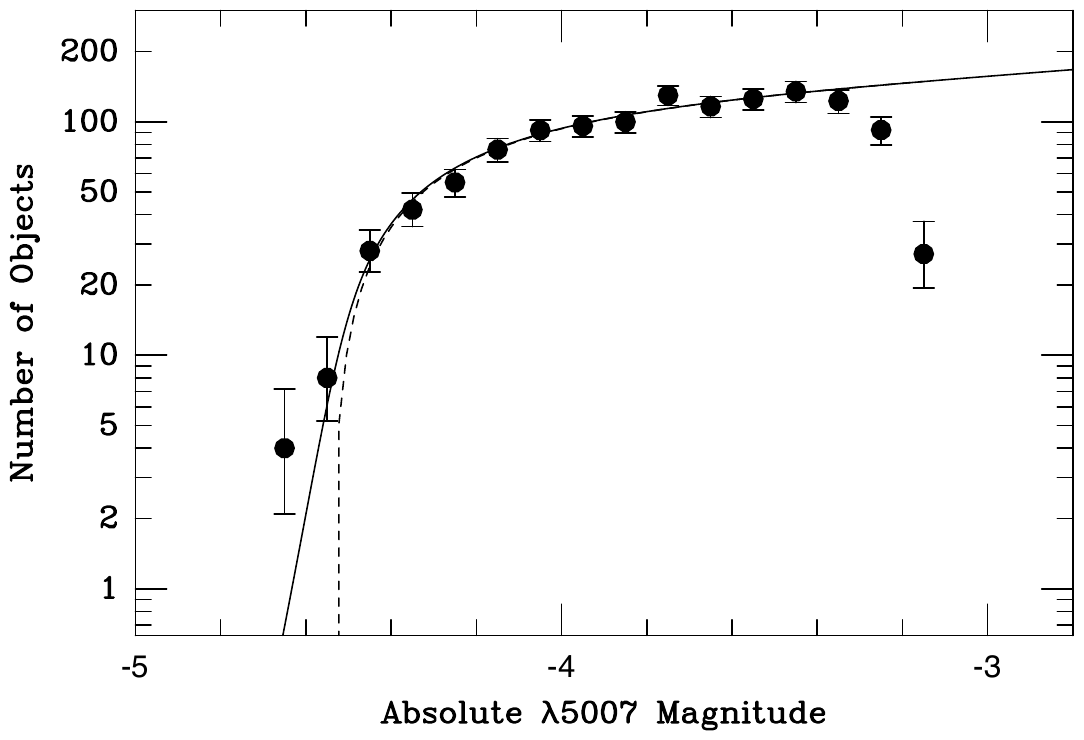}
\caption{The combined PNLF from the 14 galaxies listed in Table~\ref{tab:Grand_PNLF} as derived from 1882 PNe. The PNe's absolute magnitudes have been computed using each system's most-likely apparent distance modulus. The dashed line shows our assumed empirical PNLF (equation~\ref{eq:pnlf}); the solid curve gives the expected PNLF when accounting for the formal errors of each galaxy's distance determination. (The effect of field-to-field aperture correction errors for those galaxies observed with multiple MUSE pointings is not included in the analysis.)
\label{fig:combined_pnlf}}
\end{figure}

Table~\ref{tab:Grand_PNLF} lists the galaxies analyzed in this paper which have well-defined PNLFs.  Also listed in the table are NGC\,628 (Paper~I) and NGC\,1380 \citep[Paper I and][]{Chase+23}, both of which also have high-quality DELF measurements.  Included in the table are our estimates for the galaxies' absolute $\lambda 5007$ magnitudes for 90\% completeness (assuming their most-likely PNLF distances), the total number of PNe detected, and the number of PNe above the completeness limit.  These data are co-added into a single PNLF in Figure~\ref{fig:combined_pnlf}.  

Before discussing the luminosity function, it is imperative to be aware of the nuances involved in performing this co-addition.  The solid curve attempts to model the effect of distance uncertainties in the data set by smoothing our adopted PNLF (equation~\ref{eq:pnlf}) by kernels derived from the galaxies' likelihood contours.  While the resultant model is a better fit to the co-added PNLF, there is still a slight overdensity of objects brighter than $M^*$, and there are a number of other factors that are not included in the analysis.  For example, zero-point errors caused by uncertain MUSE flux calibrations, aperture corrections, foreground dust, and possible metallicity dependencies will not affect the co-addition for galaxies observed in a single MUSE pointing. However, these zero-point issues will distort the PNLF's shape in systems observed via a mosaic of MUSE frames (see \S\ref{subsec:NGC1433}).  The propagation of that error into the summed PNLF is not being modeled.  

Similarly, a simple co-addition does not account for the presence of PN superpositions, i.e., two (or more) point-like \OIII sources projected onto the same resolution element of the detector. When blends are present, they can distort the shape of the PNLF and produce a distance solution that is bi-modal; the likelihood contours shown in Figure~\ref{fig:NGC1512_contours} is one such example. 

In this study, we have minimized the effect of blends by excluding PNe within a few arcsec of the parent galaxy's nucleus.  As a result, even when possible super-$M^*$ objects were detected, the likelihood of their being caused by blends was relatively low (see Figures~\ref{fig:NGC1326_contours}, \ref{fig:NGC1512_contours}, and \ref{fig:NGC4418_contours}.  However, while the probability of such an occurrence in any one galaxy may be small, the likelihood of a superposition being present somewhere in the data set is significant.  The curve shown in Figure~\ref{fig:combined_pnlf} does not account for this possibility.

Another effect not modeled is the effect of dust within the target galaxies.  Internal reddening in late-type galaxies may distort the shape of PNLF by effectively sub-dividing the PN population into multiple components:  some extincted and some not.  Models suggest that the effect of this bifurcation (or trifurcation) on the PNLF distances is minor \citep[e.g.,][]{Feldmeier+97, Gonzalez-Santamaria+21, Guo+21}, but in highly-inclined or interacting systems, it may have an affect.

Finally, it is important to note that our co-addition is only sensitive to changes in the PNLF's shape amongst different galaxy populations.  Any process that shifts the location of the PNLF cutoff, but not the function's shape, will not be detected by the analysis.  In other words, Figure~\ref{fig:combined_pnlf} only shows that the shape of the PNLF depends very little on the type of galaxy being observed; it does not provide any information about whether, for example, the PNLF of spiral galaxies are systematically brighter than those of early-type systems. 

The PNLF of Fig.~\ref{fig:combined_pnlf} consists of 1016 objects brighter than their galaxies' completeness limits.  Overall the data follow the prediction based on equation~(\ref{eq:pnlf}) curve extremely well.  The two noticeable deviations are a possible slight overestimate of the number of PNe with $M_{5007}=-3.85$ and a clear underestimate of objects with magnitudes brighter than $M^*$.  

The latter feature is critical for fitting the PNLF, but the significance of the feature is difficult to determine.  As pointed out above, the curve showing the expected PNLF does not include a number of effects such as PN superpositions and zero-point differences in a mosaic of MUSE pointings.  These issues could easily produce the high-luminosity tail seen in Fig.~\ref{fig:combined_pnlf}.  

Another possible cause of the PNLF's high-luminosity tail is contamination by non-planetary nebulae.  Line diagnostics are very efficient at eliminating objects such as \ion{H}{2} regions and supernovae remnants from the PN sample. Still, some interlopers may slip through, including nebulae produced by Wolf-Rayet stars (see \S\ref{subsec:overluminousdiscussion}).  A careful investigation of the super-$M^*$ objects shown in Fig.~\ref{fig:combined_pnlf} is beyond the scope of this paper but clearly needs to be done.

Of course, the high-luminosity tail of Fig.~\ref{fig:combined_pnlf} may be real.  \citet{Davis+18b} have shown that the true \OIII luminosities of PNe in M31's bulge extend well beyond $M^*$.  They argue that the observed bright-end cutoff of the PNLF is a product of the dust produced during the AGB phase:  the more massive the core, the more massive and dusty the envelope, and the greater the circum-nebular extinction.  This hypothesis is consistent with the results of \citet{Ciardullo+99}, who found a steep correlation between the core mass of a PN and its self-extinction.  If this is true, then viewing angle may affect the observed brightness of a distant PN; for instance, a favorable orientation (e.g., along a PN's polar axis) may result in less attenuation and a brighter apparent magnitude.  Consequently, an extremely luminous PN, when viewed at just the right angle, may be recorded as overluminous.  In this cartoon model, the bright-end tail seen in Fig.~\ref{fig:combined_pnlf} is intrinsic to the PNLF itself, and equation~(\ref{eq:pnlf}) must be modified to include the existence of these objects.



\subsection{Measuring $H_0$ with Planetary Nebulae
\label{subsec:H0discussion}}


The PNLF method has been improved significantly with VLT/MUSE and the DELF techniques. With a sufficient sample of well-observed and well-selected galaxies, the method can help deliver a measurement of $H_0$, either directly via distances to galaxies in the relatively unperturbed Hubble flow, or by calibrating SN Ia luminosities in the spirit of the Cepheid and TRGB approaches.

The zero point of the PNLF distance scale, though, cannot be tied to stars in the Milky Way. Thus, their distances are not completely independent of the results of other methods (e.g., Cepheids, eclipsing binaries).  However, the PNLF works in all types of large galaxies, from the latest Pop~I systems to reddest elliptical galaxies, and the technique can reach systems outside the Local Supercluster that are ill-suited for Cepheid, TRGB, and/or SN~Ia studies.  Moreover, the requisite data can be acquired quickly and efficiently from the ground, without the need for expensive space-based observations.  Thus, a PNLF measure of $H_0$ can offer a unique perspective on the robustness of other results.

A PNLF program to measure $H_0$ would have to target galaxies that:

\begin{itemize}
\item are relatively isolated, in order to minimize the effect of peculiar motions on the estimation of the Hubble flow.   The larger the galaxy group, the larger the number of PNLF distances that would be needed to obtain a true measure of the system's Hubble velocity and distance. 

\item have distances exceeding $\sim30$ Mpc to reduce the effects of flows within the local volume of space.

\item have absolute $V$-band magnitudes of $M_V \lesssim -19.5$ (i.e., $V$ absolute luminosities of at least $\sim 5 \times 10^9 L_{\odot}$).  Any PNLF study must survey at least this amount of light to ensure that the top $\sim 1$~mag of PNLF is sufficiently populated to deliver a precision measure of the PNLF cutoff.  In practice, a target galaxy must be brighter than this, since a) observations from a limited-field instrument such as MUSE will likely not cover the entire galaxy, and b) PNe projected in the highest surface brightness regions of the system will be difficult to detect. 

\item have moderately bright point sources in the field to enable a precise measurement of the photometric aperture correction.  Foreground Milky Way stars are ideal, but for systems beyond $\sim 30$~Mpc, and bright globular clusters will suffice.  

\end{itemize}

In addition, although the PNLF can be applied to galaxies of all Hubble Types, the early-type systems are most easily analyzed, as their PN samples will not be severely contaminated by interlopers such as \ion{H}{2}, SNRs, and WR nebulae.  Uncertainties associated with internal extinction would also not be an issue.
 
With those caveats in mind, of the 24 galaxies analyzed in this paper and in Paper~I, only 6 are potentially useful for an $H_0$ study, and only two have distances greater than 30~Mpc:  NGC~6958, which is poorly measured by the present observations, and NGC\,4418. At face value, the $H_0$ values found from these two objects are $69.2\pm11.0$ and $75.0\pm9.6$~km-s$^{-1}$, respectively; their weighted average is $74.2\pm7.2$~km-s$^{-1}$-Mpc$^{-1}$.  Note that because the quoted errors are dominated by the uncertain ($\sim 10\%$) contribution of peculiar motions to the galaxies' recessional velocities, the error on $H_0$ can be reduced to $<3-4\%$ with a larger sample ($\sim25$) of galaxies.


It should be stressed that most of the galaxies mentioned in this paper were not observed in a manner optimal for distance scale studies.  NGC\,4418 happened to have a bright foreground star present in the MUSE data cube, but most of the systems discussed in this paper do not. It is also possible that some of the galaxies were observed during nights of uneven transparency.  And, most notably, the effect of metallicity shifts on the PNLF has not been considered.  For $H_0$ studies, we expect this last effect to be negligible:  although the PNLF cutoff is known to fade slightly in populations with sub-LMC metal abundances \citep[e.g.,][]{Ciardullo+92, Dopita+92, Schonberner+10}, such systems are not generally the target of a PNLF investigation.  As pointed out above, the PNLF cutoff is not well-defined in systems fainter than $M_V \sim -19.5$.  According to the mass-metallicity relation for local galaxies \citep[e.g.,][]{Tremonti+04}, this basically ensures that any system measured well enough to be useful for a study of the Hubble constant will be in the regime where metallicity effects are minimal.  


The fact that our $H_0$ result is very similar to current best estimates in the Hubble tension literature is encouraging, but likely fortuitous. Clearly, in order to advance the use of the PNLF for distance scale studies, we need a larger sample of galaxies that meet the criteria listed above, and this means new targeted observations are necessary.
\vspace{30pt}

\section{Summary}
\label{conclusions}

We selected a heterogeneous sample of 20 galaxies from the VLT/MUSE archive to study the utility and accuracy of the PNLF as a tool for determining extragalactic distances.  By combining the spectrophotometric capabilities of MUSE with the photometric procedures of \citet{Roth+21} and the fitting algorithms of \citet{Chase+23}, we were able to test our ability to obtain precision distances to galaxies spanning a wide range of morphology and star formation rate. For star-forming systems, MUSE measurements of the PNLF are quite effective unless the star formation is extreme. Although the comparison samples are small, the results for early-type galaxies are similarly encouraging, as  PNLF measurements are as consistent with other well-accepted methods ($4-8\%$) as those methods are with each other. 

Of our 20 galaxies:
\begin{itemize}
    \item 17 yielded fair-to-excellent PNLFs from which we were able to derive galactic distances with quantifiable uncertaintiesc. 
    \item 2 did not contain enough PNe for a sufficient sampling of the PNLF\null.  For these system, we could only derive upper limits to their distances.
    \item 1 is highly affected by internal extinction.  For this object, we could only derive an upper limit to the distance.
\end{itemize}
In nearly all the disappointing cases, the quality of the results were limited by inadequate sample sizes, due to poor seeing or short exposure times. In summary, we derived well-constrained distances for 10 galaxies, useful distances for 6 galaxies, upper limits to 3 galaxies, and no distance information for 1 galaxy.

In creating our galactic PNLFs, we examined the spectrum of each point-like \OIII source and used the procedures described in Paper~I to remove \ion{H}{2} regions, supernova remnants, and background Ly$\alpha$ emitting galaxies from the PN sample.  This approached worked quite well, and greatly reduced the systematic errors that could arise from the inclusion of non-PN contaminants.  

Our investigation also examined the effect of possible PN superpositions on our galactic distances \citep{Chase+23}.  In theory, the blending of two bright PNe into a single emission-line source can lead to PNLF solutions that underestimate the true distance to a galaxy and produce a systematic in the technique's distance scale. In practice, because our present study already excluded each galaxy's highest surface-brightness regions from the analysis, the effect of superpositions on the current data set was minor: in all cases the most-likely distance modulus was unaffected, and in only one case was PDF distorted enough to produce a bi-modal PDF\null.  However, since the likelihood of a PN superposition goes as the square of the distance, it will be important to include this effect in future PNLF studies. 

Two of the galaxies analyzed in this study have distances large enough to reduce their expected peculiar motions to less than 10\% of their Hubble flow velocity. For these objects, we calculated a  Hubble constant independent of the SN~Ia calibration.  Our results are encouraging, and with 25 to 30 more galaxies, the PNLF could provide an additional way to examine the current tension in the Hubble constant. Perhaps even more importantly, the PNLF allows a rare opportunity to cross-check other distance techniques (e.g, Cepheids, TRGB, SBF) that are used to calibrate SN~Ia, enabling another path for measuring $H_0$ through the PNLF. 

Consequently, we look forward to future observations that are structured specifically for PNLF distance measurements that extend into the Hubble flow. Those observations will require:
\begin{itemize}
    \item reasonably bright aperture correction reference stars in the MUSE field of view,
    \item good-to-excellent image quality (i.e., $<0\farcs 75$), and
    \item exposure times and  field coverage sufficient to identify $\sim50$ PNe in the top $\sim 1$~mag of the PNLF\.  Effectively, this means sampling $M_V \lesssim -19.5$ of galaxy light.
\end{itemize}

There is progress to be made on the shape of the PNLF as well.  The MUSE observational data are so good that we need to devise a more applicable reference PNLF than the one defined 35 years ago from PN photometry of M31's bulge. After shifting all usable PNLFs to a common distance and stacking them, we see that this higher fidelity combined PNLF fails to match the old reference PNLF in critical ways.  In particular, the distribution presents some evidence that the phenomenon of ``overluminous'' PNe may be real.  Most objects previously classified as overluminous have either been dismissed as PN contaminants or identified as the likely product of poor photometry.  But the easiest (but not the only) way to explain the behavior of the bright-end of the coadded PNLF is to modify the shape of the luminosity function's bright-end tail.  

In a future era of Extremely Large Telescopes with MUSE-like integral-field spectrographs, the PNLF's distance range, which is now limited by photon collection rates, will increase by a factor of 2 or 3.  The accuracy of the PNLF technique will then become limited by a different set of systematic errors (e.g., photometric zero-point, the near-field calibrations of the luminosity function, and the population dependence of the PNLF cutoff.  At that point, though, cosmic velocity errors will be proportionally smaller so that fewer galaxies will bee needed observed to achieve a given precision.

\section{Acknowledgments} \label{sec:acknowledgements}

We thank You-Hua Chu for helpful discussions concerning the spectra of WR nebulae.  This work was supported by the NSF through grant AST2206090. Based on data obtained from the ESO Science Archive Facility with DOI(s): https://doi.org/10.18727/archive/41 and https://doi.org/10.18727/archive/42
. This work has made use of the SIMBAD database, operated at CDS, Strasbourg, France, and of the NASA/IPAC Extragalactic Database (NED), which is operated by the Jet Propulsion Laboratory, California Institute of Technology, under contract with the National Aeronautics and Space Administration.  The Institute for Gravitation and the Cosmos is supported by the Eberly College of Science and the Office of the Senior Vice President for Research at the Pennsylvania State University.

\bibliography{MUSE-PNLF_II}{}
\bibliographystyle{aasjournal}


\appendix

\section{Notes on NGC 1385} \label{sec:notesN1385}

\subsection{The Issue}
Distances derived using the PNLF method have generally been highly consistent across authors, reflecting only the stated errors in the measurements. Thus, the discrepancy between the distance to NGC\,1385 derived in this paper and that found by \citet{Scheuermann+22} is a very unusual and warrants exploration. We first consider whether NGC\,1385's properties are 
consistent with membership in the Eridanus Cluster  20 to 25 Mpc away.  We then look in detail at the PN photometric results to uncover any clues.

\subsection{Cluster Member Distances and Kinematics in Eridanus} \label{subsec:Eridanus} 
The plausibility of NGC\,1385's distance determined in this work, compared to the much lower Tully-Fisher values from the literature, and the distance derived by \citet{Scheuermann+22}, can be assessed through its probable membership in the Eridanus cluster. Table~\ref{tab:Eridanusdistances} compares the two PNLF distances to SBF measurements of Eridanus cluster members.   Our distance of 25.0~Mpc is consistent with cluster membership, and the galaxy's radial velocity of 1499~km~s$^{-1}$ is close to the mean of the cluster \citep[see Figure~5 of][]{Willmer+89}.  Based on these data, the \citet{Scheuermann+22} distance, which is less than half our value, would seem implausible.

\begin{deluxetable*}{lcccl}[ht!]
\tablecaption{Distances of Eridanus Cluster Galaxies
\label{tab:Eridanusdistances}}
\tablehead{&&\colhead{$v_{\rm rad}$} &\colhead{Distance} &\\[-0.2cm]
\colhead{Galaxy} & \colhead{Type} & \colhead{[km\,s$^{-1}$] } &\colhead{[Mpc]} & \colhead{Method} }
\startdata
NGC 1385	 &	SBc		& 1499	&	25.0 & PNLF (this work) \\
NGC 1385	 &	SBc		& 1499	&	 9.8 & PNLF \citep{Scheuermann+22} \\
NGC 1297	&	SAB(s)0	& 1386	&	28.6 & $I$-band SBF \citep{Tonry+01} \\
NGC 1332	&	S0		& 1619	&	22.9 & $I$-band SBF \citep{Tonry+01} \\
NGC 1395	&	E2		& 1717	&	24.1 & $I$-band SBF  \citep{Tonry+01} \\
NGC 1407	&	E0		& 1779	&	25.1 & F814W SBF  \citep{Cantiello+05}	\\
NGC 1426	&	E4		& 1443	&	24.1 & $I$-band SBF \citep{Tonry+01} (2012) \\
NGC 1439	&	E1		& 1667	&	26.7 & $I$-band SBF \citep{Tonry+01} \\
\enddata 
\end{deluxetable*}

\subsection{Detection and photometry of PN candidates in NGC 1385} \label{sec:PNinNGC1385}

\begin{figure}[h!]
\includegraphics[width=0.473\textwidth]{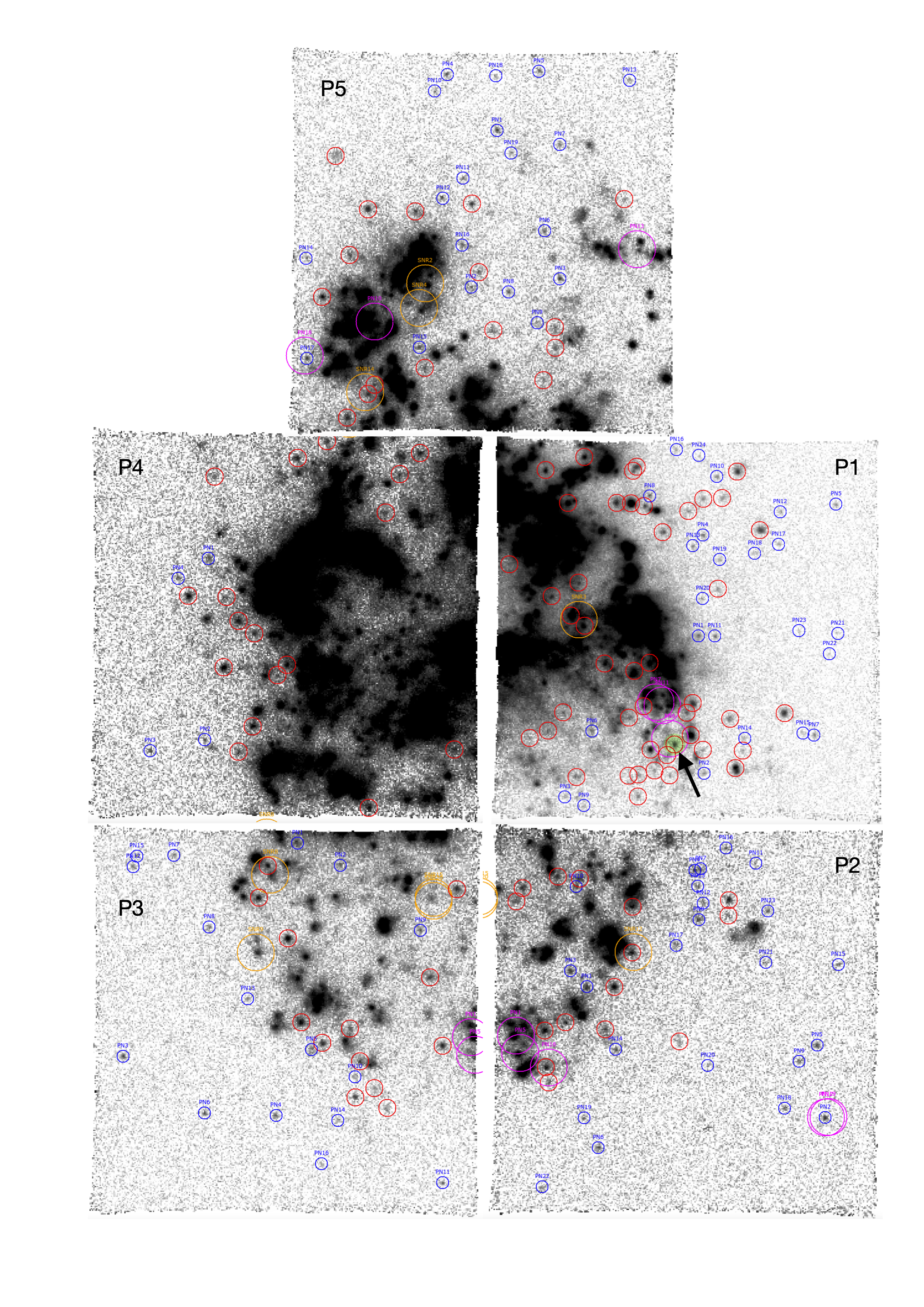}
\caption{Composite \OIII on-band map of NGC\,1385 with pointings P1 through P5 identified.  Our PN candidates are circled in blue and labeled; red circles without labels are objects classified as \ion{H}{2} regions or supernova remnants.  PNe and supernova remnants identified by \citet{Scheuermann+22} are shown in magenta and orange, respectively.
\label{fig:NGC1385_composite}}
\end{figure}

A composite of the \OIII images for the five MUSE pointings, labeled P1 through P5, is shown in Figure~\ref{fig:NGC1385_composite}. Our detections are indicated with blue markers, and the PNe are  numbered individually for each pointing, starting with PN1.  Point sources that were rejected as either \ion{H}{2} regions or SNRs are marked with red circles without labels. PNe from the sample of \citet{Scheuermann+22} are plotted in magenta and their SNRs are shown in orange.  For the following detailed comparison, it is useful to remember that our brightest PN has an \OIII magnitude of $m_{5007} = 27.55$.

\vspace{3mm}
{\noindent \bf Pointing P1:} \\
Our work yielded 24 point-like \OIII sources with spectra consistent with that of a PN; these sources have a brightness range between $27.50 \leq m_{5007} \leq 29.54$~mag. 	In contrast, \citet{Scheuermann+22} find 3 only objects in the field that they classify as PNe; their \OIII magnitudes are $m_{5007} = 26.52$, 26.65, and 27.25.  The three identifications are detailed below: 

PN6:  Our analysis suggests that the data reduction processes used by the ESO Archive are slightly different from those employed by \citet{Scheuermann+22}, and this results in a small astrometric offset between the two data sets.  Thus, we strongly suspect that the object that \citet{Scheuermann+22} calls PN6 is an unlabeled source in Fig.~\ref{fig:NGC1385_composite} which has a red marker within the magenta marker (arrow). This places the \citet{Scheuermann+22} source $0\farcs 80$ east and $0\farcs 93$ north of our position.  We classify the object as an unresolved, high excitation \ion{H}{2} region, as its [\ion{S}{2}] line ratio diagnostic is at the low-density limit.

PN7: The object is located in a giant \ion{H}{2} complex, and we classify its spectrum as that of an \ion{H}{2} region.

PN11: Like PN7, the object is located in a giant \ion{H}{2} complex, and based on its spectrum, we classified it as an \ion{H}{2} region.

In summary, we classify none of \citet{Scheuermann+22}  objects as PNe.  Conversely, none of our 24 DELF-detected PNe are reported by \citet{Scheuermann+22}. 

\vspace{5mm}
{\noindent \bf Pointing P2:} \\
We find 23 \OIII sources with spatial and spectral properties consistent with those of planetary nebulae.  The brightnesses of these objects range between $27.58 \leq m_{5007} \leq 29.14$.  In contrast, \citet{Scheuermann+22} report 5 PNe with magnitudes of $m_{5007} = 25.72$, 26.47, 27.63, 27.90, and 27.97~mag. These \citet{Scheuermann+22} identifications are:

PN1: This object, which is located within a giant \ion{H}{2} complex, is extremely bright in \OIII, but its  [\ion{S}{2}] lines indicate a low-density nebula.  If the object were a PN, it would be 1.83 mag (5.4 times) brighter than our next brightest planetary.  Moreover, if the galaxy were at the distance of the Eridanus cluster, the luminosity of the PN would imply a central star luminosity of $\log L/L_{\odot} = 4.48$, which is well above any reasonable post-AGB evolutionary track.  We conclude that this object cannot be a PN.

PN5: This source, which is projected on a complex background, is brighter in H$\alpha$ than it is in \OIII and thus violates the \citet{Herrmann+08} criterion for \OIII-bright planetary nebulae.  As indicated by the [\ion{S}{2}] doublet, the object does have a high nebular density.  Nevertheless, it is not classified as a planetary nebula.

PN16: The [\ion{S}{2}] lines of this source are half the strength of H$\alpha$.  It is a supernova remnant.

PN17: The coordinates of this source are almost identical to those of our PN19.  We agree with the \citet{Scheuermann+22} that the object is a planetary nebula.

PN19: This object corresponds to our PN2.  We also  classify it as a PN.

\vspace{3mm}
{\noindent \bf Pointing P3:} \\
We find 16 PN candidates in the brightness range between $27.76 \leq m_{5007} \leq 29.19$. \citet{Scheuermann+22} report no detections in this field.

\vspace{3mm}
{\noindent \bf Pointing P4:} \\
We find 4 PN candidates in the brightness range between $27.44 \leq m_{5007} \leq 28.32$~mag. \citet{Scheuermann+22} report no detections in this field.

\vspace{3mm}
{\noindent \bf Pointing P5:} \\
We find 19 PN candidates in the brightness range between $27.56 \leq m_{5007} \leq 29.09$. \citet{Scheuermann+22} find 3 PNe candidates with brightnesses of $m_{5007} = 27.07$, 27.37, and 27.97.  Specifically, the \citet{Scheuermann+22} identifications are:

PN10: The spectrum of this object clearly supports its classification as an \ion{H}{2} region.

PN13: Like PN10, the spectrum of this object is consistent with that of an \ion{H}{2} region.

PN18: We agree that this object is a planetary nebula. However, our photometry yields a magnitude that is $\sim 1$~mag fainter than that quoted by \citet{Scheuermann+22}, i.e., $m_{5007} = 28.84 \pm 0.28$.

\subsection{summary}
The brightest PNe in a galaxy are the most important objects for deriving a PNLF distance.  After a careful review of the brightest PNe reported by \citet{Scheuermann+22}, we find that our photometry and spectroscopy do not support those objects' classifications as PNe. Consequently, we are unable to reproduce the short (9.8~Mpc) distance to the galaxy.

\section{Notes on Aperture Correction Uncertainties} \label{sec:notesapcor}

Uncertainties associated with the detector calibration, atmospheric extinction, flux calibration, dust attenuation, population metallicity, and photon statistics all affect the precision of a PNLF measurement.  However, one of the most important terms in the error budget of a PNLF distance determination is the observation's aperture correction.  The MUSE IFS is capable of delivering a photometric precision of 0.04~mag \citep[see Section 3 and Figure~6 of][]{Roth+21}.  However, to reach this level of accuracy, it is imperative that a bright point source --- typically a foreground star --- be present within the MUSE science field-of-view. Without such a star, the relative magnitudes produced by small-aperture photometry cannot be scaled to the objects' true brightness, at least to the precision necessary for competitive extragalactic distance determinations.

This field-star prerequisite is rarely met when data are taken for another purpose.  Consequently, aperture corrections performed on archival MUSE data cubes must often be based on faint, point-like sources whose true nature is unclear.  This limits the precision of the measurements.  If the objects are truly point-like, the shot noise associated with the photometry will be non-negligible (though, in principle, the noise can be reduced by the square root of the number of PSF standards). However, if the faint point-like sources are actually marginally resolved --- for instance, if they are actually globular clusters associated with a nearby target galaxy --- then the aperture corrections for the point-source PNe will be overestimated.  This issue can be particularly vexing in early-type galaxies, where the system's globular clusters may be the best (or only) point-like sources in the field.  Moreover, it is possible that different stellar populations with different colors can introduce wavelength-dependent effects into the aperture corrections, as speculated in \cite{Roth+21}. Finally, a spatially variable background of unresolved stars in the environment of faint point-like sources can potentially introduce a systematic error that is also wavelength-dependent.  Thus, without the presence of a bright point source in the field, a truly useful error estimate is difficult to obtain.

In order to illustrate the issue, we consider two Fornax cluster elliptical galaxies observed with MUSE using similar exposure times under similar seeing conditions.  One, NGC\,1404 has a bright point source projected in the field (2MASS J03385017-3535311, G=16.2 mag). The other, NGC\,1351 has none.  Thus, their aperture corrections have very different degrees of reliability, and the overall precision of their derived distances is quite different.

\begin{figure*}[ht!]
\begin{minipage}{1.0\linewidth}
\includegraphics[width=0.44\textwidth]{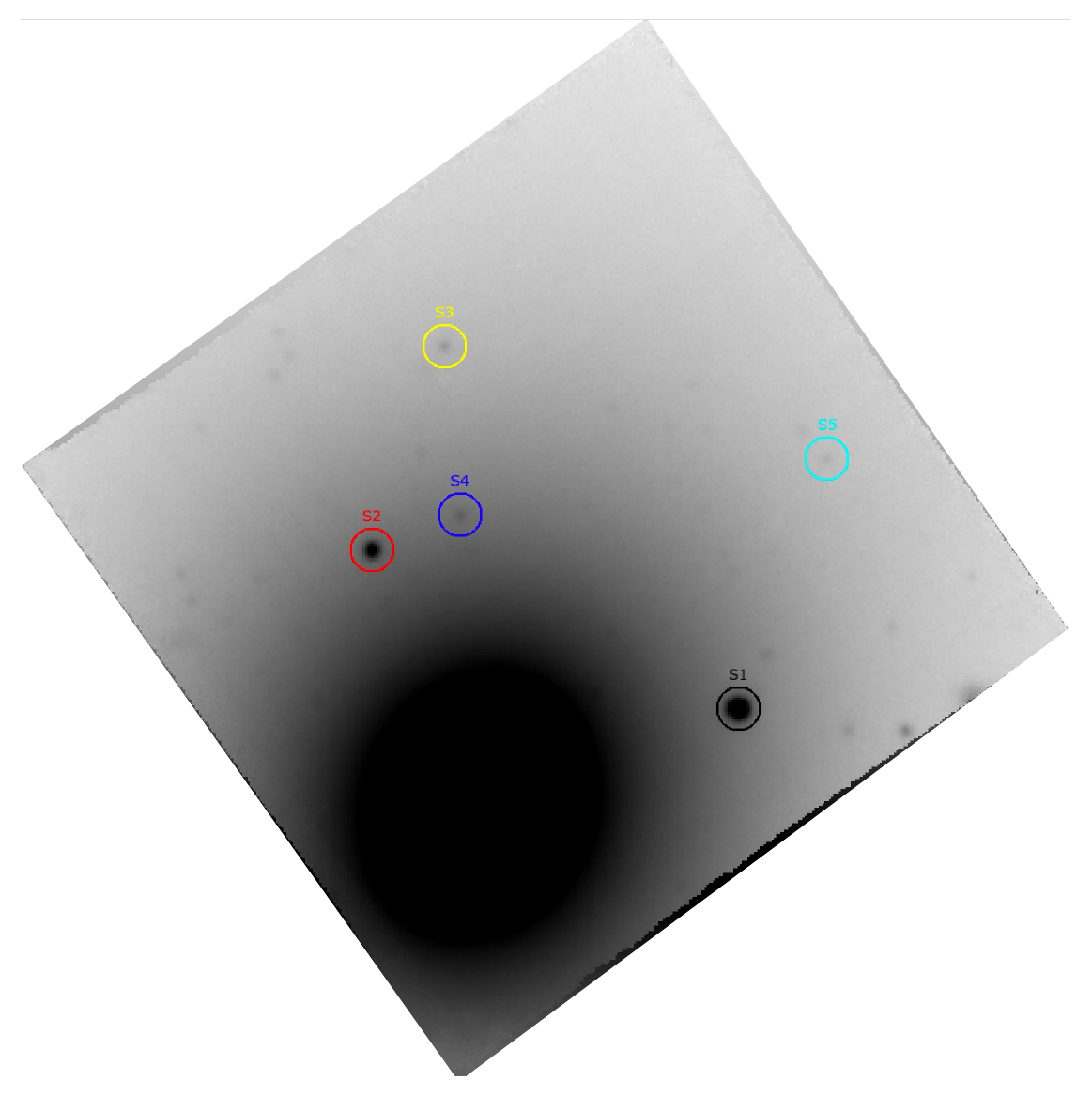} \qquad\qquad
\includegraphics[width=0.44\textwidth]{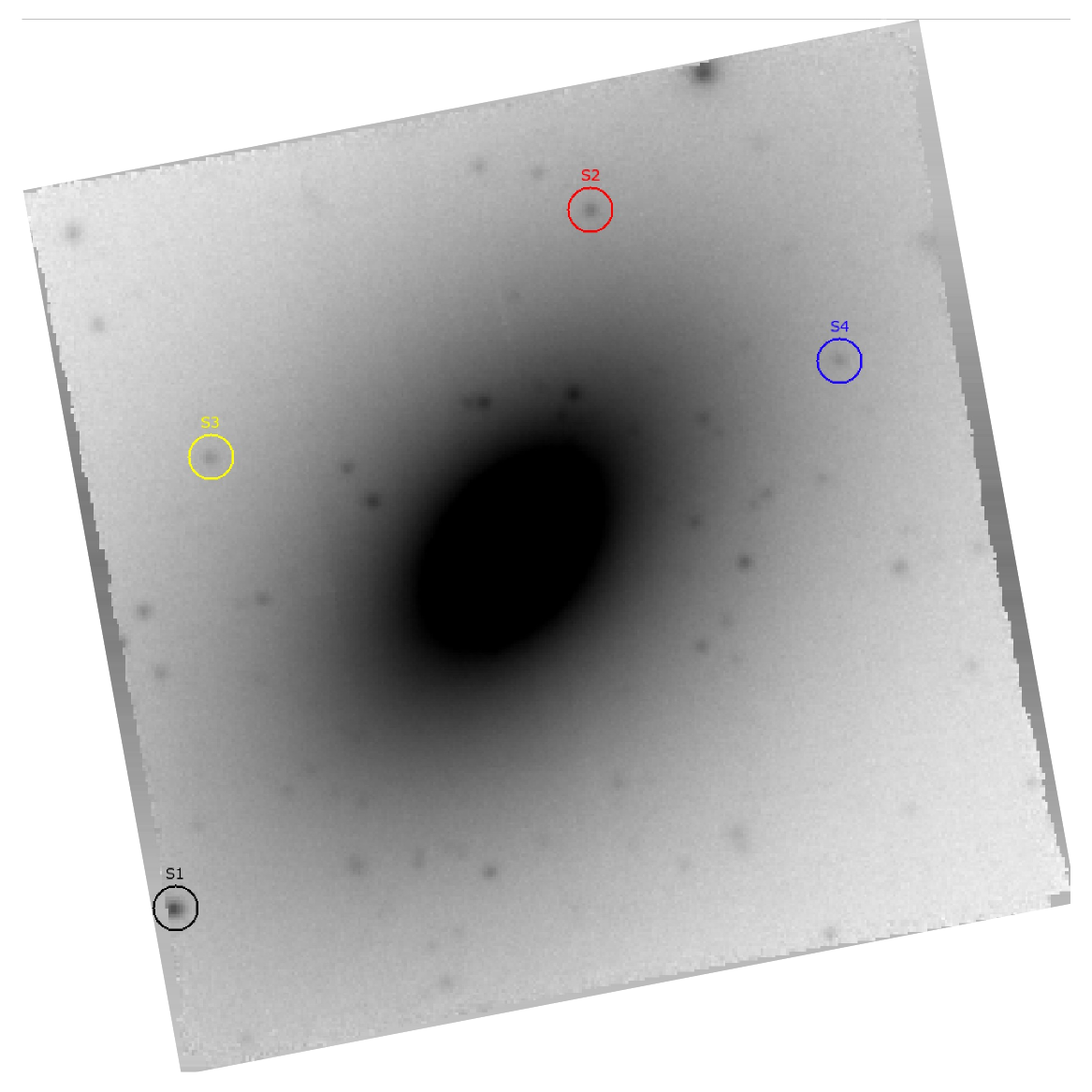}
\end{minipage}
\begin{minipage}{1.0\linewidth}
\includegraphics[width=0.44\textwidth] {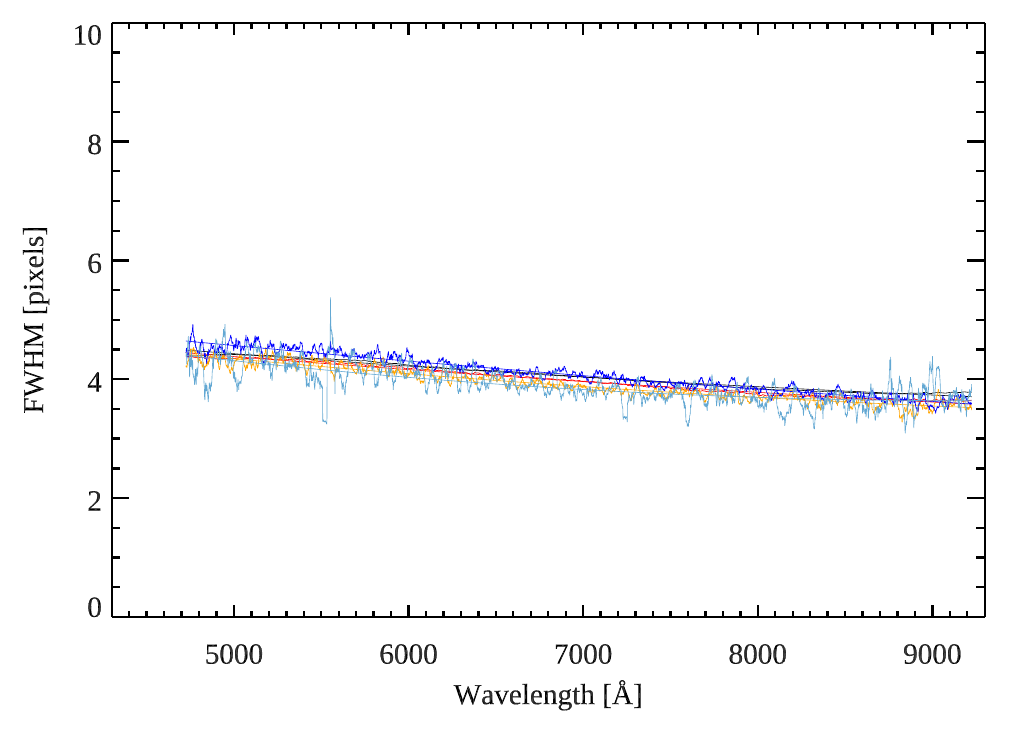} \qquad\qquad
\includegraphics[width=0.44\textwidth]{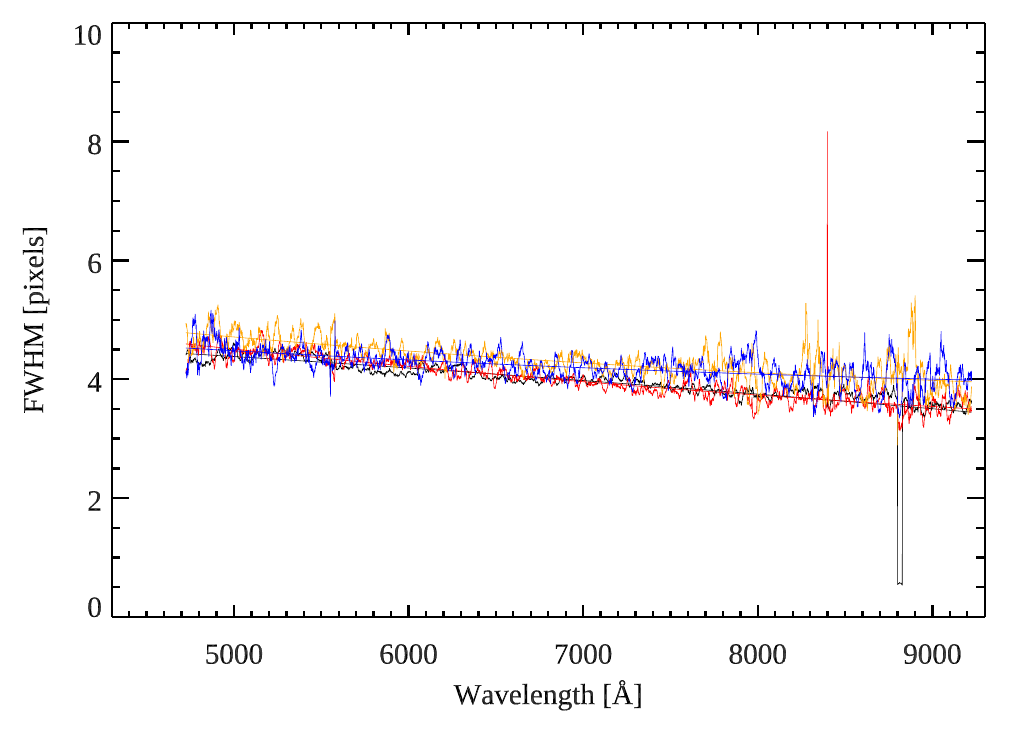}
\end{minipage}
\begin{minipage}{1.0\linewidth}
\includegraphics[width=0.44\textwidth, clip]{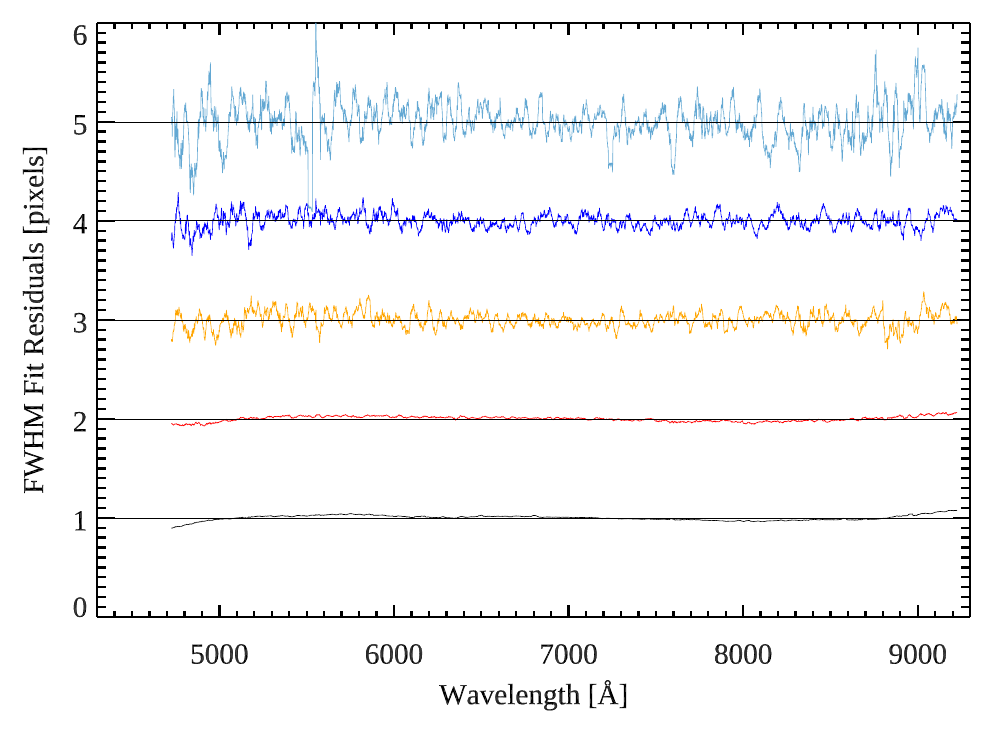} \qquad\qquad
\includegraphics[width=0.44\textwidth, clip]{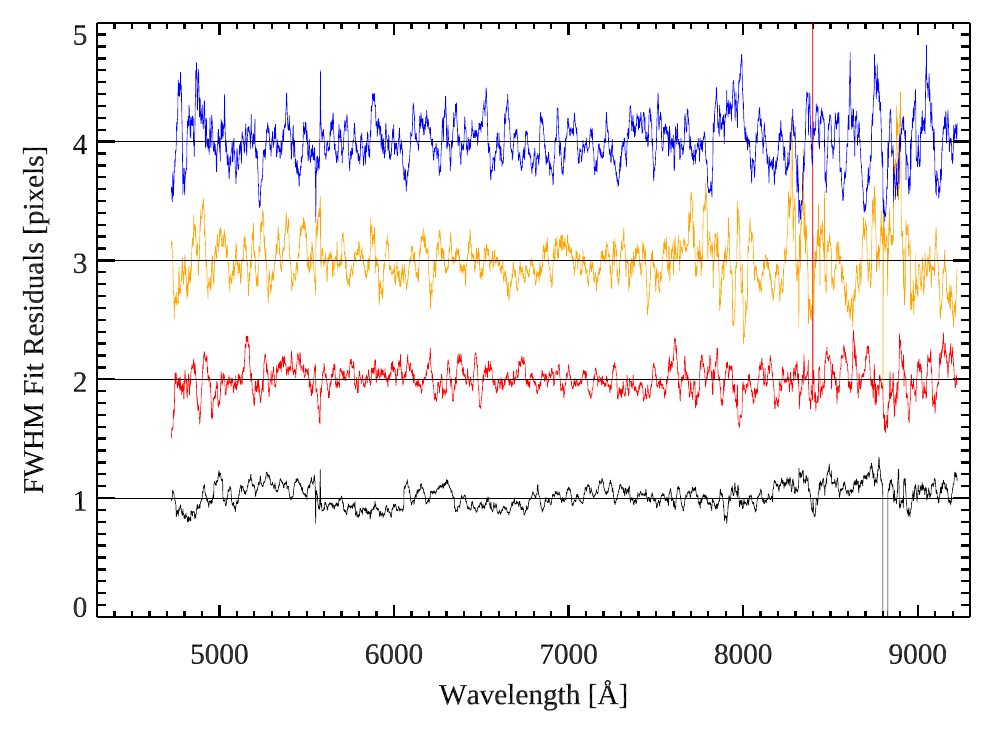}
\end{minipage}
\caption{The top panels show the continuum images for the nuclear fields of NGC\,1404 (left) and NGC\,1351 (right).  The PSF stars are identified  with colored markers.  The middle panels show the wavelength dependence of the stars' PSFs' by plotting their best-fit FWHMs.  The data are fit with a 2nd degree polynomial.  The lower panels show the residuals from this fit, with the brightest star on the bottom and the faintest on the top.  Each curve is offset by 1.0 from the previous curve for clarity.  The colors correspond to those shown in the images. Note the large residuals associated with the fainter stars.
\label{fig:NGC1404_vs_1351}}
\end{figure*}

The top panel of Figures~\ref{fig:NGC1404_vs_1351} shows continuum images of NGC\,1404 and 1351 formed from the co-addition of the 40 MUSE data cube layers between 4968~\AA\ and 5018~\AA\null.  These are the layers used to determine the cube's aperture correction, as the region is close enough in wavelength to the redshifted \OIII line so that the wavelength dependence of seeing can be ignored.  By co-adding 40 layers, we increase  the signal-to-noise over that for the \OIII $\lambda 5007$ line alone, 

The encircled objects in the figure are the point-like sources chosen to define the field's point-spread function.  Table~\ref{tab:PSF_stars} lists these sources, along with their equatorial coordinates and our derived quantities for the objects' FWHM, total flux, and inferred aperture correction.  The latter three values were derived by fitting a Gaussian to the objects' radial profiles via the Levenberg-Marquardt least squares algorithm.

To estimate the robustness of these values, we repeated the FWHM measurements all along the MUSE spectra, thereby obtaining the FWHM of the stars versus wavelength.  The middle panels of Fig.~\ref{fig:NGC1404_vs_1351} shows the results of these fits, while the bottom panel displays the residuals from a best-fit 2nd-order polynomial.   The different colors correspond to those used to identify the stars in the top panel.  



\begin{table*}[ht!]
\caption{PSF Stars}
\tiny
\begin{minipage}{0.5\linewidth}
\centering
\begin{tabular}{lccccc}
\multicolumn{6}{c}{NGC 1404 PSF Stars} \\
\hline \\ [-0.2cm]
& & &FWHM & &apcor \\ [-0.25cm]
Star  &RA(2000) &\colhead{DEC(2000)} &(pixels)
&Total Flux$^{a}$ &$\Delta$mag \\
\hline
S1 (blk) &	 3:38:50.16	&	$-$35:35:30.4 & 4.41 & 3011413 & 0.154  \\
S2 (red) &	 3:38:52.66	&	$-$35:35:17.2 & 4.41 &  650172 & 0.168 \\
S3 (yel) &	 3:38:52.16	&	$-$35:35:00.3 & 4.10 &   38331 & 0.159 \\
S4 (blu) &	 3:38:52.06	&	$-$35:35:14.3 & 4.60 &	4762 & 0.306 \\
S5 (cya)&	 3:38:49.55	&	$-$35:35:09.6 & 4.32 &	1672 & 0.005 \\
\hline
\end{tabular}
\begin{minipage}{7cm}
\vspace{0.1cm}
$^{a}$units of $10^{-20}$~ergs~cm$^{-2}$~s$^{-1}$, bandpass [4968\,\AA, 5018\,\AA] 
\end{minipage}
\end{minipage}
\begin{minipage}{0.5\linewidth}
\centering
\begin{tabular}{lccccc}
\multicolumn{6}{c}{NGC 1351 PSF Stars} \\
\hline \\ [-0.2cm]
& & &FWHM & &apcor \\ [-0.25cm]
Star  &RA(2000) &\colhead{DEC(2000)} &(pixels)
&Total Flux$^{a}$ &$\Delta$mag \\
\hline
S1 (blk) &	 3:30:36.46  &  -34:51:39.5 & 4.39 &   68552 & 0.100 \\
S2 (red) &	 3:30:34.54  &  -34:50:50.9 & 4.31 &   26782 & 0.217 \\
S3 (yel) &	 3:30:36.30  &  -34:51:08.3 & 4.72 &   18659 & 0.273 \\
S4 (blu) &	 3:30:33.38  &  -34:51:01.2 & 4.50 &   12373 & 0.181 \\
\\
\hline
\end{tabular}
\begin{minipage}{7cm}
\vspace{0.1cm}
$^{a}$units of $10^{-20}$~ergs~cm$^{-2}$~s$^{-1}$, bandpass [4968\,\AA, 5018\,\AA]  
\end{minipage}
\end{minipage}
\label{tab:PSF_stars}
\end{table*}

As can clearly be seen from residuals in the bottom panel, stars S1 and S2 in NGC\,1404, are bright enough to deliver well-behaved FWHM values that are fitted well by the polynomials.  Their FWHM values are almost identical and vary slowly with wavelength, as expected from theory \citep{Kamann+13}.  Of course, as the stellar flux decreases, the fits for the fainter stars become less and less reliable.  In NGC\,1351, all the PSF objects are faint, and their PSF fits are much noisier.


The brightest star, S1, happens to be located at the edge of the field. This does not dramatically affect the Gaussian fit, but becomes important for measuring an aperture correction, as will be shown below. 


Figures~\ref{fig:NGC1404_radial_plots} and \ref{fig:NGC1351_radial_plots} plot the radial profiles of the PSF stars in units of flux.  Each dot represents the flux of a given pixel in the continuum image shown in Fig.~\ref{fig:NGC1404_vs_1351}.  The radius from the stellar centroid is given in units of MUSE spaxels and must be multiplied by 0.2 to convert to arcseconds.  The Gaussian and Moffat fits to the data are also shown.

The bright star S1 in NGC\,1404 is located at the edge of the field.  Nevertheless, its profile is extremely well-defined. It is even possible to recognize the advantage of a Moffat fit over a simple Gaussian, specifically in the core and in the wings of the PSF\null. Star S2 in NGC\,1404 is also well-behaved, but for the fainter stars, the scatter becomes visibly larger, and the background becomes the dominant source of error. Notably, this background scatter is not all photon noise: it also arises from the spatial variation of the galaxy light underlying the star.  The strongest gradients are generally found in elliptical galaxies, but they can also be important in the spiral arms of later-type systems.  The effect is quite visible for the PSF stars in NGC\,1351 in Fig.~\ref{fig:NGC1351_radial_plots}. In future papers, we may be able to improve the stellar aperture photometry within elliptical galaxies by subtracting a smooth surface brightness model from the data, but for the sake of uniformity, we treated all galaxies in this study the same.

To obtain our aperture corrections, we first inspected the PSF fit for each star and discarded any result that was obviously an outlier.  Following Paper~I, we then measured the stars' curve-of-growths using the \texttt{APER} program in \texttt{DAOPHOT} \citep{Stetson1987}.  Figure~\ref{fig:NGC1404_apcor} shows (top to bottom) the resultant aperture corrections for the stars of NGC\,1404.  The panels plot, from top to bottom, apertures with radii of 3, 4, 5 (and, where applicable, also 6) pixels as a function of wavelength.  Unsurprisingly, the curves for S1 (black) and S2 (red) are quite similar and well-behaved.  Moreover, although the curves for S3 are much noisier, a polynomial fit is still in good agreement with the curves for the brighter stars.  Conversely, the curves for S4 deviate strongly from expectations, and the ones for S5 are completely useless.  If we ignore the latter two stars, we can formally derive the observation's mean  correction for a 4-pixel aperture to be $0.160\pm0.007$ mag 



In contrast, an assessment of aperture correction for NGC\,1351 is much more difficult.  As Figure~\ref{fig:NGC1351_apcor} reveals, the curves all deviate significantly one from another, and there is an apparent trend of  the higher aperture corrections being associated with the fainter stars.  In a case like this, it is tempting to adopt the data from the brightest star (S1) as the best estimate, but the star's proximity to the edge of the field complicates the situation. Deriving a reliable correction and statistical error for this galaxy is therefore challenging, and the best we can do is adopt a value of between 0.1 and 0.2~mag. Clearly, a more sophisticated approach for measuring the PSF, and hence the zero point of the photometry, is required to make the best use of the archival data for NGC\,1351.  However, that is beyond the scope of the current work.


\clearpage

\begin{figure}[t!]
    \begin{minipage}{1.0\linewidth}
    \centerline{
    \includegraphics[width=0.75\hsize,bb=10 70 700 550,clip]{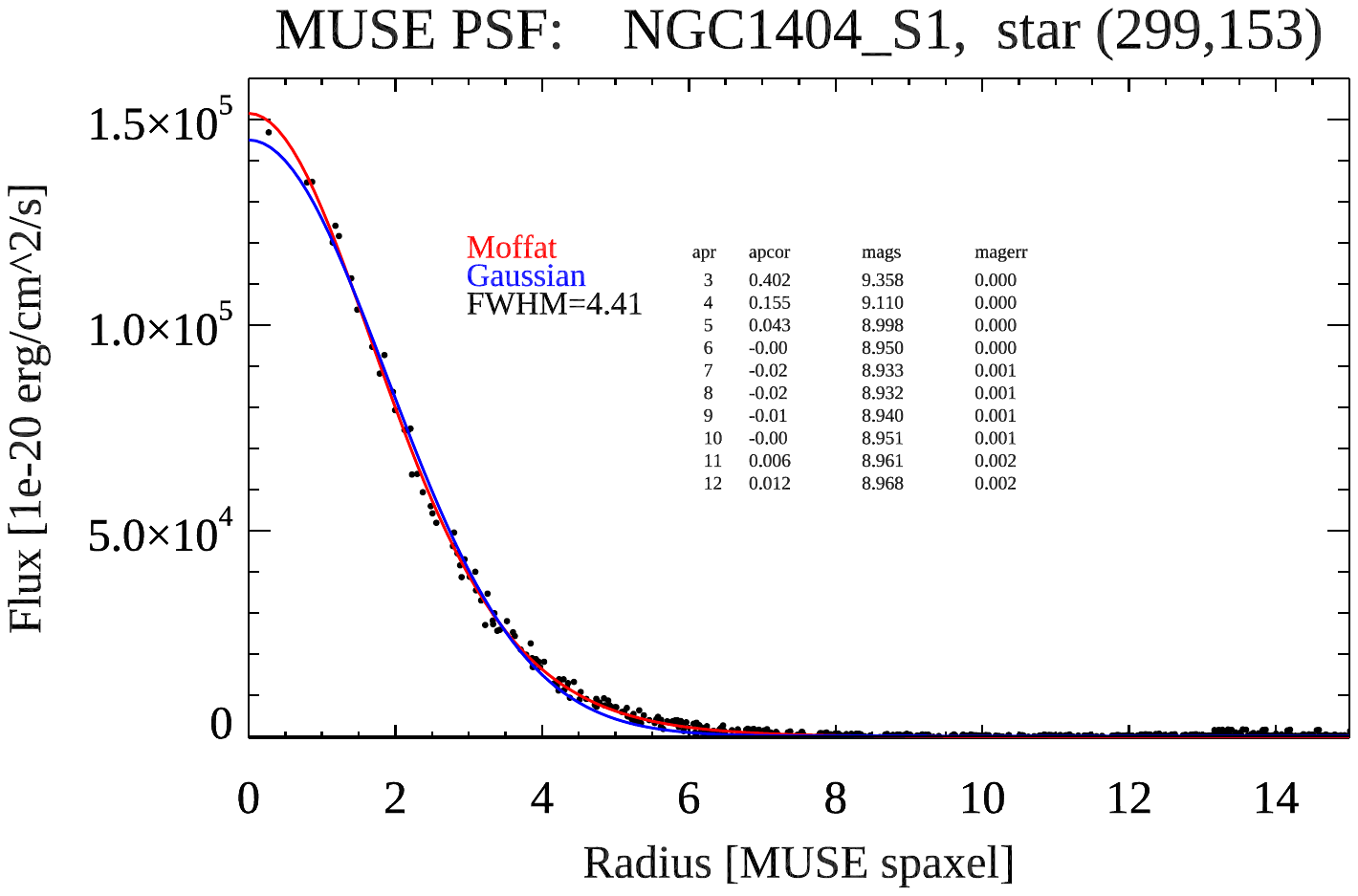}
    }
    \centerline{
    \includegraphics[width=0.75\hsize,bb=25 70 700 550,clip]{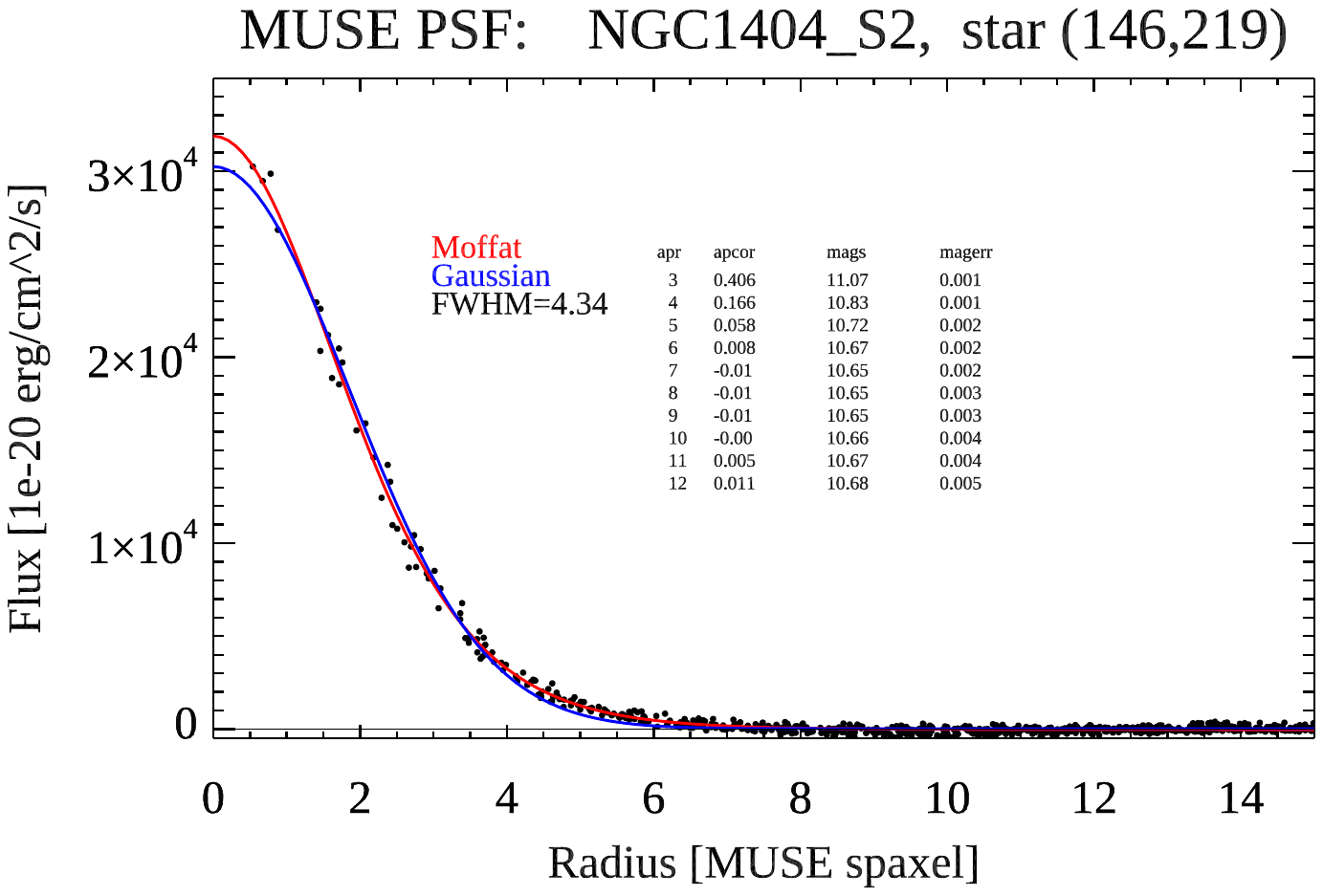}
    }
    \centerline{
    \includegraphics[width=0.75\hsize,bb=35 70 700 550,clip]{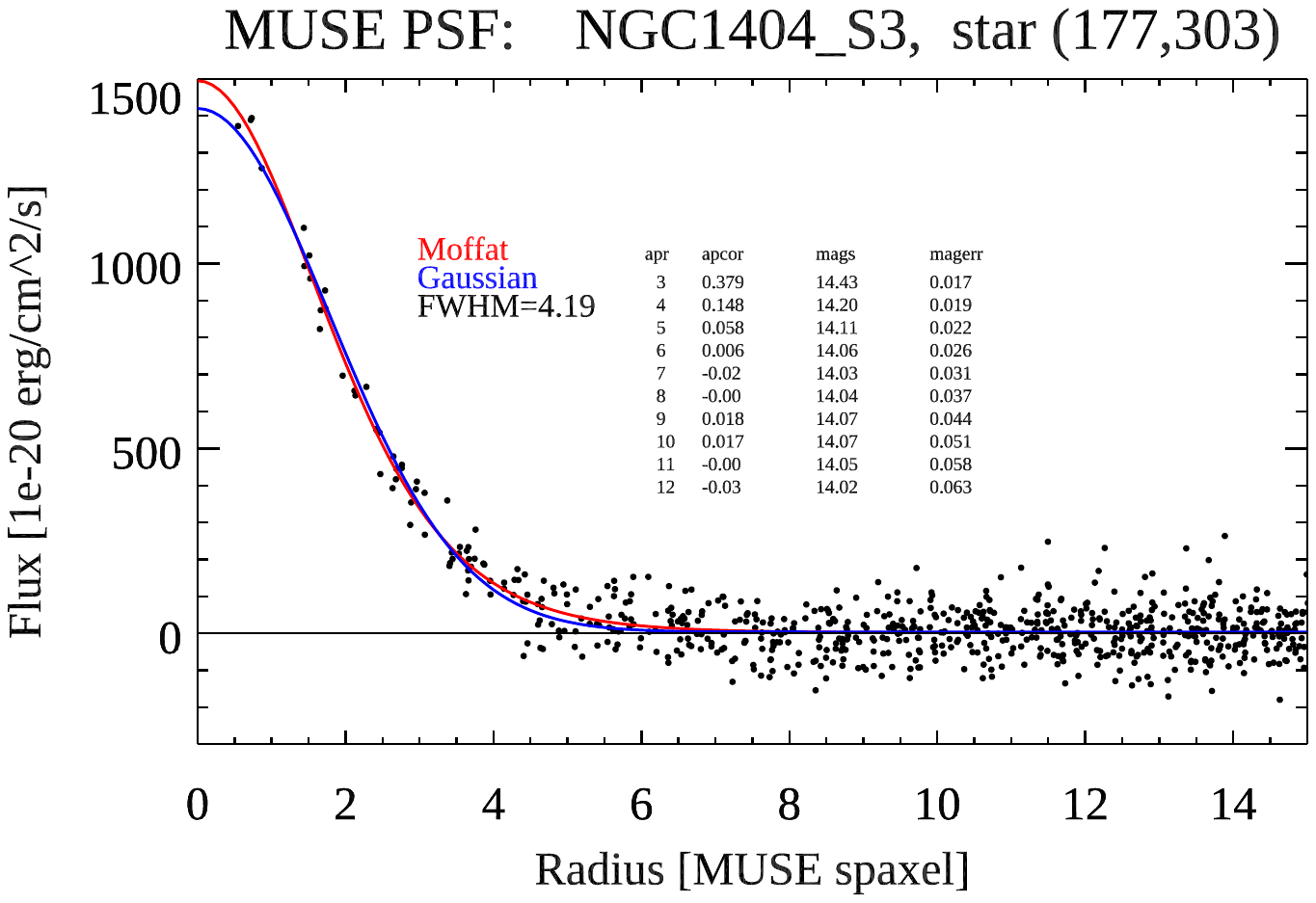}
    }
    \centerline{
    \includegraphics[width=0.75\hsize,bb=35 70 700 550,clip]{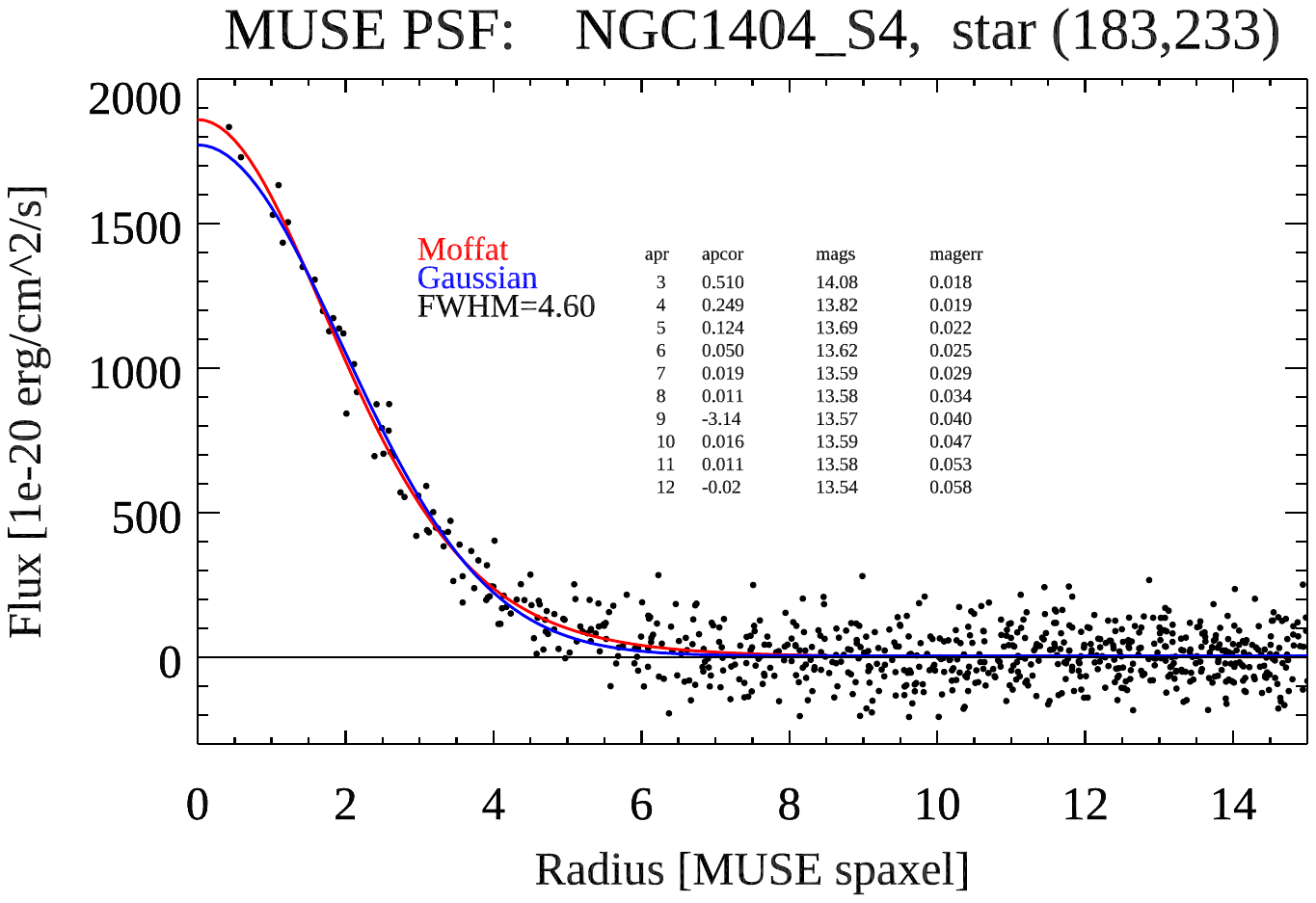}
    }
    \centerline{
    \includegraphics[width=0.75\hsize,bb=35 70 700 550,clip]{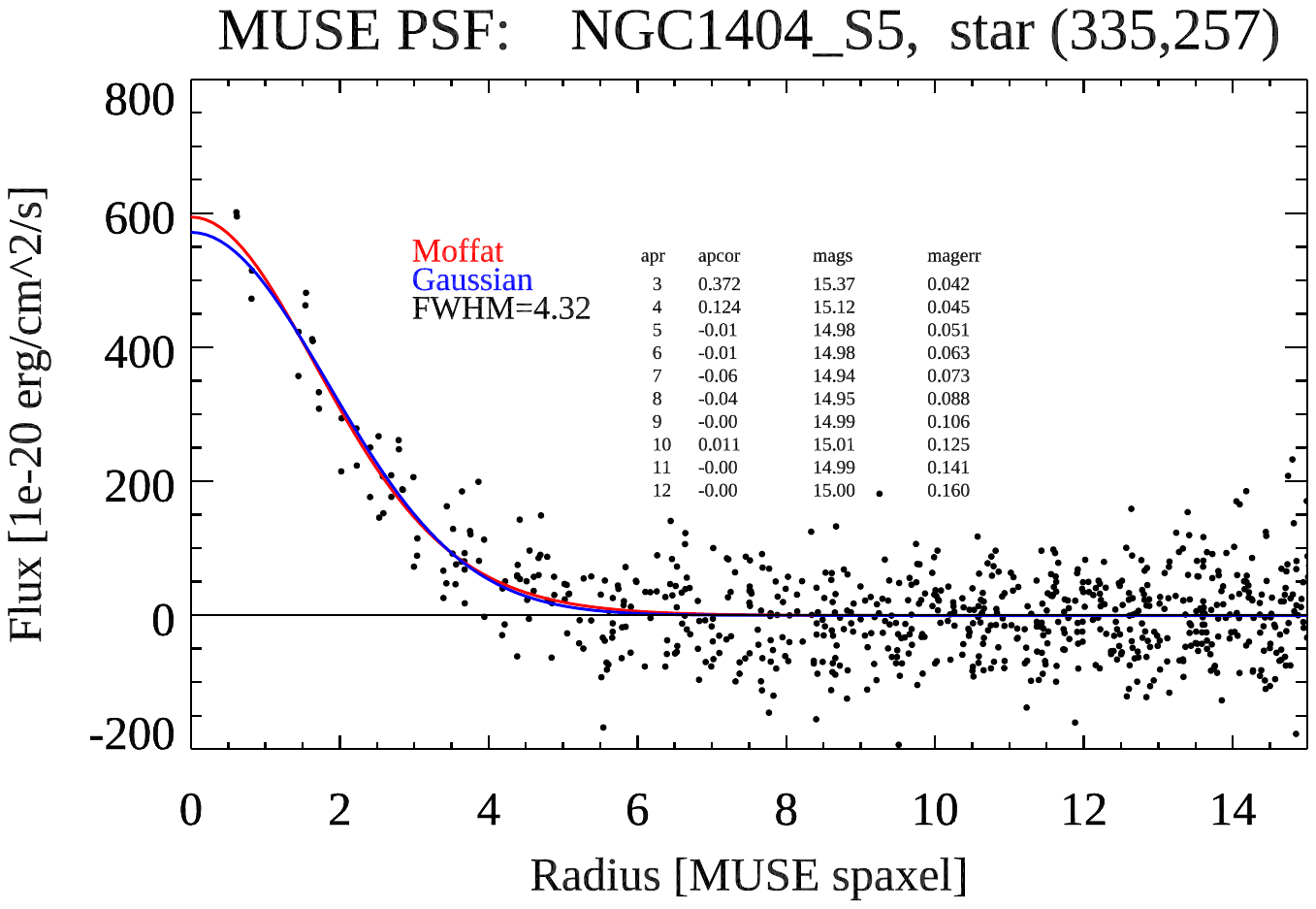}
    }
    \end{minipage}  
   \caption{NGC\;1404 PSF star radial plots, obtained from image shown in Fig.~\ref{fig:NGC1404_vs_1351}.}
 \label{fig:NGC1404_radial_plots}
\end{figure}

\begin{figure}[t!]
    \begin{minipage}{1.0\linewidth}
    \centerline{
    \includegraphics[width=0.75\hsize,bb=10 70 700 550,clip]{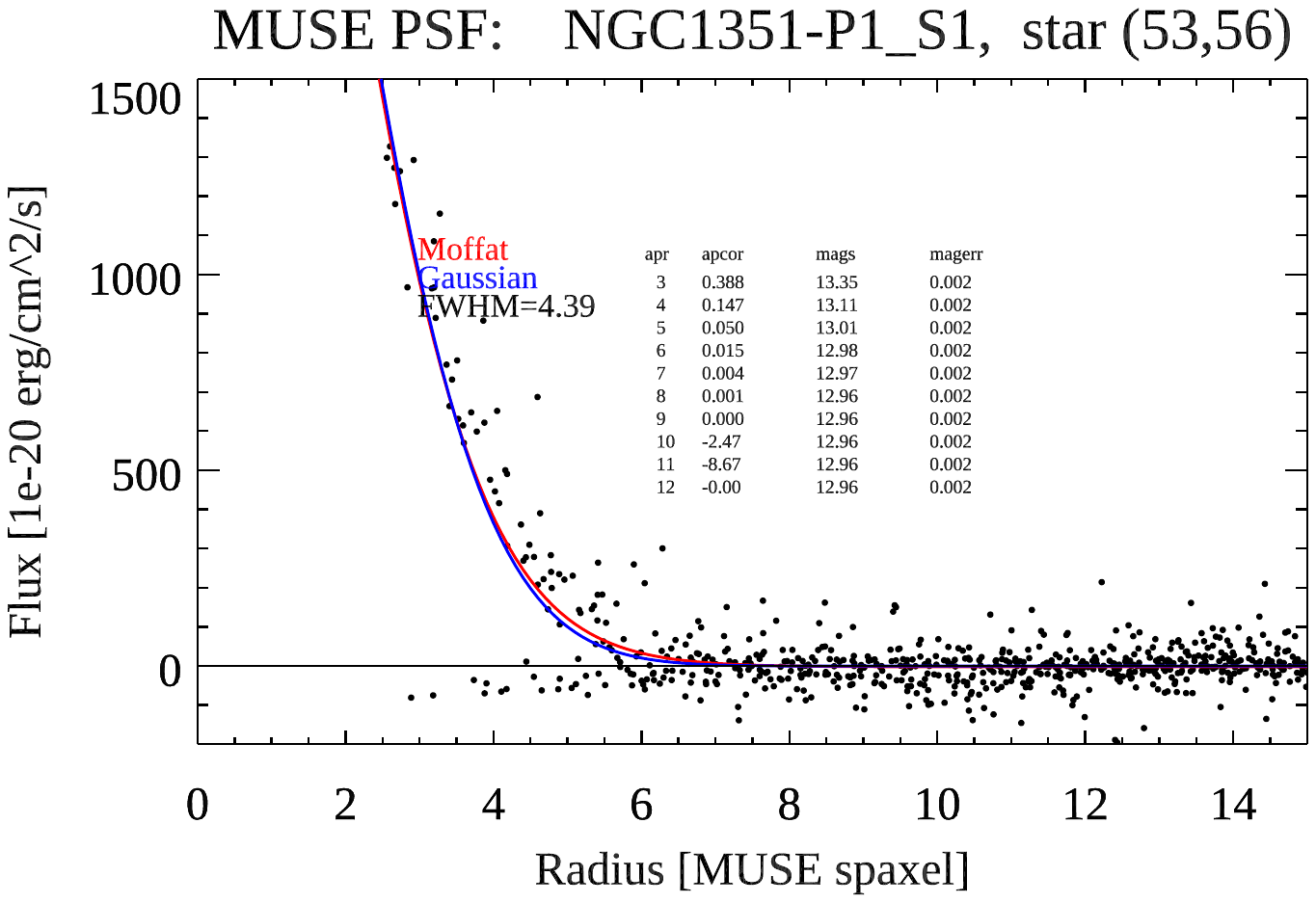}
    }
    \centerline{
    \includegraphics[width=0.75\hsize,bb=25 70 700 550,clip]{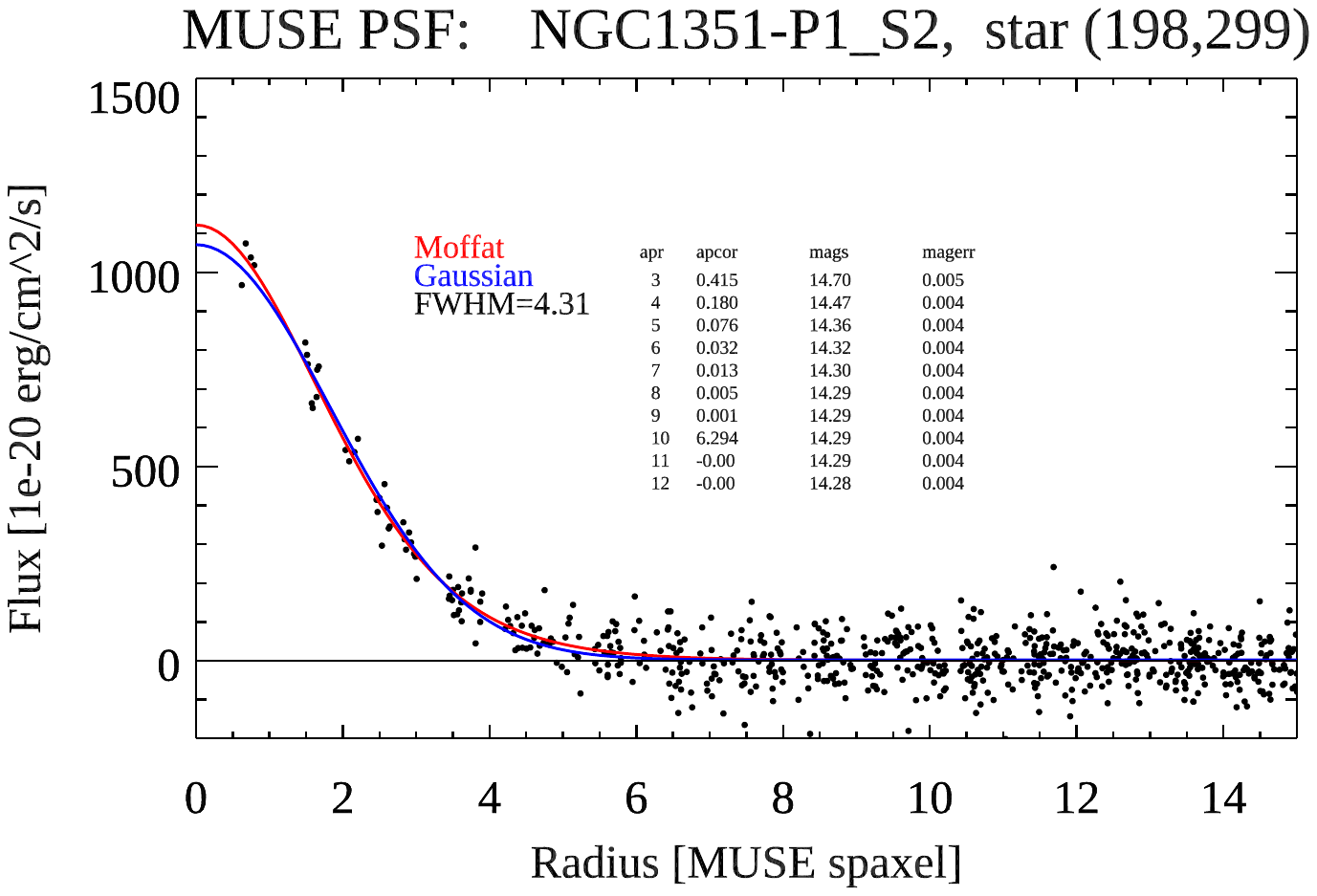}
    }
    \centerline{
    \includegraphics[width=0.75\hsize,bb=35 70 700 550,clip]{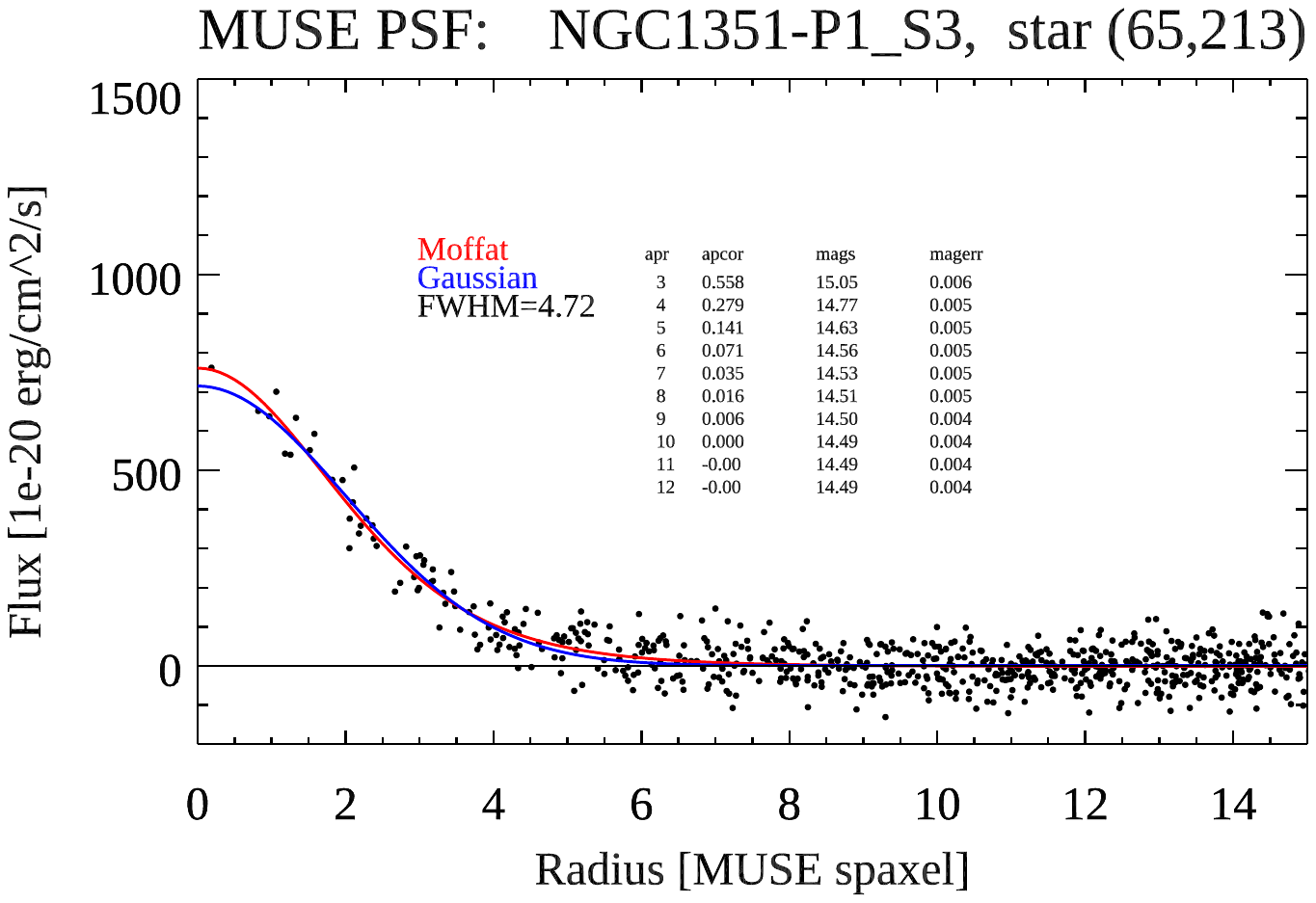}
    }
    \centerline{
    \includegraphics[width=0.75\hsize,bb=35 70 700 550,clip]{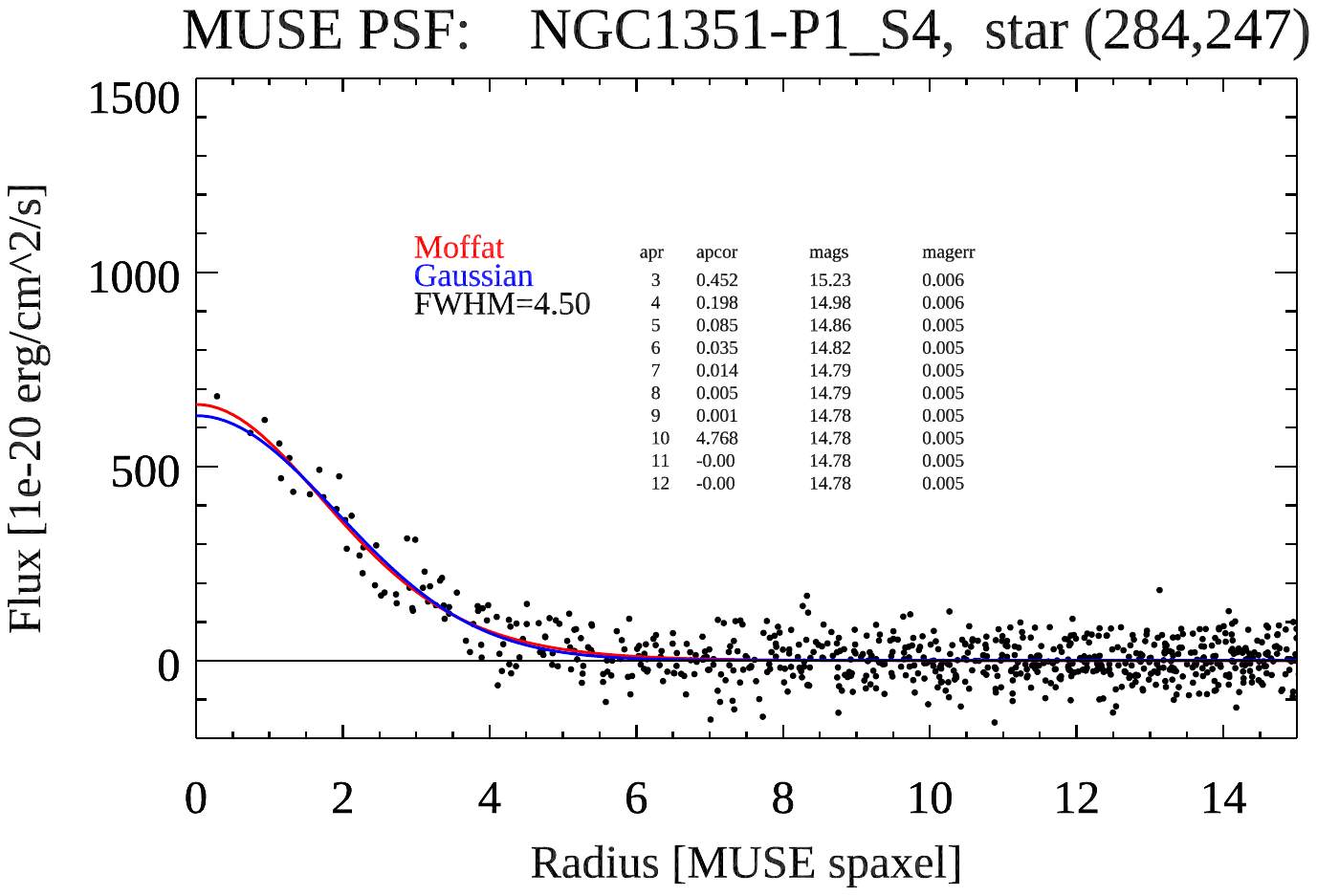}
    }
    \end{minipage}  
   \caption{NGC\;1404 PSF star radial plots, obtained from image shown in Fig.~\ref{fig:NGC1404_vs_1351}.}
 \label{fig:NGC1351_radial_plots}
\end{figure}

\clearpage

\begin{figure}[t!]
    \begin{minipage}{1.0\linewidth}
    \centerline{
    \includegraphics[width=0.82\hsize,bb=60 350 600 700,clip]{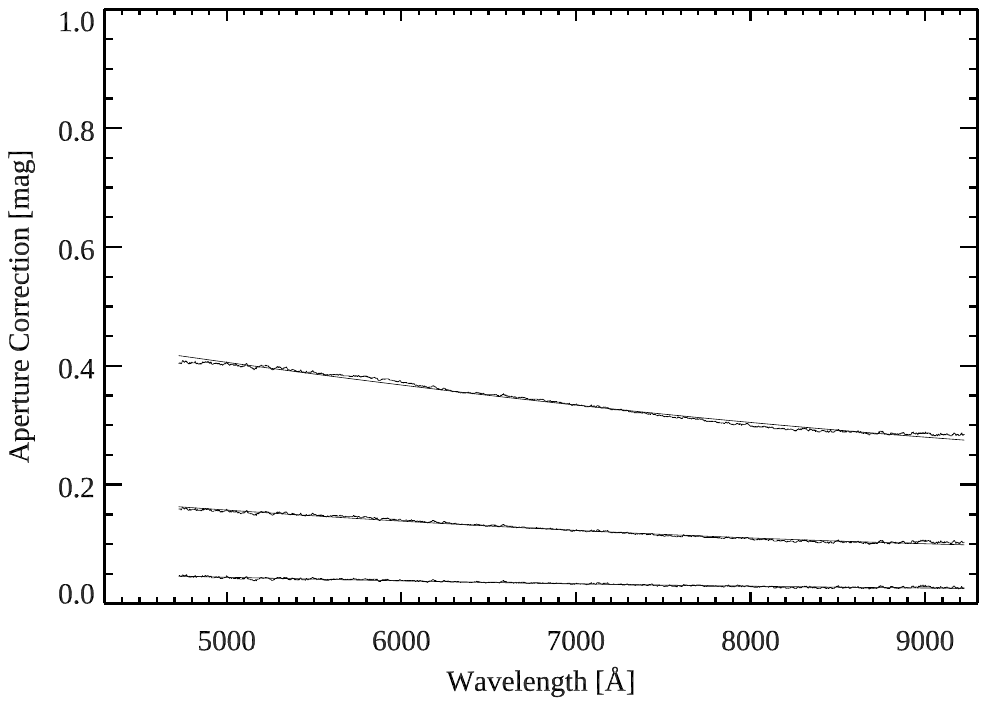}
    }
    \centerline{
    \includegraphics[width=0.82\hsize,bb=60 350 600 700,clip]{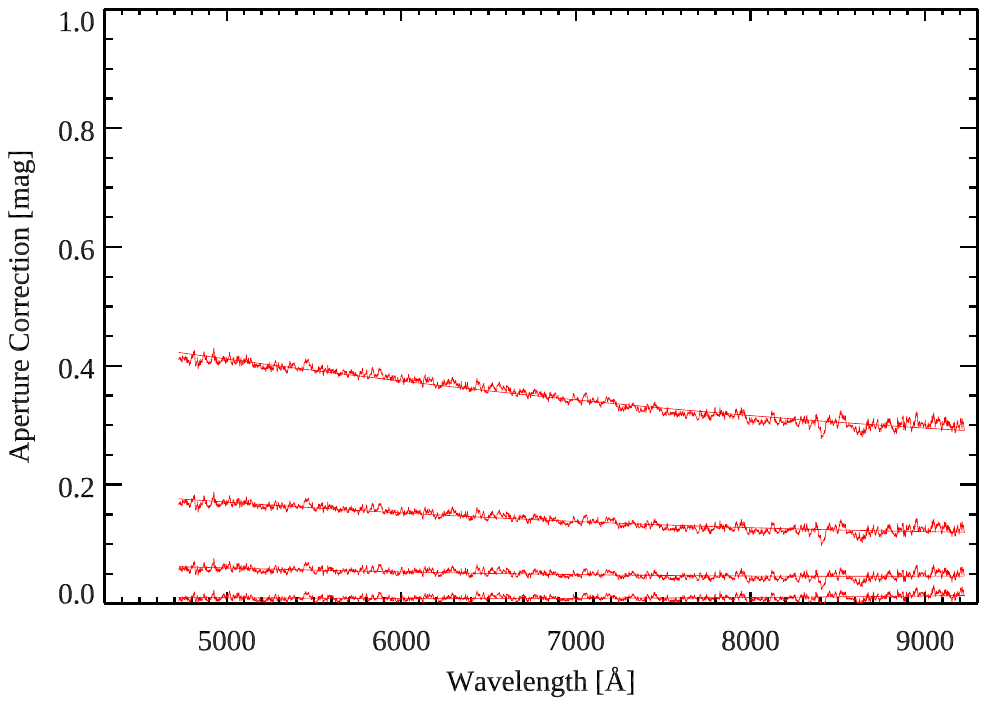}
    }
    \centerline{
    \includegraphics[width=0.82\hsize,bb=60 350 600 700,clip]{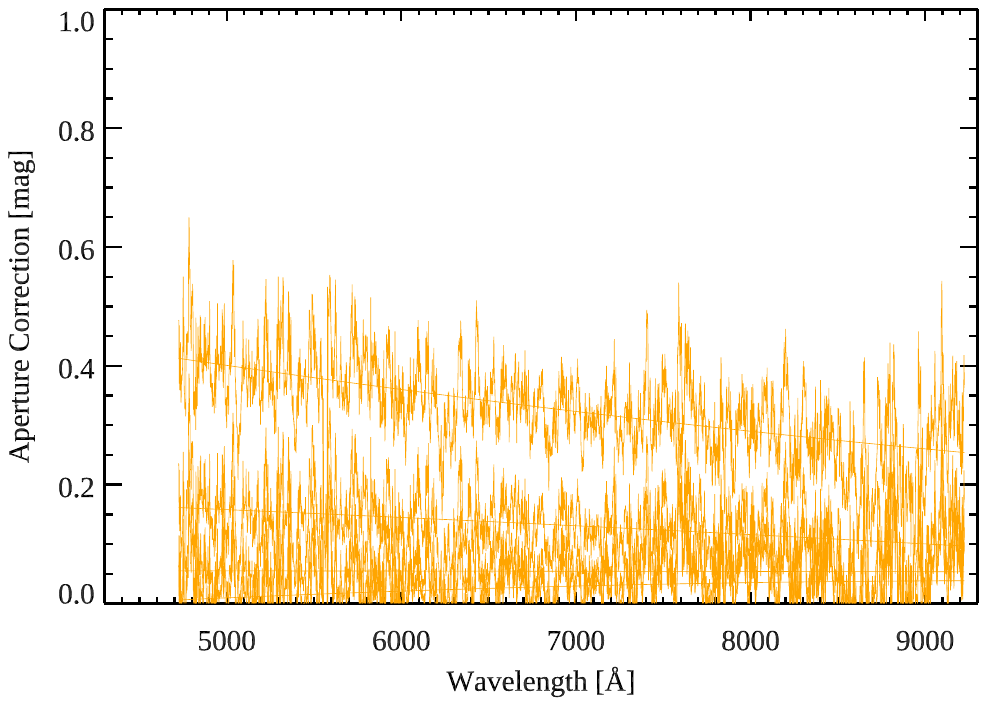}
    }
    \centerline{
    \includegraphics[width=0.82\hsize,bb=60 350 600 700,clip]{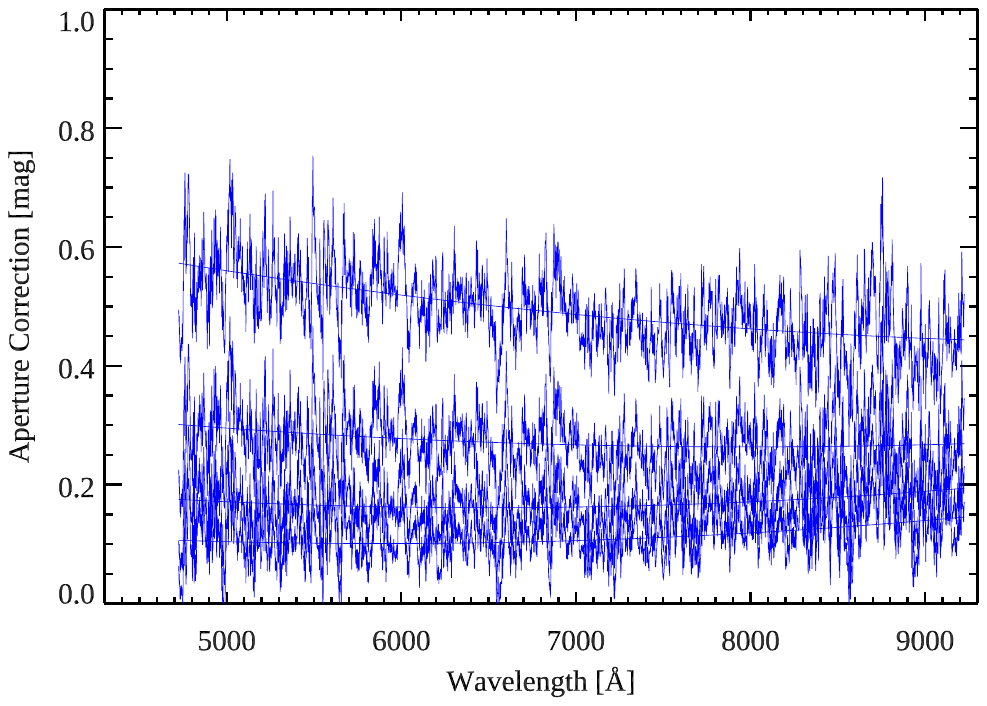}
    }
    \centerline{
    \includegraphics[width=0.82\hsize,bb=60 350 600 700,clip]{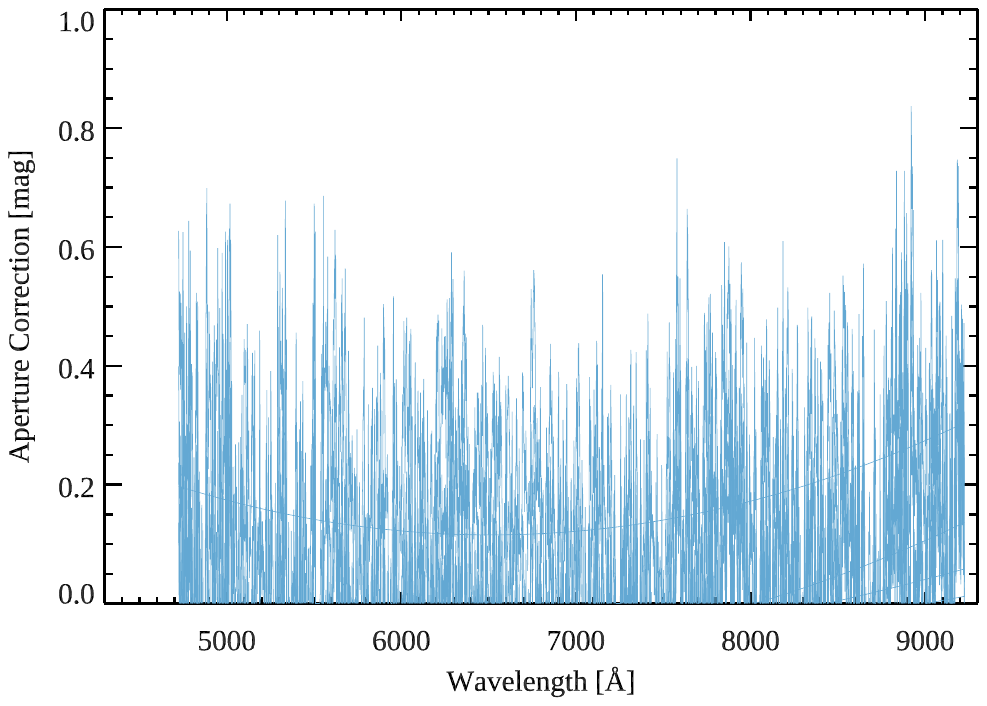}
    }
    \end{minipage}  
   \caption{NGC\;1404 PSF star aperture corrections as a function of wavelength, obtained from image shown in Fig.~\ref{fig:NGC1404_vs_1351}. PSF stars S1 through S5 are shown from top to bottom.}
 \label{fig:NGC1404_apcor}
\end{figure}

\begin{figure}[t!]
    \begin{minipage}{1.0\linewidth}
    \centerline{
    \includegraphics[width=0.82\hsize,bb=60 350 600 700,clip]{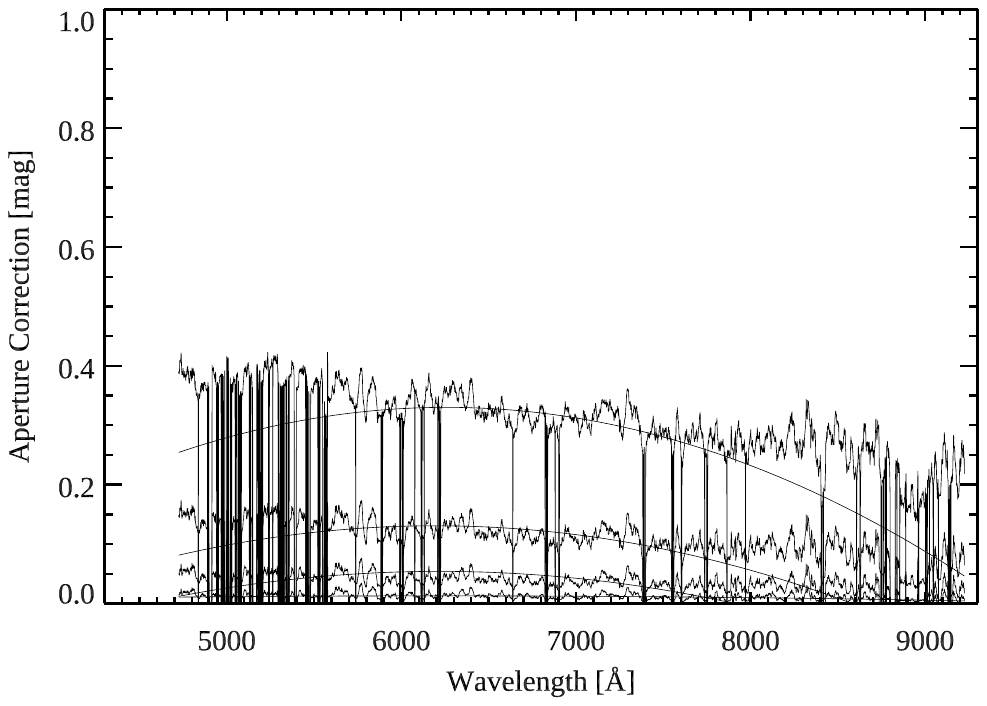}
    }
    \centerline{
    \includegraphics[width=0.82\hsize,bb=60 350 600 700,clip]{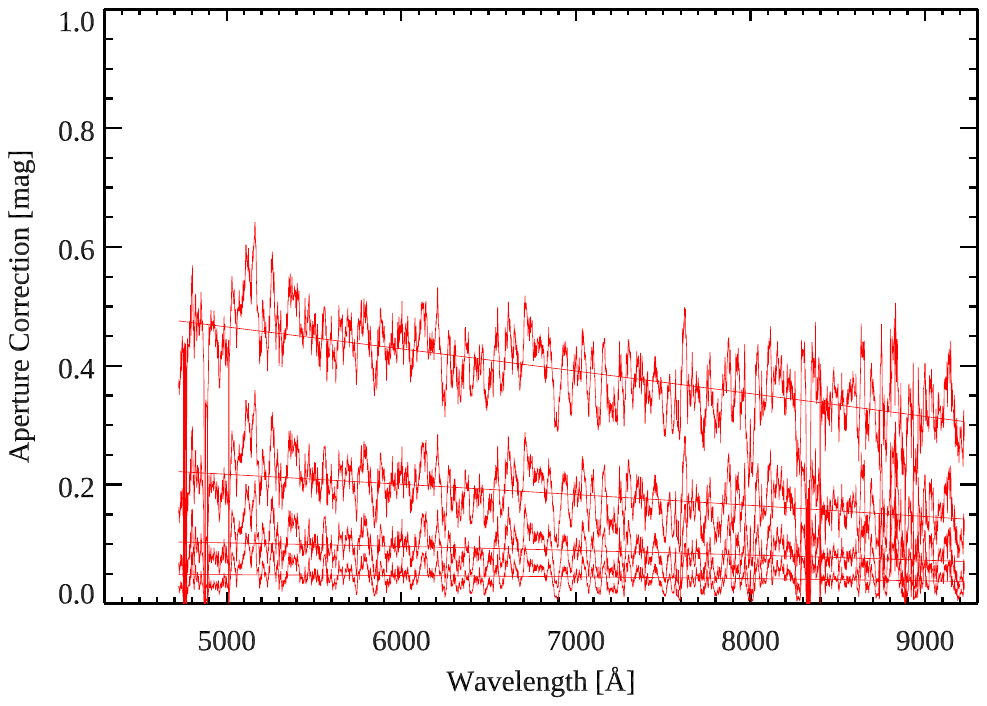}
    }
    \centerline{
    \includegraphics[width=0.82\hsize,bb=60 350 600 700,clip]{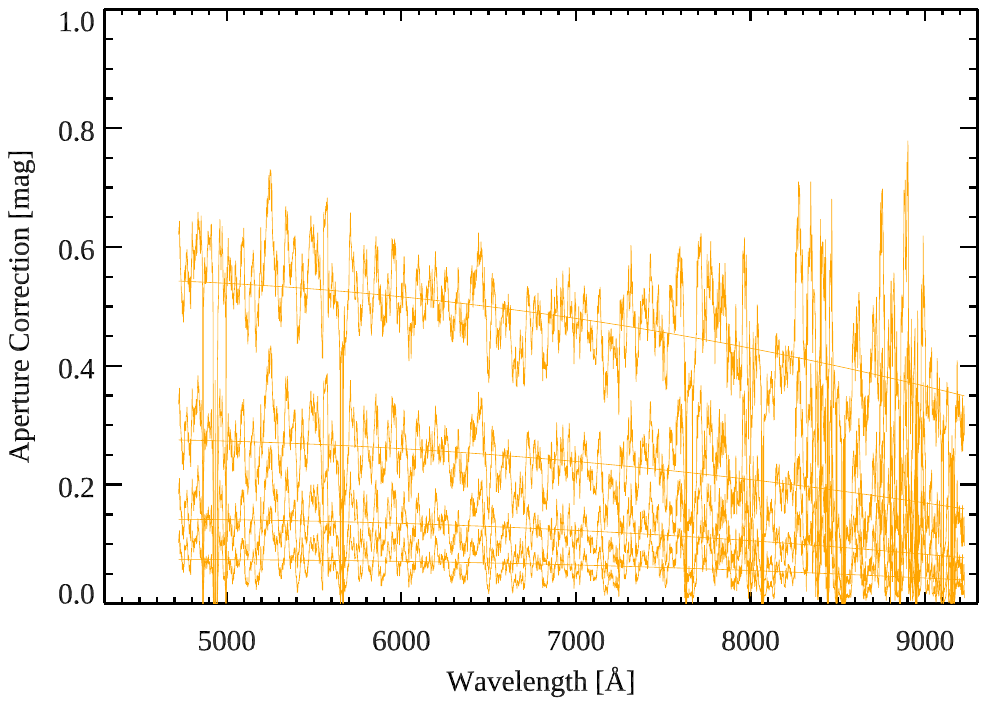}
    }
    \centerline{
    \includegraphics[width=0.82\hsize,bb=60 350 600 700,clip]{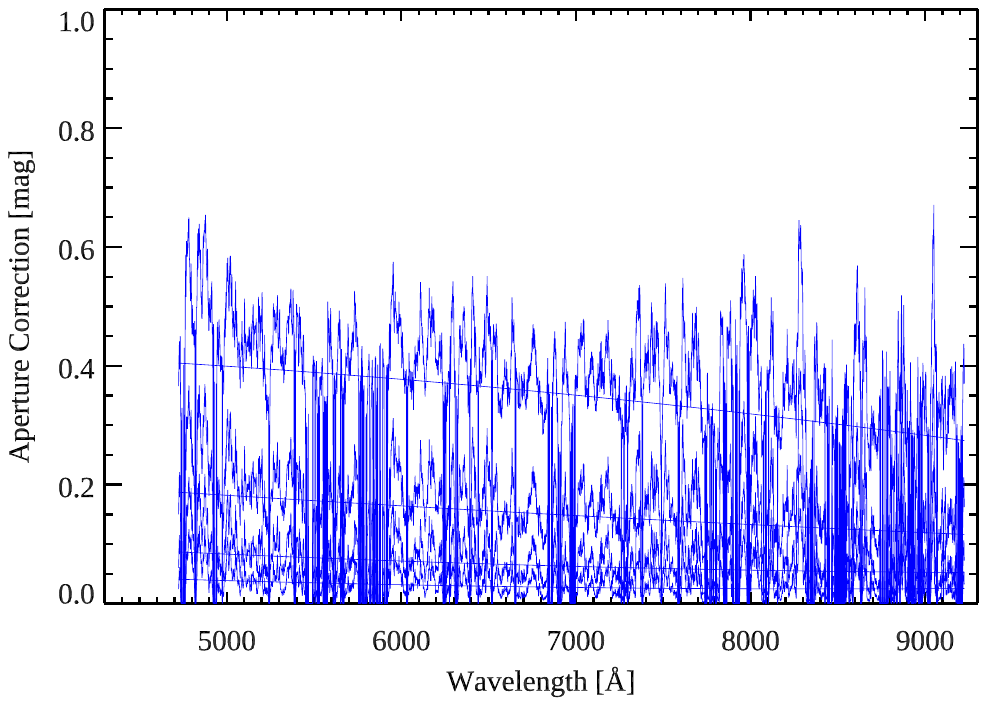}
    }
   \end{minipage}  
   \caption{NGC\;1351 PSF star aperture corrections as a function of wavelength, obtained from image shown in 
   Fig.~\ref{fig:NGC1404_vs_1351}. PSF stars S1 through S4 are shown from top to bottom.}
 \label{fig:NGC1351_apcor}
\end{figure}

\clearpage

\end{document}